\documentclass[goettingen, gauss, print]{thesis}
\usepackage{xcolor}

\title{Towards solar many-line inversions}
\author{Johannes H\"olken}
\town{Werne (Westf.)}

\thesisadvisorycommitteea{Prof. Dr. Stefan Dreizler}
\instthesisadvisorycommitteea{Georg-August-Universit\"at G\"ottingen, G\"ottingen, Deutschland}
\thesisadvisorycommitteeb{Prof. Dr. Sami K. Solanki}
\instthesisadvisorycommitteeb{Max-Planck-Institut f\"ur Sonnensystemforschung, G\"ottingen, Deutschland sowie \\ Technische Universit\"at Braunschweig, Braunschweig, Deutschland}
\thesisadvisorycommitteec{Prof. Dr. Thorsten Hohage}
\instthesisadvisorycommitteec{Georg-August-Universit\"at G\"ottingen, G\"ottingen, Deutschland sowie\\ Max-Planck-Institut f\"ur Sonnensystemforschung, G\"ottingen, Deutschland}
\thesisadvisorycommitteed{Dr. Alex Feller}
\instthesisadvisorycommitteed{Max-Planck-Institut f\"ur Sonnensystemforschung, G\"ottingen, Deutschland}

\refereea{Prof. Dr. Sami K. Solanki}
\instrefereea{Max-Planck-Institut f\"ur Sonnensystemforschung, G\"ottingen, Germany sowie \newline Technische Universit\"at Braunschweig, Braunschweig, Deutschland}
\refereeb{Prof. Dr. Stefan Dreizler}
\instrefereeb{Georg-August-Universit\"at G\"ottingen, G\"ottingen, Deutschland}
\commissiona{Prof. Dr. Hardi Peter}
\instcommissiona{Max-Planck-Institut f\"ur Sonnensystemforschung, G\"ottingen, Deutschland sowie\\ Institut für Sonnenphysik (KIS), Freiburg, Deutschland}
\commissionb{Prof. Dr. Thorsten Hohage}
\instcommissionb{Georg-August-Universit\"at G\"ottingen, G\"ottingen, Deutschland sowie\\ Max-Planck-Institut f\"ur Sonnensystemforschung, G\"ottingen, Deutschland}
\commissionc{Prof. Dr. Andreas Dillmann}
\instcommissionc{Georg-August-Universit\"at G\"ottingen, G\"ottingen, Deutschland sowie \newline Deutsches Zentrum für Luft- und Raumfahrt e.V. (DLR),\newline Institut für Aerodynamik und \mbox{Strömungstechnik}, G\"ottingen, Deutschland }
\commissiond{Prof. Dr. Stanley Lai}
\instcommissiond{Georg-August-Universit\"at G\"ottingen, G\"ottingen, Deutschland}

\submittedyear{2025}

\submitteddate{13.10.2025}
\examinationdate{24.11.2025}

\usepackage{amsmath}
\usepackage{amsfonts}
\usepackage{amssymb}
\usepackage{amsthm}
\usepackage{bbm}
\usepackage{mathtools}
\usepackage{nicefrac}
\usepackage{longtable}
\usepackage{lscape}
\usepackage{bigdelim}
\usepackage{multirow}
\usepackage{caption}
\usepackage{pdfpages}

\usepackage{orcidlink}
\usepackage{titlesec}

\newtheorem{theorem}{Theorem}[section]
\newtheorem{definition}{Definition}[section]
\newtheorem{lemma}[theorem]{Lemma}
\numberwithin{equation}{section}

\newcommand{\AnA}{\mbox{A\&A}}

\newcommand{\NN}{\mathbb{N}}
\newcommand{\RR}{\mathbb{R}}
\newcommand{\CC}{\mathbb{C}}
\newcommand{\ZZ}{\mathbb{Z}}
\newcommand{\EE}{\mathbb{E}}

\newcommand{\one}{\mathbbm{1}}

\newcommand{\EC}{\mathcal{E}}
\newcommand{\MC}{\mathcal{M}}
\newcommand{\OC}{\mathcal{O}}
\newcommand{\JC}{\mathcal{J}}
\newcommand{\TC}{\mathcal{T}}
\newcommand{\FC}{\mathcal{F}}
\newcommand{\KC}{\mathcal{K}}
\newcommand{\UC}{\mathcal{U}}
\newcommand{\VC}{\mathcal{V}}
\newcommand{\AC}{\mathcal{A}}
\newcommand{\XC}{\mathcal{X}}
\newcommand{\YC}{\mathcal{Y}}
\newcommand{\ZC}{\mathcal{Z}}

\newcommand{\transpose}{\textnormal{T}}

\newcommand{\FF}{\textnormal{FF}}
\newcommand{\DF}{\textnormal{DF}}
\newcommand{\D}{\textnormal{D}}

\newcommand{\phase}{\mathcal{P}}

\newcommand{\new}{^{\textnormal{new}}}
\newcommand{\old}{^{\textnormal{old}}}
\newcommand{\slit}{_{\textnormal{slit}}}
\newcommand{\sqr}{^{\nicefrac{1}{2}}}
\newcommand{\FFT}{\textnormal{FFT}}

\newcommand{\supcirc}{$^\circ$}

\newcommand{\x}{\(\times\)}
\newcommand{\io}[1]{\ion{#1}{}\!}
\newcommand{\vlos}{\(v_{\text{LOS}}\)}
\newcommand{\T}{\(T\)}
\newcommand{\B}{\(B\)}
\newcommand{\I}{\(I\)}
\newcommand{\Q}{\(Q\)}
\newcommand{\U}{\(U\)}
\newcommand{\V}{\(V\)}

\newcommand{\muram}{\mbox{MURaM}}
\newcommand{\hsp}{\mbox{HINODE/SP}}
\newcommand{\fiss}{\mbox{FISS}}
\newcommand{\fisssp}{\mbox{FISS-SP}}
\newcommand{\snr}{SNR}

\newcommand{\spectroflat}{\texttt{spectroflat}}
\newcommand{\specrestore}{\texttt{specrestore}}
\newcommand{\spinor}{SPINOR}

\newcommand{\nlines}{85}
\newcommand{\tlines}{171}

\newcommand{\im}{\textnormal{i}}

\newcommand{\ion}[2]{\textup{#1\,\textsc{\lowercase{#2}}}}

\newcommand{\blue}[1]{#1} 
\newcommand{\comment}[1]{}
\usepackage[german,english]{babel}
\usepackage{natbib}
\usepackage{printlen}
\usepackage{color}
\hypersetup{
    colorlinks=true,
    linkcolor=black,
    citecolor=black, 
    urlcolor=black, 
    hypertexnames=true
    }

\begin{document}
\selectlanguage{english}
\maketitle

\clearpage
\thispagestyle{empty}
\hfill
\clearpage

\newpage
\thispagestyle{empty}
\vspace*{1.2cm}
\begin{center}
    Sunspots are "[o]penings in the Sun's luminous atmosphere,\\ 
    allowing a view of the underlying, cooler surface of the Sun,\\
    which is likely inhabited".\\[10pt]
    \textit{William Herschel (1801)}
\end{center}

\tableofcontents

{
\titleformat{\chapter}[display]
  {\normalfont\huge\bfseries}{}{0pt}{\Huge}
\titlespacing*{\chapter}{0pt}{-27pt}{25pt}

\chapter*{Summary\markboth{Summary}{Summary}}
\addcontentsline{toc}{chapter}{Summary}

The Sun's atmosphere is significantly influenced by events that often measure only tens of kilometres in the horizontal direction. The phenomena that drive configuration and dynamics of the atmosphere include wave dissipation, plasma instabilities, and magnetic reconnection. They happen in both quiet and active regions and are determined by magneto-hydro-dynamic processes. Their origins often lie in the lowest layers of the solar atmosphere. To improve the understanding of chromospheric and coronal heating, the solar dynamo and its activity cycle, or to improve the prediction of space weather events we must therefore enhance our understanding of such small-scale events.

Ground-based large-aperture telescopes offer a great potential to resolve these fine details and derive the desired insights. However, to exploit their full potential we need to remove instrumental and atmospheric degradations. For this we must employ image reconstruction techniques. 
Despite the obvious benefits, image reconstruction introduces a notable increase of the noise, which presents a significant challenge in the interpretation of solar spectro-polarimetric data, which is complex and full of interdependence. Typically, the observed spectral line profiles are fitted to derive a model of the atmospheric condition. Hence, the algorithms designed to analyse and decode observations into comprehensible information about the solar atmosphere can struggle with elevated noise levels. It is, therefore, critical to address the implications of increased noise to advance our understanding of solar dynamics. 

In this work I first describe the process of calibrating such high resolution observations. 
Within this scope I develop a novel and generic approach for the so called flat-field calibration of spectro-polarimetric data.
The approach and developed code are applicable to many instruments of the same family. 
Further, I am putting together the building blocks for the mathematical proof of a widely used numerical image reconstruction algorithm.

In the second part of this Thesis I describe an instrumentation prototype which is designed to fully exploit the resolution power of a 1.6 meter telescope. This instrument is built to provide suitable data for image reconstruction of spectro-polarimetric data. In the reconstructed data from this setup the effects of residual seeing are suppressed allowing for a resolution of small-scale features down to the diffraction limit. 

Due to practical compromises the simultaneous observation and interpretation of one or two lines is still the norm for solar spectra with high spatial resolution. Previous theoretical works have shown that the simultaneous interpretation of multiple solar lines can greatly improve the robustness against noise. 
Using observations from the described prototype I advance the theoretical idea into a practical technique. 
As part of this work I develop a line-selection method to prevent influence from lines with poorly determined atomic parameters. 
With this new technique I can retrieve a coherent 3D model of the solar atmosphere, resolving structures down to a horizontal size of about 49~km.

Finally, I use those new calibration and inversion techniques to interpret an observation of a solar pore. In this last part I describe and analyse small-scale structures, which could not be spectro-polarimetrically resolved before. Based on a good qualitative match between the atmospheric configuration derived from observations and a corresponding simulated atmospheric configuration, typical pore-boundary features - namely, dark striations with bright grains moving into the pore - are identified as convective features.
\newpage

\section*{Keywords}
\markboth{Keywords}{Keywords}
\begin{itemize}
    \item Sun \begin{itemize}
        \item Photosphere
        \item Active regions
        \item Small-scale features
    \end{itemize}
    \item Techniques \begin{itemize}
        \item Data calibration
        \item Stokes inversion
        \item Image reconstruction
    \end{itemize}
    \item Instruments \begin{itemize}
        \item \fisssp{}
    \end{itemize}
\end{itemize}

\chapter*{Zusammenfassung\markboth{Zusammenfassung}{Zusammenfassung}}
\addcontentsline{toc}{chapter}{Zusammenfassung}

Die Atmosphäre der Sonne wird maßgeblich durch Ereignisse beeinflusst, die in der Sichtebene oft nur wenige Dutzend Kilometer umfassen. Zu den Phänomenen, welche die Struktur und Dynamiken der Atmosphäre bestimmen, gehören Wellendissipation, Plasmainstabilitäten und Rekonnektion des Magnetfelds. Diese treten sowohl in ruhigen als auch in aktiven Regionen der Sonne auf,  werden von Prozessen der Magnetohydrodynamik bestimmt und haben ihren Ursprung häufig in den untersten Schichten der Sonnenatmosphäre.
Um das Verständnis der chromosphärischen und koronalen Erwärmung, des Sonnendynamos und seines Aktivitätszyklus, oder die Vorhersage von Weltraumwetterereignissen zu verbessern, müssen wir daher unser Wissen über diese kleinskaligen Prozesse erweitern.

Bodengebundene Teleskope mit großer Öffnung bieten hierbei ein enormes Potenzial, die gewünschten Erkenntnisse zu gewinnen, da sie sehr kleine Details auf der Sonne auflösen können. Um jedoch ihre Möglichkeiten voll auszuschöpfen, müssen instrumentelle und atmosphärische Beeinträchtigungen entfernt werden. Dazu werden Techniken zur Bildrekonstruktion benötigt.
Neben dem offensichtlichen Nutzen wird Bildrekonstruktion auch immer von einem deutlichen Anstieg des Rauschpegels begleitet. Dies stellt eine erhebliche Herausforderung bei der Interpretation der komplexen solaren spektropolarimetrischen Daten dar. Zu ihrer Analyse werden typischerweise die beobachteten Profile der Spektrallinien gefittet, um ein Modell der Sonnenatmosphäre abzuleiten. Algorithmen, die auf diese Weise Beobachtungen analysieren und in verständliche Informationen über die Sonnenatmosphäre umsetzen sollen, können daher mit erhöhten Rauschpegeln Schwierigkeiten haben.
Es ist deshalb entscheidend, die Auswirkungen des erhöhten Rauschens zu minimieren, um unser Verständnis der Dynamiken der Sonne voranzubringen.

In dieser Arbeit beschreibe ich zunächst den Prozess der Kalibrierung hochaufgelöster Beobachtungsdaten.
In diesem Zusammenhang entwickle ich einen neuartigen und allgemeinen Ansatz für die sogenannte Flat-Field-Kalibrierung spektropolarimetrischer Datensätze. Der Ansatz, und der dafür geschriebene Code, sind für viele Instrumente derselben Klasse einsetzbar. 
Weiterhin füge ich die Bausteine für den mathematischen Beweis eines weit verbreiteten numerischen Bildrekonstruktionsalgorithmus zusammen.

Im zweiten Teil dieser Arbeit beschreibe ich einen Instrumentenprototyp, der dazu entwickelt wurde, die maximale Auflösung eines 1,6-Meter-Teleskops auszuschöpfen. Dazu erlaubt seine Konstruktion, die gemachten Beobachtungen für die Bildrekonstruktion spektropolarimetrischer Daten zu nutzen. In den rekonstruierten Daten aus diesem Aufbau werden die Auswirkungen des Seeings unterdrückt, was eine Auflösung kleinräumiger Strukturen bis hinunter zur Beugungsgrenze ermöglicht.

Aus praktischen Gründen ist die gleichzeitige Beobachtung und Interpretation von ein oder zwei Linien nach wie vor der Standard bei räumlich hochaufgelösten Sonnenspektren. Frühere theoretische Arbeiten haben gezeigt, dass die gleichzeitige Interpretation vieler Sonnenlinien die Robustheit gegenüber Rauschen zum Teil erheblich verbessern kann. Mit Beobachtungen aus dem beschriebenen Prototyp entwickle ich diese theoretische Idee zu einer in der Praxis anwendbaren Technik weiter. Im Rahmen dieser Arbeit beschreibe ich eine Auswahlmethode für Spektrallinien, um den Einfluss von Linien mit unzureichend bestimmten Atomparametern zu minimieren. Mit dieser neuen Technik kann ich ein konsistentes 3D-Modell der Sonnenatmosphäre gewinnen, das Strukturen bis zu einer horizontalen Größe von etwa 49~km auflöst.

Abschließend wendete ich diese neuen Kalibrierungs- und Inversionstechniken bei der Interpretation einer Beobachtung einer Sonnenpore an. In diesem letzten Teil beschreibe und analysiere ich Kleinststrukturen, die zuvor spektropolarimetrisch nicht aufgelöst werden konnten.
Aufgrund einer guten qualitativen Übereinstimmung zwischen dem aus der Beobachtung abgeleiteten und einem entsprechenden simulierten Modell der Atmosphäre, können typische Merkmale am Rand einer Pore – nämlich dunkle Streifen mit hellen Körnern, die in die Pore hineinwandern – als konvektive Strukturen identifiziert werden.

}
\chapter{Introduction}
\label{ch:introduction}
The preceding quote from William Herschel\footnote{For more details on Herschel's speculations on how the Sun might be populated, see \cite{Kawaler81}, \cite{Basalla06}, or \cite{French2023Sun}} is both astonishingly accurate and amusingly incorrect at the same time. Consequently, it serves as a great example of two aspects: how science generally works (observation, hypothesis, experiment, evaluation, and refinement of the hypothesis) and the significant progress we have made in the description and understanding of the Sun and sunspots in particular.

\section{Why are we interested in solar small-scale features?}
The Sun heavily influences life on Earth by providing the energy necessary for photosynthesis, driving our climate and weather systems, and thereby sustaining our biosphere. But its activity also generates space weather which can disrupt satellites, communication, and power grids, which can critically impact our modern society.
Because of its proximity, the Sun is also the only star that we can study in detail. While other stars appear as simple points of light even through the most powerful single telescopes, on the Sun we are able to resolve surface features and analyse the layers of its atmosphere. This makes the Sun our best laboratory for understanding stellar physics and hence it serves as a reference point for interpreting observations of more distant stars. 
However, achieving a very high resolution on the Sun is not particularly easy and limited by practical constraints. In this Section I strive to answer the questions why it is nonetheless important to surpass current resolution limits.

Intensity images of the Sun often show darker features that are composed of a really dark core, the umbra, surrounded by a less dark halo, the penumbra. Historically, they have been interpreted as planetary transits, clouds (and their shadows on the surface), mountains sticking out of a liquid, or openings in the atmosphere \citep{Basalla06, Arlt2020}.
Since then, the spatial resolution of solar telescopes and the sophistication of diagnostic techniques have improved. Today we know that the light we see from a dark umbra of a sunspot is coming from an area that is about 1,000 Kelvin cooler and a few hundred kilometres deeper than the surrounding photosphere. Sunspots have been identified as phenomena associated with the Sun's magnetic activity \citep{Hale08}. The strong magnetic fields in a sunspot effectively inhibit the convective processes in the Sun's upper layer \citep{Biermann41, Zhivanovich20}. In the outer third of the Sun convection is the main mechanism of heat transportation from the interior to the surface. Hence, the observable surface within a sunspot is much colder and, therefore, appears darker. At the same time, the magnetic pressure counteracts the gas pressure, preventing the hot plasma from rising as high as usual, causing what is known as the Wilson depression \citep{Wilson74}. So sunspots are indeed "[o]penings in the Sun’s luminous atmosphere,
allowing a view of the underlying, cooler surface of the Sun" as Herschel phrased it.

However, many interesting questions still remain open. Taking the strong magnetic fields into account, we can conclude that the umbra is actually too bright and hence too hot \citep{Schlueter58, Adjabshirzadeh83}. Assuming that in the presence of magnetic fields with a strength between 1,000 and 4,000 Gauss convection is completely stopped, much less heat transport should be possible via radiation alone \citep{Gough66}. To resolve this contradiction, we need a detailed description of the small-scale plasma flow and magnetic structure of sunspots beyond the currently achievable resolution limit and a description of the sub-surface field. While it is common to observe the magnetic field in the Sun's lower atmosphere (see Section~\ref{seq:decipher_the_atmos} below), it is currently not possible to directly probe the subsurface magnetic structures. Two models have been proposed to describe the invisible deeper structure of a sunspot. 
In the 1930s, Cowling, Biermann, and others proposed the "monolithic" model \citep{Cowling85}. It describes the subsurface structure as a single, coherent magnetic structure with a well-defined magnetic field. In this model, a sunspot is considered a compact, monolithic entity. In the 1970s, Parker, Spruit, and others proposed the "cluster model" \citep{Parker79, Spruit81, Choudhuri86}. Here, the sunspot is no longer regarded as a single monolithic flux tube but rather as a bundle of sheaves formed by multiple smaller flux tubes. The original models could not withstand comparison with later observations, but have been refined and updated to their modern forms \citep[See, for instance,][Section~4]{Solanki03A}. 

Unfortunately, the deeper layers of the Sun, beyond the point where the plasma becomes completely opaque, are not directly observable. 
The only observational tool we have, local helioseismology \citep{Gizon05}, is an indirect measurement technique that utilizes continuos measurements of the solar surface to detect acoustic waves and to infer the structure of the Sun's subsurface layers. Recent results seem to agree better with the cluster model, but the question of which model is more appropriate has yet to be conclusively clarified \citep[][]{Cho21}.  
The high complexity of the interactions between magnetic field, solar oscillations and other waves still leaves room for multiple interpretations of the data, leading to different predictions of the sub-surface structure. The range of interpretations can hopefully be narrowed down by future developments in this field \citep{Tripathy23} and recent, uninterrupted, multiple-hours-long, high-resolution observations from Sunrise III \citep{lagg25}. 

Simulations, on the other hand, \citep[See, for instance,][]{Schuessler06, Bharti10, Rempel11} have shed some (theoretical) light on this region, providing numerical models of the deeper layers and connecting them to observations of the solar atmosphere. Any serious scientific model must produce testable predictions and both of the above basic sunspot models predict certain small-scale structures of sunspots. However, until now, it has not been possible to achieve satisfactory agreement between observations and simulations. One reason for this is the limited resolution of available spectro-polarimetric observations, which enable us to decipher the physical characteristics of the Sun's atmosphere.
Comprehensive reviews on sunspot evolution, fine structure, and their magnetic nature have been provided by \cite{Solanki03A, Solanki06, Tritschler09, Borrero11}, and \cite{Wiegelmann14}. Therein also comparison of observations and simulations are discussed.

\begin{figure}
  \centering
  \includegraphics[width=1\textwidth]{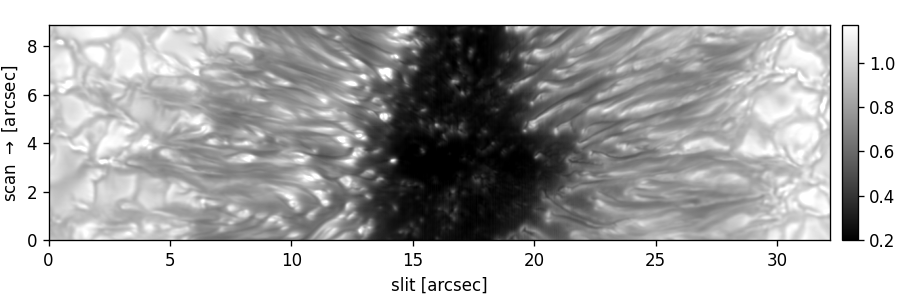}
\caption[Scan image of a sunspot]
{Scan image of a sunspot featuring some granulation cells at the sides, a fully developed penumbra with filaments stretching out along the slit direction and a dark umbra with bright UDs. The image is assembled from restored slit positions from the context imager, intensity is scaled to average quiet Sun continuum intensity.}
\label{fig:fisssp_scan_context}
\end{figure}

Traditionally, the presence of a penumbra, an area with intermediate brightness and an elongated, filamentary structure, distinguishes a sunspot from a pore. Or, in other words, a pore is an umbra without a penumbra. On closer inspection, this separation seems fairly arbitrary. Pores that develop a penumbra become sunspots and sometimes a sunspot can loose its penumbra and is then recognized as a pore. Moreover, pores are observed in various sizes, from micro-pores the size of a typical granule to pores with a fully developed umbra comparable to the size of an umbra of a small sunspot. 
A statistical comparison between pores, transitional-, and fully developed sunspots of solar cycles 24 and 25 was recently presented by \cite{Peng24}.
The umbrae of larger and more developed pores show similar small-scale features as sunspots, such as umbral dots (UDs). Figure~\ref{fig:fisssp_scan_context} presents a scan image\footnote{A scan image is an image assembled from restored context imager frames that closely matches the scene as observed with the spectrograph. For this we take a narrow cutout at the slit position from each restored context frame and append them to an extended field of view. Since the solar scene constantly changes during the scan time a single frame from the context imager would often not be representative for the scanned scene.} 
from the \fiss{} spectro-polarimeter (\fisssp{}, see  Chapter~\ref{ch:fiss_sp}) of a fully developed sunspot with a typical representation of those features.

The small brightness enhancements in umbrae of sunspots and pores were first described by \cite{Chevalier16} and referred to as UDs by \cite{Danielson64}. These small-scale features are strong candidates for a magnetoconvective energy transport, which could partly explain the observed brightness of umbrae. Simulations and nearly all observational studies agree on reduction of the magnetic field within most umbral dots, but report different reduced field values \citep[see, for example,][]{Socas-Navarro04, Schuessler06, Riethmueller08, Bharti09, Watanabe12, Riethmueller13}. Further, \cite{Watanabe12} reported counter examples with even enhanced field of up to 2k Gauss in decaying UDs. Especially the predicted small-scale plasma flows in the surroundings of UDs are still unsettled. Only \cite{Riethmueller13} could detect the average flow pattern predicted by \cite{Schuessler06}, but also in this study the details remained unresolved.

\newpage
Another convective structure that is typically observed in umbrae are light bridges (LBs). These are bright elongated structures that penetrate deeply into an umbra and sometimes even divide it in multiple parts \citep{Leka97}.
Such LBs are, as well as the boundaries of pores and surrounding granulation, dominated by small-scale structures such as dark and bright striations, which are not much wider than 100 km and are well observed with high-resolution broadband images \citep[Reported, for example, in][]{Scharmer02, Schlichenmaier16, Zhang18}. \cite{kuridze25} report even smaller striations in a network region. However, up to now, a high-resolution\footnote{By high-resolution I refer to spectro-polarimetric observations well below 0.1" resolution.} spectro-polarimetric description of such small-scale features has not yet been presented. I will further discuss such features in Chapter~\ref{ch:results}.

Going to even smaller spatial scales we find more open questions and unresolved predictions.
Movies of intensity images with the highest resolution currently available show ripples and wave patterns, hinting at the existence of very turbulent features below 50 km in size\footnote{For instance in the DKIST "fastcam" dataset (\url{https://dkist.virtualsolar.org/vanNoortfastcam/}) acquired in June 2025 with a diagnostic setup at NSO/DKIST, built in collaboration with M. van Noort.}, which are not yet really resolvable.
In magnetohydrodynamic (MHD) simulations with the \muram{} code \citep{muram}  \cite{Rempel12} and \cite{Schmassmann21} found unstable micro-convective flows in the dark umbra. They describe a shallow, unstable layer located right beneath the surface and magneto-convective filaments connected to UDs, reaching four to five Mm in depth. Recently, \cite{Bhatia25} also reported such unstable flows in \muram{} simulations of starspots. Until now, these flows could not be confirmed observationally, as their size seems below the detection capabilities of current flow-sensitive observations.

Not just since the famous detection of small-scale brightening events, popularly dubbed "campfires", in the solar corona \citep{Berghmans21, Harra21}, ranging from 400 to 4000 km in size, has it become clear that the overall energy state of the solar atmosphere is significantly influenced by small-scale events. The corresponding driving phenomena, which include wave dissipation, plasma instabilities, and magnetic reconnection, are ubiquitous across the Sun, occurring in both quiet and active regions, with their origins in the lower layers of the solar atmosphere often measuring only tens of kilometres in the horizontal direction. It is acknowledged today that understanding the structure and dynamics of solar magnetic fields down to the smallest observable scales is essential for advancing our understanding of chromospheric and coronal heating, deciphering the solar dynamo and its activity cycle, and predicting space weather events \citep[][and references therein]{Yang24}. This closely ties the small-scale observations of the lower solar atmosphere to the general understanding of our host star.

\newpage
\section{Deciphering the Sun's atmosphere}
\label{seq:decipher_the_atmos}

Quantifying the physical parameters of the solar atmosphere, most notably the temperature (\T{}), line-of-sight velocity (\vlos{}), and magnetic field, characterized by its strength (\B{}), azimuth (\(\phi\)), and inclination (\(\gamma\)), within the lower atmospheric layers down to the solar surface is key to understanding the Sun. However, it is clearly impossible to stand with a thermometer and a compass on the solar surface to measure these properties in situ. One reason for not being able to stand on the solar surface is its temperature\footnote{Indeed, it can be argued that you could (very) briefly visit the solar surface without being harmed by its temperature. See, \url{https://what-if.xkcd.com/115/}, for further discussion.}. 
Another one is that the Sun does not have a definite boundary that we could call a proper surface. We usually refer to the solar surface, known as the "photosphere", as the region where the plasma quickly becomes completely opaque. The solar surface is a layer of gas approximately 100 km thick. The gas density decreases exponentially with increasing height above the photosphere.
Fortunately, we can measure the atmospheric properties remotely, if only indirectly. 

\subsection{Observability}
The photon transport through any medium takes place by multiple repeated scatterings, absorptions, and (re-)emissions. The likelihood of these processes is determined by the composition and (optical) thickness of the medium as well as the frequency of the photon. Splitting the sunlight into its colours, or frequency components, the so called spectrum, reveals that the spectrum is not a continuous and smooth progression from blue to red. Instead, it is full of absorption and emission lines, caused by the absorptions and \mbox{(re-)emissions} from atoms or molecules in the solar atmosphere\footnote{If observed from ground the imprint of earth's atmosphere, the so called telluric absorption lines, becomes visible, too.}. We refer to the apparently line-free parts of the solar spectrum as continuum. The continuum intensity can be translated into temperature, measured in Kelvin (K), at the emitting solar surface. All spectral lines have a "natural" place within the frequency domain, usually measured in \AA{}ngstrom (\AA{}). A displacement of an observed line with respect to its nominal position can be directly translated into \vlos\, measured in km/s. 
Emission in the presence of a magnetic field, as well as scattering processes, further cause the sunlight to be polarized. Since these effects are imprinted onto the photon, and remain practically unchanged during its unobstructed journey towards the Earth, they can be measured using polarization optics and photon detectors. 

An historical overview on the discovery of magnetic signatures in the sunlight and an introduction to the process of inferring the atmospheric parameters from observations can be found in \cite{IntroductiontoSpectropolarimetry}. 
 A good basic introduction, with focus on observability and practical instrumentation, can be found in \cite{TheSun}. A deeper discussion of the solar case was presented in \cite{IntroductiontoSpectropolarimetry}, while in \cite{Frisch22} the general theory of radiative transfer beyond the solar application can be found. Here, I only want to briefly motivate how we can measure and use the polarization to estimate the magnetic field, as it is the most complex of the physical parameters to infer.

\begin{figure}
  \centering
  \includegraphics[width=1\textwidth]{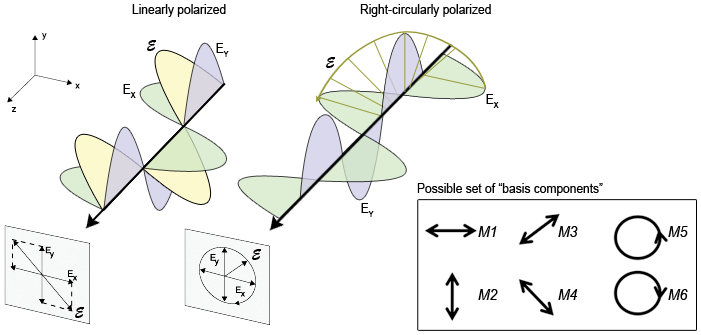}
\caption[Schematic drawing of fully polarized light]{Fully linearly and right-circularly polarized monochromatic light propagating along the line of sight axis \(z\) \citep[left, adapted from][Figure 2]{Goncalves02} and one possible set of "basis components" with which different states of polarization can be described (right). }
\label{fig:polarized_light}
\end{figure}

In the idealized case of fully polarized monochromatic light\footnote{In reality, natural light is never fully polarized and monochromatic. I have chosen this simplification for the practical reason that this idealization allows for a rather compact presentation of the core concepts of polarimetry.} we can describe the electric field vector in the relevant transversal \((x,y)\)-plane in a simplified form as
\begin{equation} 
\label{eq:electric_Field_vector}
\mathcal{E} \; := \; (E_x,E_y) \; = \; \big(A_x\,\cos(\omega t), A_y\,\cos(\omega t - \phi)\big),
\end{equation}
where \(A_j\) is the magnitude of the peak amplitude of the wave, \(\omega := 2\pi\nu\) the angular frequency (for frequency \(\nu\)), and  \(\phi\) the phase. 
Further, all possible polarization states can be described with the selection of "basis components" 
\mbox{\(\mathcal{M} := \big\{M_i\,|\,i\in\{1,\ldots,6\}\big\}\)}. Figure~\ref{fig:polarized_light} exemplifies two polarization states and a possible set of "basis components".  
Unfortunately, the components of the electric field vector cannot be measured directly with imaging photon-counting detectors. But with the chosen set of "basis components" we can describe linear combinations of them, the so-called Stokes parameters. In the idealized case this can be described by
\begin{eqnarray}
    \label{eq:stokes_parameters}
    I \;:= & M_1 + M_2 \;=\; M_3 + M_4 \;=\; M_5 + M_6  & = \; A_x^2 + A_y^2 \nonumber \\
    Q \;:= & M_1 - M_2 & = \; A_x^2 - A_y^2  \nonumber \\
    U \;:= & M_3 - M_4 & = \; 2A_xA_y \cos(\phi) \nonumber \\
    V \;:= & M_5 - M_6 & = \; 2A_xA_y \sin(\phi).
\end{eqnarray}

In contrast to the components of the electrical field vector \(\mathcal{E}\), the Stokes parameters can be detected. In our case of six basis components we can use a setup as follows: The incoming light must be sorted or filtered by basis component, which is causing detectable fluctuations in the image intensity. This is done with two optical components: A retarder ($\mathcal{R}$) and a polarizer ($\mathcal{P}$). The first device shifts the phase of one component of the electric field by a given \(\delta\), for instance for the \(y\)-component:
\begin{equation}
    \mathcal{R}\big((E_x,\,E_y)\big) \;=\; (E_x,\,E_y e^{i\delta}).
\end{equation}
The second device only transmits linearly polarized light in one direction, the direction of angle \(\theta\) in the transversal plane, such that the exiting light becomes completely $\theta$-linearly polarized. The resulting field can be described at
\begin{equation}
    \mathcal{P}\big((E_x,\,E_y)\big) \;=\; (E_x \cos\theta,\,E_y \sin\theta).
\end{equation}
Both devices together will cause a measurable difference in the intensity ($I_{meas}(\theta, \delta)$) that depends on the angles \(\theta\) and \(\delta\).
When rotating $\theta$ to 0$^\circ$ and 90$^\circ$ we can create the basis components $M_1$ and $M_2$, rotating to 45$^\circ$ and 135$^\circ$ creates \(M_3\) and \(M_4\), and adding further a quarter wave retardance to the last two creates $M_5$ and $M_6$. This yields another description of the four Stokes parameters in terms of six measurable intensities:
\begin{eqnarray}
    \label{eq:stokes_measurement}
    I \;= & M_1 + M_2 & = \; I_{meas}(0,0) + I_{meas}(\nicefrac{\pi}{2},0) \nonumber \\
    \;= & M_3 + M_4 & = \; I_{meas}(\nicefrac{\pi}{4},0) + I_{meas}(\nicefrac{3\pi}{4},0) \nonumber \\
    \;= & M_5 + M_6 & = \; I_{meas}(\nicefrac{\pi}{4},\nicefrac{\pi}{2}) + I_{meas}(\nicefrac{3\pi}{4},\nicefrac{\pi}{2}) \nonumber \\
    Q \;= & M_1 - M_2 & = \; I_{meas}(0,0) - I_{meas}(\nicefrac{\pi}{2},0)  \nonumber \\
    U \;= & M_3 - M_4 & = \; I_{meas}(\nicefrac{\pi}{4},0) - I_{meas}(\nicefrac{3\pi}{4},0) \nonumber \\
    V \;= & M_5 - M_6 & = \; I_{meas}(\nicefrac{\pi}{4},\nicefrac{\pi}{2}) - I_{meas}(\nicefrac{3\pi}{4},\nicefrac{\pi}{2}).
\end{eqnarray}

In our case of fully polarized monochromatic light the Jones representation (Equation~\ref{eq:electric_Field_vector}) can be identified with the measurable Stokes representation (Equation~\ref{eq:stokes_measurement}) via Equation~\ref{eq:stokes_parameters}. 
However, the Jones representation is not adequate for natural light, which is always partially polarized, and has a non zero frequency bandwidth. The Stokes formalism does not need such restrictions. A detailed comparison of Stokes and Jones representations can be found in \cite{shurcliff62}.
We can define the degree of polarization \(\varrho\) by 
\begin{equation}
\varrho \quad:=\quad \frac{\sqrt{Q^2+U^2+V^2}}{I}.
\end{equation}
Natural light always satisfies \(\varrho \in (0,1)\).

As we will discuss later\footnote{See discussion on seeing and image restoration in the following Section~\ref{sec:intro:dround_based}.} the time it takes to complete a modulation cycle is critical. Modern instruments therefore often use a setup that only needs four instead of six modulation states. A common setup of this type consists of two retarders, typically one of a quarter wave (\(\phi_1:=\nicefrac{\lambda}{4}\)) and the other of a half wave (\(\phi_2:=\nicefrac{\lambda}{2}\)), at variable rotations of \(\alpha_{1,2}\) and a polarizing beam splitter (PBS) that acts as a linear analyser \citep[See, for instance,][]{Pillet99, Gandorfer99, Keller03, Feller14, Iglesias16, MiHI}. 
The transformation on the Stokes parameters caused by any optical element can be described as a \(4\times{}4\)-matrix, the so called  Mueller matrix \citep{collett05}. 

\newpage
The Mueller matrices for the components above are easily defined as
\begin{equation*}
P\quad:=\quad \frac{1}{2}\,\left(\begin{array}{cccc}
         1 & 1 & 0 & 0 \\
         1 & 1 & 0 & 0 \\
         0 & 0 & 0 & 0 \\
         0 & 0 & 0 & 0 
    \end{array}\right)
\end{equation*}
for the polarizer, 
\begin{equation*}
V(\phi)\quad:=\quad \left(\begin{array}{cccc}
         1 & 0 & 0 & 0 \\
         0 & 1 & 0 & 0 \\
         0 & 0 & \cos\phi & -\sin\phi \\
         0 & 0 & \sin\phi & \cos\phi 
    \end{array}\right)
\end{equation*}
for the retarder, depending on the retardence \(\phi\), and
\begin{equation*}
R(\alpha)\quad:=\quad \left(\begin{array}{cccc}
         1 & 0 & 0 & 0 \\
         0 & \cos 2\alpha & \sin 2\alpha & 0 \\
         0 & -\sin 2\alpha & \cos 2\alpha & 0 \\
         0 & 0 & 0 & 1 
    \end{array}\right)
\end{equation*}
for the rotation \(\alpha\) of the elements along the optical axis. Typically, the rotation changes by \(45^\circ\) and both retarders have a rotational offset to the optical axis, thus \(\alpha_{1,2} = \theta_{1,2} + \beta_{1,2}\) with \(\beta_{1,2}\in\{0^\circ, 45^\circ\}\) variable and \(\theta_{1,2}\in[0^\circ,360^\circ)\) fixed. Assuming the incoming light vector \(S_0=(I_0,Q_0,U_0,V_0)^T=(a,b,c,d)^T\) not to change during one modulation cycle,
We can describe the resulting four modulation states, defined by the combinations of possible \(\alpha_{1,2}\), as 
\begin{equation}
MS(\alpha_1,\alpha_2,\phi_1,\phi_2) \quad=\quad P\cdot R(-\alpha_2)\cdot V(\phi_2)\cdot R(\alpha_2)\cdot R(-\alpha_1)\cdot V(\phi_1)\cdot R(\alpha_1)\cdot S_0.
\end{equation}
Each state again describes a full Stokes vector, but with our photon detectors we can only measure the \I{} component, which corresponds to the first component of the resulting Mueller matrix of the system, and which is different for each combination of \(\phi_{1,2}\) and \(\alpha_{1,2}\). With \(\phi_{1,2}\) fixed and two possible values for each \(\alpha_i\) this results in a set of four equations of the form
\begin{eqnarray}
    \label{eq:modulation_measurement}
    2\, I(\alpha_1,\alpha_2,\phi_1,\phi_2)  &=& a 
     + \big(b\cos 2\alpha_1 + c\sin 2\alpha_1\big)\cos 2(-\alpha_1)\cos 2\alpha_2\cos 2(-\alpha_2) \nonumber\\
    && + \sin 2(-\alpha_1)\cos 2\alpha_2\cos 2(-\alpha_2) \nonumber \\
    && \quad \cdot \, \Big(\cos\phi_1 \big(-b\sin 2\alpha_1 + c\cos 2\alpha_1\big) \nonumber \\
    && \qquad     + \sin\phi_1\big(-b\sin 2\alpha_1 + c\cos 2\alpha_1\big)\Big) \nonumber\\ 
    && + \sin 2\alpha_2\cos 2(-\alpha_2)) \nonumber \\
    && \quad \cdot \,\bigg(\big(b\cos 2\alpha_1 + c\sin 2\alpha_1\big)\sin 2(-\alpha_1) + \cos 2(-\alpha_1) \nonumber\\ 
    && \qquad \cdot \, \Big(\big(-b\sin 2\alpha_1 + c\cos 2\alpha_1\big)\cos\phi_1  \nonumber \\
    && \qquad \quad + \sin\phi_1(-b\sin 2\alpha_1 + c\cos 2\alpha_1)\Big)\bigg).
\end{eqnarray}
Since \(\alpha_1,\,\alpha_2,\,\phi_1,\) and \(\phi_2\) are all known parameters, we can retrieve \(S_0\) from the four measured intensity values by solving a system of linear equations. 
The modulation scheme and its efficiencies can be optimized by choosing different retardence values \(\phi_i\) and different offsets \(\theta_i\) for the rotation along the optical axis.
Details of such an optimization are, for instance, described in \cite{Iglesias16}. 

Combining the first rows of each modulation state matrix \(MS(\alpha_1,\alpha_2,\phi_1,\phi_2)\) into a new matrix results in the so called modulation matrix (\(\mathcal{M}\)) of the system. This allows us to determine how suitable our system is to measure the individual Stokes parameters. 
For our system with \(4\) modulation states, the efficiency \(\epsilon_s \in [0,1]\) for Stokes parameter \(s\) can be calculated as:
\begin{equation}
     \epsilon_s \quad=\quad \frac{1}{4} \sqrt{ \sum_{k=1}^4 \mathcal{M}_{k,s}^2 }.
\end{equation}
Values closer to one correspond to a better observability of the corresponding parameter. 
In an actual telescope observation also the polarization created by the telescope, the instrument and also the sensor response itself play an important role and measurements need to be calibrated for those effects. The modulation scheme can be optimized accounting for real world imperfections, when \(\theta_i\) remain adjustable. Such a calibration and optimization method for measured polarimetric modulated intensities is described in \cite{2008A&A...489..429V}.

\subsection{Instruments for solar observations}
Instruments that aim to collect the spectral and polarimetric information that we need to decipher the Sun's atmosphere must record three data dimensions, namely two spatial and one spectral dimension. Naturally, it is difficult to collect a 3D data structure with a 2D sensor. There are basically three types of instruments for the observation of high spatial resolution spectra: (i) Filtergraphs that capture narrow band images at various wavelengths, scanning the spectral dimension over time, (ii) Spectrographs that capture images with one spectral and one spatial dimension, scanning the other spatial dimension over time, and (iii) Integral field instruments that capture full spectral and spatial data simultaneously. All three types are described in more detail and the currently available instruments summarized and contrasted in \cite{Iglesias19}. The main conceptual difference of these instrumentation types lies in the dimension that is sampled with time: As the Sun changes constantly during observation, the signal recorded along the time dimension originates from a slightly altered feature. Filtergraphs produce coherent 2D images, while the observed feature evolves during a scan along the wavelength dimension, which can effect the observed line shape.
Slit spectrographs, on the other hand, sample coherent wavelength information, while the feature evolves during the collection of one of the spatial dimensions.
Integral field units, which sample coherent information in all three dimensions, are today still limited to a relatively small field of view (FOV) that does not allow to observe larger structures in their entirety.

\begin{figure}
  \centering
  \includegraphics[width=1\textwidth]{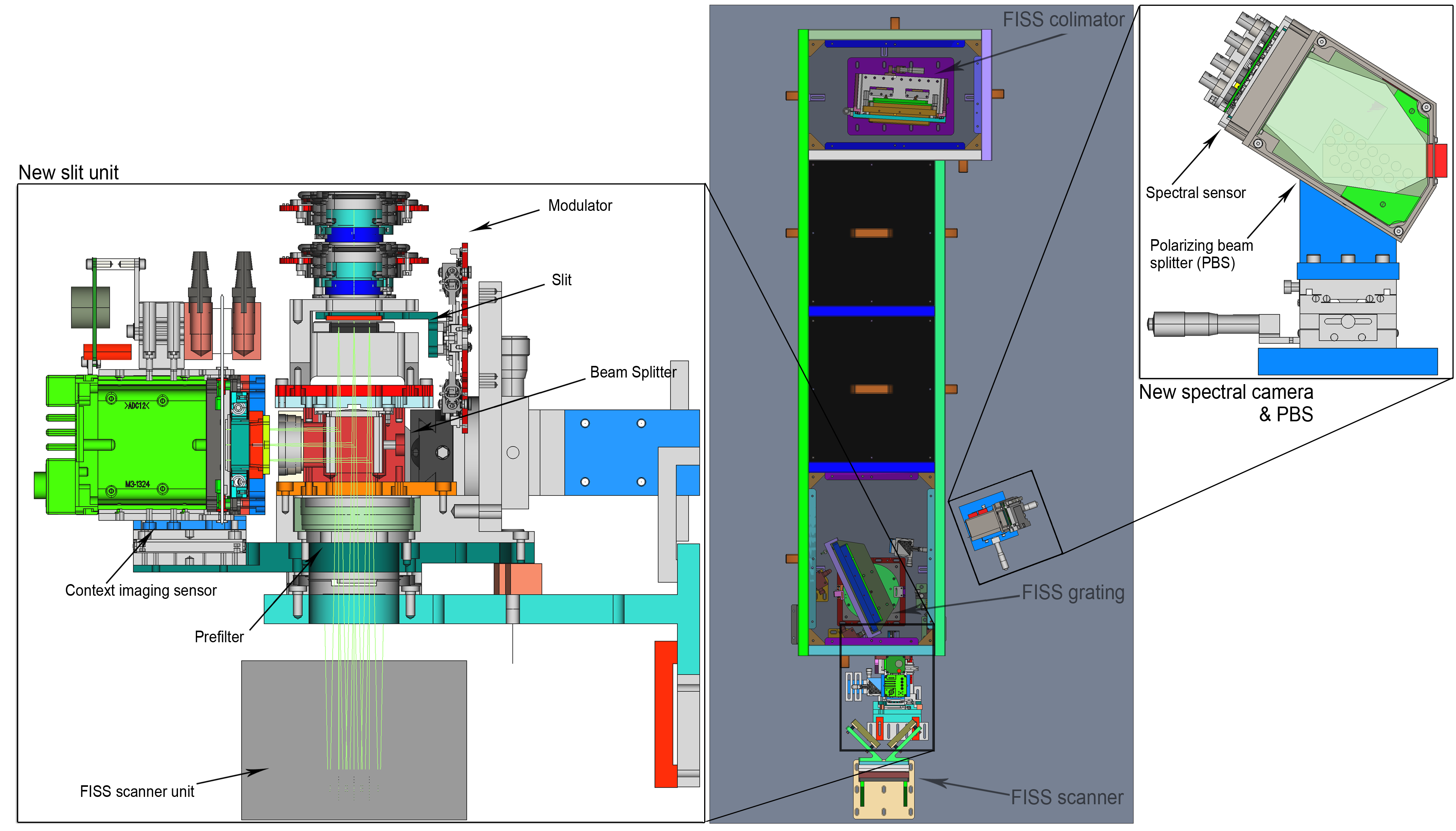}
\caption[CAD drawing of \fisssp{}]
{Computer Aided Design (CAD) drawing of \fisssp{}, the spectro-polarimetric extension of \fiss{}. The two inlays show the extension modules consisting of the new slit unit (left) and the new spectral camera unit (right).}
\label{fig:fiss-sp_cad}
\end{figure}

All these instruments can be equipped with polarimetric capabilities to measure the Stokes parameters. We refer to such instruments that measure the polarization of light across various wavelengths and on spatially resolved features as spectro-polarimeters (SPs). In practice multiple possible setups exist, an introduction to observational polarimetry can be found in \cite{TheSun} and in \cite{IntroductiontoSpectropolarimetry} Chapter~5.
In this Thesis I will focus on a subset of (ii): grating-based slit-scanning spectrographs with polarimetric capabilities. In this case a narrow slit is placed in the focal plane of the telescope and the beam is then directed onto a grating to disperse the spectrum. The solar image is moved across the slit to scan a region of interest (ROI). 

The main instrument used for this thesis is a spectro-polarimetric extension of the Fast Imaging Solar Spectrograph \citep[FISS,][]{Chae13}. Two extension modules where added to the existing instrument to allow for spectro-polarimetric observations. These modules are (1) a new slit unit, consisting of a context imager, a slit selection library, a pre-filter, and the polarimetric modulator and (2) a new large sensor spectral camera and a PBS. 
A design drawing of the resulting \fisssp{} setup including the two extension modules is presented in Figure~\ref{fig:fiss-sp_cad}. The SP extension utilises a dual beam modulation scheme with \(2\times4\) modulation states based on two ferro-electric liquid crystals (FLCs) as retarders and a PBS analyser. As already mentioned, the time it takes to complete a modulation cycle is critical, especially in ground based observations. A FLC based modulator allows for a four state modulation scheme with a temporal modulation frequency of up to 1 kHz \citep{Gisler03} and hence allows for the desired high frame rates. The setup,  calibration, and first-light observations of \fisssp{} will be detailed in Chapter~\ref{ch:fiss_sp}.

\subsection{Interpretation of the observations}
Observations from such SP instruments enable us to investigate the physical parameters of the solar atmosphere using a process termed "Stokes inversion" which employs the polarization profiles of spectral lines. However, converting these observations into usable information is highly complex, as their interpretation relies on an intricate relationship between atmospheric parameters and the way light is absorbed and emitted as a function of wavelength and height, described by the optical thickness and source function. This function describes the ratio of emission, absorption and re-emission along the line of sight and is a function of the matter properties (density, temperature, composition). To simplify this, let us assume that at any point in the solar atmosphere, all microscopic processes (like collisions, excitations, and ionizations) are in equilibrium, and the local temperature accurately describes the system's thermodynamic state. This state is referred to as Local Thermodynamic Equilibrium (LTE). Under this assumption the source function \(S_{\!\nu, T}\) can be described by the Planck function \(B_{\nu}(T)\) of the local temperature \(T\) at a given frequency \(\nu\). Focusing on the beam itself and neglecting influences from adjacent material and adjacent or crossing beams, that is, assuming a one dimensional atmosphere, the observed intensity \(I_{\nu}\) can be described by the simplified radiative transfer equation (RTE) as follows \citep{TheSun}:

\begin{equation}
    \label{eq:rte}
    I_{\nu}(s) \;=\; I_{\nu}(s_0) \cdot e^{-\tau_{\nu}(s_0,s)} \,+\, \int_{s_0}^s B_{\nu}\big(T(s')\big) \cdot \alpha_{\nu}(s') \cdot e^{-\tau_{\nu}(s',s)}\, ds',
\end{equation}
where \(\tau_{\nu}(s_1,s_2)\) is the scalar optical depth between \(s_1\) and \(s_2\), and \(\alpha_{\nu}\) the absorption coefficient. While the contribution of the individual parts, especially the incident beam \(I_{\nu}(s_0)\), are nicely visible in the integral form, for a compact notation its differential form is often preferred:
\begin{equation}
    \frac{dI_\nu}{ds} \quad=\quad \alpha_\nu(s)\,\Big[\, B_\nu(T(s)) - I_\nu(s)\,\Big].
\end{equation}
If the composition and temperature of the solar atmosphere and the velocities along the line of sight are known, the solution of the RTE results in synthetic Stokes-\I{} profiles for a selected wavelength range and spectral lines.

\begin{figure}
  \centering
  \includegraphics[width=1\textwidth]{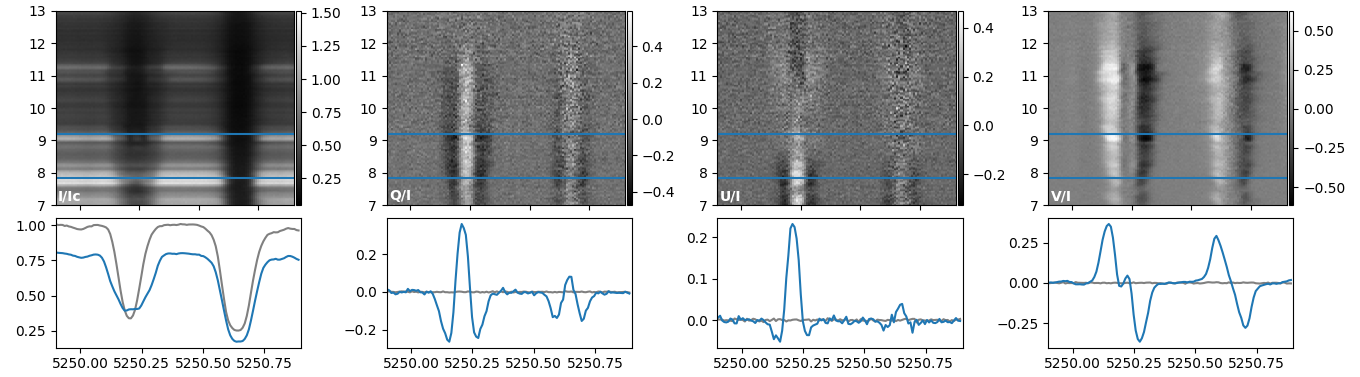}
\caption[Example of observed Zeeman signals]{Example of Zeeman polarization signals in \I{}, \Q{}, \U{} and \V{} from the 5250~\AA{} \ion{Fe}{I}-lines in a pore. The top row shows the spectra with vertical axis expanding along the slit (arc-seconds) and horizontal axis in wavelength direction (\(\AA{}\)). The bottom row shows the spatially averaged profiles from the area between the blue lines in the spectra in blue and a reference average quiet Sun profile without a Zeeman signature in gray. The spectra have been recorded with \fisssp{} and are restored.}
\label{fig:zeeman}
\end{figure}

\begin{figure}
\begin{minipage}[b]{0.62\textwidth}
  \includegraphics[width=1\textwidth]{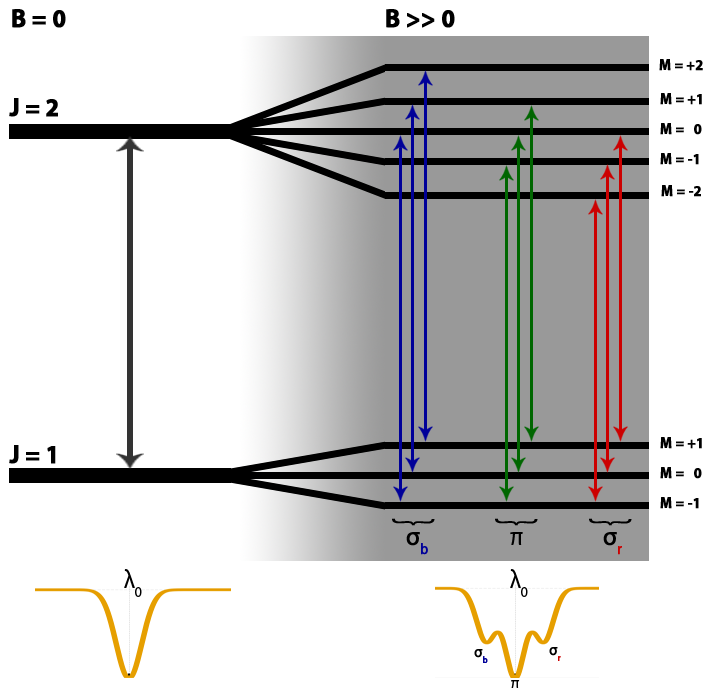}%
\end{minipage}\quad\begin{minipage}[b]{0.35\textwidth}%
  \caption[Idealised scheme of the normal Zeeman effect]{
  Idealised transition scheme of the normal Zeeman effect for a transition \mbox{\(J:2\rightarrow1\)}. On the left transition without field (\(B=0\)) is presented, while on the right a relevant field (\(B >> 0\)) is present. The red, green, and blue arrows represent the transitions with \(\Delta M\) of -1, 0, and +1 respectively.
  The transitions of each group have the same energy potential (indicated by the same length of the arrows) and hence result in the same line component in the splitting pattern, exemplified below.}%
  \label{fig:normal_zeeman}%
\end{minipage}
\end{figure}

In the presence of a magnetic field the aforementioned polarization processes add further complexity. The biggest influence on the line shape is known as Zeeman effect\footnote{The discovery of this effect and its influence on solar physics is described in \cite{DELTOROINIESTA96}.}, as shown in Figure~\ref{fig:zeeman}. This effect describes the splitting of an absorption line due to splitting of energy levels of an atom into sub-levels in the presence of a magnetic field \(B\). 
An energy level with total angular momentum quantum number \(J\) splits into \(2J+1\) sub-levels with different magnetic quantum numbers (\(M\)). The energy spacing between these sub-levels is proportional to the unsigned magnetic field strength \(|B|\) and the Landé factor\footnote{The Landé factor (\(g\)) is the dimensionless proportion between the total magnetic and total angular momentum of an atom, and hence describes its probability to split. A first energy level specific version was presented by \cite{Lande23} and later generalized by \cite{Shenstone29}.}. When a magnetic field is present, each of the electronic energy levels involved in an atomic transition can split, resulting in multiple possible transitions and thus the original spectral line splits into multiple so called Zeeman components. Which transitions between sub-levels are allowed is governed by selection rules:
(i) Magnetic quantum number transitions are only allowed between states where the magnetic quantum number changes by zero or by one \mbox{(\( \Delta M \in\{ 0, \pm 1\} \))}. And (ii) also total angular momentum quantum transitions must satisfy the change by 0 or one \mbox{(\( \Delta J \in \{ 0, \pm 1\} \))}, but with the additional constraint that transitions between two states both having \( J = 0 \) are forbidden. We can group the Zeeman components by their magnetic quantum number transition:
\begin{itemize}
\itemindent=2.5em
\item[\(\Delta M = 0\)\;\;] transitions produce linearly polarized components, which are not shifted in wavelength (\(\pi\)-components).  
\item[\(\Delta M = +1\)] transitions produce elliptically polarized components, which are relative to the nominal line position shifted to shorter wavelengths (\(\sigma_b\)-components).  
\item[\(\Delta M = -1\)] transitions also produce elliptically polarized components, but with opposite spin direction, which are relative to the nominal line position shifted to longer wavelengths (\(\sigma_r\)-components).
\end{itemize}
A schematic drawing of components and transitions of the \textit{normal Zeeman effect} is presented in Figure~\ref{fig:normal_zeeman}. The actual observed splitting and polarization of the line profile then depends of the amount of field, which allows us to estimate the field strength in the inversion process.  Further details, especially on the more general \textit{anomalous Zeeman effect}, where transitions no longer necessarily have the same energy potential, can be found in \cite{IntroductiontoSpectropolarimetry}, Chapter 8.

To compute the full Stokes vector \mbox{\(\mathbf{I}_{\nu} := (I_{\nu}, Q_{\nu}, U_{\nu}, V_{\nu})\)} a vector radiative transfer equation (VRTE) is needed. The differential form for polarized radiation can be given as \citep{IntroductiontoSpectropolarimetry}:
\begin{equation}
    \label{eq:vrte}
    \frac{d\mathbf{I}_{\nu}}{d\tau_C} \quad=\quad \mathcal{K}_{\nu}\; \Big[\mathbf{S}_{\nu, T} - \mathbf{I}_{\nu}\Big],
\end{equation}
where \(\mathbf{S}_{\nu, T} := (B_{\nu, T}, 0, 0, 0)^T\) is the source function vector and \(d\tau_C\ = -\chi^C \,ds\) the optical depth of the continuum with \(-\chi^C\) the frequency-independent absorption coefficient for the continuum. This coefficient translates the geometrical depth \(s\), that we used before, to the optical depth, which is a function of opacity and hence more suitable for an observation. Further, \(\mathcal{K}_{\nu}\) denotes the \(4\times4\) propagation matrix that contains the combined effects of absorption coefficients \(\alpha_{\nu}\), magneto‑optical (dispersive) coefficients, and the Zeeman components. It therefore determines how polarization states are absorbed and interconverted as the beam propagates.

So far we have assumed that  temperature, velocities, and magnetic field configuration are known and we can thus solve the (V)RTE. Unfortunately, those are exactly the values that we wish to extract from the observations. 
Least-squares inversion methodologies, involving either analytical or computational solutions of the (V)RTE, are utilized to tackle this issue. These techniques compare observed Stokes profiles to those synthesized from a presumptive model atmosphere. The disparity is then employed to adjust the atmospheric parameters until the match between the synthetic and observed profiles is satisfactory. 
This process leads to the creation of a model atmosphere capable of representing the spectro-polarimetric observations, considering the model's assumptions and constraints.

To briefly summarize the process of Stokes inversion, it involves three steps: (1) Computing synthetic Stokes profiles for a given configuration of the solar atmosphere by solving the (V)RTE. (2) Comparing the observed Stokes profiles with the synthesized ones. (3) Evaluating the disparity between synthetic and observed spectra and adjusting the atmospheric parameters, and iterating steps 1-3 until a satisfactory match is found.
A comprehensive description of solar Stokes-inversions can be found in \cite{IntroductiontoSpectropolarimetry} and \cite{ToroIniesta16}. 

\newpage
\section{Diffraction-limited, ground-based solar observations}
\label{sec:intro:dround_based}

The current defacto standard for high-resolution, slit-scanning spectro-polarimetric observations is set by the spectro-polarimeter aboard the  \mbox{HINODE} satellite \citep[HINODE/SP,][]{Lites01, Kosugi07, Tsuneta08}. 
While the HINODE solar optical telescope (SOT) boasts a 0.5-meter aperture, the next generation of ground-based solar telescopes with primary apertures exceeding 1.5 meters has arrived. 
Unfortunately, when observing the Sun it is impossible to obtain a "true" solar image.  
Any recorded image suffers from instrumental and external degradations. Sophisticated data reduction and calibration methods need to be applied to remove as much as possible the instrumental effects from the data. In Chapter~\ref{ch:data_reduction} I will discuss some of these techniques in more detail. But even if we consider the collected frames as calibrated with respect to common effects like dark current, bias, sensor non-linearity, flatfield, and gain some instrumental and environmental effects remain.

\begin{figure}
  \centering
  \includegraphics[width=210pt]{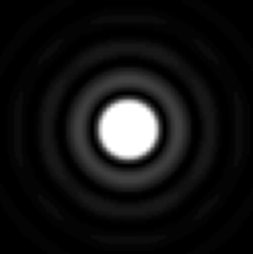}
  \includegraphics[width=210pt]{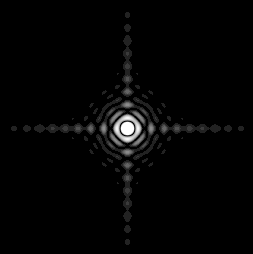} 
\caption[Telescope aperture diffraction patterns]{Artificial images of diffraction patterns caused by telescope apertures: Airy Disk pattern of an ideal unobstructed circular aperture (left) and "Christmas Star" pattern of a circular aperture with central obstruction and a four-leg spider (right). Generated using \url{http://aberrator.astronomy.net}.}
\label{fig:diffraction}
\end{figure}

During the journey from the solar surface to the detector plane of an instrument, the solar image is modified by the diffraction caused by the telescope aperture and other optical aberrations. If the telescope is placed within the Earth atmosphere the effects of the Earth's atmospheric turbulence further degrade the recorded image. The primary effect of both of these processes on the solar image is to distort and smear the recorded image, which strongly effects the small scale structures. In technical terms, this process reduces the amplitudes of the high-spatial-frequency components of the Fourier transformed image, and scrambles their phase.
The static effect of the telescope and other optical elements downstream is very well understood and can be modelled to the extent of real-world imperfections in the physical setup. See Figure~\ref{fig:diffraction} for examples of idealized diffraction patterns and \cite{AstronOptics} for an overview on telescope optics.

However, for ground-based observations the main contributor to image degradation is the external effect of atmospheric seeing. Light is emitted spherically, but as the Sun is about \(1.5\times 10^8\) km (or \(1\) AU) away from the Earth, the curvature of the wavefront can be neglected. We therefore assume the sunlight to arrive as a plane wave.
When this plane wavefront reaches Earth's atmosphere the light is refracted from one patch of air, with a given temperature and density, to another patch with different temperature and density. Moreover, the small-scale structure of the atmosphere changes quickly and randomly. 
These deviations from the ideal plane wavefront cause the spatial distribution of light intensity originating from the same source. In the case of a point source the information would simply be smeared out over a larger area. In the case of an extended source the effect is more subtle: Here, the seeing scrambles the signal from adjacent structures and the image becomes blurry. We describe the spatial distribution of light intensity of an image of a point source as point spread function (PSF), or its Fourier transform, the optical transfer function (OTF). The combined wavefront aberrations, diffraction effects, and optical aberrations from seeing, telescope, and instrument cause the effective PSF of our images.
Following \cite{Noll76}, \cite{Wang80}, and \cite{Wilson96}, the PSF can be efficiently described using an infinite linear combination of Zernike polynomials \citep{ZERNIKE1934}.  This will be discussed in more detail in Chapter~\ref{ch:reconstruction}.

\begin{figure}
  \centering
  \includegraphics[width=400pt]{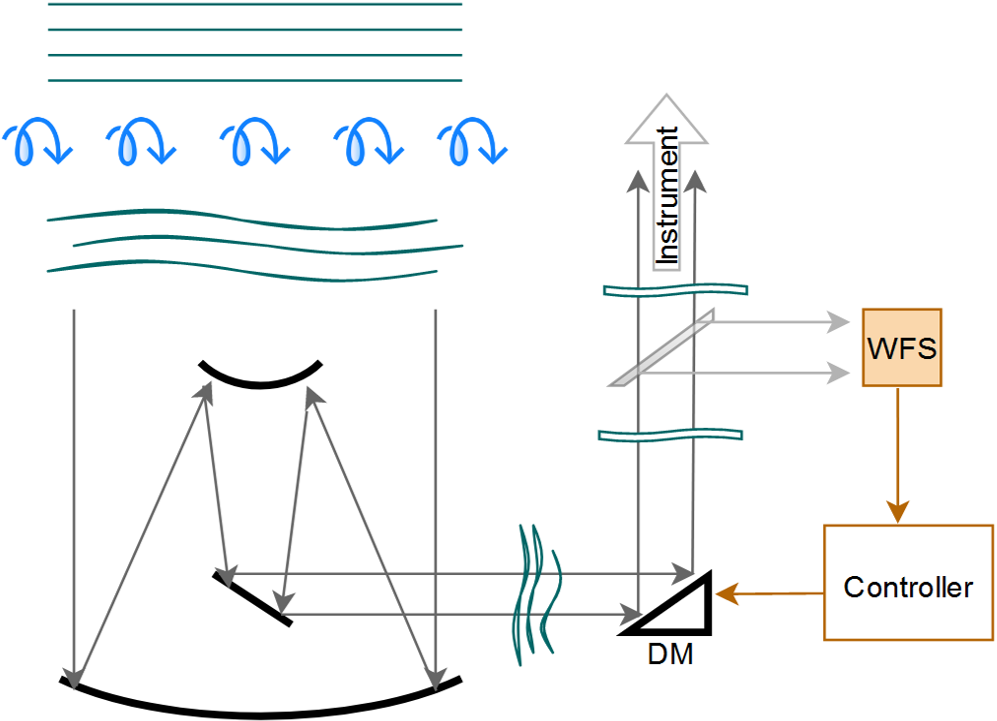} 
\caption[Schematic drawing of an adaptive optics system]{Schematic drawing of an adaptive optics system. The plane wavefront is distorted by the atmosphere. The distorted wavefront reaches the deformable mirror (DM) and a less distorted wavefront is sent through a beam-splitter. Most of the light is directed towards the instrument(s) while a small portion is fed into the wave front sensor (WFS) to compute the online correction via the DM.}
\label{fig:ao}
\end{figure}

According to Taylor's Hypothesis \citep{Taylor38} the timescale on which the seeing condition changes is bound to the wind speed. Even if this hypothesis has been criticized lately \citep{SaintJacques2000, Hilland24}, the hypothesis that seeing has a limited frequency is not at debate. Following \cite{Wilson96}, also the number of of Zernike polynomials needed is limited. 
Consequently, real time sensing and correction of wavefront errors became possible, with the availability of fast computers and fast cameras. Figure~\ref{fig:ao} shows a schematic overview of such an adaptive optics (AO) system. However, for a perfect description of a PSF caused by a wavefront error still a large number of Zernike polynomials and a very small sampling time of the sensor and negligible image readout time combined with a high duty cycle would be needed. Also, the deformable mirror (DM) is limited in response time, amplitude, and shapes that can be modeled \citep{Glindemann2000}. Thus, the real-time correction delivered by an AO system can only correct for some finite orders and frequencies of an aberration that was measured some (short) time ago. With larger fields of view (FOV) or larger apertures the best correction is also limited to an isoplanatic patch (usually in the center of the FOV), that is the area of sky where atmospheric turbulence affects all light rays similarly.

Mathematically, the degradation of the true solar scene, hereafter just scene, can be described as a convolution:
\begin{equation}
    \label{eq:convolution}
    f(x) \;=\; p \circledast s\,(x) \;:=\; \int p(x-\hat{x}) \cdot s(\hat{x})\, d\hat{x},
\end{equation}
where \(f\) is the image we observe, \(s\) the scene, \(p\) is the PSF normalized to \mbox{\(\int p(x) \, dx= 1\)},  and symbols \(x, \hat{x} \in \mathbb{R}^2\) denote the positions in the  observed image and scene respectively. We can interpret this process as follows: Each point on the object is substituted with a PSF that has a strength proportional to the intensity value of the point and is centered on its position. In the case of an extended object the image we observe is created by the sum of all these partially overlapping PSFs. 
This process is obviously a linear operation. It is therefore relatively easy to compute the PSF from the recorded image if we know the scene. Once the PSF is known, we can undo its effect. This process is referred to as deconvolution and best described within the Fourier domain, as convolution translates to multiplication:
\begin{equation}
    f(x) \;=\; \big(p \circledast s\big)(x) 
    \qquad \xrightarrow{\textnormal{Fourier transform}} \qquad 
    F(u) \;=\; P(u) \cdot S(u),
\end{equation}
where \(u\in\mathbb{C}\) is the spatial frequency coordinate. In Fourier space we can now divide by the OTF \(P\) to restore the scene: 
\begin{equation}
    \label{eq:deconvolution}
    S(u) \;=\; \frac{F(u)}{P(u)}.
\end{equation}
Using the inverse Fourier transform we can now retrieve the undegraded scene. It is important to note that, since we divided by \(P\), any measurement error in \(f\) (i.e., noise), will be amplified in the undegraded image.

In this context it is unfortunate that, opposed to any other observable star, the Sun is not a simple point source and we cannot know the scene. We therefore need a metric that allows for a "blind" reconstruction, that is a metric where no a priory knowledge of the scene is needed. 
Multi Frame Blind Deconvolution (MFBD) \citep{Schulz93, Miura94, Paxman96}, is a post-facto estimation technique of the PSF. Their underlying assumption is that over a given time span the only change is the PSF that describes the seeing, while scene and instrumental aberrations remain constant. In this Thesis I will discuss an improved version, the Multi Object MFBD (MOMFBD) approach, as described by \cite{Noort05}. This version combines the information from multiple objects (recordings of the same solar scene at different wavelengths), and very many frames (typically frames are recorded with several hundred Hertz) per object. With this setup we have hundredths of measurements of the same scene distorted by different PSFs which allows us to estimate the PSFs and therefore restore the undegraded solar image.
Further, in this post-faco approach the analysis can be split to multiple patches of the image (each within the size of the isoplanatic patch, which is inversely related to the telescope's aperture and the atmospheric altitude layers that dominate the seeing). Given sufficient overlap in the patches, the obtained PSFs can be stitched together. This effectively allows to de-convolve the full FOV of the recorded frames. As the operation is not performed in real time, the time delay problem of AO systems does not apply here.

\begin{figure}
  \centering
  \includegraphics[width=1\textwidth]{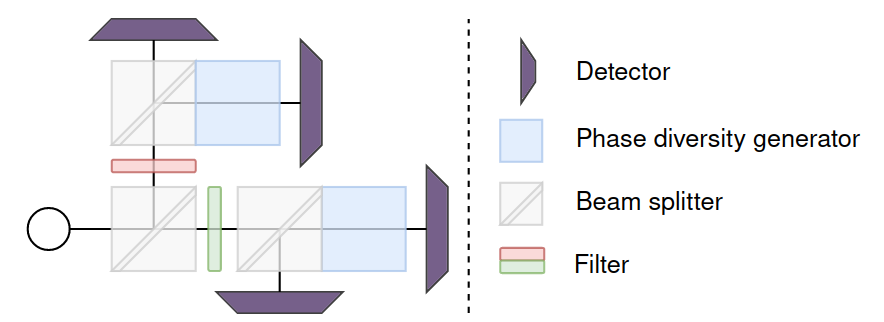} 
\caption[Data collection scheme for MOMFBD]{MOMFBD data collection scheme \citep[simplified from][]{Noort05}. All detectors are strictly synchronized, such that all frames from all sensors are collected at a particular time \(t\). Each filter allows the passage of light in a different wavelength window, allowing to observe multiple "objects" at the same time. The phase diversity channels of each object add additional information to sense the wavefront error (see \cite{Meynadier99} for details).}
\label{fig:data_collection}
\end{figure}

As all objects are distorted by the same seeing, the PSF is, apart from a wavelength dependent scaling factor, the same. Thus, we can use all objects as input and solve the deconvolution problem simultaneously. 
In practice we can approximate the PSF as a finite linear combination of Zernike or Karhunen-Lo\`eve modes \citep{Wilson96}.
Figure \ref{fig:data_collection} shows a schematic overview of data collection with two different objects and two channels each. 

Solar evolution time is the time it takes until a solar scene is significantly changed. For instance, the time span of granular evolution is about 5 Minutes and after 15 Minutes the solar scene on the photosphere has completely changed (see for example \cite{Dialetis86}). However, the length of the time scale where the scene can be assumed to be truly unchanged depends on the telescope's aperture \citep[See][Figure~1]{Iglesias19}. By restricting frame selection to a small enough time scale we can assume that the solar scene has not changed from frame to frame and the variation is exclusively related to different realizations of the PSF. As a first approximation we can estimate the "seeing quality", that is how much the atmosphere degraded our images, by measuring the root-mean-square (RMS) contrast of our frames: The higher the contrast for the same scene, the better the seeing. 

The minimization idea is the same for all objects and if computed simultaneously the individual objects just provide additional constraints for the metric. To sketch the 
idea of the MOMFBD algorithm we can thus ignore the multi object part and focus on a single object.
Let \(N\in\NN\) and \(f_i, i\in [0,N]\) the frames taken into account. Let \(s\) again denote the scene and \(p_i\) the PSF corresponding to each frame \(f_i\) described by the set of coefficients \(\mathbf{\alpha}_i \subset \RR\) for the linear combination of the modes. As convolution is a linear operation and we are in the case of discrete detector pixels, the following equation holds
\begin{equation} \label{eq:psf}
    s \circledast \frac{1}{N} \sum_{i=0}^N p_i \quad=\quad \frac{1}{N} \sum_{i=0}^N f_i.
\end{equation}
We can now proceed with an iterative approach. In every step do
\begin{enumerate}
    \item Compute the PSFs (\(\hat{p}_i\)) from a proposed set of coefficients \(\hat{\mathbf{\alpha}}_i\)
    \item De-convolve each frame (\(f_i\)) with the proposed PSF (\(\hat{p}_i\)) to estimate the scene (\(\hat{s}_i\)).
    \item Weight the estimated scenes by the seeing quality (\(\omega_i \)) and compute the common estimate as weighted average (\(\hat{s} = \sum \omega_i\, \hat{s}_i/N\)).
    \item Compare each raw frame (\(f_i\)) with the common scene estimate  re-convolved with the corresponding PSF (\(\hat{s} \circledast \hat{p}_i\)).
    \item Stop the iteration if the disparity between re-convolved scene estimate and raw frames is satisfactory, else generate a new set of coefficients \(\hat{\mathbf{\alpha}}_i\) and start over.
\end{enumerate}
It is important to note two things here: (a) the estimated scene \(\hat{s}\) does not depend on the individual frame but is a common estimate and (b) also no knowledge of the actual scene is needed to compute \(\hat{s}\). The idea is to minimize the disparity of Step 4, again weighted by the estimated seeing quality of each frame (\(\omega_i\)).

Of course also the MOMFBD technique cannot restore the true solar scene completely. It truncates the estimation at a finite number of modes, determined by numerical feasibility, and a non-negligible sensor-integration time is needed to collect enough photons. Therefore, an underestimation of degradation caused by higher frequencies and modes is systematic.
I will provide the detailed mathematical description of the MOMFBD algorithm and error metric in Chapter~\ref{ch:reconstruction}.

Let us now assume the image reconstruction problem to be solved. We also need to discuss the degradation of spectral data. Neglecting internal spectrograph-aberrations, it does not matter if we record the photons at the focal plane of our instrument directly, or feed it into another instrument. Thus, the same convolution mechanisms apply to photons which are collected via post focus instruments, as slit (scanning) spectrographs.
Unfortunately, spectrographic data do not have two spatial dimensions. Instead, we have a spatial and a spectral dimension on our sensor. Also, integration time on spectral sensors is usually considerably longer to compensate for the much lower photon flux (compared to the wide band imagers used for MOMFBD). Thus, the available spectral information alone is not sufficient to estimate the effective PSF. 

\begin{figure}
  \centering
  \includegraphics[width=1\textwidth]{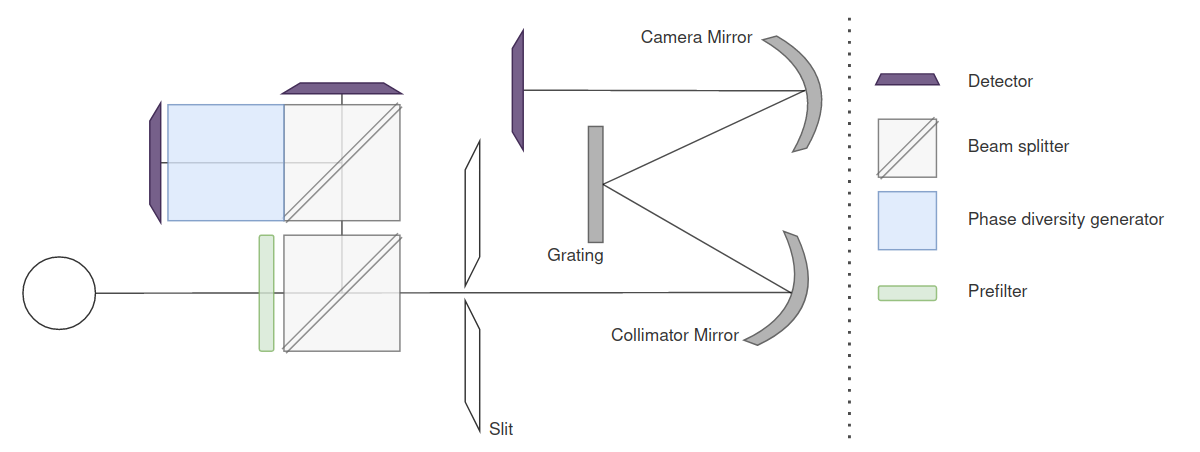}
\caption[Data collection scheme for spectral image reconstruction]{Data collection scheme for spectral image reconstruction with a context imager. All detectors are synchronized, but allow for the context sensors to run at a frame-rate that is a multiple of the spectrograph sensor. The slit area needs to be within the FOV of the context imager.
\citep[Simplified schematic following][]{Noort17}}
\label{fig:spectral_data_collection}
\end{figure}

Following \cite{Noort17}, we extend the data acquisition setup for our spectrograph by a slit-jaw or context camera. The reconstruction of the images from the context camera using MOMFBD provides an estimation of the PSF that we can apply to the spectral images. Figure~\ref{fig:spectral_data_collection} provides a schematic drawing of the data collection in case of a slit spectrograph with a context imager. The slit-jaw case is similar. Instead of a beam splitter in front of the slit, the reflective slit-jaws are tilted and a camera (with or without phase diversity channel) is recording the reflected scene. 

The application of the PSF estimated from the context images to the simultaneously recorded spectra allows the spatial reconstruction of spectroscopic data and will be discussed at the end of  Chapter~\ref{ch:reconstruction} and examples of restored spectrographic data are given in Chapters~\ref{ch:reconstruction}, \ref{ch:fiss_sp}, \ref{ch:many_line}, and \ref{ch:results}.

\section{The problem with signal and noise}
We can divide the signal contained in our recorded frames in two additive parts: The scientifically relevant signal and any other signal. The latter we often refer to as "noise".
Assuming perfectly calibrated frames, where all systematic instrumental degradations have been removed, our images are not pure signal. Each discrete frame that is recorded with photo-electric imaging sensors is inherently affected by a pixel and time dependent uncertainty. This random intensity variation is known as photon or shot-noise.
Each frame we take is therefore "contaminated" with random contributions, that are added to the scientifically relevant signal. 
The emission of a photon is a statistical process that can be described as a Poisson distribution (see Section~\ref{sec:poisson}). 
All camera sensors consist of an array of pixels. The sensor converts light (in terms of collected photons) from each pixel into a digital matrix of integers. The numbers in this matrix are referred to as analogue-to-digital units (ADU, often simply called “counts”). The conversion factor from the number of photo-electrons [\(e^{-}\)] registered by the sensor to ADU is called the gain-factor, or simply gain.
Ignoring any other influences, the noise can be estimated by the square root of the total number of registrered electrons and we can define the signal-to-noise ratio (SNR) as

\begin{equation}\label{eq:SNR}
\textnormal{SNR} 
\; := \; \frac{science\,signal}{\sqrt{total\,signal}} 
\; = \; \frac{science\,signal}{\sqrt{science\,signal + noise}}.
\end{equation}

In the previous sections we have established image degradation by seeing and telescope diffraction, which causes distortion and blurring of small structures by mainly reducing the amplitude of high-frequency components in the Fourier domain. 
The shot noise results in a measurement error in the Fourier phases and amplitudes. 
As there is no non-statistical way to separate the science signal from the random shot noise, image restoration processes must be executed on the total signal.
Due to the linearity of the deconvolution operation (Equation~\ref{eq:convolution}) and the additive nature of science signal and noise (Equation~\ref{eq:SNR}), this operation must amplify noise and science signal alike (Equation~\ref{eq:deconvolution}). In its pure mathematical form it can therefore not alter the SNR.

However, image reconstruction is not simply a mathematical manipulation of random data. Its goal is to reverse systematic degradations introduced by optical effects. For example, an isolated point source appears spread out in the recorded data due to atmospheric seeing and instrumental diffraction. In the process of reconstruction, we aim to gather this distributed signal back into its original location. 
For this we make use of the fact that the true signal is systematically spread across multiple pixels by the optical system, while random noise is distributed, well, randomly. When deconvolving the data we effectively recombine the signal into its correct pixel(s), but the noise from the surrounding pixels is also accumulated. Importantly, because noise is random and not systematically aligned with the signal, its contributions tend to partially cancel each other out when summed. Thus, the contribution of random noise to the final reconstructed signal is less significant than the contribution of the signal itself. 
As a result, depending on the shape and extend of a PSF, the signal-to-noise ratio (SNR) in each re-accumulated pixel improves by up to the square root of the PSFs area. In reality, we need to consider the shape of the PSF, which is never a perfect box. Hence, the improvement in SNR is much less than this upper limit.

In case of image reconstruction of extended objects the situation becomes more complex: Here, we have multiple overlaying contributions from  a continuously extending source. In this case the re-accumulation will always contain signal from several features and this "contamination" effectively cancels the benefits we have found in the case of the isolated point source.
Since deconvolution does not amplify the SNR, image reconstruction highly depends on the quality of the input frames. The more compact the effective PSF is, the more signal stands out of the noise, which results in better starting conditions. This nicely illustrates that it is highly desirable to tune the telescope and instrument optics for maximum transmission and minimal degradation and to use an AO system.
The smaller the spatial scale of a feature, the higher its Fourier frequency. Naturally, the large scale (low-frequency) structures are less affected by a compact PSF. Even though we restore all frequencies, we need to understand solar image reconstruction as an operation explicitly aimed to restore the amplitude of high-frequency components in the Fourier domain, while preserving the SNR.

In Fourier space, diffraction at the edge of a telescope aperture causes a gradual but monotonic reduction in the amplitudes with increasing frequency. At the maximum frequency that the telescope can transmit, the theoretical diffraction limit, all image signal must vanish completely. Increasing the aperture increases the diffraction limit, because the bigger the telescope's aperture the weaker the amplitude reduction across all frequencies. Unlike these optical effects, shot noise maintains a constant power density, unaffected by frequency. This creates a resolution limit where noise amplitudes surpass signal amplitudes. The goal of image reconstruction is to shift this resolution limit towards the diffraction limit. Since the SNR is not altered by the deconvolution of extended objects, the question, how the SNR is ultimately amplified, is non trivial. The two main mechanisms for this are filtering and precise feature averaging.

The idea of the filtering is not to amplify frequencies beyond the diffraction limit. The Rayleigh criterion describes the practical diffraction limit as the radius of the first dark ring in the Airy pattern (see Figure~\ref{fig:diffraction}) of the ideal telescope \citep{AstronOptics}. This can be computed as 
\begin{equation}
    \label{eq:rayleigh}
    \alpha \;=\; 1.22\cdot \frac{\lambda}{D},
\end{equation}
where \(\alpha\) denotes the angular radius, \(\lambda\) the wavelength, and \(D\) the diameter of the aperture. This formula translates the diffraction limit from apertures size to angular resolution, and therefore to the pixel scale of our detectors. 
Using a discrete Fourier transform we can determine the frequency beyond which there can only be noise and we can effectively filter it during the restoration process and prevent this part of the noise to be amplified. 
Both restoration methods discussed in this thesis, MOMFBD and the spectral image restoration, employ such filters (see, \cite{Noort05}, \cite{Noort17}, and Chapter~\ref{ch:reconstruction}).

Accumulating multiple raw frames will only smear out the signal further, as the signal peak is usually located at different pixel positions, due to erratic image movement (the so called tip and tilt components). The easiest way to mitigate this is a global cross-correlation of the raw frames prior to accumulation, since a correlated accumulation will bring the noise down by a factor of up to \(\sqrt{N}\), where \(N\in\NN\) denotes the number of frames accumulated. 
However, especially the seeing induced distortion effects different parts of the image differently and a global correlation will not account for those local effects.
Image restoration includes the local removal of the tip and tilt displacements and also the minimization of local random image distortions, where here local refers to the patch size. The accumulation of restored frames hence truly adds the signal from same sources and avoids further signal smearing. As reconstruction depends on the input SNR, not all frames can be restored to the same quality. In MOMFBD the noisiest frames are added with a very low weight, to suppress the noise even further (see, \cite{Noort05} and Chapter~\ref{ch:reconstruction}).

In conclusion, although deconvolution increases signal and noise alike, the use of filtering and coherent accumulation enhances the SNR also in the case of image reconstruction of extended objects. This becomes fully visible in the case of image restoration of solar spectra (See Figure~\ref{fig:power_spectrum_IU} and Figure 16 in \cite{Noort22}).

Unfortunately, the increased absolute noise level remains problematic for interpretation of image reconstructed spectroscopic data based on Stokes inversion. As discussed, during the Stokes inversion process a synthetic noise-free signal is contrasted with the noise contaminated observation and the metric to deduce the quality of the fit is a \(\chi^2\)-based merit function. Because of the enhanced SNR in the restored data, the detection of true positive signals is improved. As desired, this is especially the case for signals corresponding to small scale features.
But in regions where there is no relevant signal the amplified absolute noise allows to fit a wide range of possible synthetic signals with a similar fit quality (see, Chapter~\ref{ch:many_line}). 
This effect is the strongest in the magnetic parameters, as they depend mostly on Stokes \Q, \U, and \V{}. Since temperature and velocities mainly depend on Stokes \I{}, where there is always relevant signal from the solar line, the effect on those parameters is much weaker \citep{Noort12, Danilovic16}. 
However, due to the coupled influence of the atmospheric parameters on the Stokes profiles (See Section~\ref{sec:background}), all parameters are affected. \cite{Noort12} established that \(10^{-2}\) noise can have a relevant negative effect on the deduced atmospheric parameters.
Currently established practical noise values for unrestored spectro-polarimetric data are ranging from \(10^{-3}\) to \(10^{-5}\) of the continuum intensity \citep[See for instance][]{Noort12, Iglesias16}. After the application of spectral image reconstruction, the absolute noise number of diffraction limited spectro-polarimetric observations typically resides in the \(10^{-2}\) regime \citep[See for instance][and Chapter~\ref{ch:fiss_sp} of this thesis]{Noort22}. 

\section{Goal and structure of this thesis}
We can summarize the above as follows:
\begin{enumerate}
    \item We must resolve the Sun at the smallest scales possible to understand local and global processes in our host star.
    \item Large aperture telescopes are now available. But without image reconstruction we do not exploit their full potential. 
    \item Image reconstruction amplifies signal and SNR but also the absolute noise level.
    \item High absolute noise in spectro-polarimetric data can be problematic for our inversion codes as it degrades an already not particularly well posed problem even further.
\end{enumerate}
Points (1) and (2) demand for image reconstruction of solar spectra while (3) and (4) raise questions on the reliability of the deduced atmospheric configuration. \textit{This is the dilemma to be tackled in this thesis}.
The herein proposed solution is to utilize the combined signal from many spectral lines in a process termed solar many-line inversion \citep{Riethmueller19}. A process that is well known in the analysis of stellar spectral profiles \citep{Donati97} and highly integrated solar spectral profiles \citep[see, for instance,][]{Stenflo83, Solanki84, Solanki85, Allende98, Borrero02}, but new to spatially highly resolved solar spectro-polarimetric observations.

The high resolution spectro-polarimetric observation of the Sun inherently requires multi-dimensional data with  spatial, spectral and time dimensions. Technical constraints such as telescope aperture, detector size and readout time, data rates and storage capacities, as well as available computing power impose practical limits on observations.
In the design of every instrument, compromises between cadence, FOV, and resolution in all three dimensions (spatial and spectral) must be made.
For those reasons, the simultaneous high-resolution observation of only a few (usually one or two) spectral lines is the norm for high spatial resolution spectra. 
Advancements in detector and computer technology have relaxed these constraints considerably and the simultaneous observation of many solar lines is now possible. 
Consequently, \cite{Riethmueller19} studied the merits of the simultaneous interpretation of many lines on a synthetic observation. Their encouraging results not only show a better robustness against noise, and hence a possible way out of this dilemma. On top, they suggest the possibility of a better hight coverage of the turbulent atmosphere. 
Today, multiple instruments that deliver many-line observations are available. The two most noteworthy allow the simultaneous observation of hundredth of spectral lines and are the Sunrise III UV Spectro-polarimeter and Imager \citep[SUSI, ][]{lagg25, Feller25} and the \fiss{} spectro-polarimeter (\fisssp{}, Chapter~\ref{ch:fiss_sp}) and I had the honor to be involved in both of them. 

Usually, the way from a novel idea to something actually usable is long. In this thesis I collect the building blocks to advance the theoretical idea of solar many-line inversion into a new practical technique.  

The first one of those blocks is a new calibration approach. 
In grating spectroscopy the recorded spectral lines appear curved on the detector. This spectrographic curvature effect is an optical distortion, commonly known as line smile \citep{Zhao03}. For the extraction of precise atmospheric properties, most notably the Doppler-velocities, this imprint has to be removed from the data with the highest precision possible. During the process of spectral image restoration the degradation from the PSF is minimized, and contributions to a pixel are shifted along the spatial and temporal dimensions. Hence, this process also heavily depends on the orthogonality of the observed spectra.
This effect is wavelength dependent and due to unavoidable optical imperfections it can vary quite drastically over the spectral FOV. For traditional instruments with only a few lines in a narrow spectral regime of interest, a simple tracing of one line and a polynomial fit is usually enough \citep{Woehl02}. For the interpretation of many-lines from a large spectral range, a new approach was needed. 
In Chapter~\ref{ch:data_reduction} I describe and verify the generic software library "\spectroflat{}" for full field dependent calibration and reduction of spectro-polarimetric data from slit scanning spectrographs. This work was originally published as \cite{Hoelken24}, updates since then can be found in Appendix~\ref{app:spectroflat}.

The second block is the image reconstruction of solar spectra. 
The method to reconstruct an extended 2D FOV from scanned solar spectra utilized here is descried in \cite{Noort17} and \cite{Noort22}. It is based on the MOMFBD image reconstruction algorithm which is described in \cite{Noort05}. Unfortunately, the theorems and proofs for this algorithm are scattered across multiple research papers and textbooks. 
In Chapter~\ref{ch:reconstruction} all concepts and mathematical utilities needed are provided and the relevant theorems and proofs are collected from various publications and brought into a common set of symbols and notations, and therein omitted proofs are given here.

The third block concerns the collection of such diffraction-limited many-line spectro-polarimetric data. 
The design, setup, and data reduction for the spectro-polarimetric extension of the Fast Imaging Solar Spectrograph \citep[\fiss,][]{Chae13}, which is installed at the 1.6-meter clear aperture Goode Solar Telescope \citep[GST,][]{Goode10, cao2010b, Goode12} at the Big Bear Solar Observatory (BBSO) in California, USA, is discussed in Chapter~\ref{ch:fiss_sp}. The extended instrument is referred to as \fisssp{}. This text is already submitted as an original paper to \AnA{} and presents the first diffraction-limited spectro-polarimetric measurements at a large-aperture telescope.

With the data recorded, calibrated, and restored, we can now test the new many-line inversion idea on real solar observations. 
Like \cite{Riethmueller19}, I used the \spinor{} inversion code \citep{Frutiger00phd, Frutiger00} to do so. Unfortunately, just inverting spatially resolved data with all the lines that are somewhat known and can be reproduced in an average quiet-Sun spectrum, did not bring the expected improvement in terms of coherency, physicality and height coverage. 
It was expected that uncertainties in line parameters and blend configurations would average out, but I found that some "bad apples" can spoil the results. Hence, a deliberate selection process of lines and line blends was required. 
When analyzing stellar spectra \cite{Bigot06} and \cite{Heiter21} where confronted with similar challenges. They compiled lists of "well-behaved" lines, but since these lists focus on non-blended lines in stellar spectra they are much too sparse for the solar dataset analysed here. The development and verification of a selection method for solar data is described in Chapter~\ref{ch:many_line}, and the details of the selected lines are provided in Appendix~\ref{sec:atom_dat}. This text is also already submitted as an original paper to \AnA{}. 

In Chapter~\ref{ch:results}, I analyse \fisssp{} scans inverted with this new many-line technique. Here, I describe features of the solar atmosphere near a pore that can now, with the spatial resolution of 0.068" and combining the information from \nlines{} solar absorption lines, be successfully resolved and interpreted. I revisit former high resolution broad band observations and spectro-polarimetric observations with lower resolution. 
The comparison of the inverted observations with a high-resolution \muram{}-simulation allows me to present the first analysis of the 3D fine-structure of small-scale striations and moving bright grains in the vicinity of a pore to describe their
magneto-convective nature.
This text is a paper draft which is targeted for timely submission.

Chapter~\ref{ch:conclusion} offers a summarizing conclusion of the main findings and possibilities to followup and extend on this work. 
Two appendices are noteworthy: In the aforementioned Appendix~\ref{sec:atom_dat}, the atomic line parameters used for this work are presented in a tabular form, Appendix~\ref{app:notation} lists the most important notations and acronyms used throughout this thesis.

All analyses were done in Python. Extensive use was made of 
\texttt{numpy}\footnote{\url{https://numpy.org/}},
\texttt{scipy}\footnote{\url{https://scipy.org/}}, and
\texttt{astropy}\footnote{\url{https://www.astropy.org/}} for the data analysis, 
\texttt{anaio}\footnote{\url{https://pypi.org/project/anaio/}} for accessing rice compressed ANA \texttt{*.f0} files,
and \texttt{matplotlib}\footnote{\url{https://matplotlib.org/}} to create the plots.
\chapter{Data reduction} 
\label{ch:data_reduction}

\textit{The contents of this section correspond to the published article "\textbf{Spectroflat: A generic spectrum and flat-field calibration library for spectro-polarimetric data}" (J. Hölken, \mbox{H. P. Doerr}, A. Feller, and F. A. Iglesias: 2024, \AnA{} 687, A22, \cite{Hoelken24}). 
I wrote, documented, tested, and evaluated the software, ran the inversions, performed the analysis, created all figures, and wrote the majority of the text.}

\subsection*{Abstract}
\begin{footnotesize}
\textit{Context.} Flat-fielding spectro-polarimetric data with one spatial and one spectral dimension is inherently difficult as the imprint of the spectral lines needs to be separated from other wavelength-dependent instrumental effects (e.g., fringes or prefilter profiles) and wavelength-independent effects (e.g., dust and sensor response). Current approaches for spectrometers are often based on moving the grating or they depend on optical models and/or on lab calibration data. They are also limited to small spectral regions and are instrument-specific. Approaches that would be suitable for polarimeters have not been reported yet.

\noindent
\textit{Aims.} We present an approach that allows for flat-field calibration data to be to obtained for diffraction-grating-based, long-slit spectrographs combined with temporally modulated polarimetry from high-resolution solar telescopes. This approach is based on nominal flat-fielding procedures performed during the instrument's science operations.

\noindent
\textit{Method.} We performed a precise and field-dependent correction of the spectrographic distortion effect (resulting in curved spectral lines, typically denoted as a "smile" effect) to ensure the orthogonality of the spectral and spatial dimensions. We identified distortions by tracking the position of multiple spectral lines within the full spectral field of view. From the raw modulated flats, we then removed the solar line imprints and derived separate flat-fields for sensor and slit dust features. Optionally, wavelength calibration and continuum correction can be included in this process.

\noindent
\textit{Results.} We have created generic Python libraries that can be plugged into existing Python-based data reduction pipelines or used as a standalone calibration tool. We show that for spectrographs covering many spectral lines, a correction of the smile distortion based on optical models alone is not sufficient. Our results demonstrate a suppression of fringes, sensor artifacts, and fixed-pattern imprints in demodulated data by one order of magnitude. For intensity images, the photon noise level can be closely attained after calibration. Our correction works across the full spectral range. The algorithm was tested for different wavelength regimes with emission (EUV range) or absorption (near-UV, VIS, IR range) spectra, on data acquired with ground-based (SST/TRIPPLE-SP, GREGOR/GRIS), balloon-borne (SUNRISE-III/SUSI), and space-based (SolO/SPICE) instruments. The data calibrated with our method offer robust and precise inversion results.

\noindent
\textit{Conclusion.} We have extended existing spectroscopic flat-field techniques to modern instruments with large imaging sensors covering many spectral lines simultaneously, and with polarimetric capabilities, where methods described so far are not adequate. We believe that our method is applicable as a standard calibration approach for most modern high resolution large-FOV, long-slit spectrographs -- both with and without polarimetric capabilities.

\end{footnotesize}

\section{Introduction}
\label{sec:dr:introduction}
The flat-field calibration process characterizes and corrects images for the non-homo- geneous response of imaging sensors, such as those based on charge-coupled devices (CCD) and complementary metal-oxide-semiconductor (CMOS) detectors, across their light-sensitive area. The flat-field also typically includes the contribution of the optical components in the beam path, from the main telescope mirror to the imaging detector. In order to obtain images closer to the real object, all recorded data must be corrected for the effects accounted for by the flat-field (e.g., dust specs, vignetting, fringes, etc). The quality of this calibration process strongly influences the scientific utility of the observational data. Flat-field calibration is thus compulsory for any kind of astronomical observation and it affects ground-, balloon- and space-based observations alike. This paper considers the specific difficulties inherent in flat-fielding solar spectrographs (producing images with one spectral and one spatial dimension) combined with polarimetry.
As the accumulation of dust particles is a recurrent process and the stability of other flat-field features depends strongly on the structural stability and configuration of the telescope and post-focus instrument, flat-field data are to be obtained close in time before and/or after the observation. In practice, multiple flat-fields are recorded during an observation day \citep[See e.g.,][]{Chae13, Kotrc16, Wang19, Song22}.

Lab flats, which are generated using controlled setups based on artificial light sources, typically cannot be obtained close in time with respect to the scientific data acquisition. They also tend not to represent the solar observations with sufficient accuracy (e.g., in terms of beam geometry, light level, or temperature distribution inside the instrument). Indeed, the calibration source that is closest to the scientific observing conditions is the Sun itself. As the Sun is neither spectrally nor spatially uniform, we need methods to remove solar imprints in both dimensions from the calibration data.

Typically, an imaging flat-field is retrieved from an extended, non-uniform source by shifting the image on the sensor and using the mean of multiple such displaced images \citep{Tyson86}. Other methods, such as  proposed by \cite{KLL91}, derive the flat-field from a small series of displaced images via a least-squares estimate. Most commonly, the spatial uniformity is enforced by moving the telescope in a random-walk manner or following a (rotating) ellipsis or Lissajous-pattern in a quiet Sun region near disk center, while constantly collecting and averaging frames \citep{Denker99}. 
For spectrographic observations this approach alone is not sufficient, since the solar line features will not be displaced enough on the spectrograph sensor by pointing the telescope to a slightly different location; at the same time, the telluric lines will not be displaced at all. Although averaging spectral images (collected along a path around the solar disk center) does produce uniform quiet-Sun-like line profiles, they still need to be removed in order to obtain the desired flat calibration data.

Current flat-fielding procedures for diffraction-grating-based, long-slit\footnote{By long-slit we refer to a slit in the typical field of view at a high resolution solar telescope, which is up to 70 arcseconds at the time of writing.} spectrographs often include moving the grating (either to a nearby continuum section~or constantly back and forth to displace line imprints), defocussing the image, and/or ignoring vignetting and other spectral features alltogether \citep[Examples in][]{Keller03, Chae13, Wueser18}.
\cite{Jones03J} and \cite{Wang19} pointed out that for spectroscopic observations, any change to the instrument configuration modifies the imaging of optical impurities by either shifting them or changing their imprints amplitude and/or shape on the sensor. Furthermore, \cite{Jones03J} demonstrated that such configuration changes modify the fringe patterns as well.  

Spectroscopic data are inherently affected by an optical distortion, which causes the spectral lines to appear bent on the detector, commonly known as line smile \citep{Zhao03}. The smile distortion has to be corrected before the desired flat-field information can be successfully extracted. Typical spectrographic requirements for a maximum residual smile distortion range between 10 \% and 20 \% of a pixel size \citep{Kaiser08, Reimers15}. 
The smile correction is currently done in various ways. One method estimates the expected curvature function using an optical model and use it to correct the data, as described in \cite{Qiu22}, for instance. Another method adjusts the optical design to meet the aforementioned requirements on residual smile distortion without any postcorrection \citep{Wei16}. However, this is not always possible or desirable, as the additional lenses or mirrors required may lead to higher instrumental complexity and losses in efficiency and/or imaging quality. Techniques to determine the smile curvature from flat-field measurements exist; for instance fitting a low-order polynomial to the detected line core positions of a single line \citep[See for instance][]{Bommier02, Woehl02, Chae13}. 
\cite{Cai17} first presented an approach to extract the smile characteristic from multiple lines. They used the median of all telluric lines in the wavelength window to correct the full frame. This makes their approach more precise than previous methods with a single line smile characteristic, but it does not allow for a differential smile correction of the full field of view (FOV). Further, several scientifically interesting spectral regions without any telluric lines exist. Data from space based or balloon borne instruments do also not show telluric lines for obvious reasons.
Moreover, the spectrographic curvature effect can vary quite drastically for large spectral FOVs (see e.g., Section~\ref{sec:desmiling}), limiting the applicability of these smile correction methods to instruments that cover only narrow spectral regions of a few \AA{}ngstroms.

In the case where the mean images can be considered spatially uniform and "free from smile", other approaches for removing spectral lines and producing a flat-field exist: \cite{Jones03J} removed the line imprints in the image by dividing the 1D spatial information for each spectral point by a least-squares fit to this spectral point along the spatial dimension (parallel to the slit). A more common approach is the  mean profile method \citep[see][]{Bommier02, Woehl02}, where every spectral profile of the frame (i.e., a cut perpendicular to the spatial dimension for a given spatial resolution element) is divided by the spatially averaged spectral profile.

In the case of spectropolarimetric data calibration, the correction of both spatial flat-field effects and smile becomes even more challenging. This is because high-precision polarimetry requires the detection of small differential signals and, thus, differential distortion effects between modulation states may have a significant influence on the resulting Stokes parameters. For high-precision polarimetry with accuracy requirements of 0.1\% or less in the measured degree of polarization, the requirements for the maximum differential smile distortion after correction is at the level of a few percent of a pixel only (comparable to effects and scales of seeing introduced cross-talk; e.g., \cite{Judge04} and \cite{Iglesias16}). Also, inversion codes are sensitive to flat-fielding effects and especially fixed pattern noise after demodulation \citep[see, e.g.,][]{Rubio03, Longcope09} therefore, they are heavily dependdentt on properly calibrated data. Until now, to the best of our knowledge, no detailed procedure for spectropolarimetric data has been published so far.

With the availability of modern large-size detectors, instruments that allow the observation of wide spectral windows - covering several nanometers and therefore registering tens to hundreds of spectral lines simultaneously - have become more relevant and widespread. \cite{Riethmueller19} investigated the spectropolarimetric Stokes inversions used in remote sensing of the Solar atmosphere. They compare the performance of the traditional approach of inverting only a few spectral lines registered in multiple wavelength regimes to the joint inversion of many lines covering a wide spectral range and observed simultaneously. They found more accurate inversion results and an extended height coverage in the "many lines" case. 
On the other hand, the successful application of numerical methods used to restore images subject to variable instrumental or ambient optical aberrations, such as the spectral restoration technique developed by \cite{Noort17}, also require an accurate smile correction over large (spectral) FOVs.

In this work, we introduce a novel algorithm, \spectroflat{}, and we provide a publicly available generic library for the high-precision calibration of spectro-polarimetic observation data. Our technique accounts for detection and correction of the smile distortion over larger spectral FOVs and generates flat-field images in an automated manner from calibration data captured close in time to the observation and in the operational instrument configuration. 
The rest of the paper is organized as follows. 
Section~2 explains the method in depth, including notation, input data requirements, smile correction, and the extraction of multiple flat-field images. It also summarizes the optimal data reduction approach with calibration data obtained through \spectroflat{}.
Section~3 validates the algorithm using solar data from four instruments including the infrared (IR), visible (VIS), and near- (NUV) and extreme-ultra violet (EUV) spectral regimes. These instruments are: (1) the spectro-polarimetric extension of the TRI-Port Polarimetric Echelle-Littrow (TRIPPEL) spectrograph \citep{Kiselman11} attached to the Swedish Solar Telescope \mbox{(SST/TRIPPLE-SP)}\footnote{Unfortunately, this prototype is not yet fully described. Some details of the extension are given in \cite{Saranathan21}, Section~6.}; (2) the GREGOR Infrared Spectrograph (GREGOR/GRIS) \citep{Collados12} installed at the Observatorio del Teide, Tenerife; (3) the Sunrise Ultraviolet Spectropolarimeter and Imager (SUNRISE-III/SUSI) \citep{Feller20}. We note that an overview of the SUNRISE I+II missions can be found in \cite{Solanki10, Solanki17}, while a description of the \mbox{SUNRISE-III} mission is in preparation\footnote{Author's note: This article was published in 2024, prior the second flight of SUNRISE-III. In the meantime the SUNRISE-III mission and instrumentation overview is published as \cite{lagg25}.}; (4) the Spectral Imaging of the Coronal Environment instrument on board of Solar Orbiter (SolO/SPICE) \citep{SPICE}. 
In Section~4 we summarize our main conclusions.

\newpage
\section{Method}
 The input data is assumed to be previously calibrated for the camera zero-input response (dark current and bias level) and non-linearity. Furthermore, we assumed the individual frames to be already sorted with regard to their modulation state (if any) and combined into mean images for each modulation state.
 The targeted flatness, for instance, in terms of the root-mean-square (RMS) variation of the mean flat-field frame, largely depends on the science case and targeted sensitivity. 
Let us define the spectro-polarimetric input data cube as having dimensions of $s\times y \times \lambda$, where the symbols denote the modulation state, spatial location along the spectrograph slit, and wavelength, respectively. We refer to the individual mean modulation state data ($y \times \lambda$) in such a cube as state images, the following method is applicable to every state image independently. 
The result of this approach are flat-field matrices for each modulation state. We denoted this set of flat-fields as a "modulated flat". Intensity-only spectrographic data is treated as data with only one modulation state, i.e. a cube with $s$-dimension equal to one. 
In addition, we assume the spectral dimension is extending along to the imaging sensor rows (\(x\)-axis) and the spatial dimension extends along the sensor columns (\(y\)-axis). Therefore, we refer to individual spectra simply as rows. Since transposing a matrix is a numerically conservative operation, this requirement is purely formal.

The image rotation with respect to the sensors \(x\)-axis is best measured with dedicated data from a suitable calibration target. If no distortion effects are apparent the measured rotation can be given as a configuration value, and will then be corrected by the library. Higher level distortions (e.g., keystone) are out of the scope of the current implementation and need to be corrected before feeding the data to \spectroflat{}. The deviation along the \(y\)-axis is removed by the smile correction. We call the shift of a pixel in the spectral direction with respect to its reference position (i.e., its deviation from a parallel line to the \(y\)-axis), its offset.

Overall, \spectroflat{} is capable of correcting the full FOV of the input or only a region of interest (ROI). As the functionality is the same in both cases, we can assume and showcase full FOV correction. The implementation makes heavy use of concurrent processing. Computing time depends on the number of cores and amount of RAM available but is typically in the order of minutes.
\spectroflat{} follows the basic idea of estimating separate flat-fields for different fixed pattern imprints sources, for instance, the sensor, slit, and illumination field, as presented by \cite{Woehl02}. However, our smile detection is done for the full FOV and we extended the idea to also extract the flat-fields for each modulation state individually. Further, the flat-fields themselves are extracted using new algorithms.

\subsection{Preparation}
\label{sec:preflat}

\begin{figure}
    \centering
  \includegraphics[width=1\textwidth]{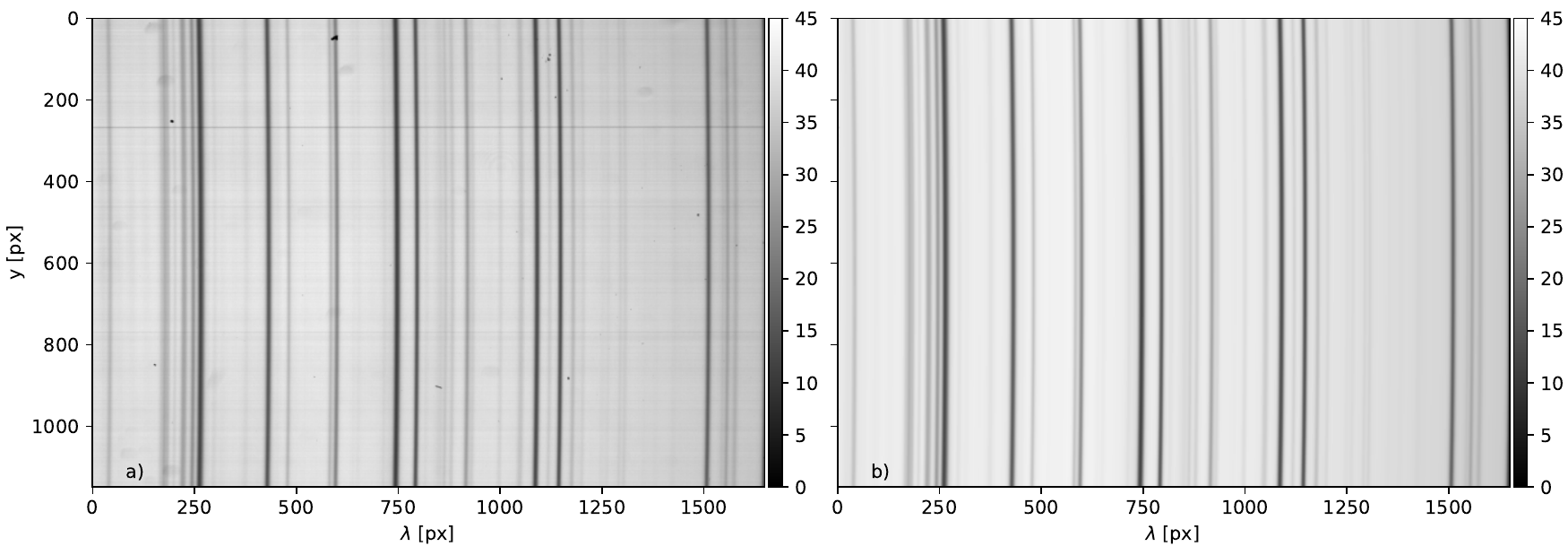}
  \caption[\spectroflat{} pre-flat correction]{
  TRIPPEL-SP input state image of the first modulation state is shown in panel (a). 
  Panel (b) shows the corresponding ad-hoc corrected image. Both color bars denote relative intensities in ADUs.}
  \label{fig:pre-flat}
\end{figure}

The first step is an adhoc removal of the dust imprints and any other spatial variation from the state images to increase the quality of the smile correction. The corrected image is obtained by fitting a high order (default  13) polynomial to the pixel values of each image column \citep[compare][]{Jones03J}. 
If the sensor presents a column to column variation, for instance, due to conversion gain, the algorithm can be configured to remove this from the corrected image. In this case, the average column to column variation of all columns is used to adjust the gains in all columns multiplicatively. If such a component exists, it has to be taken out in this step since (a) the stripy pattern might interfere with the line core detection of the smile detection (see below) and (b) after smile correction and rotation this pattern will not be parallel to the image \(y\)-axis anymore and removal would become unnecessarily complex.
An example of the results of this procedure is shown on the right hand side of Figure~\ref{fig:pre-flat}. The example flat-field data is taken from the 2017 TRIPPEL-SP campaign on the Swedish Solar Telescope (SST).

\subsection{Desmiling} \label{sec:desmiling}
The smile detection algorithm is executed on the corrected state images obtained in the preparation step. Where dust imprints and column to column sensor effects are temporarily removed. 
As correction reference, the mean spectrum of the central five rows is used to further reduce the influence of noise and row to row variations. From this 1D reference spectrum (\(R\)), a set of spectral lines is selected automatically based on configuration values (relative line depth, distance between line centers).
For each selected line, a Lorentzian is fitted to detect the line center with sub-pixel resolution. For each row, the same set of lines is selected and their corresponding sub-pixel center positions are detected using the same fit routine.

The relative displacement between line centers of a given row (\(C\)) and the reference spectrum (\(R\)), is determined by fitting several polynomials (\(p\), of degrees two to a configurable maximum) to the detected differences in the line center positions. For each fit polynomial, the new position of each pixel is obtained using a cubic spline (CS) interpolation, resulting in a smile-corrected row spectrum (\(\Tilde{C}_p\)). In a brute force approach the polynomials and corrected spectra are computed for each polynomial-degree in the configured range. The polynomial that minimizes the error function,
\begin{equation}\label{eq:smile_error}
E(p) \;:=\; \sum_{x} \EC(p, x) \;:=\;  \sum_{x} \frac{\Big(\Tilde{C}_p(x) - R(x)\Big)^2}{\Tilde{C}_p(x)^2},
\end{equation}
is selected. The offsets of the individual pixels in spectral direction are computed using the selected polynomial. This is done for every image row individually.

\begin{figure*}
    \centering
  \includegraphics[width=1\textwidth]{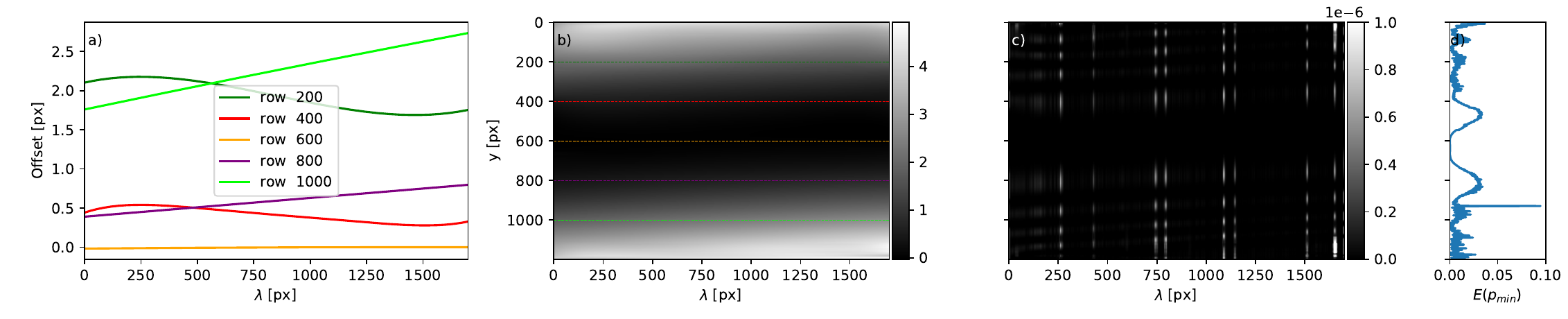}
  \caption[\spectroflat{} offset map and residual error]{
  Offset map profiles shown in panel (a) for the colored rows indicated in panel (b).  
  Map of detected offsets in panel (b). Color indicates the amount of shifting needed in units of pixels for each pixel along the x-axis to remove the smile distortion. 
  Panel (c) shows a map of the pixel-wise error \(\EC(p_{min},x)\), and panel (d) the result of the error function as of Equation~\ref{eq:smile_error} for the selected polynomials.}
  \label{fig:offset-map}
\end{figure*}

\begin{figure}
\begin{minipage}[b]{0.55\textwidth}
  \includegraphics[width=1\textwidth]{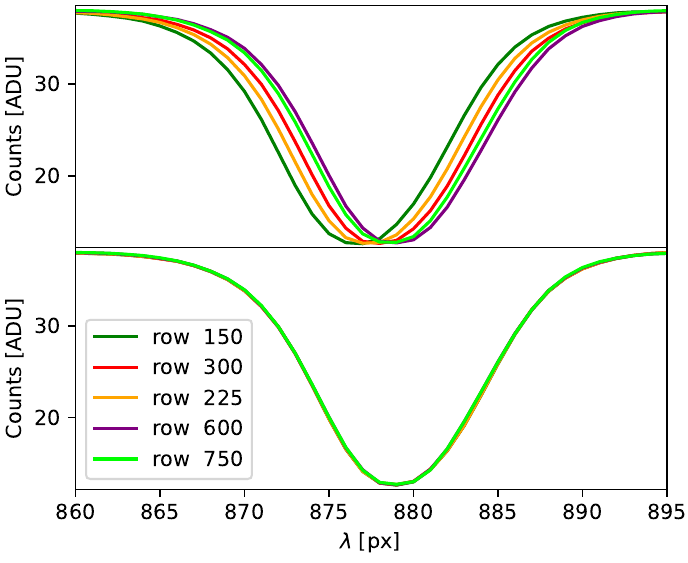}  
\end{minipage} \quad 
\begin{minipage}[b]{0.42\textwidth}
  \caption[Smile correction example]{
  Intensity profile of different rows around the 5250.6 \AA{} spectral line. The prepared input data is shown in the upper panel, while the lower shows the intensity profiles after correcting the smile distortion.}
  \label{fig:offset-correction}
\end{minipage}
\end{figure}

After the offsets of all rows have been detected, an offset and an error map are generated. After all modulation states have been processed, the default behavior is to average the individual offset maps and use the mean map to correct all modulation states, in order to avoid differential polarization artifacts.\footnote{
It is possible to extract an offset map per modulation state. As none of our datasets suffer from modulation state-dependent smile effects introduced by the modulation package, we have not been able to test this mode rigorously until now. Simple simulations show that it might be possible to mitigate differential artifacts in some cases, but testing is incomplete at time of writing. We therefore mark this mode as experimental.} To correct the smile distortion of any applicable frame, each row is CS-interpolated according to the detected offsets.
Figure~\ref{fig:offset-map} shows the global offset and error maps corresponding to Figure~\ref{fig:pre-flat}. The cuts in panel (a) demonstrate that a global smile curvature determined from a model or a single line is not enough for instruments targeting multiple spectral lines. Panel (c) shows the pixel-wise error values \(\EC(p_{min}, x)\) for the selected polynomial. Figure~\ref{fig:offset-correction} shows spectral profiles of the image around the \ion{Fe}{I}, 5250.6 \AA{} line. In the lower panel, the line centers are shifted and differ by less than 10 \% of a pixel.

\newpage
 For all the instruments analyzed in this work, unless the instrument configuration does not change, the resulting smile distortion pattern is stable over time. It can therefore be applied to frames that are already temporally averaged, for instance, in the slit scanning direction. Not correcting the individual frames saves computation time during data reduction.

\subsection{Sensor and slit flats}
For preparation (see Section~\ref{sec:preflat}), we used an ad-hoc correction to remove dust from the image in order to improve the smile detection. However, this is an imprecise method to extract the dust imprints from the flat images. 
To create a precise flat-field matrix, a 1D spectrum can be generated by averaging all rows of the smile-corrected image. 
This 1D spectrum is then repeated in spatial dimension to create another artificial image. This artificial image is then "re-smiled", namely, shifted according to the offset map to re-introduce smile. The original input image is divided by it to retrieve what we call a dust-flat. 
Optionally, this process can be repeated multiple times. A good metric to determine how many iterations are needed is to stop when the difference between consecutive dust-flats is small. We have achieved relative differences with a standard deviation in the \(10^{-4}\)-range after three to four iterations.

\begin{figure*}
    \centering
  \includegraphics[width=1\textwidth]{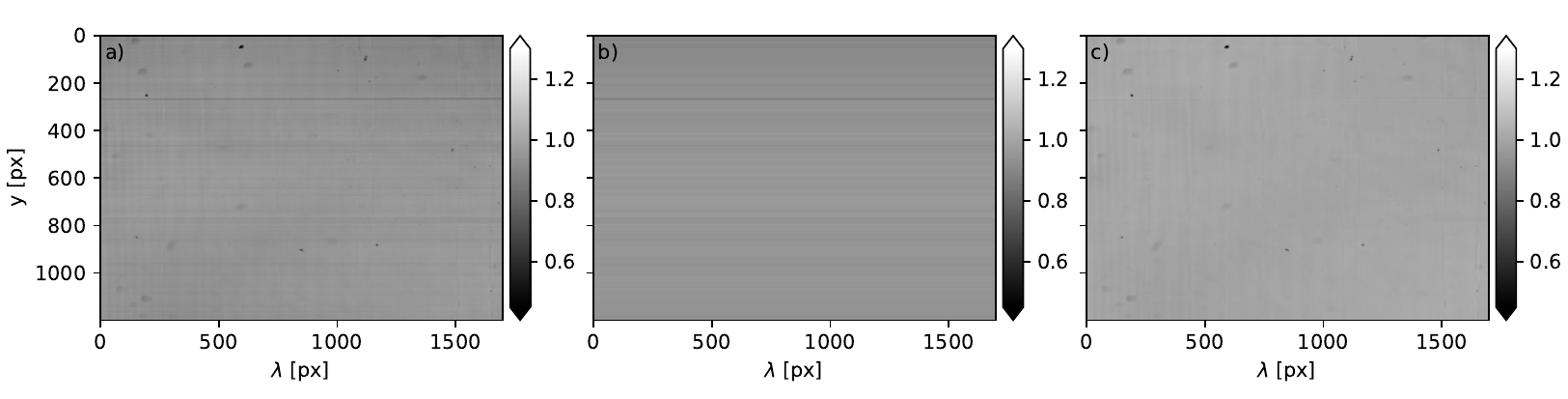}
  \caption[\spectroflat{} dust, slit and sensor flat examples]{
  TRIPPLE-SP Dust-flat (a),
  separated slit flat with the horizontal slit features (b) and
  separated sensor flat (c). All color bars show normalized values. Range in all plots adjusted to increase visibility of features.}
  \label{fig:slitsensor-flat}
\end{figure*}

A slit flat is generated from the dust-flat. For this, all columns are averaged to a 1D spatial representation, removing all features that vary in the spectral dimension. This column is again repeated to the full image dimensions to create the slit flat. Subtracting the slit-flat from the dust-flat yields the sensor flat. Figure~\ref{fig:slitsensor-flat} shows the final dust-flat and the separated sensor and slit flats of the example in Figure~\ref{fig:pre-flat}. 
The separation of the slit flat from the sensor flat allows us to track and correct "drifting" slit features (e.g., due to changes in instrument temperature and/or mechanical configuration) independently of the dust imprints on the sensor in the science data. Furthermore, it can work vice versa to remove slit features independent of  changing features in the sensor flat, such as moving dust grains.
Since the slit and sensor dust flat(s) are extracted from the input flat images without smile correction they can also be applied to science frames without the need of desmiling and interpolation.

\subsection{Illumination field}
The illumination field represents the smooth low order intensity variation. With \texttt{spectro- flat}, this information is mainly included in the dust flat. Accordingly, the variation of the final \spectroflat{} illumination fields are always in the \(10^{-6}\) range. Therefore, we did not apply these. However, we kept this step for compatibility with the \cite{Woehl02} approach, and for border cases where it might be needed.

\begin{figure*}
    \centering
  \includegraphics[width=1\textwidth]{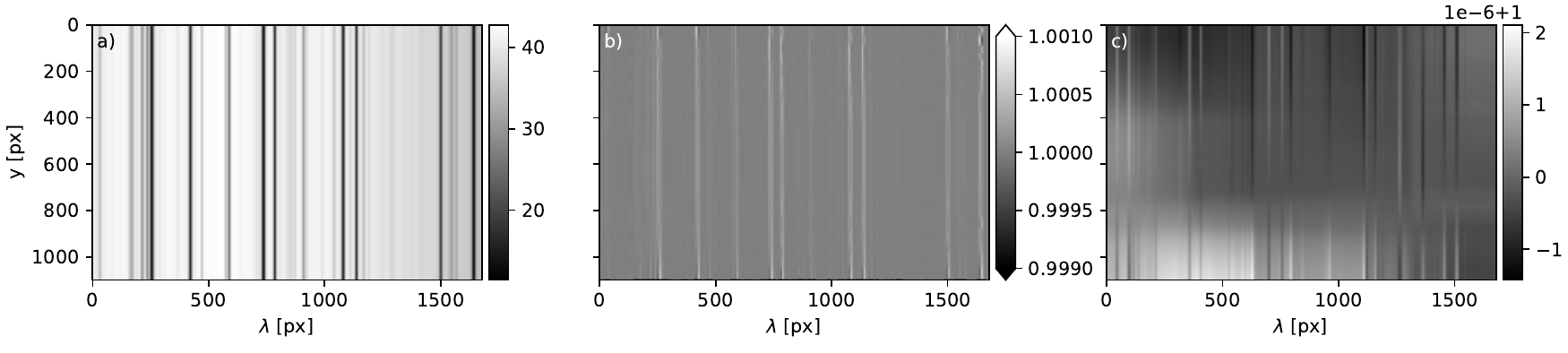}
  \caption[\spectroflat{} illumination field extraction example]{
  Steps of the illumination field extraction: a) Desmiled input image, 
  b) illumination field, i.e., the desmiled input image after removing the spectral lines, showing vertical line residuals
  and c) clean illumination field after reducing the line residuals. Note: panels b) and c) have different color scales to improve the visibility of the residuals.}
  \label{fig:illumination-pattern}
\end{figure*}

From the dust-flat-corrected and desmiled input image, we then removed the line imprints using the standard mean profile method.
We found it common to be left with some undesired, sharp line residuals due to spatial gradients within the lines. Similarly to \cite{Woehl02}, we removed them by smoothing. This was done by first linearly interpolating over these residuals in spectral direction and then applying a Gaussian filter to the full frame. The result is an illumination field on the sensor. Figure~\ref{fig:illumination-pattern} shows the corrected input image corresponding to Figure~\ref{fig:pre-flat}, before and after line removal, and after residuals cleaning. 
The time stability of the illumination field is comparable to the stability of the smile-distortion pattern, thus, it can also be applied to science frames that have already been temporally averaged. 

\subsection{Wavelength calibration \& continuum correction}
As an optional step, the detected smile offsets can be amended by a wavelength calibration. The desmiling and the wavelength calibration of the science data can then be performed in a single interpolation step. 

The input data can be calibrated using several solar atlases. The Hamburg Fourier transform spectrograph (FTS) atlas \citep[see][]{Neckel99} and the the Second Solar Spectrum (SS2) atlas of \cite{stenflo14} are currently supported. Other atlases can be provided as a plug-in to the \texttt{atlas-fit} routine, for instance, to support other spectral regimes.
First, a spectral re-sampling and convolution with a Gaussian of the atlas is carried out to match the input data sampling and resolution.
Then, the sub-pixel spectral difference between the line centers in the data and the atlas is fitted using the same procedure described in Section~\ref{sec:desmiling}. The resulting additional offset information is added to the previously generated smile offset map.

\begin{figure*}
    \centering
  \includegraphics[width=1\textwidth]{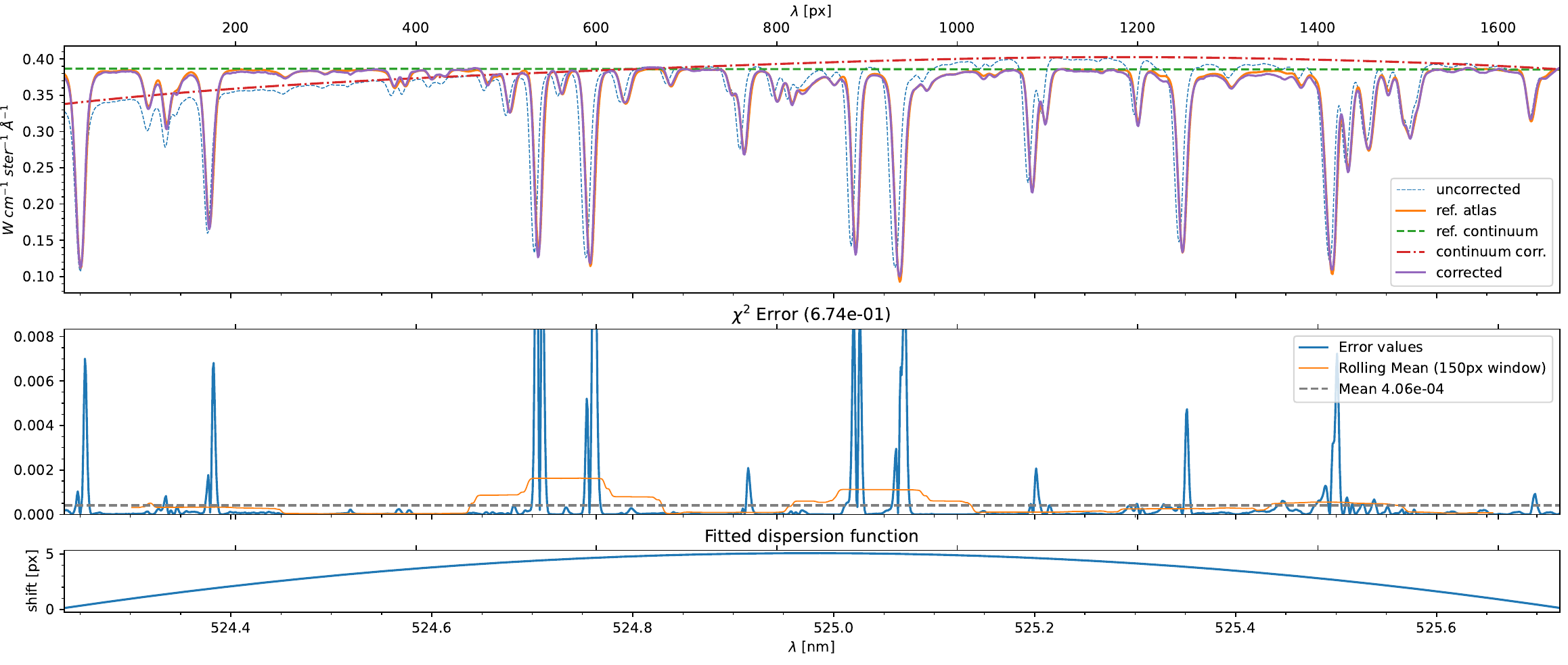}
  \caption[\texttt{atlas-fit} wavelength calibration]{
  Spectral and continuum calibration using the FTS quiet-Sun, disk-center atlas from \cite{Neckel99}.
  The top panel shows the uncorrected and corrected spectrum, the used continuum correction curve, the atlas reference continuum level and the atlas (see legend). The middel panel presents the difference between the atlas and the calibrated data in units of the above panel. The lowest panel shows the fitted dispersion offset in pixels.}
  \label{fig:wl-calibration}
\end{figure*}

\newpage
If the selected atlas provides information on the continuum intensity, the continuum calibration of the data is done as follows (see Figure~\ref{fig:wl-calibration}):
First, the difference between the maximum intensity and the continuum value specified in the reference atlas is estimated. For this, the atlas spectrum is divided into a configurable number of equidistant sections and the maximum point in every section is selected. Next, a high-order polynomial (configurable, default: 14, should never exceed the number of sections) is fitted to the selected maximum points. Outliers of the maximum points are filtered out if they differ more than 20\,\% of the fitted polynomial. Then, we fit a fifth-order spline through the remaining maximum points. Finally, the difference between the spline interpolation and the reference continuum is computed and used as reference spectral continuum variation. 

In a second step, for each individual row in the input data spectrum, the same approach specified above is used to find a spline interpolation of the measured continuum intensity profile. The computed reference continuum variation is added to the solution found for the input data spectrum. The correction of the solution with the reference continuum variation from the atlas prevents the overcorrecting of strong and blended lines. 
The resulting continuum correction solution is a 2D map and can be used in addition to or instead of the \spectroflat{} illumination field. As for other flat-fields, the correction is multiplicative and normalized to one.

To allow for better fits of the wavelength solution and the continuum correction, the algorithm performs a least-squares fit of the effective full width half maximum (FWHM) and the gray stray light component of the Atlas to the data. Lines of known variable depth (i.e., telluric lines) can be masked out for this purpose. As a first order approximation the spectrographic PSF can be assumed to be a global Gaussian. Hence, we convoluted the Atlas with a Gaussian function with FWHM equal to the differential FWHM between instrument and atlas. The atlas FWHM is assumed to be known, while the data FWHM is a free parameter. The other free parameter in the fit is the intensity offset introduced by a stray-light component \citep{MartinezPillet92}. Figure~\ref{fig:wl-calibration} shows an example calibration result for the TRIPPLE-SP flat-field dataset.

\subsection{Data reduction process}
\label{sec:steps}
Once all calibration data is obtained using the \spectroflat{} algorithm the reduction of a batch of science frames \(\{F_t\}_{t\in T}\) is usually carried out according to the method described below.
To describe the integration into a full calibration workflow, we also include steps with calibration data obtained via other methods that are not within the scope of this work. For instance, non-linearity (NL) correction, tracking moving dust parts on the slit and sensor, and obtaining demodulation matrices from polarimetric calibration.

\begin{enumerate}
    \item Apply NL and zero-point correction (e.g. dark field) to each individual frame:
    \begin{equation}F_t'\;:=\; \DF(F_t)\end{equation}
    \item Apply \spectroflat{} dust flats. Identify \(t\) with a modulation state, and select the corresponding modulated flat for each frame. If needed shift slit and sensor flat for each frame such that it matches the dust patterns best:
    \begin{equation}F_t'' \;:=\; \FF_{dust}(F_t') \;=\; \frac{F_t'}{\FF_{slit} \,\cdot\, \FF_{sensor}}\end{equation}
    \item Sort data by modulation state $s$ and bin \(N\) frames per state in temporal (i.e., scanning) direction:
    \begin{equation}F_{s,\Delta t} \::=\: \frac{1}{N} \sum_{s(t)} F_t''\end{equation}
    \item Apply \spectroflat{}  smile distortion correction to averaged frames:
    \begin{equation}F_{s,\Delta t}^{(1)} \;:=\; \textnormal{desmile}(F_{s,\Delta t})\end{equation}
    \item Optionally, apply the \spectroflat{}  illumination field:
    \begin{equation}F_{s,\Delta t}^{(2)} \;:=\; \FF_{illum}(F_{s,\Delta t}^{(1)}) \;=\; \frac{F_{s,\Delta t}^{(1)}}{\FF_{illum}}\end{equation}
    \item Optionally, apply the \texttt{atlas-fit} continuum correction matrix:
    \begin{equation}F_{s,\Delta t}^{(3)} \;:=\; \FF_{cont}(F_{s,\Delta t}^{(i)}) \;=\; \frac{F_{s,\Delta t}^{(i)}}{\FF_{cont}}, \qquad i \in \{1, 2\}\end{equation}
    \item Optionally, apply the spectral image reconstruction \citep[see][]{Noort17}:
    \begin{equation}F_{s,\Delta t}^{(4)} \;:=\; \textnormal{Specrestore}(F_{s,\Delta t}^{(i)}), \qquad i \in \{1, 2, 3\}\end{equation}
    \item For the polarimeters: demodulate using the demodulation matrix \(M_{demod}\), to obtain Stokes images $S := \{I,Q,U,V\}$:
    \begin{equation}S^{(i)} \;:=\; \D(F^{(i)}) \;:=\; M_{demod} \,\cdot\, F^{(i)}, \qquad i \in \{1, 2, 3, 4\}\end{equation}
    where \(F := \{F_{s,\Delta t}\}_{s\in\{1,...,n\}}\) and \(n\) is the number of modulation states.
\end{enumerate}
If the illumination field and continuum correction are both applied, they can be combined into a single "soft flat" matrix which will reduce steps (5) and (6) to a single operation. Since multiplication is numerically less expensive than division, it is advised to save all flat-field matrices as gain tables, namely, as \(\frac{1}{\FF_*}\). 

\section{Results and discussion}
The \spectroflat{} algorithm yields flats for sensor and slit dust imprints for each polarimetric modulation state. It has been tested with data from several long-slit spectrographs combined with temporally modulated polarimetry installed at high resolution solar telescopes. 
It further characterizes the smile distortion pattern precisely and provides smile correction methods for science frames. It is able to do spectral and continuum level calibrations by comparing with a reference atlas.

We applied our method to several datasets obtained with the facility instrument GREGOR/GRIS, the SST/TRIPPEL-SP prototype instrument, with \mbox{SUNRISE-III/SUSI} during pre-flight validation tests, and with SolO/SPICE. 
The wavelength regimes cover infrared (IR), visible (VIS), near-ultraviolet (UV) and extreme ultraviolet (EUV) datasets. 
The IR, VIS, and UV datasets feature absorption lines, while the EUV presents an emission spectrum. 
The \spectroflat{} algorithm and library are suitable for instruments using CCD and CMOS sensors alike. 
The test datasets span cases from pure spectrographic (no polarimetric modulation) observations to observations with up to 12 states per polarization modulation cycle. 

\begin{figure}
\begin{minipage}[t]{.48\textwidth}
  \includegraphics[width=1\textwidth]{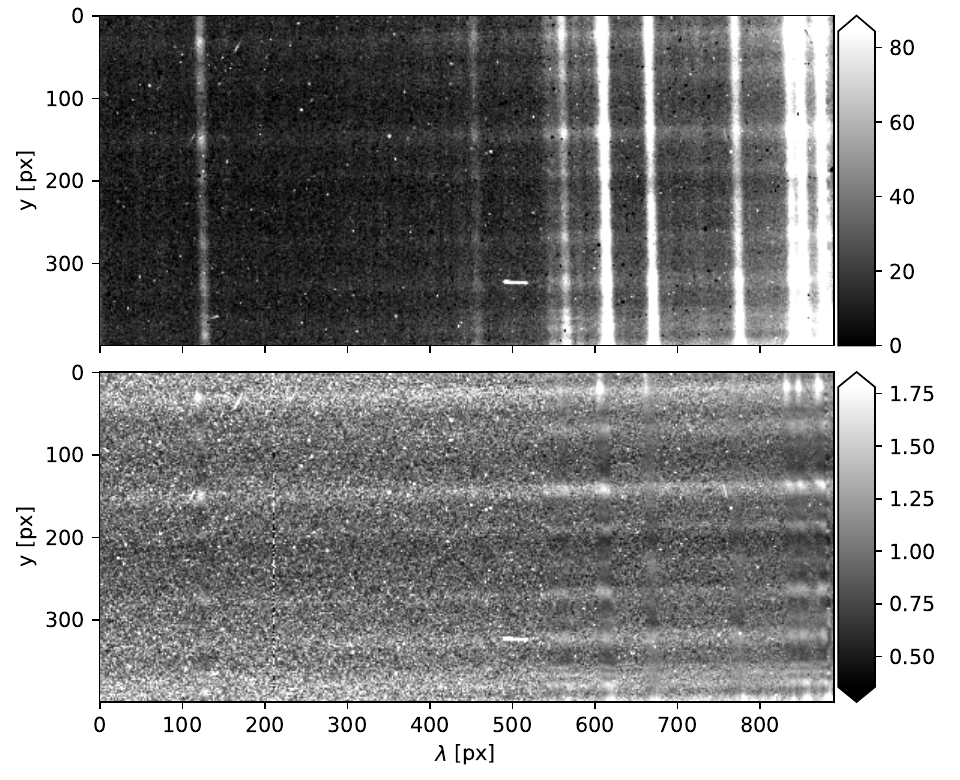}
  \caption[SPICE L1 dataset and extracted flat field]{
  SolO/SPICE L1 test dataset created from multiple quiet Sun science observations in the upper panel and extracted pseudo-flat in the lower one. The \spectroflat{} algorithm cannot distinguish between actual flat-field artifacts and residual solar structures. Thus, these residual horizontal structures are regarded as flat-field effects. The vertical features in the flat are not line residuals but traces of the burn-in effect on the sensor.
  }
  \label{fig:spice-l1}
\end{minipage} \quad
\begin{minipage}[t]{.48\textwidth}
  \includegraphics[width=1\textwidth]{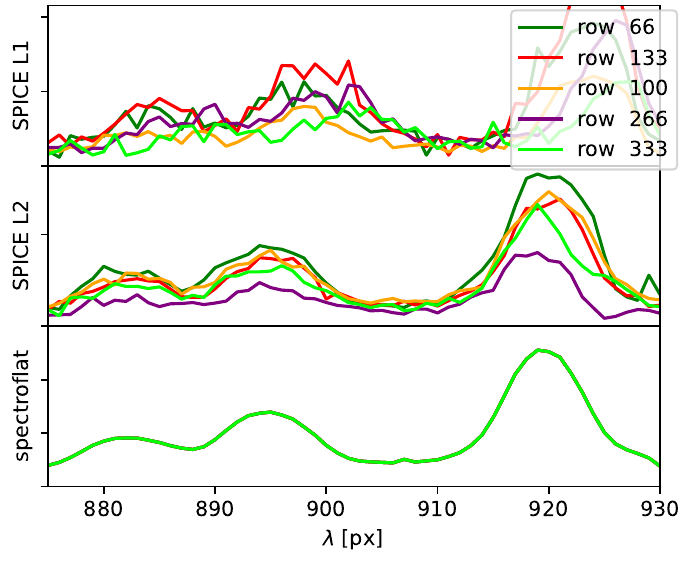}
  \caption[SPICE L1 smile correction result]{
  Comparison of smile correction on the SolO/SPICE dataset using intensity profiles of different rows. L1 is without distortion correction and L2 is with flat and distortion corrections (including desmiling) applied. Note: the outstanding smoothness in the spectroflat dataset is caused by the fact that residual spatial solar features from the not-a-flat-field test dataset are regarded as instrumental flat-field features and thus removed.
  }
  \label{fig:spice-smile}
\end{minipage}
\end{figure}

The test dataset from SPICE is not a regular flat-field; rather it was constructed from a quiet Sun raster map with full spectral coverage taken from the SPICE 3.0 data release \citep{spice.3.0}\footnote{The SPICE data from 8 Mar 2022 (ObsID 100663741) are available through the Solar Orbiter Archive at \url{https://soar.esac.esa.int/soar/}}. It is therefore not sufficiently spatially flat to perform a full analysis. However, we were able to perform a full smile correction and showcase the general applicability of the correction approach. Figure~\ref{fig:spice-l1} shows the constructed pseudo-flat measurement and extracted dust flat. The result of \spectroflat{} smile correction, compared to level one and two data products of the SPICE pipeline, is shown in Figure~\ref{fig:spice-smile}.
In all other datasets, \spectroflat{} produced reliable and stable results, efficiently removing fixed pattern imprints from the science data, with only minimal configuration adjustments per instrument.

The nominal flat-fielding procedure for GRIS is not suitable for our new approach of modulated flats. We were able to generate a non-modulated flat-field from the data and demonstrate the applicability of our method. However, for our new approach of having an individual flat-field map per modulation state, only a quarter of the 50 frames can be used per modulation state. To ensure sufficient spatial and spectral flatness we combined two subsequent flat-field recordings.

\begin{figure*}
    \centering
  \includegraphics[width=1\textwidth]{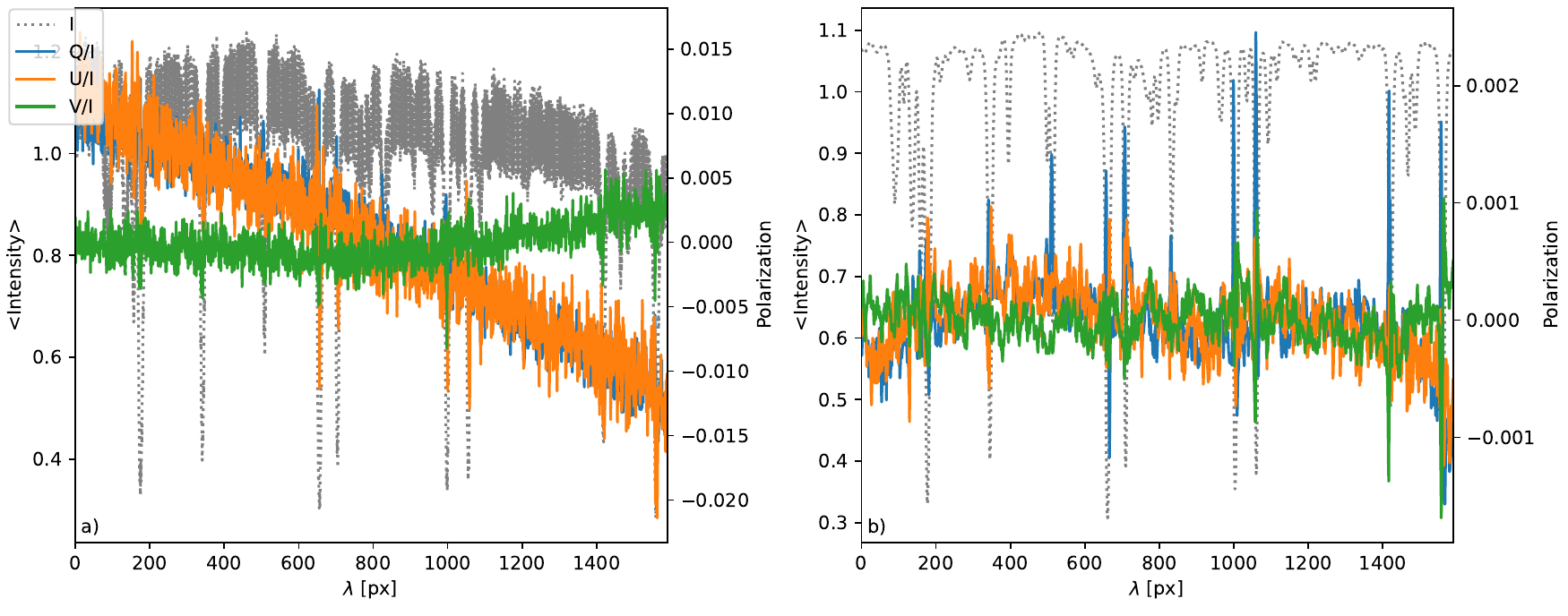}
  \caption[Comparison of raw and \spectroflat{}-corrected flat field frames]{
  Spectral profiles of a demodulated TRIPPLE-SP flat-field without (\(\D(\FF')\) in panel a) and with (\(\D(\FF^{(1)})\) and \spectroflat{} applied in panel b). Left axes denote normalized Stokes I intensity while axes on the right hand side denote the relative polarization signal for Q/I, U/I, and V/I.
  }
  \label{fig:sst-flat-demod}
\end{figure*}

After applying the sensor and slit flats the fixed pattern imprints and the artifacts in the demodulated data are effectively reduced by up to an order of magnitude. Figure~\ref{fig:sst-flat-demod} shows the spectral profiles of the demodulated flat-field frame from Figure~\ref{fig:pre-flat} before and after correction. The strong global slopes in the linear polarization states Q/I and U/I are removed. 
The residual I to Q/I, U/I and V/I cross talk and V/I to Q/I and U/I cross-talk around the stronger lines are not tackled by the algorithm presented here and need to be corrected on a per instrument basis. 
The sensor used in the TRIPPLE-SP setup features a strong column-to-column response variation, which is almost completely removed by the \spectroflat{} algorithm. Such a heavily biased dataset is considered to be a worst case and its results should be evaluated separately from the rest.

Demodulating the averaged flat-fields allows us to evaluate the polarimetric performance of the procedure. In short, no polarization is expected and the resulting Stokes components should be a flat line at zero with added photon noise. Thus, the smaller the variation and residual peaks, the better the result. 
With \(\FF := \{\FF_s\}_{s\in\{1,...,n\}}\) we denote the per modulation state averaged flat-field calibration frames, namely, the input image. With \(S = \{I,Q,U,V\}\), we denote the demodulated and averaged Stokes science images. To prepare our data we remove the instrumental polarization background (mean offset from zero) after demodulation.
The demodulation residuals, measured as the mean peak to peak value computed for several continuum regions (i.e., with no spectral lines present) before and after applying \spectroflat{} to all datasets are given in Table~\ref{table:flat_artifacts}.
Another estimation of the residuals is given in Table~\ref{table:flat_rms}, which presents the root-mean-square (RMS) variation of the full spectral profiles of the central row including absorption line regions. 
Both metrics show a reduction of fixed pattern imprints by up to an order of magnitude.\footnote{
Note: all results are single beam results and noise might be further reduced by instrument specific approaches, such as cross-talk correction and dual-beam demodulation.}  

\begin{table}
\begin{minipage}[t]{.48\textwidth}
\centering 
\begin{tabular}{l c r r} 
\hline\hline
Instrument &  & \(\D(\FF')\) & \(\D(\FF^{(1)})\)\\ 
\hline 
TRIPPLE-SP  & \(\langle I\rangle\)  & 25.26 & 3.52 \\
(SST)       & \(Q/I\)               & 0.58 & 0.07 \\
            & \(U/I\)               & 0.85 & 0.09 \\
            & \(V/I\)               & 0.66 & 0.07 \\
\hline 
SUSI           & \(\langle I\rangle\)   & 27.20 & 25.01 \\
(SUNRISE-III)  & \(Q/I\)                & 0.63 & 0.06 \\
               & \(U/I\)                & 0.45 & 0.15 \\
               & \(V/I\)                & 0.54 & 0.24 \\
\hline 
GRIS     & \(\langle I\rangle\) & 49.23 & 5.22 \\
(GREGOR) & \(Q/I\)              & 0.49 & 0.18 \\
         & \(U/I\)              & 0.32 & 0.17 \\
         & \(V/I\)              & 0.71 & 0.18 \\
\hline 
\end{tabular}
\end{minipage}\qquad%
\begin{minipage}[c]{.48\textwidth}
\caption[Residuals in demodulated flat-field frames per Stokes parameter]{
Residuals in demodulated flat-field frames per Stokes parameter as mean peak to valley variation within a continuum region after dark field correction and after applying the \spectroflat{} flat-field corrections to the flat-field data themselves. Reduction symbols as in Section~\ref{sec:steps}, values are \(10^{-2}\).\\[4.8em]
}
\label{table:flat_artifacts}
\end{minipage}
\end{table}

\begin{table}
\begin{minipage}[t]{.48\textwidth}
\centering 
\begin{tabular}{l c r r} 
\hline\hline
Instrument &  & \(\D(\FF')\) & \(\D(\FF^{(1)})\)\\ 
\hline 
TRIPPLE-SP  & \(Q/I\)          & 0.66 & 0.04 \\
(SST)       & \(U/I\)          & 0.65 & 0.03 \\            
            & \(V/I\)          & 0.15 & 0.02 \\
\hline 
SUSI            & \(Q/I\)      & 0.40 & 0.05 \\
(SUNRISE-III)   & \(U/I\)      & 0.18 & 0.04 \\
                & \(V/I\)      & 0.17 & 0.09 \\
\hline 
GRIS     & \(Q/I\)             & 0.15 & 0.07 \\
(GREGOR) & \(U/I\)             & 0.10 & 0.05 \\
         & \(V/I\)             & 0.22 & 0.08 \\
\hline 
\end{tabular}
\end{minipage}\qquad%
\begin{minipage}[c]{.48\textwidth}
\caption[RMS variation in demodulated flat-field frames per Stokes parameter]
{Same as Table~\ref{table:flat_artifacts} but for RMS variation of the full central row. Reduction symbols as in Section~\ref{sec:steps}, values are \(10^{-2}\).\\[5.8em]}
\label{table:flat_rms}
\end{minipage} 
\end{table}

\begin{figure}
\begin{minipage}[b]{0.64\textwidth}
  \includegraphics[width=1\textwidth]{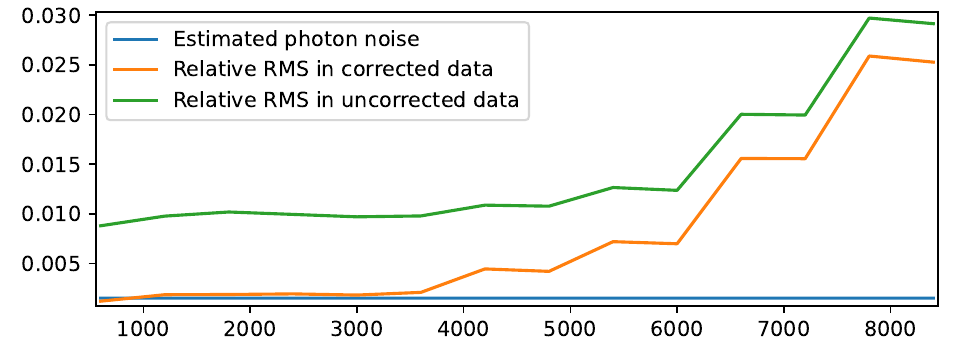}
\end{minipage} \quad 
\begin{minipage}[b]{0.32\textwidth}
  \caption[Photon noise estimation vs. RMS in data]{
  Estimated photon noise versus relative flat-field RMS for different ROI sizes of a SUSI flat-field recording. See text for details.}
  \label{fig:photon_noise}
  \end{minipage}
\end{figure}

For the SUSI dataset the detector conversion gain used is 1.8\,DN/e$^{-}$. Assuming a Poisson distribution for the photo-electrons generated in the detector, the relative photon noise can be estimated via
\begin{equation}
\textnormal{pn}(\FF) \;:=\; \left( \sqrt{ \sum_{i=1}^N \frac{F_i}{\textnormal{gain}}} \right)^{-1},
\end{equation}
where \(F_i\) denotes the flux in individual flat-field frames in units of DN, and \(N\) the number of frames averaged. 
Comparison with the relative flat-field RMS, given by
\begin{equation}
\textnormal{RMS}_{rel} \;:=\; \frac{\textnormal{RMS}}{\textnormal{Mean intensity}},
\end{equation}
yields an estimate on fixed pattern imprints still present after correction. Since there is basically no line-free continuum region in the SUSI near UV dataset recorded at 412\,nm, we analyzed ROIs with fixed height of 600 px (spatial dimension) and different widths (1 to 14 px, spectral dimension) in a quasi-continuum section. For a ROI width up to 6 px width (3\,600 px total) the relative RMS is on the same scale as the estimated photon noise. For larger widths weaker spectral line imprints contribute to the RMS measurement. Thus, we conclude that no relevant fixed pattern imprint is left above the noise level, after correction. Figure~\ref{fig:photon_noise} shows the relative RMS of corrected and uncorrected data in comparison to the estimated photon noise level. 

To highlight the benefits of \spectroflat{}, we compare it with the \cite{Woehl02} method applied to the same dataset. Some implementation details are not specified in \cite{Woehl02}; thus, we did our best to select values that produce an optimal correction. 
The original algorithm presented by \cite{Woehl02} does not account for strong column to column variations. As noted above, the TRIPPLE-SP data is heavily influenced by such a column comp pattern. We have therefore added the same pre-correction step as described in Section~\ref{sec:preflat} to our implementation of \cite{Woehl02} algorithm, to allow for a fairer comparison. 
We note that \spectroflat{} treats the modulation states of the averaged flat-field frames independently, while  \cite{Woehl02} does work with a single mean image composed of all modulation states. Moreover, \cite{Woehl02} performed the division by the mean spectrum  after smoothing and interpolation. In \spectroflat{}, we do not apply any smoothing prior to the extraction of the flat-field information. This helps avoid residuals that originate from artificially broadened spectral lines. 

\begin{figure*}
    \centering
    \includegraphics[width=1\textwidth]{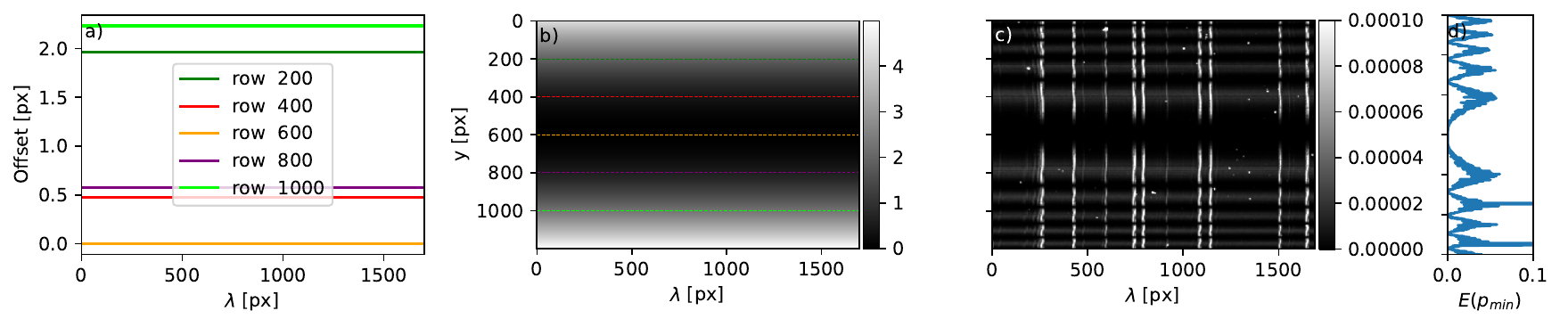}
  \caption[W\"ohl representative offset map and residual error]{
  Same as Figure~\ref{fig:offset-map}, but for the correction method from \cite{Woehl02}.
  }
  \label{fig:woehl-smile}
\end{figure*}

A key difference between the methods is that \spectroflat{} employs multiple lines for the smile correction, while \cite{Woehl02} uses a single one. In our implementation of \cite{Woehl02}, we selected the \ion{Fe}{I} line at 5250.6~\AA{} for smile detection, which is strong and located near image center. Moreover, \cite{Woehl02} do not remove dust imprints prior to the smile detection. This may cause errors in the line center detection. Figure~\ref{fig:woehl-smile} shows the offset and error-maps of the smile correction for \cite{Woehl02}.
In comparison to Figure~\ref{fig:offset-map}, the multi-line approach of \spectroflat{} shows a better performance (by two orders of magnitude) in terms of residual error (see Equation~\ref{eq:smile_error}).

\begin{figure}
    \centering
  \includegraphics[width=1\textwidth]{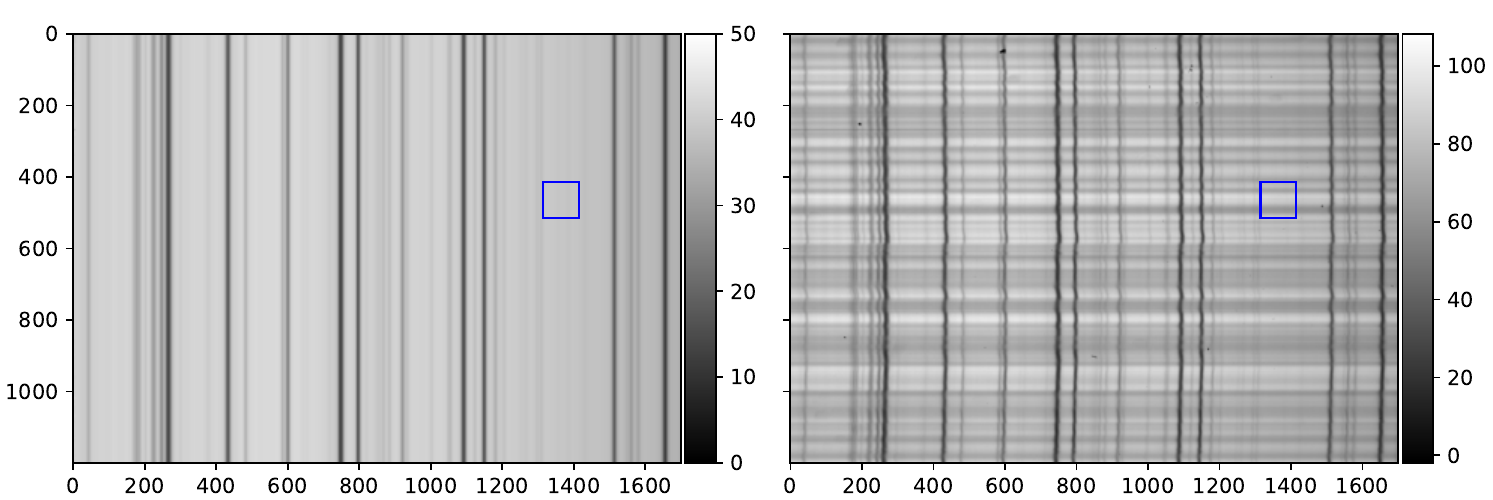}
  \caption[Selected area for RMS computation]{
  Position of the box to compute the RMS variation (Table~\ref{table:stds}) in one of the flat-field state images (left panel) and Stokes I of a science frames (right panel). Color bars give relative intensity in ADUs.
  The box is chosen such that spectral lines and strong dust impurities are avoided.
  }
  \label{fig:std_box}
\end{figure}

A widely used metric for flat-field performance is the residual RMS variation in a relatively uniform area of the corrected flat-field frames \citep[see][their Table~2]{Woehl02}. Table~\ref{table:stds} shows the RMS as a percent of the mean signal, computed within the rectangular box (as shown in Figure~\ref{fig:std_box}).
This metric translates to the signal-to-noise ratio (S/N) defined as the mean intensity divided by the RMS in the selected region. On the flat-fielded TRIPPLE-SP flat images (\(\FF_s\)) this yields values of 14.6 for the uncorrected data, 69.9 after applying the \cite{Woehl02} method and 118.3 after \spectroflat{} correction.
And similarly, comparing spectroflat alone against the full GRIS pipeline, we get S/N values of 5.7 for the uncorrected data, 94.6 after applying the GRIS pipeline (including dual beam recombination), and 92.1 after \spectroflat{} correction (single beam) for the \mbox{GREGOR/GRIS} dataset recorded in the 10\,820~\AA{} window.

\begin{table}[hp]
\caption[Comparison of RMS values from correction with different methods]
{RMS variation in percent of the mean count in a sub region of TRIPPLE-SP data chosen such that strong spectral lines and strong dust impurities are avoided.
}
\label{table:stds}
\centering 
\begin{tabular}{l | c c c c} 
\hline\hline
Method & \(\sum_s \FF_{s}/n\) & \(\FF_{dust}\) & \(\FF_{illum}\) & \(I_{science}\) \\
\hline 
Uncorrected & 7.17 \% & -       & -       & 10.98 \% \\ 
\hline 
\cite{Woehl02} & 1.04 \% & 5.32 \% & 0.27 \% & 9.89 \% \\
\hline 
Spectroflat & 0.85 \% & 6.84 \% & 0.00 \% &  9.08 \% \\
\end{tabular}
\end{table}

\begin{figure}[hp]
    \centering
    \includegraphics[width=1\textwidth]{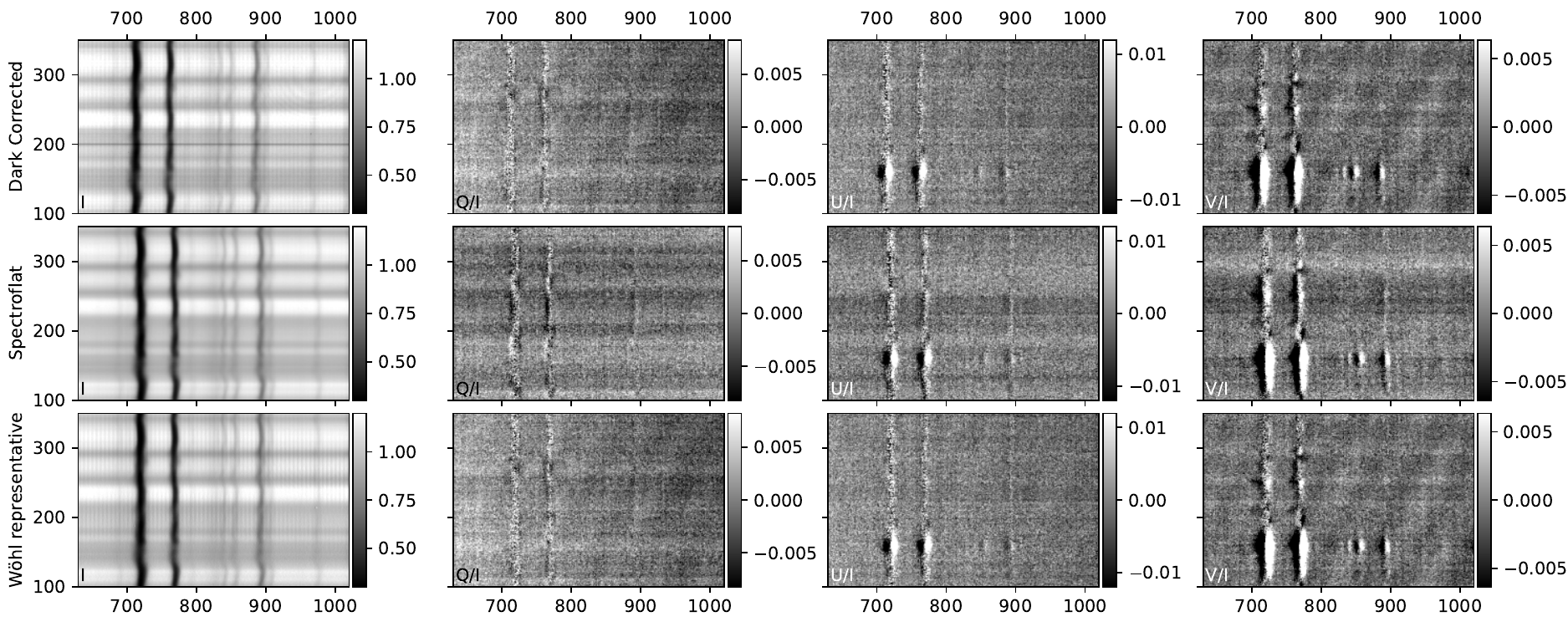}
  \caption[Comparison of reduced data sets: 2D ROI]{
  ROI from a five-second (500 frames) average of quiet Sun Stokes spectra from the 2017 SST/TRIPPLE-SP campaign. Panels in the upper row present data with no correction applied, while the middle row shows the same after \spectroflat{} and our wavelength calibration are applied. 
  The lower row panels show the same ROI, after the \cite{Woehl02} method is applied. Note: any instrumental cross-talk was not corrected in any of these datasets.
  }
  \label{fig:sst-demod}
\end{figure}

\begin{figure}[hp]
    \centering
    \includegraphics[width=1\textwidth]{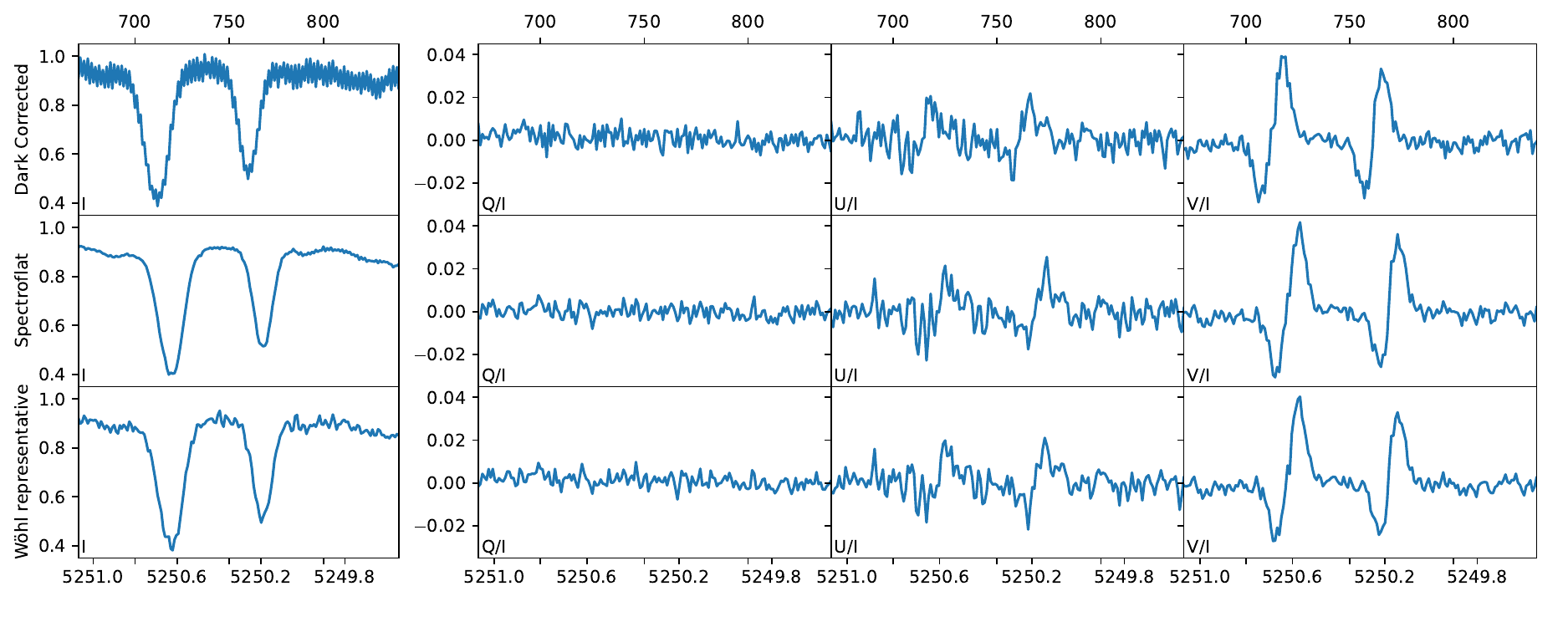}
  \caption[Comparison of reduced data sets: 1D profile]{
  Spectral profiles at row number 150 of the images shown in Figure~\ref{fig:sst-demod}. Top and bottom \(x\)-axis scales indicate the pixel position and the wavelength in~\AA{}, respectively. Note: the spectrum is flipped in wavelength direction with respect to the standard convention.
  }
  \label{fig:sst-demod-cuts}
\end{figure}

For further comparisons of the \spectroflat{} and \cite{Woehl02} methods, we examined reduced \mbox{SST/TRIPPLE-SP} science data. Figure~\ref{fig:sst-demod} shows the I, Q/I, U/I and V/I spectra of a five-second averaged quiet Sun measurement. The upper row presents data that was only corrected for the camera zero-response as a reference. The strong sensor response pattern imprint and slit-dust impurities are clearly visible in Stokes I. All Stokes parameters show diagonal fringes and gradients across the FOV. 
These effects are drastically reduced by the \spectroflat{} correction shown in the middle row. In comparison, \cite{Woehl02} correction (bottom row) does not fully remove the fringes and intensity gradients from the Q/I, U/I and V/I images. Also, we observe a residual column to column response pattern in I.
Figure~\ref{fig:sst-demod-cuts} shows spectral profiles at row number 150 of the images in Figure~\ref{fig:sst-demod}. There is a reduction in fixed pattern imprints in all Stokes parameters for both correction methods. However, the residual imprints from the column to column variation remain visible after the \cite{Woehl02} correction. In the other Stokes parameters, the differences between both correction methods are minimal after division by I.

\begin{figure}[!p]
    \centering
    \includegraphics[width=1\textwidth]{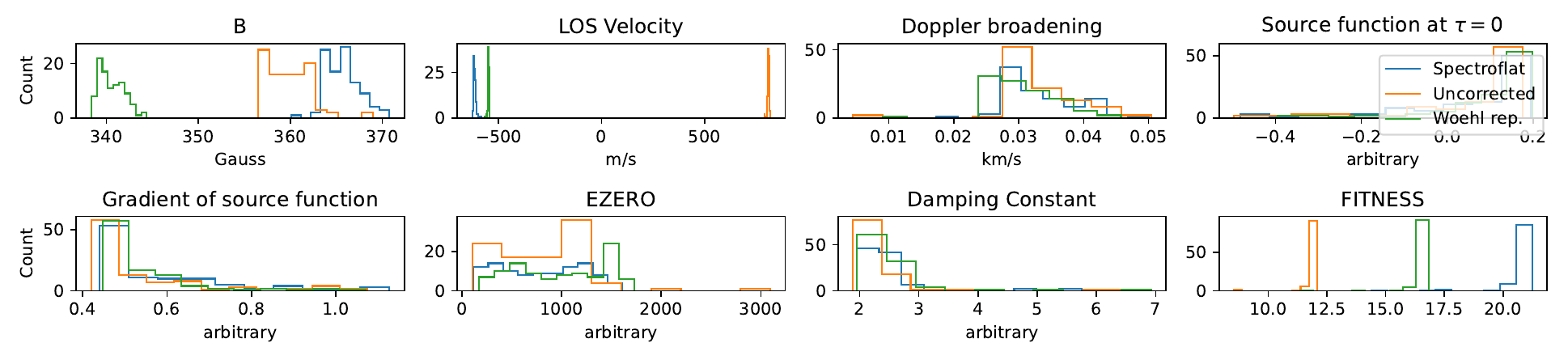}
  \caption[Histograms of 100 inversions on \ion{Fe}{I} 5250.6~\AA{} line at one spatial pixel]{
    Histogram of 100 inversions run on the \ion{Fe}{I} 5250.6~\AA{} line at row 150 of the datasets shown in Figure~~\ref{fig:sst-demod}.
    }
  \label{fig:sst-inversion-5250-hist}
\end{figure}

\begin{figure}[!p]
    \centering
    \includegraphics[width=1\textwidth]{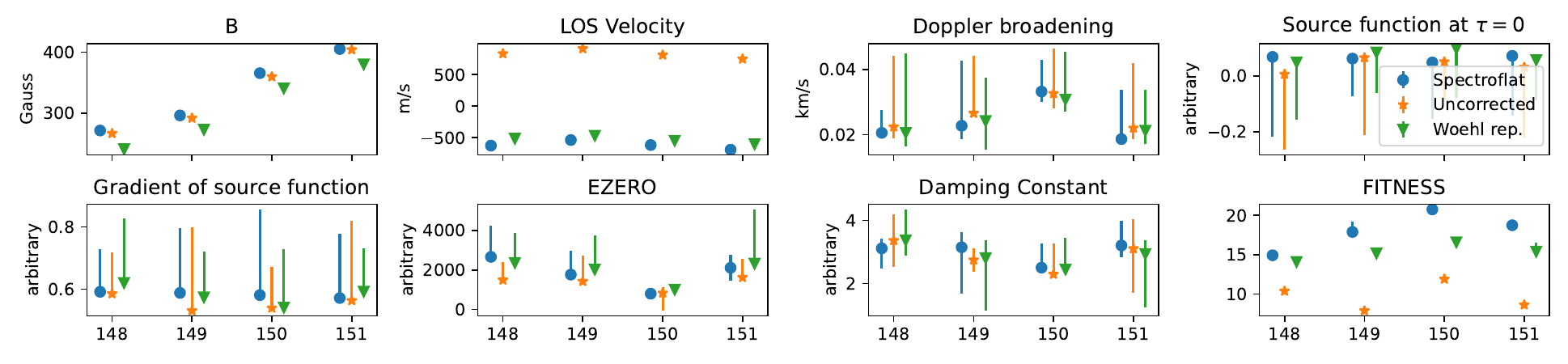}
  \caption[Comparison of inversion results of adjacent spatial pixels for \ion{Fe}{I} 5250.6~\AA{}]{
    Results of the HeLIx\(^+\) inversion code, run on the \ion{Fe}{I} 5250.6~\AA{} line, at rows 148 to 151 of the datasets shown in Figure~~\ref{fig:sst-demod}. The symbols indicate the histogram peak value, while the error bars the full range of the resulting distributions of the 100 inversion runs. 
    }
  \label{fig:sst-inversion-5250}
\end{figure}

To test the influence of the flat-field and spectral calibrations on the value of the atmospheric parameters retrieved by the inversion algorithms typically employed in remote sensing, we performed a comparison using the Milne-Eddington (ME) HeLIx\(^+\) inversion code \citep[see][]{Lagg04, Lagg09}. We applied it to the measurements shown in Figure~\ref{fig:sst-demod}. 
To determine the atmospheric parameters,  HeLIx\(^+\)  maximizes the FITNESS metric, namely, the inverse of the difference between measured and synthetic Stoke parameters, using a genetic optimization algorithm. See \cite{Lagg04} for details on the HeLIx\(^+\) algorithm and fitting procedure. 
The exact propagation of residual flat-field errors to the inverted atmospheric parameters depends on the inversion code utilized. Therefore, we use a simplistic ME-model and a statistical approach to test for the influence on the inversion robustness, rather than on individual results. 
We invert some selected spectral profiles from each case shown in Figure~\ref{fig:sst-demod} in 100 independent HeLIx\(^+\) runs with random start parameters (within the defined bounds). The genetic code will determine slightly different best fits for each run. Hence, the comparison of peaks and footprints of the result distribution for basically uncorrected and the two reduced datasets can be used as a sanity and robustness check respectively.
After removing the V cross-talk the linear polarization signal present in this dataset is below the noise level (see Figure~\ref{fig:sst-demod-cuts}). Hence, we selected weights to ignore Q/I and U/I to avoid fitting artifacts triggering offsets in the magnetic flux B. All inversion results shown rely solely on I and V/I, which prevents the deduction of azimuth and inclination angles of the magnetic field. 
The histograms of the distribution of the results are given in Figure~\ref{fig:sst-inversion-5250-hist}, in terms of the distribution of the inversions of the \ion{Fe}{I} 5250.6~\AA{} line in row \(y=150\). Likewise, Figure~\ref{fig:sst-inversion-5250} presents the results for the adjacent rows, which have similar Stokes profiles. 
For all runs the quality of the fit (FITNESS) of spectroflat is better compared with the dataset calibrated with \cite{Woehl02} method. However, given that the differences in the corrected spectra of both methods are minimal, they provide similar inversion results, also in general agreement with the noisy raw data. The most prominent difference is the line of sight (LOS) velocity. The wavelength calibration was executed on the central image row (\(y=600\)) for all datasets. For the uncorrected data, the smile was not removed, thus at row \(y=150\) we have a several pixel offset from the calibrated wavelength.

\begin{figure}[!p]
    \centering
    \includegraphics[width=1\textwidth]{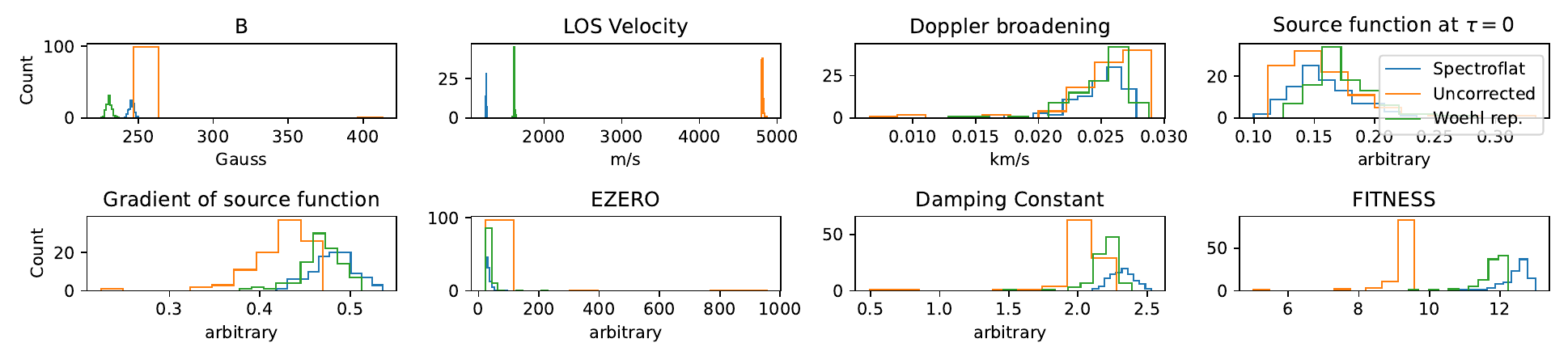}
  \caption[Histograms of 100 inversions on the \ion{Fe}{I} 5247.0~\AA{} and \ion{Cr}{I} 5247.5~\AA{} lines at one spatial pixel]{
      Same as Figure~\ref{fig:sst-inversion-5250-hist} except that here the \ion{Fe}{I} 5247.0~\AA{} and \ion{Cr}{I} 5247.5~\AA{} lines are inverted.
    }
  \label{fig:sst-inversion-5247-hist}
\end{figure}
\begin{figure}[!p]
    \centering
    \includegraphics[width=1\textwidth]{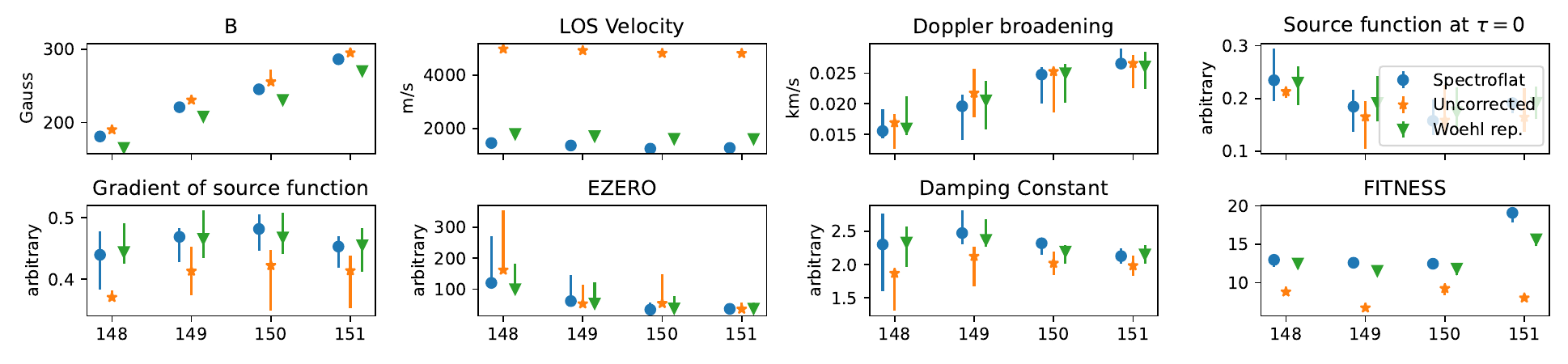}
  \caption[Comparison of inversion results of adjacent spatial pixels for \ion{Fe}{I} 5247.0~\AA{} and \ion{Cr}{I} 5247.5~\AA{}]{
     Same as Figure~\ref{fig:sst-inversion-5250} except that here the \ion{Fe}{I} 5247.0~\AA{} and \ion{Cr}{I} 5247.5~\AA{} lines are inverted.  
    }
  \label{fig:sst-inversion-5247}
\end{figure}

\newpage
To further investigate the influence of the smile correction, we also studied lines where the smile solution of \cite{Woehl02} and spectroflat no longer agree, and their differences are more pronounced. Figure~\ref{fig:sst-inversion-5247-hist} shows the distribution of 100 individual inversion runs for row \(y=150\) for the \ion{Fe}{I} 5247.0~\AA{} and \ion{Cr}{I} 5247.5~\AA{} lines. As both lines are formed under roughly the same atmospheric conditions, simultaneous ME inversions can be executed with out penalties.
Figure~\ref{fig:sst-inversion-5247} shows the result for the adjacent rows which have similar Stokes profiles.

\begin{figure}
    \centering
  \includegraphics[width=1\textwidth]{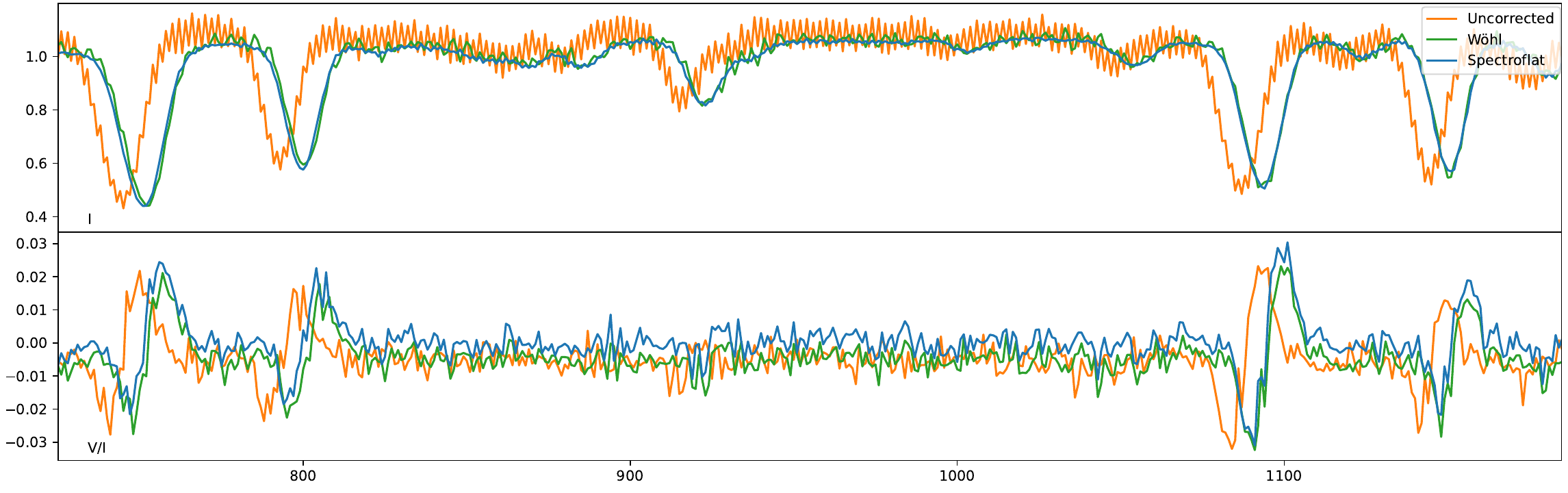}
  \caption[Spectral profiles after different smile correction and flat fielding methods]{
  Row \(y=148\) of the demodulated Quiet Sun frame corrected with different algorithms after wavelength calibration on the central row \(y=600\). The orange line shows the dataset without any smile or flat-field correction applied. For the W\"ohl representative dataset (green) the smile was detected using the \ion{Fe}{I} line at 5250.6~\AA{} (at x=1150 px) only. The spectroflat method (blue) uses all strong lines in the FOV simultaneously. While both methods agree on the 5250~\AA{} lines, they already start to notably diverge on the 5247~\AA{} lines (350 pixels apart).}
  \label{fig:diff_smile_corr}
\end{figure}

The most dominant differences are again for LOS velocity which can be traced down to smile correction and wavelength calibration: We show earlier in this work that the smile distortion is not uniform (see Figure~\ref{fig:offset-map}). For the W\"ohl representative dataset the smile correction was performed on the \ion{Fe}{I} line at 5250.6~\AA{}. For the results on that line (see Figure~\ref{fig:sst-inversion-5250}) we have a mean difference of less than 100~m/s. However, in the 5247~\AA{} region the smile solutions already show relevant offsets (see Figure~\ref{fig:diff_smile_corr}) leading to a mean difference exceeding 500~m/s. The difference in LOS velocity in the inversion results showcases the impact of residual error in the single line smile correction for larger spectral FOVs.

\newpage
\subsection*{Limitations}
As for flat-fields for pure imaging instruments, the input data to this algorithm has to be sufficiently uniform in spatial direction. Any remaining (solar) structure will be regarded as a flat-field effect and therefore be "corrected" out of the science frames. This would change the science data in an undesired way. Moreover, if the data is too noise dominated, the noise will affect the smile detection error metric (Equation \ref{eq:smile_error}). As the \spectroflat{} algorithm cannot distinguish between actual flat-field artifacts originating from the instrument and residual solar structures, the observer and user of the \spectroflat{} algorithm have to ensure that the recorded flat-field data are averaged along a path around the disk center and sufficiently flat, depending on the requirements of a given science case.
Both issues can be mitigated by integrating enough and sufficiently displaced frames. 

The smile detection and wavelength calibration are aimed at calibrating the full FOV by applying row-wise changes. In order to do so, several spectral lines have to be tracked simultaneously. Hence, it requires multiple absorption or emission lines in the correction area; in particular, lines near the left and right borders of the correction area improve the result.

Finally, the \spectroflat{} desmile algorithm is focused on detecting and removing the spectrographic curvature effect. Thus, other geometrical distortions (especially in spatial direction) need to be tackled separately. Also instrument specific cross-talk effects are beyond the scope of this algorithm.

\section{Conclusion}
We have introduced two publicly available python libraries. \spectroflat{} to generate spectro-polarimetric flat-fields and \texttt{atlas-fit} to amend the calibration data generated by \spectroflat{} with wavelength calibration, along with continuum correction information against a given solar atlas \citep{spectroflat, atlas-fit}. Both can be installed from the generic python \texttt{pip} repository, while the sources can be inspected and downloaded from a public Gitlab repository. Both libraries are agnostic to instrument specifics and can be included into existing python-based data-reduction pipelines. Technical description is available in the cited references.

Our algorithms allow to obtain precise flat-field maps from nominal flat-field measurements recorded close in time to the science observations and using a representative instrument configuration with Sun pointing instead of an artificial light source. Thus, we avoid the shortcomings mentioned in Section~\ref{sec:dr:introduction}. When applying our method we see a significant reduction of fixed pattern noise and residual artifacts. Furthermore, smile distortion effects are precisely registered and corrected. We have shown its applicability on a wide range of instruments, wavelength regimes and sensor types.

A variation of smile-distortion as a function of wavelength is expected. However, we have been able to show that the smile distortion may vary less predictably along the spectral dimension, exceeding second-order distortions. Higher-order distortion effects, caused by optical manufacturing and alignment tolerances, are typically not reproduced by an optical model. Therefore, the currently prevalent approach (obtaining this information around a single line or from an optical model) is not suitable for large sensors targeting multiple solar lines simultaneously. 
Since multiple spectral windows without any telluric lines exist and telluric lines are not present at all in datasets from balloon-borne or space-based instruments, an approach using these lines for smile correction, as presented by \cite{Cai17}, was not suitable for this generic library. However, it might be of value to explore such a mode in a later release.

To prove the general performance of \spectroflat{}, we examined the residual artifacts and the noise level of demodulated flat-fields from different instruments. We found the fixed pattern noise, measured as RMS or peak-to-peak variation in Stokes Q, U, and V, to be reduced by up to an order of magnitude by our method. 
Further, for a given science observation, we compared the calibration result obtained with our method with the result obtained with the method by \cite{Woehl02}: 
\begin{itemize}
    \item The fixed pattern imprint, quantified as the S/N of the Stokes I data, is significantly better suppressed by our method. 
    \item For the full frame we were able to achieve a better fringe and gradient suppression in Q/I, U/I and V/I. 
    \item The differences in the 1D Q/I, U/I, and V/I profile correction results of both methods are minimal. 
    \item The inversion results after both calibration methods agree within inversion error expectations, indicating that our method does not introduce systematic errors.
    \item Higher fitness values for the inversions of the data obtained with our method indicate a better compatibility of the Stokes profiles with the atmospheric model used by the inversion algorithm.
    \item The results on LOS velocity benefit strongly from a full FOV smile solution.
\end{itemize}
For an in-depth analysis of the influence of the calibration methods on inversion results, a comparison against synthetic data is needed. This is planned for a follow up study.

Furthermore, our algorithm does not depend on an illumination field and works with sensor and slit flat alone. The smile correction can thus be applied on a temporally averaged frame, which significantly reduces the numerical costs of the flat-field correction, as compared to the original \cite{Woehl02} method. 
When applying the generic \spectroflat{} to a \mbox{GREGOR/GRIS} single beam flat-field, the resulting S/N value is almost as high as after running the full instrument-specific GRIS pipeline. A detailed study on possible improvements of the GRIS pipeline using spectroflat is in preparation. 
For the SUSI test data we where able to show that the fixed pattern imprints are suppressed down to the photon noise level. 

The few configuration values required by our algorithm were found to be robust. In all tests based on a given instrument we were able to use the same instrument-specific configuration on all corresponding flat-field datasets.
Due to its broad applicability and its model independence with respect to specific instrument properties and wavelength regimes, we believe that our method is applicable as a standard calibration approach for most modern large FOV long slit spectrographs with or without polarimetric capabilities. 
We welcome the solar-instrumentation community to use the code as is or to contribute to the code for even broader applicability.\footnote{Author's note: Since October 2025, \spectroflat{} is an officially affiliated package of the SunPy project \citep{sunpy_community2020}.}

Because of its robustness and clear benefits, including its relative numerical inexpensiveness, an on-board implementation on future space missions (with the goal of performing flat-fielding on board) appears feasible. This is a promising outlook since downstreaming fully reduced science data helps in reducing telemetry bandwidth significantly.
Such a standardized pipeline approach will be explored with the (post-flight) data calibration pipeline of the Sunrise Ultraviolet Spectropolarimeter and Imager (SUNRISE-III/SUSI) on the Sunrise re-flight planned for summer 2024\footnote{Author's note: After the successful re-flight of SUNRISE-III in summer 2024, \spectroflat{} is now indeed an integral part of the SUSI data reduction pipeline. For details on the pipeline implementation and the integration of \spectroflat{} kindly refer to \cite{Feller25}. Further updates on \spectroflat{} can be found in Appendix~\ref{app:spectroflat}.}.

\chapter{Image Restoration}
\label{ch:reconstruction}

\textit{This chapter is focused on the mathematical background of the MOMFBD image reconstruction algorithm, used to estimate the PSF that degraded a recorded image. This PSF can then later also be used to reconstruct solar imaging spectro-polarimetric data in a pseudo-deconvolution process named \specrestore{}.
The form of the thesis changes to a more mathematical work in this chapter and is structured in definitions, lemmata, theorems, and proofs. }

\section{Background}
In this section I will provide the mathematical toolbox needed to derive the image  reconstruction algorithm. I will state general results on Zernike Polynomials and Karhunen-Lo\`eve Expansions which are used to describe the PSF and a quick summary of Poisson processes. Further, the Expectation Maximation (EM), and Gerchberg-Saxton (GS) algorithms are introduced. 
Proofs from textbooks are omitted while proofs from articles are reproduced where needed. 

\subsection{Mathematical Utilities}
\subsubsection{Zernike Polynomials} \label{sec:zernike}
Zernike polynomials are suitable to describe optical aberrations and to some degree also distortions caused by atmosphere (\cite{ZERNIKE1934, Noll76}). Optical pupils usually are circular domains, it is hence natural to define the Zernike polynomials in polar coordinates. 
For this section let \mbox{\(\rho \in [0,1]\)} be the radial distance and \(\phi \in [0,2\pi]\) the azimuth angle.

\begin{definition}
\label{def:zernike} \textbf{(Zernike polynomials)}\\
Let \(n,m \in \NN_0\) with \(n \geq m\). 
The even Zernike polynomials are defined as
\begin{equation}
    Z_n^m(\rho,\phi) \quad:=\quad R_n^m(\rho) \cdot \cos(m\phi)
\end{equation}
and the odd ones as 
\begin{equation}
    Z_n^{-m}(\rho,\phi) \quad:=\quad R_n^m(\rho) \cdot \sin(m\phi).
\end{equation}
The radial polynomials \(R_n^m(\rho)\) are defined by
\begin{equation} \label{eq:zernike_radial_poly}
    R_n^m(\rho) \quad:=\quad 
    \begin{cases}
    \sum\limits_{k=0}^{(n-m)/2} \frac{(-1)^k\,(n-k)!}{k!\,\left(\frac{n+m}{2}-k\right)!\,\left(\frac{n-m}{2}-k\right)!}\rho^{n-2k}, & \text{if \(n-m\) even}\\
    0, & \text{if \(n-m\) odd.}
  \end{cases}
\end{equation}
\end{definition}

\begin{figure}
  \centering
  \includegraphics[width=0.95\textwidth]{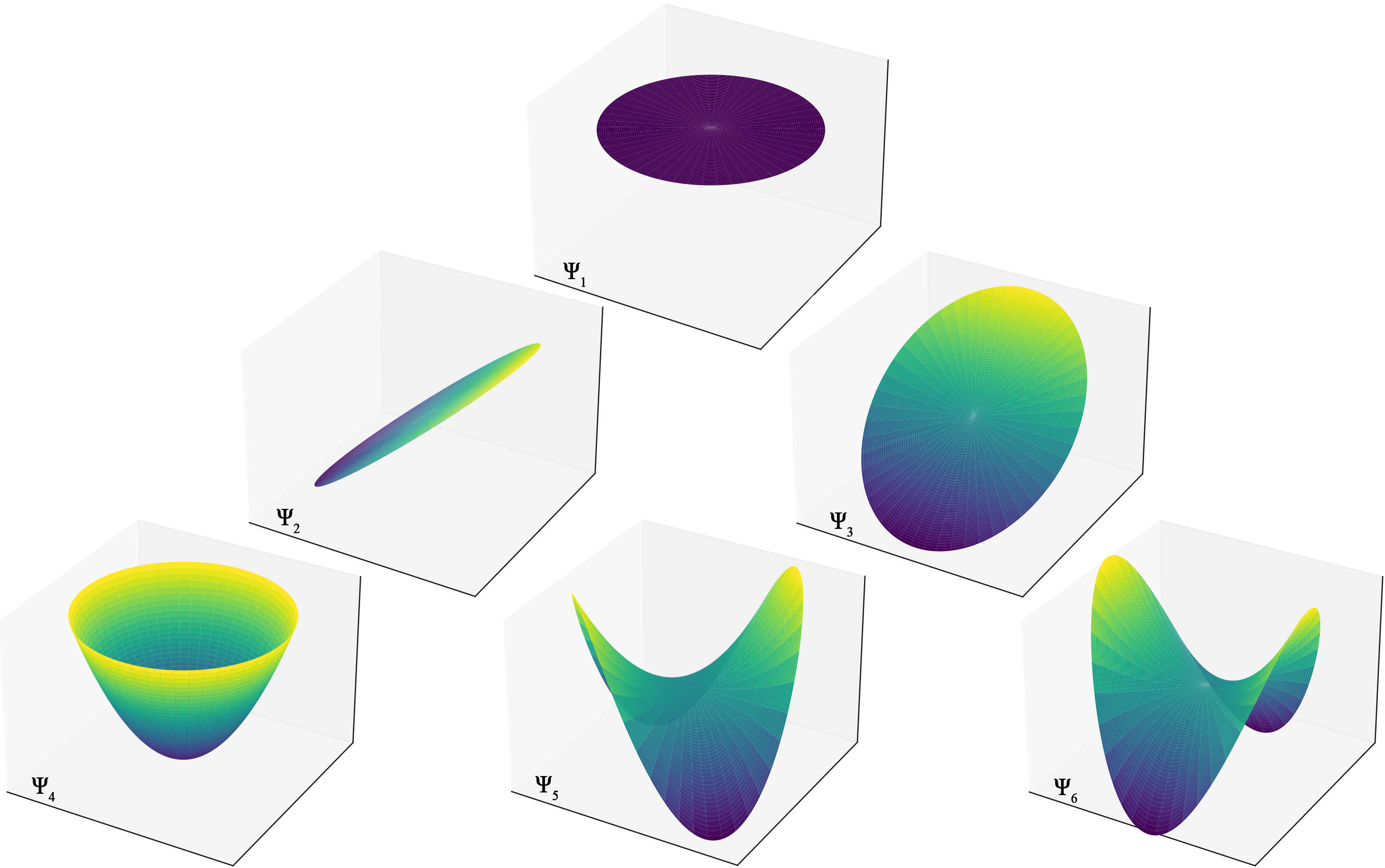}
\caption[3D visualization of Zernike Polynomials]{3D visualization of the first six Zernike polynomials. See Table~\ref{tab:Zernikes} for the corresponding equations.}
\label{fig:zernikes}
\end{figure}

\noindent
Following \cite{Noll76}, we can normalize the Zernike polynomials by multiplication with 
\begin{equation} \label{eq:hmn}
    h_m(n)  \quad:=\quad 
    \begin{cases}
    \sqrt{2(n+1)}, & \text{if } m\neq 0\\
    \sqrt{n+1}, & \text{else}.
  \end{cases}
\end{equation}
Several competing indexing schemes exists, we are using the radially ordered scheme introduced by  \cite{Noll76}, where 
\[
i \;:=\, i(n,m) \;:=\; \frac{n(n+1)}{2} + |m| + 
 \begin{cases}
    0, & \text{if } m > 0 \textnormal{ and } n \equiv \{0,1\} \, (\textnormal{mod} 4);\\
    0, & \text{if } m < 0 \textnormal{ and } n \equiv \{2,3\} \, (\textnormal{mod} 4);\\
    1, & \text{if } m \geq 0 \textnormal{ and } n \equiv \{2,3\} \, (\textnormal{mod} 4);\\
    1, & \text{if } m \leq 0 \textnormal{ and } n \equiv \{0,1\} \, (\textnormal{mod} 4).
  \end{cases}
\]
We write \(\Psi_i\) for the \(i\)-th Zernike polynomial as indicated in Table \ref{tab:Zernikes} below, and visualizations in Figure~\ref{fig:zernikes}. We call the indices \textit{modes}. A finite set of modes is often denoted with \(\MC \subset \NN\). Modes \(2\) and \(3\) are referred to as the tip-tilt components of the PSF.

\begin{table}[!ht]
\begin{center}
    \begin{tabular}{c|c|c|c|c|l}
        \hline
        \textbf{Mode} & Symbol & \(n\) & \(m\) & \(h_m(n)\, Z_n^{\pm m}(\rho,\phi)\) & Optical Name \\ \hline
        \textbf{1} & \(\Psi_1\) & 0 & 0 & \(1\) & Piston \\
        \hline
        \textbf{2} & \(\Psi_2\) &  1 & 1 & \(2\rho \, \cos(\phi)\) & Tip (\(y\)) \\
        \textbf{3} & \(\Psi_3\) &  1 & -1 & \(2\rho \, \sin(\phi)\) & Tilt (\(x\)) \\
         \hline
         \textbf{4} & \(\Psi_4\)  &  2 & 0 & \(\sqrt{3}\,(2\rho^2-1)\) & Defocus \\
         \textbf{5} & \(\Psi_5\) &  2 & -2 & \(\sqrt{6}\,\rho^2 \, \sin(2\phi)\) & Astigmatism (oblique) \\
         \textbf{6} & \(\Psi_6\) &  2 & 2 & \(\sqrt{6}\,\rho^2 \, \cos(2\phi)\) & Astigmatism (vertical)\\ \hline 
    \end{tabular} 
    \caption[Index number, equation, and name of the first six Zernike Polynomials]
    {Noll mode index number, equation, and name of the first six Zernike Polynomials. 
    Note that polynomials where \(R_n^m\) vanishes are not indexed. 
    See Figure~\ref{fig:zernikes} for a visualization.}
    \label{tab:Zernikes}   
\end{center}
\end{table}

\begin{lemma} \textnormal{(Orthogonality of the Zernike Polynomials)}\\
The Zernike Polynomials are orthogonal. In particular, the following equation holds:
\begin{equation}
   \frac{1}{\pi}\; \int_0^{2\pi}\int_0^1 \Psi_i(\rho, \phi) \, \Psi_j(\rho, \phi) \,\rho \; d\rho \, d\phi \quad = \quad \delta_{i,j}.
\end{equation}
\end{lemma}
\begin{proof}
We can split the claimed property in the orthogonality of the radial and angular parts. For the angular part the following cases exist: 
The mixed term is always zero: 
\begin{align*}
    \int_0^{2\pi} \,\sin(m\phi)\,\cos(m'\phi) d\phi \quad=\quad&  \frac{m \sin(2m\pi)\sin(2m'\pi) + n (cos(2m\pi)\cos(2m'\pi) -1)}{m^2 - m'^2 } \\
    =\quad& 0 \qquad  \forall\; m,m'\in\NN_{0},
\end{align*}
since \(\sin(2z\pi) = 0\) and \(\cos(2z\pi) = 1\) for all \(z\in\ZZ\). 
If we assume \(m=0\) and \(m'\in\NN_0\), we get 
\[ \int_0^{2\pi} 0 \cdot \sin(m'\phi) \,d\phi \;=\; 0 \quad\textnormal{ and }\quad \int_0^{2\pi} \cos(m'\phi) \,d\phi \;=\; 2\pi\,\delta_{m,m'},\]
and the case where \(m=m'=0\) is always the \((\cos,\cos)\) case. 
Let's assume that \(m\in\NN\), then
\[
  \int_0^{2\pi} \,\sin(m\phi)\,\sin(m'\phi)\, d\phi \;=\; \pi\,\delta_{m,m'} \;=\; \int_0^{2\pi} \,\cos(m\phi)\,\cos(m'\phi) \,d\phi, 
\]
as we always get a result of the form "\(a\sin(2z\pi)-b\sin(2z'\pi)\)", with \(a,b\in\RR\) and \(z,z'\in\ZZ\), if \(m\neq m'\). 
And the integral over \(\cos^2\) and \(\sin^2\) is \(\pi\).

For the radial part we now just need to check for cases where \(m=m'\). The Zernike polynomials were designed to be orthogonal in this case
by subtracting appropriate low order terms. Thus,
\[
    \int_0^1 R_n^m(\rho)\,R_{n'}^m(\rho)\,\rho\,d\rho \quad=\quad  \frac{ \delta_{n,n'}}{2(n+1)}.
\]
With the choose of \(h_m(n)\) as in Equation~\ref{eq:hmn} we have shown the claimed condition.
\end{proof}

\noindent
Expanding the Zernike polynomials by substituting \(x := \rho\cos(\phi)\) and \(y := \rho\sin(\phi)\) we obtain polynomials in Cartesian coordinates, for example, 
\begin{equation}
\Psi_6 \;=\; Z_2^2(\rho,\phi) \; = \; \sqrt{6}\,\rho^2\cos(2\phi) \quad\xrightarrow{\textnormal{corresponds to}}\quad  \sqrt{6} \, (x^2 - y^2).    
\end{equation}
Apart from the natural circularity of our problem, it has further benefits to define Zernikes in polar coordinates: (i) As just shown, Zernike polynomials are orthogonal over the unit disk when expressed in polar coordinates. This is an important characteristic for expanding the PSF into a sum of Zernike polynomials.
(ii) In polar coordinates, Zernike polynomials separate into a radial part (\(R_n^m(\rho)\)) and an angular part (\(\cos(m\phi)\) or \(\sin(m\phi)\)) which is matching the physics of modes in an optical system.
And finally, (iii) the expression of Zernike polynomials is generally much simpler in polar coordinates.

\newpage
\subsubsection{Karhunen-Lo\`eve Expansion}
While Zernike polynomials are ideal for optical aberrations, Karhunen-Lo\`eve (KL) modes are more suitable at estimating wavefront errors created by atmospheric turbulences \citep{Noll76, Roddier90}.
KL modes can be constructed by using KL-expansion of the Zernike polynomials.
Usually, this is used for modes above three (i.e., non tip-tilt modes) but the use of KL expansion is optional and can be selected per execution of the MFBD algorithm. We therefore just briefly discuss the topic here for completeness. 
In a nutshell: The idea of the KL-Expansion is similar to the idea of selecting a Hilbert-basis from the Fourier-series representation.

I will follow \cite["Data-Driven Computational Methods", Chapter 5]{Harlim18} for the general theory part and \cite{Orlov97} for the description of our use case. In the former a detailed introduction to KL-expansions can be found. In particular the relevant proofs can be found ibid.

\begin{definition}
\textnormal{(Hilbert-Schmidt kernel \& integral operator)} \\
Let \(T\subset\RR\) be a bounded domain. The function \(k:T\times T\rightarrow\RR\) is called a \textnormal{Hilbert-Schmidt kernel} if \(k\in L^2(T\times T)\), that is 
\[\int_{T\times T} \big| k(t,s) \big|^2 \,dt\,ds \quad <\quad \infty.\] 
Further we define the \textnormal{Hilbert-Schmidt integral operator} as 
\[\big(Ku\big)(t) \;:=\; \int_T k(t,s)\,u(s)\,ds, \quad\forall\,u\in L^2(T).\]
\end{definition}

\noindent
Using the Cauchy-Schwarz Inequality for Integrals we see
\begin{eqnarray}
\int_T \Big|\big(Ku\big)(t)\Big|^2\, dt & = & \int_T \left|\int_T k(t,s)\,u(s)\,ds\right|^2\, dt\nonumber\\
&\leq& \int_T \left(\int_T\Big|k(t,s)\Big|^2 ds\right) \cdot \left(\int_T\Big|u(s)\Big|^2\,ds\right) \, dt\nonumber\\
&=& \lVert k \rVert^2_{L^2(T\times T)} \cdot \lVert u \rVert^2_{L^2(T)} \quad < \quad \infty
\end{eqnarray}
And hence, it follows that \(Ku \in L^2(T)\) is well defined and \(K:L^2(T)\rightarrow L^2(T)\) is bounded with \(\lVert K\rVert \leq \lVert k\rVert_{L^2}\).

\begin{definition}
\textnormal{(Self-adjoint, positive operator)}\\
Let \(H\) be a Hilbert space. We say a bounded linear operator \(K:H\rightarrow H\) is \textnormal{self-adjoint} if
\[\langle Ku,v\rangle \;=\; \langle u,Kv\rangle \quad\forall\,u,v\in H.\]
Further, we say \(K\) is \textnormal{positive semidefinite} if \(\langle Ku,u\rangle \geq 0\) for all \(u\in H\). 
\end{definition} 
As a lemma one can derive that a Hilbert-Schmidt integral operator \(K\) with kernel \(k\) is self-adjoint if and only if \(k\) is symmetric, that is \(k(s,t) = k(t,s)\) for all \(s,t\in T\).  
With these definitions at hand we can give the main result that is needed to understand KL expansion:
\begin{theorem}
\textnormal{(Mercer's Theorem)} \label{th:mercers_theorem}\\
Let \(T\subset\RR\) be a compact interval and \(k:T\times T \rightarrow\RR\) be a continuous function. Further, let \(k\) be a symmetric Hilbert-Schmidt kernel and the corresponding Hilbert-Schmidt integral operator \(K\) be positive semidefinite.
Then there exist real \(\lambda_j\geq 0\) and continuous functions \(\varphi_j \in L^2(T)\) that satisfy \(\big(K\varphi_j\big)(t) = \lambda_j\varphi_j(t)\) such that
\[ k(t,s) \quad=\quad \sum_{j\in\NN}\;\lambda_j\,\varphi_j(t)\,\varphi_j(s)\]
uniformly on \(T\times T\). Moreover, \(\{\,\varphi_j\,|\,j\in\NN\,\}\) forms an orthonormal basis of \(L^2(T)\).
\end{theorem}
Recall that \(\{\,\varphi_j\,|\,j\in\NN\,\}\) being an orthonormal basis means that any function in \(L^2(T)\) can be described as an infinite sum of these basis functions. 

For the application on KL-expansion let \(T\subset\RR\) a bounded interval, \(\Omega\) a probability space, and \(X:\Omega\times T \rightarrow \RR\) a centered mean-square-continuous\footnote{
A stochastic process \(\{X_t\}_{t\in T}\) is called mean-square-continuous, if 
\[\lim_{\epsilon\rightarrow0}\; \EE\big[(X_{t+\epsilon} - X_t)^2\big] \;=\;0.\]
} stochastic process in \(L^2(\Omega\times T)\). 
Define \(K:L^2(T) \;\rightarrow\; L^2(T)\) via
\begin{equation}
    \big[Ku\big](t) \quad:=\quad \int_T k(t,s)\,u(s)\,ds,
\end{equation}
where \(k = R_X\) is the autocovariance function of \(X\) given by
\begin{equation}
    R_X(t,s)\quad:=\quad \EE\big[\big(X_t - \mu_X(t)\big)\big(X_s-\mu_X(s)\big)\big],
\end{equation}
where \(\mu_X(t) := \EE[X_t]\) is the mean function. For centered processes \(\mu_X = 0\). Obviously \(R_X\) is symmetric, therefore \(K\) is self-adjoint.
Since \(X\) is mean-square-continuous, \(k = R_X\) is continuous on \(T\times T\) and 
\(R_X\in L^2(T\times T)\) because \(X\in L^2(\Omega\times T)\).
Further, \(K\) is positive semidefinite because,
\begin{equation}
\langle Ku, u \rangle \;=\; \mathbb{E}\left[ \left( \int_T X_t \, u(t)\, dt \right)^2 \right] \;\geq\; 0    \qquad \forall\, u \in L^2(T)
\end{equation}
since \(\mathbb{E}[Y^2]\) is always nonnegative for any real-valued random variable \(Y\). 
Thus, \(R_X\) is a Hilbert-Schmidt kernel and the Hilbert-Schmidt operator \(K\) with kernel function \(R_X\)  satisfies the requirements (self-adjoint, continuous, positive semidefinite) of Mercer's Theorem and therefore has real eigenvalues \(\lambda_j\geq0\) and eigenfunctions \(\varphi_j\in L^2(T)\) that form an orthonormal basis of \(L^2(T)\). 
Since \(X\in L^2(\Omega\times T)\) we can expand \(X\) as
\begin{equation}
    \label{eq:x_expansion}
    X(\omega,t)\quad=\quad \sum_{j\in\NN} x_j(\omega)\,\varphi_j(t),
\end{equation}
with expansion coefficients 
\begin{equation}
    \label{eq:kl_exp_coefficients}
    x_j(\omega) \;=\; \langle X(\omega,\cdot),\varphi_j\rangle \;=\; \int_T X(\omega,s)\,\varphi_j(s)\,ds.
\end{equation}

\begin{theorem} 
\label{th:kl_expansion}
\textnormal{(Karhunen-Loève Expansion)}\\
The expansion defined by Equations~\ref{eq:x_expansion} and \ref{eq:kl_exp_coefficients} converges and is uniformly in \(T\). That is, it does not depend on the particular value chosen in \(T\). Moreover, the coefficients are of mean zero and uncorrelated. In formulas 
\[\EE[x_j]\;=\;0,\quad\EE[x_jx_k]\;=\;\lambda_j\,\delta_{jk}\textnormal{ \; and } \textnormal{Var}[x_j^2]\;=\;\lambda_j,\]
where \(\delta_{jk}\) is the Kronecker-Delta. 
\end{theorem}

\noindent
\textbf{In our case} we have a two dimensional turbulence induced phase distortion \(p:T\rightarrow\mathbb{C}\) over a (circular) aperture \(T\subset\mathbb{R}^2\), the telescope pupil. By scaling everything with the pupil radius we can assume that \(T\) is the unit circle. Denote the auto-correlation function of the phase \(p\) with 
\begin{equation}
R_p(s,t) \quad:= \quad \big\langle p(s+t)\,p^*(s)\big\rangle,    
\end{equation}
for \(s := (\cos\phi, \sin\phi)\) and \(t := (\cos\theta, \sin\theta)\), with \(\phi, \theta \in [0, 2\pi)\), then we can formulate a two-dimensional KL equation as an homogeneous integral equation 
\begin{equation} \label{eq:kl_integral_eq}
    \lambda_j^2\varphi_j(t) \quad=\quad \int_T R_p(s,t)\, \varphi_j(s)\, ds.
\end{equation}
The correlation function of a turbulence induced phase distortion is symmetric. Thus, we want to apply Theorem \ref{th:kl_expansion} to the right hand side of the equation to get
\[p(t) \quad=\quad \sum_{j\in\NN} x_j \, \varphi_j(t),\]
with 
\[x_j \quad:=\quad \frac{1}{\pi} \int_T p(t)\, \varphi_j(t) \, dt.\]
However, computing a description of a seeing-induced PSF for the two dimensional case is numerically a non trivial task and the generalization of the KL theory to a multi-dimensional case is also more complex. 
Since seeing introduces phase distortions are isotropic \citep{Tatarski61} the correlation function \(R_p\) satisfies 
\begin{equation}
    R_p(t,s) \quad=\quad R_p\big(|t-s|\big)
\end{equation}
and we can reduce the two dimensional case to a one dimensional one (and use the 1D results from above).
Similar to the definition of Zernike polynomials (see Section \ref{sec:zernike}) we want to find a 
radial component \(L_j^m\) such that
\begin{equation} \label{eq:kl_radial_angluar_split}
    \varphi_j(t) \quad\sim\quad L_j^m(\rho) \cdot 
     \begin{cases}
   \cos(m\phi), & \text{ if \(m\neq0\) and \(j\) is even}\\
    \sin(m\phi), & \text{ if  \(m\neq0\) and \(j\) is odd}\\
    1, &\text{ if } m=0,
  \end{cases}
\end{equation}
where \(\rho \in [0,1]\) is again the radial distance and \(\phi \in [0,2\pi]\) the azimuth angle. According to \cite{Voitsekhovich95}, the correlation function \(R_p\) can be written as 
\begin{equation}
    R_p(s,t) \quad=\quad 0.49\cdot r_0^{-\nicefrac{5}{3}} \int_T \Phi_n(v) \,\exp\big(\im \, v\, (s-t)\big)\, dv,
\end{equation}
where \(\Phi_n\) is the fluctuation spectrum for the refractive index \(n\) and \(r_0\) the Fried parameter\footnote{
The \textit{Fried parameter} \citep[see][]{Fried66} is defined as 
\[r_0 \quad:=\quad \left( 0.523\, k^2\, \textnormal{sec}(\zeta)\; \int_0^{\infty} C_n^2(z)\, dz \right)^{-\nicefrac{3}{5}}\]
Where \(k = \nicefrac{2\pi}{\lambda}\) is the wave number and \(\zeta\) the zenith-angle. \(C_n\) describes the turbulence strength for refractive index \(n\) above the telescope as function of height. The Fried parameter can be interpreted as the \(0\)-th moment of turbulence. It defines the scale length for which the wavefront phase errors are of the order of 1 radian.
}. If we use the assumed split in radial function \(L_j^m\) and angular part from 
Equation~\ref{eq:kl_radial_angluar_split} we can translate Equation~\ref{eq:kl_integral_eq} to the one-dimensional integral (\cite{Orlov97})
\[
    \lambda_j^2\, L_j^m(\rho) \;=\; \int_0^1 L_j^m(r)\, k(r,\rho)\,dr,
\]
with Hilbert-Schmidt kernel function 
\[
    k(x,y) \quad:=\quad 1.96 \,\pi\, r_0^{-\nicefrac{5}{3}} \int_0^{\infty} z\, \Phi_n(z) \, J_m(zx) \, J_m(zy) \,dz,
\]
where \(J_m\) denotes the \(m\)-th Bessel-function\footnote{
The \(n\)-th Bessel function is defined as 
\[
  J_n(x) \quad:=\quad \sum_{r=0}^{\infty} \frac{(-1)^r (\nicefrac{x}{2})^{2r+n}}{\Gamma(n+r+1)n!},
\]
where \(\Gamma(n) := (n-1)!\) is the Euler Gamma Function.
}. Now that we are in 1D, we wish to express the radial part of the KL expansion \(L_n^m\) by the radial part of the Zernike polynomials \(R_n^m\) as defined in Equation~\ref{eq:zernike_radial_poly}.
From the reformulated Equation~\ref{eq:kl_integral_eq} we can derive the homogeneous system 
\begin{equation}
    \sum_{j\in\NN} \beta_{j,n}^m\, A_{j,k}^m \quad=\quad \lambda_m^2\,\beta_{k,n}^m,
\end{equation}
using the ansatz 
\begin{equation}
    L_j^m(\rho) \quad=\quad \sum_{k=m}^{\infty} \beta_{k,n}^m\, \sqrt{2j+2}\,R_j^m(\rho).
\end{equation}
The components of the matrix \((A_{j,k}^m)\) are given by 
\begin{align}
    A_{j,k}^m \;=\;& 3.92\,\pi r_0^{-\nicefrac{5}{3}}(-1)^{\nicefrac{(k+j+2m)}{2}}\, (1-\delta_{0k})\,(1-\delta_{0j}) \sqrt{(k+1)(j+1)}\nonumber \\
    & \quad\cdot\; \int_0^{\infty} z^{-1}\,\Phi_n(z)\,J_{j+1}(z)\,J_{k+1}(z)\, dz.
\end{align}
One can easily see that the computation now heavily depends on the chosen model for the fluctuation spectrum  \(\Phi_n\) for the refractive index. Among the more widely used models are the 
\textit{Kolmogorov-Model}, the \textit{Exponential-Model} and the \textit{Von-Karman-Model}. Tailored expressions for the matrix-components \(A_{j,k}^m\) for these models can be found in \cite{Orlov97}.

\subsubsection{Translated Poisson Process} \label{sec:poisson}
The model of a translated Poisson process describes the situation when acquiring images through an distorted atmosphere. I will only give a short overview here, a general introduction can be found in \cite["Random Point Processes in Time and Space", Chapter 3]{Snyder91}. Especially all proofs for the results presented here can be found there.
Before we discuss the translated Poisson process we need to introduce some concepts:

\begin{definition} \textnormal{(counting process)}\\
Let \(\XC\) be a measurable space (e.g., \(\XC\in\RR^d\)) and \(\omega = \{x_1, x_2, \ldots\}\) where each \(x_i\) denotes a coordinate in \(\XC\). Let \(A\subset\XC\) and 
\begin{equation*}
N(A:\omega) \quad:=\quad \sum_i I_A(x_i)
\end{equation*}
the number of points in \(\omega\) that lie in \(A\), with 
\begin{equation*}
      I_A(x) \quad := \quad  
     \begin{cases}
    1, & \text{ if } x \in A\\
    0, &\text{ if } x \notin A,
  \end{cases}
\end{equation*}
then we call \(N(A:\omega)\) a \textnormal{counting process}.
\end{definition}

\noindent 
Usually, \(\omega\) is fixed in a given context and we hence omit it.
When seen as a function of \(A\), \(N:\XC\rightarrow\mathbb{N}_0\) defines a positive semidefinite random process on \(\XC\) and we call the \(N(A)=k\) the count of \(A\) under \(N\), or simply count.

\begin{definition} \textnormal{(Poisson counting process)}\\
Let \(\XC\) be a measurable space. A \textnormal{Poisson counting process} (or simply \textnormal{Poisson process}) on \(\XC\) with a monotonically increasing intensity function \(\lambda:\XC\rightarrow\NN_0\) is a counting process \(N(\cdot)\) on \(\XC\) such that, for any measurable subset \(A \subseteq \XC\):
\begin{itemize}
\item[(a)] \(N(A)\) is Poisson-distributed with mean \(\lambda(A)\), that is
   \[
   P\big[N(A) = k\big] = \frac{\lambda(A)^k}{k!} \exp(-\lambda(A))
   \]
   for \(k\in\NN_0\).
\item[(b)]
   For any collection of disjoint, measurable subsets \(A_1, A_2, \ldots, A_n \subset \XC\) the counts 
   \(N(A_1),\, N(A_2),\, \ldots,\, N(A_n)\) are independent random variables.
\end{itemize}    
\end{definition}

\noindent
Note that \(\mathbb{E}[N(A)] = \lambda(A)\). 
A typical example for \(A\) and \(\XC\) would be to identify \(\XC\) with time and \(A\) with the number of events that have occurred up to time \(A\). A more common notation in this case would be \(\{N(t) : t \geq t_0\}\) and the conditions translate to 
\begin{enumerate}
    \item \(P\big[N(t_0) = 0\big] = 1\), that is we almost surely start with zero counts.
    \item The increment \(N(s,t) = N(t) - N(s)\) for any \(t_0\leq s < t\) is Poisson distributed with mean \(\lambda(t)-\lambda(s)\).
    \item The increments are independent random variables.    
\end{enumerate}
Properties (b) or (3) are the characteristic property of the Poisson process: It simply means that the counts in separate, nonoverlapping intervals are statistically independent, regardless of how big the intervals are or where they are located.

Now we can discuss the translated Poisson process. In general it is a model for a measured point process that is bound to another underlying unobservable point process. The former one is referred to as \textit{output point process} in \(\XC\) the latter as \textit{input point process} in \(\YC\).

\begin{theorem} \textnormal{(A translated Poisson process is a Poisson process)} \label{th:translated_poisson}\\
Let \(\{N(A)\,|\,A\subset\XC\}\) be a Poisson counting process
with the integrable intensity function \(\lambda(x)\)  in the input space \(\XC\). 
If an input location \(x\in\XC\) is randomly translated to an output location \(y\in\YC\) via a conditional probability density function \(\rho(y|x)\) to form an output point-process \(\{M(B)\,|\,B\subset\YC\}\), then the latter is also 
a Poisson counting process with intensity function
\begin{equation}
  \mu(y) \quad:=\quad \int_{X}\rho(y|x) \, \lambda(x) \, dx.
\end{equation}
\end{theorem}

Following \cite{Snyder91} (Chapter~2) we can compute the log likelihood of a Poisson process as:
\begin{equation} 
     \mathbf{L}_{\XC}(N,\lambda) \quad=\quad  - \int_{\XC} \lambda(x)\,dx \,+\, \int_{\XC} \ln\big(\lambda(x)\big) \, N(dx) \,-\, \ln\big(N(\XC)!\big),
\end{equation}
Where \(N(dx)\) acts like a Dirac counting measure at the observed points, that is
\begin{equation*}
      N(dx) \quad := \quad  
     \begin{cases}
    1, & \text{ if \(x\) is a point where an event has occurred} \\
    0, &\text{ otherwise. }
  \end{cases}
\end{equation*}
With the notation above, the log likelihood for the translated process is:
\begin{equation} 
     \mathbf{L}_{\YC}(M,\mu) \quad=\quad  - \int_{\YC} \mu(y)\,dy \,+\, \int_{\YC} \ln\big(\mu(y)\big) \, M(dy) \,-\, \ln\big(M(\YC)!\big).
\end{equation}
Because \(N(dx)\) evaluates to \(0\) or \(1\) in a discrete manner, the last integral in the log likelihood function can be simplified to a sum \citep[compare,][]{Hohage2016}. If the process observed points at positions \(x_1, ..., x_n\) the integral translates to
\begin{equation*}
\int_{\XC} \ln\big(\lambda(x)\big)\, N(dx) \quad=\quad  \sum_{i=1}^n \ln\big(\lambda(x_i)\big)).
\end{equation*}
This is especially the case in the domain of discrete pixels, where we soon will operate.

\begin{theorem}
\textnormal{(Conditional mean for count-record data)} \label{th:conditional_mean}\\
Let \(\{N(A)\,|\,A\subset\XC\}\) a Poisson process with an integrable intensity function \(\lambda(x)\) and  \(\{M(B)\,|\,B\subset\YC\}\) an output point-process with a translating conditional probability density function \(\rho(y|x)\). Denote the count-record data by \(M = \{\YC_1,...,\YC_{M(\YC)},\,M(\YC)\}\), where \(\YC_m\) is the location of the observed point in the output space. 

Then the conditional expectation given the count-record data \(M\) is given by 
\begin{equation}
    \EE\big[N(A)\,|\,M\big] \quad=\quad 
    \int_{\YC} \frac{\int_A \rho(y|x) \lambda(x) dx}{\int_{\XC} \rho(y|x) \lambda(x) dx} M(dy).
\end{equation}
\end{theorem}

The input space \(\XC\) and output space \(\YC\) may lay in the same space. In this case we can think of points being "shifted around", which is a fair assumption in our case of an image degradation. In this context I will also refer to \(\rho\) as the point spread function (PSF).

\subsection{Algorithms}
So far I have introduced Zernike Polynomials and KL-modes to describe the seeing induced PSF and the log likelihood metric for Poisson processes. Next I will discuss two algorithms that are needed to bind these two blocks together in discrete pixel space. The expectation-maximization (EM) algorithm is an iterative approach to find the (local) maximum likelihood of a statistical process while the Gerchberg-Saxton (GS) algorithm is a rapid computational method to recover the phase of two intensity measurements. 

\subsubsection{Expectation Maximization Algorithm} \label{sec:em}
In this subsection I will briefly discuss the idea of the EM algorithm and translate it to our case of Poisson distributed processes in discrete pixel space. I refer to \cite{Snyder91} (Chapter 3) for details. 

To describe an EM algorithm, the concepts of a complete-data space \(\ZC\) and an incomplete-data space \(\YC\) are used. Further, a  mapping \(h:\ZC \rightarrow\YC\) between them  must exist. In our case the concepts can be translated to:

\begin{itemize}
    \item \textbf{Incomplete-data space} (\(\YC\)): This usually is the space where we take our measurements (i.e., images).
    \item \textbf{Complete-data space} (\(\ZC\)): This is a hypothetical space with no physical meaning. It is needed to make the expectation and maximization steps of the EM algorithm traceable and make the computation needed numerically feasible. Note that the complete-data space depends on a choice and is not unique.
    \item \textbf{Mapping} (\(h\)): The complete-data space is \textit{larger} than the incomplete one, thus we usually have a point-wise many-to-one mapping. Mapping and incomplete data form a constraint (\(h(z) = y\)) that the complete data may have.
\end{itemize}
Let \(\theta\) be a vector of unknown parameters and \(\mathbf{L}(\theta,\YC)\) the corresponding (log) likelihood function. The maximization-algorithm consists of two steps which can be boiled down to:
\begin{itemize}
    \item[\textbf{E}] Define \(Q(\theta,\hat{\theta}\old)\) as the expected value of the complete-data log likelihood function \(\mathbf{L}_{cd}(\theta,\ZC)\), conditioned on the observed parameters in \(\YC\) and using the estimated \(\hat{\theta}\old\) from the iteration before. 
    \item[\textbf{M}] Find the parameters, such that 
    \[\hat{\theta}\new \quad=\quad \underset{\theta}{\textnormal{argmax}} \;\; Q(\theta,\hat{\theta}\old).\]
\end{itemize}
Note that obviously the \textit{E-Step} cannot be done in the fist iteration. Thus, an initial guess or starting condition is needed. 
Also, the definition of \(Q\) as conditioned expected value of a log likelihood function on the unobservable data in the complete-data space \(\ZC\) needs a bit more explanation. The complete-data log likelihood function is given by 
\begin{equation}
    \mathbf{L}_{cd}(\theta,\ZC) \quad:=\quad \ln\big(\rho_{\ZC}(\ZC|\theta)\big),
\end{equation}
where \(\rho_{\ZC}(\ZC|\theta)\) denotes the density of the complete data as function of the parameters. Note that in general the desired \(\hat{\theta}\) does not maximize \(\mathbf{L}_{cd}\). The complete-data log likelihood is a helper function to define the important \(Q\) function as follows: 
\begin{equation}
    Q(\theta,\hat{\theta}\old) \quad:=\quad \EE\old\big[\, \mathbf{L}_{cd}(\theta,\ZC)\,|\, \YC\,\big],
\end{equation}
as conditioned expectation given the incomplete-data \(\YC\) and using the previous \(\hat{\theta}\old\) in the following sense:
\begin{equation} \label{eq:exp_of_log}
\EE\old\big[\, \mathbf{L}_{cd}(\theta,\ZC)\,|\, y=\YC\,\big] \quad=\quad 
     \int_{\ZC} \rho_{\ZC|_{\YC}}(\ZC|_{\YC} \,|\, \hat{\theta}\old) \ln\big(\rho_{\ZC}(\ZC|\theta)\big) d\ZC,
\end{equation}
where 
\[
 \rho_{\ZC|_{\YC}}(\ZC|_{\YC} \,|\, \theta) \quad=\quad 
 \begin{cases}
    \frac{\rho_{\ZC}(\ZC|\theta)}{\rho_{\YC}(\YC|\theta)},& \text{if } \ZC\in\ZC(\YC)\\
    0,              & \text{otherwise}
\end{cases}
\]
and \(\ZC(\YC) = \{z\in\ZC\,|\, h(z) = \YC \}\). The denominator in this equation is the density of the incomplete data that satisfies:
\[ \int_{\ZC(\YC)} \rho_{\ZC}(\ZC|\theta) d\ZC \quad=\quad \rho_{\YC}(\YC|\theta). \]
With this notation we can establish:

\begin{theorem} (EM-Algorithm) \label{th:em-algorithm}\\
Let the sequence \(\{\theta^{(k)}\,|\, k\in\NN\}\) be defined according to the \textbf{E} and \textbf{M} steps above, then the corresponding sequence of 
\(\{\mathbf{L}(\theta^{(k)},\YC)\,|\, k\in\NN\}\) is nondecreasing. 
\end{theorem}
\noindent
For the proof of this theorem I refer to \cite{Snyder91} (p. 137f). \\[5 pt]

\textbf{In our case} of a translated Poisson process the (log) likelihood function \(\mathbf{L}(\theta,\YC)\) is simply the one that I established in Section \ref{sec:poisson}. With the notation of the previous subsection let \(N(\XC)\) be the number of observed points in \(\XC\). We assume that the input space \(\XC\) is partitioned into disjoint subsets \(\XC_1,...,\XC_{N(\XC)}\). For the output-space we assume \(\YC_1,...,\YC_{N(\YC)}\) as partitioning (i.e., think of sensor pixels). 
Then the intensity function on \(\XC\) can be written as
\begin{equation}
    \lambda(x) \quad=\quad \sum_{k=1}^{N(\XC)} \lambda_k \, \one_{x\in \XC_k}.
\end{equation}
Thus, the intensity function is patchwise constant. Since the intensity function \(\lambda\) is strictly non-negative, it is clear that \(\lambda_k \geq 0\) for all \(k = 1,...,K\). Without translations, the maximum likelihood estimate \(\hat{\lambda}\) would be given by 
\begin{equation}
    \hat{\lambda}(x) \;=\; \sum_{k=1}^{N(\XC)} \hat{\lambda}_k \, \one_{x\in \XC_k} \textnormal{, \quad  where }\,
     \hat{\lambda}_k \;:=\; \frac{N(\XC_k)}{\|\XC_k\|},
\end{equation}
with \(\|\XC_k\|\) denotes the size of the \(k\)-th sub region. If we think of the input space as the complete-data space, and ignore all terms in the log likelihood that do not depend on the intensity, we get
\begin{equation}
    \Tilde{\mathbf{L}}_{cd}(\lambda) \quad=\quad 
    - \sum_{k=1}^{N(\XC)} \lambda_k \, \one_{x\in \XC_k} + \sum_{k=1}^{N(\XC)} N(\XC_k) \ln(\lambda_k).
\end{equation}
This is exactly the \(\lambda\) dependent part of the log likelihood of the input process. With that we can translate 
\begin{equation} \label{eq:discrete_q}
    \Tilde{Q}(\lambda,\hat{\lambda}\old) \quad=\quad 
     - \sum_{k=1}^{N(\XC)} \lambda_k \, \one_{x\in \XC_k} + \sum_{k=1}^{N(\XC)} \EE\old\big[ N(\XC_k)\,|\,M\big] \ln(\lambda_k),
\end{equation}
where \(M\) is the measured incomplete data as in Section \ref{sec:poisson}. Applying the conditional mean Theorem \ref{th:conditional_mean}, we get
\begin{equation} \label{eq:discrete_q_expanded}
    \Tilde{Q}(\lambda,\hat{\lambda}\old) \quad=\quad 
     - \sum_{k=1}^{N(\XC)} \lambda_k \, \one_{x\in \XC_k} + \sum_{k=1}^{N(\XC)} 
        \left[ \int_{\YC} \frac{\hat{\lambda}_k\old f_k(y) }{\sum\limits_{l=1}^{N(\XC)}\hat{\lambda}_l\old f_l(y)} M(dy) \right] \ln(\lambda_k),
\end{equation}
with 
\[ f_k(y) \quad := \quad \int_{\XC_k} \rho(y|x) \, dx.\]

\subsubsection{Gerchberg-Saxton Algorithm} \label{sec:gs}
I will now follow \cite{Gerchberg72} for the general presentation of the phase-recovery algorithm for intensity images. The discussed method is an iterative algorithm that depends on the Fourier Transform relation between the image and diffraction plane and can be formulated in a setup and two computation steps:\\

\textbf{Setup:} The input data are the diffraction plane intensity images measured and the amplitudes of the sampled images. The latter ones are proportional to the square roots of the measured intensities. 
Further, a random number generator is used to generate a matrix of random numbers \((a_{i,j})\in[-\pi,\pi]^{n\times m}\) where \(n,m\in\NN\) are the image dimensions. This Matrix is used as initial guess for the phases that correspond to the image amplitudes.\\

\textbf{Step 1 (Image Plane):} Multiply the current phase estimate with the sampled image plane amplitudes. Perform a Fast Fourier Transform (FFT) of the discrete function and compute the phases of the complex result. \\

\textbf{Step 2 (Diffraction Plane):} Multiply the current phase estimate with the sampled diffraction plane amplitudes. Perform a Fast Fourier Transform (FFT) of the discrete function and compute the phases of the complex result. \\

\begin{figure}
  \centering
  \includegraphics[width=430pt]{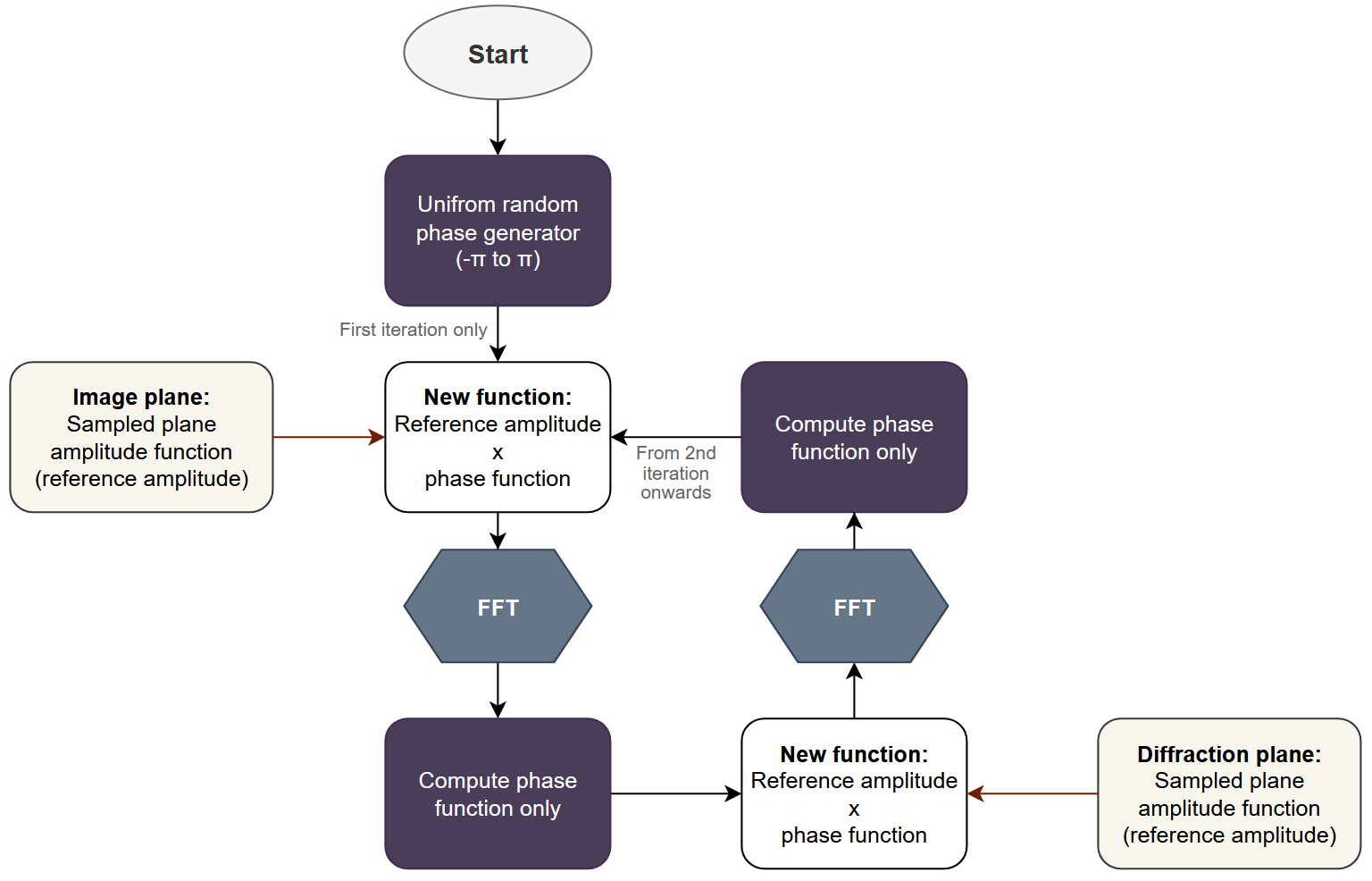} 
\caption[Flowchart of the Gerchberg-Saxton Algorithm]{Flowchart of the Gerchberg-Saxton phase determining algorithm, adapted from \cite{Gerchberg72}.}
\label{fig:gs_scheme}
\end{figure}

The setup step is only done once to start the algorithm. Computing steps 1 \& 2 are repeated per iteration. The basis of this method is the fact, that changing only the amplitude distribution results in a change of both, amplitude and phase, after Fourier Transform.  Figure \ref{fig:gs_scheme} shows a schematic overview of the GS algorithm.  

The algorithm works as follows: Let \(\phi_k\) be the current phase in the image plane (at the beginning this will be the initial random phase guess \((a_{i,j})\)). Let \(u_0, U_0\) denote the measured amplitude in image and diffraction planes respectively. 
In \textbf{step one} we compute 
\begin{equation} \label{eq:gs1}
    U_{k+1} \quad:=\quad |U_{k+1}|\cdot\exp(\im\psi_{k+1}) \quad=\quad  \FFT\big(|u_0|\cdot\exp(\im \phi_k)\big).
\end{equation}
In \textbf{step two} we take the phase part \(\exp(\im\Phi_{k+1})\) and multiply it with the measured amplitude of the diffraction plane \(U_0\). 
From this we get the next phase estimate \(\phi_{k+1}\) via 
\begin{equation} \label{eq:gs2}
u_k \quad:=\quad  |u_k|\cdot\exp(\im\phi_{k}) \quad=\quad  \FFT\big(|U_0|\cdot\exp(\im \psi_{k})\big).
\end{equation}
The amplitude in the diffraction plane is the Fourier Transform of the image plane function.  Let \(N \in \NN\) be the number of all measured points then
Parsevals's Theorem can be directly translated to
\begin{equation}
    \sum_{n=0}^{N-1} |u_0[n]|^2 \quad = \quad \frac{1}{N} \sum_{n=0}^{N-1} |U_0[n]|^2.
\end{equation}
We define two error functions
\begin{equation} \label{eq:gs_fourier_error_0}
    E_k \;:=\; \left(N \cdot \sum_{n=0}^{N-1} \big(|U_k[n]| - |U_0[n]|\big)^2\right)\sqr
\end{equation}
and 
\begin{equation} \label{eq:gs_fourier_error}
    e_k \;:=\; \left(\sum_{n=0}^{N-1} \big(|u_k[n]| - |u_0[n]|\big)^2\right)\sqr.
\end{equation}
We multiplied the diffraction plane error with \(\sqrt{N}\) for compatibility with the image plane error. If this errors become zero it is clear that we have found the correct phase for the points in both planes.

First we note that solutions found by this algorithm are not unique. Any constant arbitrary phase added to any amplitude\(\times\)phase-function, that matches the intensity conditions in image and diffraction plane, will generate a new solution function. Thus, only relative phases of the solution are meaningful. 

In the case where the intensities in both planes are symmetric on the image center, the complex conjugate of the solution would also form a solution. However, this case is neglectable for solar imagery. 
For the error convergence we see that if 
\[E_k \xrightarrow[\;k\;]{} M \in \NN \quad\textnormal{and}\quad e_k \xrightarrow[\;k\;]{} m \in \NN\]
the phase in the image and the diffraction plane must be the same for \(k \geq \hat{k}\in\NN\).
The original paper \cite{Gerchberg72} does not contain a formal proof that the errors always have a finite limit, but the authors claim that this always happens even if the algorithm does not find a desired solution and thus that the algorithm is always finished in finitely many steps. 
For a formal proof of the error convergence I follow \cite{Thimons18}.

\begin{theorem}
\textnormal{ \cite[Theorem 1]{Thimons18}} \\
The diffraction plane and image plane errors do not increase, as the algorithm progresses. In particular the following inequality holds
\begin{equation}
    E_{k+1}\quad\leq\quad e_k\quad\leq\quad E_k \qquad \forall\, k\in\NN_0.
\end{equation}
\end{theorem}
\begin{proof}
Define \(W_k := |U_0|\cdot\exp(\im \psi_{k})\) and \(w_k := |u_0|\cdot\exp(\im \phi_{k-1})\). We start with a proof of the right inequality. 
\begin{align} \label{eq:gs_mult_exp}
    \left(N\sum_{n=0}^{N-1} \big|U_k[n] - W_k[n]\big|^2\right)\sqr  =& \left(N\sum_{n=0}^{N-1} \Big|\exp(\im \psi_{k}) \cdot \big(|U_k[n]| - |W_k[n]|\big)\Big|^2\right)\sqr \nonumber \\
    =& \left(N\sum_{n=0}^{N-1} \Big(\big|\exp(\im \psi_{k})\big| \cdot \big||U_k[n]| - |W_k[n]|\big|\Big)^2\right)\sqr  \nonumber  \\
    =& \left(N\sum_{n=0}^{N-1} \big(|U_k[n]| - |W_k[n]|\big)^2\right)\sqr.
\end{align}
Since \(|W_k[n]| = |U_0[n]|\) this is \(E_k\).
From the step definition in Equation~\ref{eq:gs2} we see that \(\FFT(U_k) = w_k = |u_0|\exp(\im\phi_{k-1})\) and \(\FFT(W_k) = u_k =  |u_k|\exp(\im\phi_{k}) \).
As FFT is a linear operation and with Parseval's Theorem we can transfer the left hand side of the equation above to the image domain. We get 
\begin{equation*}
    E_k \;=\; \left(N\sum_{n=0}^{N-1} \big|U_k[n] - W_k[n]\big|^2\right)\sqr \;=\;\left(\sum_{n=0}^{N-1} \Big|w_k[n] - u_k[n]\Big|^2\right)\sqr. 
\end{equation*}
For the ease of notation let's focus on the \(n\)-th term \(t_k[n] := \big|w_k[n] - u_k[n]\big|^2\) in the sum above.
\begin{align*}
   t_k  =\;&  \Big||u_0|\exp(\im\phi_{k-1}) - |u_k|\exp(\im\phi_{k})\Big|^2  \\
    \stackrel{(*)}{=}\;& \Big||u_0|\cos(\phi_{k-1}) + \im|u_0|\sin(\phi_{k-1})  - |u_k|\cos(\phi_{k})  - \im|u_k|\sin(\phi_{k})\Big|^2  \\
    \stackrel{(**)}{=}&\Big[|u_0|\cos(\phi_{k-1}) - |u_k|\cos(\phi_{k})\Big]^2 + \Big[|u_0|\sin(\phi_{k-1})  -  |u_k|\sin(\phi_{k})\Big]^2  \\
    =\;& \big|u_k\big|^2 + \big|u_0\big|^2 - 2 \big|u_k\big|\big|u_0\big|\big[\cos(\phi_{k-1})\cos(\phi_{k}) + \sin(\phi_{k-1})\sin(\phi_{k})\big] \Big. \\
    =\;& \big|u_k\big|^2 + \big|u_0\big|^2 - 2 \big|u_k\big|\big|u_0\big|\cos(\phi_{k-1}- \phi_{k}) \Big.,
\end{align*}
where \((*)\) is the definition of complex exponentials and \((**)\) the definition of the complex magnitude. Since \(\cos \leq 1\) we get 
\[a \geq a \cdot \cos(\theta), \qquad \forall\, a\in[0,\infty),\;\forall\,\theta\in\RR,\]
regardless of the sign of \(\cos(\theta)\). And therefore
\begin{align*}
    E_k \;\geq\;& \left(\sum_{n=0}^{N-1}  \big|u_0[n]\big|^2 + \big|u_k[n]\big|^2 - 2 \big|u_0[n]\big|\big|u_k[n]\big| \right)\sqr \\
    \;=\;&  \left(\sum_{n=0}^{N-1} \big(|u_k[n]| - |u_0[n]|\big)^2\right)\sqr \\
    \;=\;& e_k.
\end{align*}
Which is what we wanted to show in the first part. For the other inequality we start by multiplying \(|\exp(\im\phi_k)|\) analog to Equation~\ref{eq:gs_mult_exp} and see that
\[e_k \quad = \quad  \left(\sum_{n=0}^{N-1} \big|u_k[n] - w_{k+1}[n]\big|^2\right)\sqr. \]
We use Parseval's Theorem once again and translate this to the diffraction plane:
\[\left(\sum_{n=0}^{N-1} \big|u_k[n] - w_{k+1}[n]\big|^2\right)\sqr  \quad = \quad  \left(N \sum_{n=0}^{N-1} \big|W_k[n] - U_{k+1}[n]\big|^2\right)\sqr. \]
This result can be derived with the same rationals as above to 
\begin{align*}
    e_k =\;& \left(N \sum_{n=0}^{N-1} \big|W_k[n] - U_{k+1}[n]\big|^2\right)\sqr \\
    \geq\;& \left(N \sum_{n=0}^{N-1} \big|U_0[n]\big|^2 + \big|U_{k+1}[n]\big|^2 - 2 \big|U_0[n]\big|\big|U_{k+1}[n]\big| \right)\sqr \\
    =\;& E_{k+1}.
\end{align*}
The combination of both results yields the proof of the theorem.
\end{proof}

\noindent 
\cite{Thimons18} further suggest to start the algorithm with a constant initial phase instead of a random one, given that at least one of the magnitudes is non-centrosymmetric. 

\section{The MOMFBD Algorithm}
\label{sec:momfbd}
In Section~\ref{sec:intro:dround_based} I have introduced the MOMFBD \citep{Schulz93, Miura94, Paxman96, Noort05} algorithm in five steps:
\begin{enumerate}
    \item Compute the PSFs (\(\hat{p}_i\)) from a proposed set of coefficients \(\hat{\alpha}_i\)
    \item De-convolve each frame (\(f_i\)) with the proposed PSF (\(\hat{p}_i\)) to estimate the scene (\(\hat{s}_i\)).
    \item Weight the estimated scenes by the seeing quality (\(\omega_i \)) and compute the common estimate as weighted average (\(\hat{s} = \sum \omega_i \hat{s}_i/N\)).
    \item Compare each raw frame (\(f_i\)) with the common scene estimate  re-convolved with the corresponding PSF (\(\hat{s} \circledast \hat{p}_i\)).
    \item Stop the iteration if the disparity between re-convolved scene estimate and raw frames is satisfactory, else generate a new set of coefficients \(\hat{\alpha}_i\) and start over.
\end{enumerate}
In this section I will discuss the metric to judge the quality of the result and how to estimate the set of coefficients \(\hat{\alpha}_i\) in a way that ensures convergence of the algorithm. 
For this I will follow the argument from \cite{Noort05}, but will collect all referenced results, exemplify missing proofs, and introduce my own, common notation where necessary. 

\subsection{Definitions}
\label{sec:momfbd_def}
In this section let \(\OC = [0,...,O], O\in\NN\) the set of objects, each recorded at a specific wavelength \(\lambda_o\), and \(\KC_o = [0,...,K_o]\) with \(K_o\in\NN\) and \(o\in\OC\) the channels per object. 
Further, let \(\TC = [0,...,T], T\in\NN\) the set of frames that are recorded strictly in sync for all objects. Refer to Figure \ref{fig:data_collection} for a schematic drawing of the data collection process. 
Typically, we deal with one to ten objects and one or two channels each. However, with cameras recording at 400 Hz and a sample time of a minute, the number of processed frames can be quite large. 

To simplify the notation we collapse frame number and channel into a single index \(\JC_o := \TC \times \KC_o\). 
Also, we again first focus on a single object and ignore any \(\OC\) related indices. We will see shortly, why this can be done without a loss in generality.

On the order of the isoplanatic angle, we can assume a space invariance of the PSF. However, for larger fields of view (FOV) this is violated and we have to process subfields of a suitable size separately. For simplicity we assume that the whole frame lies within an isoplanatic region. As Frames are composed of pixels we assume coordinates \mbox{\(v\in\VC\subset\NN^2\subset\RR^2\)} from a discrete region \(\VC\). With this each frame \(f_j, j\in\JC\) can be described as 
\begin{equation}
    f_j(v) \quad=\quad (s \circledast p_j)(v) + n_j(v),
\end{equation}
where \(s\) is the true solar scene (the same for all frames), \(p_j\) the PSF and \(n_j\) a Gaussian noise term. In the Fourier domain this translates to 
\begin{equation} \label{eq:fourier_frame}
    F_j(u) \quad=\quad S(u) \cdot P_j(u) + N_j(u),
\end{equation}
where \(u\in\UC\subset\CC\) is the spatial frequency coordinate in a discrete region \(\UC\) and the upper case symbols denote the Fourier transforms of their lower case equivalents. Per convention, symbols with a hat ( \(\hat{}\) ) denote an estimate of the denoted entity. 

\begin{figure}
  \begin{minipage}[t]{0.48\textwidth} 
  \includegraphics[width=0.47\textwidth]{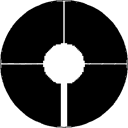} 
  \includegraphics[width=0.47\textwidth]{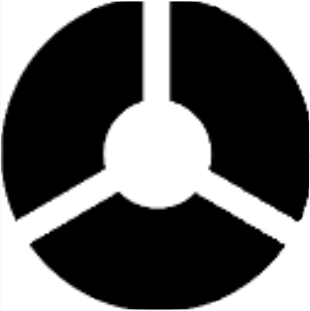} 
  \end{minipage}\quad%
  \begin{minipage}[b]{0.5\textwidth} 
\caption[Telescope pupil examples]{Telescope pupil examples: Negative of the pupil footprints from the \mbox{Sunrise III} \citep[left,][]{lagg25} and HINODE SOT \citep[right,][]{Kosugi07} telescopes in discrete pixel space
\mbox{\(b:\VC\subset\RR^2\rightarrow \{0,1\}\).}}
\label{fig:pupil}
\end{minipage}
\end{figure}

With this we can estimate the optical transfer function (OTF) as Fourier transform of the PSF as
\begin{equation} \label{eq:OTF}
    \hat{P}_j \;=\; \FC\big(|\FC^{-1}(\hat{Q}_j)|^2\big) \quad\textnormal{and}\quad \hat{Q}_j \;:=\; B \cdot \exp(\im \cdot \hat{Z}_j),
\end{equation}
where \(\FC\) is the Fourier transform operator and \(B:\UC\rightarrow\{0,1\}\) is a binary operator that specifies the footprint of the telescope pupil.
See Figure \ref{fig:pupil} for examples of a pupil footprint.

\noindent
With the notation established in Section \ref{sec:zernike} we parametrize the unknown phase as
\begin{equation} \label{eq:zernike}
    \hat{Z}_j \quad=\quad \theta_j + \sum_{m\in\MC} a_{j,m} \cdot \Tilde{\Psi}_{m},
\end{equation}
where \(\Tilde{\Psi}_{m} := \Tilde{\Psi}_{m,o} := \nicefrac{\Psi_m}{\lambda_o}\) denotes the mode normalized with respect to the wavelength of the object and \(\theta_j \in \RR\) is a factor for known channel differences (e.g. de-focus, sub pixel registration, known aberrations). With the normalization by wavelength we can enforce that for different objects \(o\) the coefficients \(a_j\in\RR\) are identical for the same instance \(t\). 

Using only finitely many modes is numerically necessary but comes at the cost of a systematic underestimation of the degradation caused by high-order aberrations in the wavefront.
For the mathematical theory it is irrelevant if \(\Psi_m\) are Karhunen-Loève Modes or plain Zernike Polynomials. In reality, one often uses plain tip-tilt Zernikes (\(m=2,3\), see Table~\ref{tab:Zernikes}) and Karhunen-Lo\`eve Modes for \(m>3\), in order to ease numerical computation and deal with the central obstruction of the telescope pupil.

From Equation~\ref{eq:zernike} each PSF is effectively given by a vector of coefficients and the PSFs for the complete frame set can be described by \(\theta_j\) and a matrix 
\begin{equation} \label{eq:alpha}
    \alpha \quad:=\quad (a_{j,m})_{j\in\JC,m\in\MC}^{\transpose} \in\RR^{M\times T}.
\end{equation}

We established the log likelihood function for a translated Poisson process from \(X\) to \(Y\) in Section~\ref{sec:poisson} as:
\begin{equation} \label{eq:org_log_likelihood}
     \mathbf{L} \quad=\quad  - \int_{Y} \mu(y)\,dy \,+\, \int_{Y} \ln\big(\mu(y)\big) \, M(dy) \,-\, \ln\big(M(Y)!\big),
\end{equation}
with \(\{M(B)\,|\,B\subset Y\} \) the corresponding Poisson-counting process on \(Y\) and original integrable intensity function \(\lambda \in L_1(X)\)  is transferred via a PSF \(\rho\) as: 
\[\mu(y) \quad:=\quad \int_{X}\rho(y|x) \, \lambda(x) \, dx. \] 

We transfer this to our notation with \(M \sim f_j\), \(\rho \sim p\), \(\lambda \sim s\). Further we define \(\hat{d}_j = s\circledast\hat{p}_j\) the translated intensity function, i.e. the noise free part of the recorded frame. Take into account that \(\VC\) is discrete and also, that we want to have the total log likelihood for all transferred processes \(\{f_j\}_{j\in\JC}\) then 
Equation~\ref{eq:org_log_likelihood} translates to
\begin{equation}
   \mathbf{L}(\hat{d}) \quad=\quad \sum_{j\in\JC} \left[ -\sum_{v\in\VC} \hat{d}_j(v) + \sum_{v\in\VC} f_j(v)\cdot\ln\big(\hat{d}_j(v)\big)\right] + T,
\end{equation}
where \(T\) represents a term that does not affect maximization and is therefore ignored in the rest of this section. 
We note that \(\hat{d} := (\hat{d}_j)_{j\in\JC}\) only depends on the parameters of the estimated PSF, and therefore we write \( \mathbf{L}(\alpha, \theta) = \mathbf{L}(\hat{d})\) with \(\theta = (\theta_j)_{j\in\JC}\).
In terms of maximization we are then looking for an \(\Tilde{\alpha}\in\RR^{M\times T}\) and \(\Tilde{\theta} \in \RR^{\JC}\) such that 
\begin{equation} \label{eq:orig_condition}
    \mathbf{L}(\Tilde{\alpha}, \Tilde{\theta}) \;\geq\; \mathbf{L}(\alpha, \theta) \qquad \forall\, \alpha\in\RR^{M\times T} \;\,\forall\, \theta:\in \RR^{\JC}.
\end{equation}
 Since convolution is linear, scaling in the form of \(\sigma s \circledast \frac{\hat{p}_j}{\sigma}\) does not change the likelihood. Thus, we can assume that  
\begin{equation}\label{eg:s_eq_one}
    \sum_{v\in\VC} s(v) \;=\; 1.
\end{equation}

\subsection{Expectation-Maximization} \label{sec:mfbd_em}
Now we will follow \cite{Schulz93} to construct an expectation-maximization (EM) algorithm, introduced in Section \ref{sec:em}, to find such \(\Tilde{\alpha}\) and  \(\Tilde{\theta}\) to satisfy 
Equation~\ref{eq:orig_condition}. As a first step we give an maximization estimate for the \(\theta_j\) parameter of Equation~\ref{eq:zernike}.

\begin{theorem} \label{th:theta} \textnormal{ \cite[compare][Theorem 5.1]{Schulz93}}\\
The Maximum Likelihood (ML) estimate for \(\theta_j\) is 
\begin{equation} \label{eq:theta}
   \Theta_j := \big|\exp(\im\Tilde{\theta}_j)\big|^2 \quad=\quad \frac{\sum\limits_{v\in\VC} f_j(v)}{G_j}
\end{equation}
with 
\begin{equation}
    G_j \quad:=\quad \sum_{v\in\VC}\left|\sum\limits_{u\in\UC} \, B(u) \cdot \exp\Big(\im \cdot \sum\limits_{m\in\MC} a_{j,m}\cdot\Psi_m(u)\Big) \cdot \exp(-\im2\pi vu)\right|^2.
\end{equation}
\end{theorem}

\noindent
\begin{proof}
From Equations~\ref{eq:OTF} and \ref{eq:zernike} we can write \(\hat{p}_j\) as 
\begin{align} \label{eq:def_phat}
    \hat{p}_j(v) &=\quad \left|\int_{u\in\UC} \,B(u) \cdot \exp(\im\theta_j)  \cdot \exp\Big(\im \cdot \sum\limits_{m\in\MC} a_{j,m}\cdot\Psi_m(u)\Big) \cdot \exp(-\im2\pi vu) \; du\right|^2 \nonumber\\ 
    &=: \quad  \big|\exp(\im\theta_j)\big|^2 \cdot g_j(v),
\end{align}
using discrete Fourier Transform on \(\UC\). We also note that \(G_j=\sum_{v} g_j(v-\tau)\) is not a function in \(\tau\). With that we can re-write the log likelihood as
\begin{align}
     \mathbf{L}(\alpha, \Tilde{\theta}) &=  \sum_{j\in\JC} \left[ -\sum_{v\in\VC}\sum_{\tau\in\VC} \Theta_j\hat{g}_j(v-\tau)s(\tau) + \sum_{v\in\VC} f_j(v)\cdot\ln\big(\Theta_j \cdot \big(s\circledast{}g_j\big)(v)\big)\right] \nonumber\\
     &= - \sum_{j\in\JC} \Theta_j \left(\sum_{v\in\VC} \sum_{\tau\in\VC} \hat{g}_j(v-\tau) s(\tau) \right) \nonumber\\ 
     &\qquad + \sum_{j\in\JC} \sum_{v\in\VC} f_j(v) \ln(\Theta_j) + \sum_{j\in\JC} \sum_{v\in\VC} f_j(v)\cdot \ln\big(\big(s\circledast{}g_j\big)(v))\big) \nonumber \\
     &= - \sum_{j\in\JC} \Theta_j G_j + \sum_{j\in\JC} \sum_{v\in\VC} f_j(v) \ln(\Theta_j) + \sum_{j\in\JC} \sum_{v\in\VC} f_j(v)\cdot \ln\big(\big(s\circledast{}g_j\big)(v)\big).
\end{align}
Partial differentiation for any given \(k\in\JC\) yields 
\begin{equation}
    \frac{\partial\mathbf{L}(\alpha, \Tilde{\theta})}{\partial\Tilde{\theta}_k}   \quad=\quad -G_k +  \frac{\sum\limits_{v\in\VC} f_j(v)}{\Theta_k}. \quad 
\end{equation}
Thus, the derivative is zero for any \(\Theta_k\) as defined in Equation~\ref{eq:theta} and the second derivative is negative, since all \(f_j\) are strictly non-negative.
\end{proof}

As a corollary we see that the optimal choice of \(\theta\) only depends on \(\alpha\), hence we will omit the parameter for the likelihood function and write 
\begin{equation}
      \mathbf{L}(\alpha) \quad:=\quad \mathbf{L}(\alpha,\theta).
\end{equation}
Next we want to give an iterative expectation-maximization (EM) algorithm to choose \(\alpha\in\RR^{M\times T}\).

\begin{theorem} \label{th:GEM} \textnormal{\cite[compare][Theorem 5.2]{Schulz93}}\\
A sequence of estimates of the phase parameters updated as 
\begin{equation} \label{eq:def_shat}
    \hat{s}\new(v) \quad=\quad \sigma \cdot \hat{s}\old(v) \sum_{j\in\JC}\sum_{\tau\in\VC} \frac{\hat{p}_j\old(\tau-v)}{\hat{d}_j\old(\tau)}f_j(\tau),
\end{equation}
where \(\sigma := \left(\sum_{j\in\JC}\sum_{v\in\VC} f_j(v)\right)^{-1}\) is the scaling factor to ensure Equation~\ref{eg:s_eq_one} holds and \(\hat{p}_j\new\) chosen such that
\begin{equation} \label{eq:condition_to_choose_pj}
\sum_{v\in\VC} \xi(\hat{p}_j\old(v)) \cdot \ln\big(\hat{p}_j\new(v)\big) \quad \geq \quad \sum_{v\in\VC} \xi(\hat{p}_j\old(v)) \cdot \ln\big(\hat{p}_j\old(v)\big),
\end{equation}
with 
\begin{equation} \label{eq:def_xi}
\xi(\hat{p}_j(v)) \quad:=\quad \hat{p}_j(v) \left[\sum_{\tau\in\VC}  \frac{\hat{s}\old(\tau-v)}{\hat{d}_j\old(\tau)}f_j(\tau) \right]
\end{equation}
is an EM algorithm with
\begin{equation} \label{eq:req_update}
    \mathbf{L}(\alpha\new) \quad\geq\quad \mathbf{L}(\alpha\old).
\end{equation}
\end{theorem}

\begin{proof} For the same scaling argument as above, we can restrict our estimates of the true solar scene \(\hat{s}_j\) to the set \(\AC := \{\,s\,| \sum_{v\in\VC}s(v) = 1\}\). 
Further we express the observed data \(\{f_j(v)\}\) via an arbitrary set of Poisson-distributed random variables \(\Tilde{f}_j\) via 
\begin{equation} \label{eq:ftilde}
    f_j(v) \quad=\quad \sum_{\tau\in\VC} \Tilde{f}_j(\tau|v).
\end{equation}
The expected value of this data is 
\begin{equation} \label{eq:exp_tilde_f}
    \EE[\Tilde{f}_j(\tau|v)] \quad=\quad s(\tau) \cdot p_j(\tau-v).
\end{equation}
Hence,
\begin{equation} \label{eq:exp_f}
    \EE[f_j(v)] \quad=\quad \sum_{\tau\in\VC}  \EE[\Tilde{f}_j(\tau|v)] \quad=\quad \sum_{\tau\in\VC}  s(\tau) \cdot p_j(\tau-v) \quad=\quad d_j(v).
\end{equation}
The new set does not reveal any new physical meaningful information.  Important is that it is statistically consistent with the observed data as a mathematical model. With this we now can apply the terminology of Section \ref{sec:em} of complete data \(\{\Tilde{f}_j(\tau|v)\} = \ZC\) and incomplete data \(\{f_j(v)\} = \YC\). The many-to-one relation of complete to incomplete data \(h\) is given by Equation~\ref{eq:ftilde}.
As above we can derive a log likelihood function for the complete data ignoring all terms, that do not depend on \(s\) or \(p_j\)
\begin{align}
    \Tilde{\mathbf{L}}(s, p_j) \;=& - \sum_{j\in\JC}\sum_{v\in\VC}\sum_{\tau\in\VC} s(\tau) p_j(\tau-v) \nonumber\\
    & \quad + \sum_{j\in\JC}\sum_{v\in\VC}\sum_{\tau\in\VC} \Tilde{f}_j(\tau|v) \ln\big( p_j(\tau-v)s(\tau) \big).
\end{align}
Now we follow the EM-algorithm description from Section \ref{sec:em} and define 
\begin{equation}
    Q(s, p_j) \quad:=\quad \Tilde{Q}(s, p_j,\hat{s}\old,\hat{p}_j\old) \quad=\quad \EE\old\Big[\Tilde{\mathbf{L}}(s, p_j) \Big|\Big. \{f_j(v)\}\Big],
\end{equation}
where \(\EE\old[\,\cdot\,|\{f_j(v)\}]\) denotes a conditioned expectation given the observed Poisson-distributed data \(\{f_j(v)\}\) and using the previous functions \(\hat{s}\old\) and \(\hat{p}_j\old\) for computing \(f_j\) and \(\Tilde{f}_j\). With this notation we have changed the requirement for updating the estimates from Equation~\ref{eq:req_update} to 
\begin{equation}
    Q(\hat{s}\new, \hat{p}_j\new) \quad\geq\quad Q(\hat{s}\old, \hat{p}_j\old).
\end{equation}
From Equation~\ref{eq:discrete_q} we get a concrete formulation of \(Q\) and since \(s,\hat{s}_j \in \AC\) we can remove one \(s\) from the formula. Further, we separate \(Q\) to a sum of two independent functions of \(s\) and \(p\) respectively:
\begin{align}
Q(s, p_j)    \;=& - \sum_{j\in\JC}\sum_{v\in\VC} p_j(v) \nonumber\\
    & \quad + \sum_{j\in\JC}\sum_{v\in\VC}\sum_{\tau\in\VC} \EE\old\big[\,\Tilde{f}_j(\tau|v)\,\big|\,f_j(\tau)\,\big] \ln\big( p_j(\tau-v)s(v) \big) \nonumber \\
    =& \underbrace{- \sum_{j\in\JC}\sum_{v\in\VC} p_j(v) + \sum_{j\in\JC}\sum_{v\in\VC}\sum_{\tau\in\VC} \EE\old\big[\,\Tilde{f}_j(\tau|v)\,\big|\,f_j(\tau)\,\big] \ln\big( p_j(\tau-v) \big)}_{=:\,Q_p(p_j)}  \nonumber \\
    & \underbrace{\quad + \sum_{v\in\VC} \left[\sum_{j\in\JC}\sum_{\tau\in\VC} \EE\old\big[\,\Tilde{f}_j(\tau|v)\,\big|\,f_j(\tau)\,\big]\right] \ln\big(s(v) \big)}_{=:\,Q_s(s)} \nonumber\\
    =& \;Q_p(p_j) + Q_s(s).
\end{align}
Now, since all \(\Tilde{f}\) are mutually independent random Poisson distributed variables, we see from Equations~\ref{eq:discrete_q_expanded}, \ref{eq:exp_tilde_f},  and \ref{eq:exp_f} that 
\begin{equation} \label{eq:exp_tilde_f_cond}
     \EE\old\big[\,\Tilde{f}_j(\tau|v)\,\big|\,f_j(\tau)\,\big] \;=\;  \frac{\EE\old[\Tilde{f}_j(\tau|v)]}{\EE\old[f_j(\tau)] }  f_j(\tau)
     \;=\; \frac{p_j\old(\tau-v)\hat{s}\old(v)}{d_j\old(\tau) } f_j(\tau).
\end{equation}
By summing over all observed points and all frames we get 
\begin{align}
    \sum_{j\in\JC}\sum_{\tau\in\VC} \EE\old\big[\,\Tilde{f}_j(\tau|v)\,\big|\,f_j(\tau)\,\big] \;=\;& 
            \hat{s}\old(v) \sum_{j\in\JC}\sum_{\tau\in\VC} \frac{p_j\old(\tau-v)}{d_j\old(\tau) } f_j(\tau) \nonumber \\
            \;=\;& \frac{\hat{s}\new(v)}{\sigma}.
\end{align}
This allows us to re-write \(Q_s\) as 
\begin{align}
    Q_s(\hat{s}) \;=\;& \sum_{v\in\VC} \left[\sum_{j\in\JC}\sum_{\tau\in\VC} \EE\old\big[\,\Tilde{f}_j(\tau|v)\,\big|\,f_j(\tau)\,\big]\right] \ln\big(\hat{s}v) \big) \nonumber \\
    =\;& \sum_{v\in\VC} \frac{\hat{s}\new(v)}{\sigma} \ln\big(\hat{s}(v) \big).
\end{align}
For this reformulation it is easy to apply the Laplace operator 
\begin{equation}
    \nabla O_s(\hat{s}) \;=\; \left( \frac{\hat{s}\new(v)}{\sigma \,\hat{s}(v)} \right)_{v\in\VC}.
\end{equation}
Since all scene function estimates must be positive, it is obvious that \(\nabla O_s \neq 0\) and \mbox{\(\Delta Q_s < 0\)} for all \(\hat{s}\). 
Define  
\begin{equation}
\Lambda(\hat{s},\lambda) \quad:=\quad Q_s(\hat{s}) + \lambda \left(\sum_{v\in\VC} \hat{s}(v)\right) - \lambda,
\end{equation}
for \(\lambda\in\RR\). For critical points of \(\Lambda\) we get the conditions
\begin{equation} \label{eq:lap_op_cond1}
    \frac{\partial\Lambda}{\partial \hat{s}(w)} \;=\; \frac{\hat{s}\new(w)}{\sigma \,\hat{s}(w)} + \lambda\;=\; 0 
\end{equation}
and
\begin{equation} \label{eq:lap_op_cond2}
    \frac{\partial\Lambda}{\partial \lambda} \;=\; \sum_{v\in\VC} \hat{s}(v) - 1 = 0.
\end{equation}
For \(\lambda\neq0\) Condition~\ref{eq:lap_op_cond1} can be rearranged to 
\begin{equation} \label{eq_sw}
    \hat{s}(w) \quad=\quad -\frac{\hat{s}\new(w)}{\sigma\lambda}.
\end{equation}
Plugging that into Condition~\ref{eq:lap_op_cond2} yields 
\begin{alignat}{2}
        & 0 &\;=\;& 1 +  \sum_{v\in\VC} \frac{\hat{s}\new(v)}{\sigma\lambda} \nonumber \\
\Leftrightarrow\qquad & - \lambda &\;=\;& \sum_{v\in\VC} \frac{\hat{s}\new(v)}{\sigma} \quad =\quad \frac{1}{\sigma}.
\end{alignat}
Now we plug this result into Equation~\ref{eq_sw} and found the maximum at \(s\new\).

\noindent
Revisiting Equation~\ref{eq:exp_tilde_f_cond} for \(\Tilde{f}_j(\tau|\tau-v)\) we get
\begin{align}
     \sum_{\tau\in\VC} \EE\old\big[\,\Tilde{f}_j(\tau|\tau-v)\,\big|\,f_j(\tau)\,\big] 
     \;=\;&  \sum_{\tau\in\VC}  \frac{\hat{p}_j\old(v)\hat{s}\old(\tau-v)}{d_j\old(\tau) } f_j(\tau) \nonumber\\
     =\;& \hat{p}_j\old(v) \sum_{\tau\in\VC}\frac{\hat{s}\old(\tau-v)}{d_j\old(\tau) } f_j(\tau) \nonumber \\
     =\;& \xi(\hat{p}\old_j(v)).
\end{align}
With this  above we can rearrange \(Q_p\) in the form below:
\begin{align}
    Q_p(p_j) \;=\quad& \sum_{j\in\JC}\left[ - \sum_{v\in\VC} p_j(v) + \sum_{\tau\in\VC}\sum_{v\in\VC}  \EE\old\big[\,\Tilde{f}_j(\tau|v)\,\big|\,f_j(\tau)\,\big]  \ln\big( p_j(\tau-v) \big)\right] \nonumber \\
    \stackrel{v=\tau-w}{=}& \sum_{j\in\JC} \left[ - \sum_{v\in\VC} p_j(v) +  \sum_{\tau\in\VC}\sum_{w\in\VC}  \EE\old\big[\,\Tilde{f}_j(\tau|\tau-w)\,\big|\,f_j(\tau)\,\big]  \ln\big( p_j(w) \big)\right] \nonumber \\
    =\quad& \sum_{j\in\JC}\left[ - \sum_{v\in\VC} p_j(v) + \sum_{w\in\VC} \xi(\hat{p}\old_j(w)) \ln\big(p_j(w)\big) \right].
\end{align}
We see that for \(\Tilde{p}_j := \xi(\hat{p}_j\old)\) we find a maximum, since 
\[
\frac{\partial Q_p(p_j)}{\partial p_l} \quad=\quad - \sum_{v\in\VC} 1 + \sum_{v\in\VC} \frac{\xi(\hat{p}\old_l(v))}{p_l(v)},
\]
and the second derivative is always negative. 
\end{proof}

\noindent \\
The theorem gave us the flexibility to "choose" the PSF (via \(\hat{p}_j\)) under a specific condition. 
Next, we want to give an explicit algorithm to construct \(\hat{p}\new_j\). We again follow \cite{Schulz93} for this.
The idea is to use the Gerchberg-Saxton algorithm, as introduced in Section \ref{sec:gs}. However, the GS algorithm requires
measurements in the image and diffraction plane. We therefore need to show how this algorithm is applicable here. 

Revisiting Equation~\ref{eq:def_phat}, which gives \(\hat{p}_j\), we can think of 
\(\{\sqrt{\hat{p}_j(v)}\,|\,v\in\VC\}\) as image- and of \(\{B(u)\,|\,u\in\UC\}\) as corresponding diffraction-plane magnitudes.
Assume \(\{\sqrt{\xi(\hat{p}_j(v))}\,|\, v\in\VC\}\), with \(\xi(\cdot)\) as defined in 
Equation~\ref{eq:def_xi}, to be a measurement of the image domain magnitude.
Remember the definition of the phase function in Equation~\ref{eq:zernike}:
\[
    \hat{Z}_j(u) \quad:=\quad \theta_j +  \sum_{m\in\MC} a_{j,m} \cdot \Psi_{m}(u),
\]
where \(\Psi\) denotes the Zernicke or KL modes. As already noticed, \(\hat{Z}_j(u)\) is given by the vector \((a_{j,m})\).
To construct \(\hat{p}_j\new\), we need to construct \(\hat{Z}_j\new\) or \((a_{j,m})\new\) respectively. 
The GS iteration step to do so could then condensed into one equation:
\begin{equation} \label{eq:gs_update}
    \hat{Z}_j\new(u) \quad=\quad \phase\left(
    \FC^*\left[ \sqrt{\xi(\hat{p}\old_j(v))} \cdot \exp\left(\im \cdot \phase\big[\Tilde{p}_j\old(v)\big]\right) \right]
    \right).
\end{equation}
Where \(\phase(x)\) denotes the phase of a complex \(x\) and 
\[
    \Tilde{p}^{\_\_}_j(v) \quad:=\quad \FC\Big( B(u)\,\exp\big(\im\hat{Z}^{\_\_}_j(u)\big)\Big).
\]
The Image plane convergence criteria, derived from the image plane error of the GS algorithm as given in Equation~\ref{eq:gs_fourier_error}, now translates to 
\begin{align}
    \sum_{v\in\VC} \left| \Big[ \xi\big(\hat{p}_j\old(v)\big)\Big]\sqr\right.-&\left. \Big[ \hat{p}\old_j(v) \Big]\sqr \right|^2 
         - \sum_{v\in\VC} \left| \Big[ \xi\big(\hat{p}_j\old(v)\big)\Big]\sqr - \Big[ \hat{p}\new_j(v) \Big]\sqr \right|^2 \nonumber\\
         =\; & 2 \sum_{v\in\VC} \Big[\xi\big(\hat{p}_j\old(v)\big)\Big]\sqr \, \left(  \Big[ \hat{p}\new_j(v) \Big]\sqr -  \Big[ \hat{p}\old_j(v) \Big]\sqr \right) \nonumber \\ 
         =\; & 2 \sum_{v\in\VC} \Big[\xi\big(\hat{p}_j\old(v)\big)\Big]\sqr \,  \Big[ \hat{p}\new_j(v) \Big]\sqr \, \left( 1 -  \left[\frac{\hat{p}_j\old(v)}{\hat{p}_j\new(v)} \right]\sqr \right) \nonumber \\
         \geq\;& 0.
\end{align}
We can also reformulate the condition on the PSF from the EM algorithm, given in 
Equation~\ref{eq:condition_to_choose_pj}, to
\begin{align}
\sum_{v\in\VC} \xi(\hat{p}_j\old(v)) \cdot \ln\big(\hat{p}_j\new(v)\big) -& \sum_{v\in\VC} \xi(\hat{p}_j\old(v)) \cdot \ln\big(\hat{p}_j\old(v)\big) \nonumber\\ 
    =\;& \sum_{v\in\VC} \xi(\hat{p}_j\old(v)) \;\ln\left( \frac{\hat{p}_j\new(v)}{\hat{p}_j\old(v)} \right) \nonumber\\ 
    =\;& 2 \sum_{v\in\VC} \xi(\hat{p}_j\old(v)) \;\ln\left( \left[\frac{\hat{p}_j\new(v)}{\hat{p}_j\old(v)} \right]\sqr \right) \nonumber \\
    \stackrel{(*)}{\geq}\;& 2 \sum_{v\in\VC} \xi(\hat{p}_j\old(v)) \;\left(1 -  \left[\frac{\hat{p}_j\old(v)}{\hat{p}_j\new(v)} \right]\sqr \right) \nonumber \\
    \stackrel{(!)}{\geq}\;& 0.
\end{align}
Here \((*)\) holds since \(\ln(x) \leq x-1\) and \(r\ln(x) = \ln(x^r)\) imply \(\ln(x) \geq 1-x^{-1}\) and inequality \((!)\) is the reformulation of the condition. Because of the similarity of these conditions, we can expect that an update of \(\hat{p}_j\) according to Equation~\ref{eq:gs_update} fulfills the requirement on the EM algorithm in Equation~\ref{eq:condition_to_choose_pj}. We therefore define 
\begin{equation}
    \hat{p}_j\new \quad := \quad 
     \begin{cases}
   \Tilde{p}_j(v), &\text{if } \EC(\Tilde{p}_j) \geq \EC(\hat{p}_j\old) \\
   \hat{p}\old_j(v), &\text{ else,}
  \end{cases}
\end{equation}
with
\[ \EC(x) :=  \sum_{v\in\VC} \xi(\hat{p}_j\old(v)) \cdot \ln\big(x(v)\big).\] 
For the cases where we keep \(\hat{p}_j\old\) we rely on the update of \(\hat{s}\new\), as defined in Equation~\ref{eq:def_shat}, to maximize the likelihood function in the current iteration.

\subsection{Object free error metric}
As a last step we now need to reduce the metric to a form that does not depend on the true solar scene, 
which cannot be known, but only on estimated quantities and the recorded data frames as constructed above. 
Let's remind ourselves that the recorded images can be described as 
\mbox{\(f_j(v) \;=\; (s \circledast p_j)(v) + n_j(v)\)}, where \(p_j\) is the PSF, \(s\) the true solar scene and \(n\) a Gaussian noise term.
Thus, we are in the case of Additive White Gaussian Noise (AWGN) where
maximizing the likelihood (as we did in the previous chapter) is equivalent 
to finding the "minimum distance" (as means square error (MSE)\footnote{Let \(n\in\NN\) and \(X_0,\ldots,X_n\in\RR\) be the true data and \(\hat{X}_0,\ldots,\hat{X}_n\in\RR\) the corresponding predictions, then
\[\textnormal{MSE} \;:=\; \frac{1}{n}\sum_{i=0}^{n}\big(X_i - \hat{X}_i\big)^2.\]
}) to the truth data. 

As convolution is easier described and understood in Fourier space we use the translated equation \(F_j = S\,P_j + N_j\) and minimize the equivalent error metric in the frequency domain. 
\begin{equation} \label{eq:Lj}
L(\alpha_j) \quad:=\quad \sum_{u\in\UC} \Big| F_j(u) - \big(\hat{S}\hat{P}_j\big)(u)\Big|^2.
\end{equation}

We use the Wiener deconvolution to reconstruct the solar scene. The corresponding formula is well known in the literature
\citep[see for instance]["Digital Image Processing Using Matlab", p. 173f]{Gonzales03}:
\begin{equation}
    \hat{S}_j \quad=\quad \frac{\hat{P}_j^* \EE[S^*S] F_j}{|\hat{P}_j|^2 \EE[S^*S] + \EE[N_j^*N_j]},
\end{equation}
where the superscript asterisk (\(^*\)) denotes complex conjugation. As we assumed \(N_j\) to be white noise we can consider the spectral densities of \(S\) and \(N_j\) to be constant. Define \(\gamma := \nicefrac{\EE[N_j^*N_j]}{\EE[S^*S]}\), then we can re-write the above equation as
\begin{equation} \label{eq:wiener_s}
    \hat{S}_j \quad=\quad \frac{\hat{P}_j^* F_j}{|\hat{P}_j|^2 + \gamma}.
\end{equation}
The main assumption of the MOMFBD algorithm is, that all frames from all objects and all time steps show the same solar scene, degraded by different realizations of a random PSF. And indeed, the main work in Section \ref{sec:mfbd_em} was to show that we can maximize the likelihood for the same object for all frames simultaneously. Thus, using the linearity of the convolution (see Equation~\ref{eq:psf}), we can write
\begin{equation}
    \frac{1}{M}  \frac{\sum\limits_{j\in\JC} \hat{P}_j^* F_j}{\sum\limits_{j\in\JC}|\hat{P}_j|^2 + \gamma} \;=\; \hat{S} \;=\; \sum_{j\in\JC} \frac{\hat{S}}{M}
        \;=\; \frac{1}{M} \sum_{j\in\JC} \frac{\hat{P}_j^* F_j}{|\hat{P}_j|^2 + \gamma},
\end{equation}
where \(M := \sharp(\JC\times\KC)\). Plugging this into Equation~\ref{eq:Lj} yields
\begin{equation}
L(\alpha_j) \quad=\quad  \sum_{u\in\UC}  \left| F_j(u) - \frac{\left[\left(\sum\limits_{i\in\JC} \hat{P}_i^* F_i\right) \hat{P}_j\right](u) }{\sum\limits_{i\in\JC}|\hat{P}_i(u)|^2 + \gamma} \right|^2.
\end{equation}
As \(M\) is constant for any given data set, it has no effect on minimization and was ignored. To expand the \(|\cdot|^2\) expression we have
\[|a-b|^2 \quad=\quad |a|^2 + |b|^2 - 2\textnormal{Re}(ab^*)\qquad \forall\;a,b\in\CC.\]
With \(a = F_j(u)\) the interesting parts are the \(b\) terms. For readability let's ignore the point-wise evaluation for a moment. For \(ab^*\) we get:
\begin{equation*}
   F_j \left( \frac{\left(\sum\limits_{i\in\JC} \hat{P}_i^* F_i\right) \hat{P}_j }{\sum\limits_{i\in\JC}|\hat{P}_i|^2 + \gamma}  \right)^* 
   \quad=\quad 
   \frac{\left(\sum\limits_{i\in\JC} \hat{P}_i F_i^*\right) \hat{P}_j^*F_j }{\sum\limits_{i\in\JC}|\hat{P}_i|^2 + \gamma}.
\end{equation*}
Since \((x+y)^* = x^*+y^*\) and \((xy)^* = x^*y^*\) for all \(x,y\in\CC\). Now if we sum over \(j\in\JC\) we see that 
\begin{equation}
    \sum_{j\in\JC} \frac{\left(\sum\limits_{i\in\JC} \hat{P}_i F_i^*\right) \hat{P}_j^*F_j }{\sum\limits_{i\in\JC}|\hat{P}_i|^2 + \gamma} 
    \quad=\quad 
    \frac{\left(\sum\limits_{i\in\JC} \hat{P}_i F_i^*\right) \left(\sum\limits_{j\in\JC}\hat{P}_j^*F_j\right) }{\sum\limits_{i\in\JC}|\hat{P}_i|^2 + \gamma} 
    \quad=\quad 
    \frac{\left|\sum\limits_{i\in\JC} \hat{P}_i F_i^*\right|^2 }{\sum\limits_{i\in\JC}|\hat{P}_i|^2 + \gamma}. 
\end{equation}
Which is already real. For the \(b^*b\) part we get
\begin{equation*}
   \left( \frac{\left(\sum\limits_{i\in\JC} \hat{P}_i^* F_i\right) \hat{P}_j }{\sum\limits_{i\in\JC}|\hat{P}_i|^2 + \gamma}  \right)^* \left( \frac{\left(\sum\limits_{i\in\JC} \hat{P}_i^* F_i\right) \hat{P}_j }{\sum\limits_{i\in\JC}|\hat{P}_i|^2 + \gamma}  \right)
   \quad=\quad 
   \frac{\left|\sum\limits_{i\in\JC} \hat{P}_i F_i^*\right|^2 |\hat{P}_j|^2 }{\left(\sum\limits_{i\in\JC}|\hat{P}_i|^2 + \gamma\right)^2.} 
\end{equation*}
Summing over \(j\in\JC\) leads to 
\begin{equation*}
   \frac{\left|\sum\limits_{i\in\JC} \hat{P}_i F_i^*\right|^2 \left(\sum\limits_{j\in\JC}  |\hat{P}_j|^2\right) }{\left(\sum\limits_{i\in\JC}|\hat{P}_i|^2 + \gamma\right)^2} 
   \quad\stackrel{(*)}{\leq}\quad 
   \frac{\left|\sum\limits_{i\in\JC} \hat{P}_i F_i^*\right|^2 \ }{\sum\limits_{i\in\JC}|\hat{P}_i|^2} 
   \quad\stackrel{(**)}{\leq}\quad 
   \frac{\left(\sum\limits_{i\in\JC} |\hat{P}_i|^2\right) \cdot  \left(\sum\limits_{i\in\JC} |F_i|^2\right)}{\sum\limits_{i\in\JC}|\hat{P}_i|^2} 
   \quad=\quad \sum_{i\in\JC} |F_i|^2.
\end{equation*}
Where \((*)\) holds since \(\gamma\geq0\) and \((**)\)  is a special case of the Cauchy-Schwarz inequality. Putting it all together we get
\begin{equation}
    \mathbf{L}(\alpha) \;=\; \sum_{j\in\JC} L(\alpha_j) \quad\leq\quad 
    \sum_{u\in\UC} \left[ 2\, \sum_{j\in\JC} |F_j(u)|^2 \,-\, 2\,\frac{\left|\sum\limits_{j\in\JC}(F_j^*\hat{P}_j)(u)\right|^2}{\sum\limits_{j\in\JC}|\hat{P}_j(u)|^2 + \gamma} \right].
\end{equation}
As the constant factor of \(2\) does not effect the Minimization, it can be ignored. And we end up with our final error metric \citep{Noort05}:
\begin{equation}
    \mathbf{L}(\alpha) \quad=\quad \sum_{u\in\UC} \left[  \sum_{j\in\JC} |F_j(u)|^2 - \frac{\left|\sum\limits_{j\in\JC}(F_j^*\hat{P}_j)(u)\right|^2}{\sum\limits_{j\in\JC}|\hat{P}_j(u)|^2 + \gamma} \right].
\end{equation}

\begin{figure}
  \centering
  \includegraphics[width=1\textwidth]{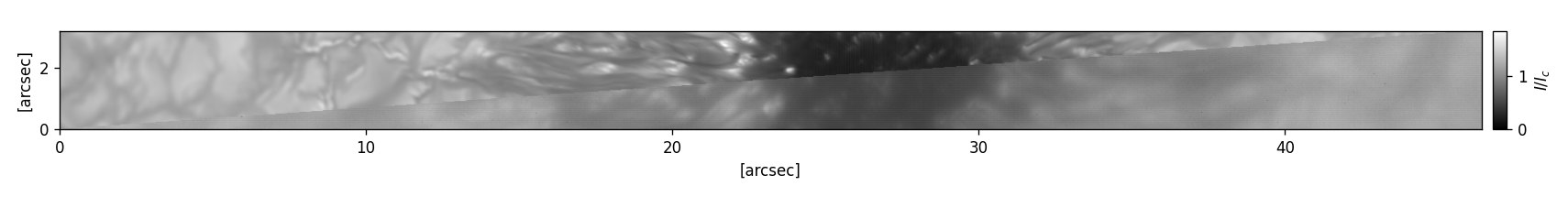} 
    \caption[Comparison of restored and raw context images]{
        Comparison of a raw input frame and the MOMFBD reconstructed result. In the lower-right triangle of the image an excerpt of one original frame is shown. Small-scale features are not visible and the complete information is smeared out. In the upper-left triangle an excerpt of the corresponding restored image is shown. In both cases the intensity is scaled to average quiet Sun continuum intensity (\(I_c\)), the data were recorded with \fisssp{}.
    }
\label{fig:momf_result}
\end{figure}

Figure~\ref{fig:momf_result} presents an example comparison of a single input frame with the corresponding MOMFBD restored frame using a frame recording time-span of 1.5 seconds (540 frames) and 90 KL modes from two channels and a single object. The frames where sub-divided in patches that lie within the isoplanatic size and later re-assembled to a single image. The frames have been recorded with \fisssp{} (see next Chapter).

\section{Spatial Restoration of Spectra}
\label{sec:spec_restore}
In this section I will give a brief mathematical outlook on the restoration algorithm for solar spectra as described in \cite{Noort17}. 
Due to the broadband pre-filter in our data collection setup (see Figure \ref{fig:spectral_data_collection}), we can assume that the PSF is the same for all wavelengths covered by the spectral sensor. 

Denote the coordinates on our context sensor by \((x,y)\in\NN_0^2\) and the coordinates on the Sun by \(u,v\).
For spectral data the "scene" or intensity function \(s^{\textnormal{spectral}}\) takes three arguments: The spatial coordinates \(u,v\) and the spectral coordinate \(\lambda\). The two dimensional (spatial) scene (that we have used in the previous subsection) can be written as 
\[
s^{\textnormal{spatial}}(u,v) \quad = \quad \sum_{\lambda} s^{\textnormal{spectral}}(u,v,\lambda).
\]
Therefore we will drop the superscripts "spectral" and "spatial" for notation purposes and rely on context and/or the number or arguments to differentiate between those two. 

Let \(j\in\JC=\KC\times\TC\) be again the index of frames (compare Section \ref{sec:momfbd_def}). Let \(p_j(x,y|u,v)\) denote the PSF of the \(j\)-th frame, e.g. as extracted from MOMFBD. We can describe the intensity value for each point \((x,y)\) in every frame \(f_j\) of our context imager as 
\begin{equation} 
    f_j(x,y) \quad=\quad \sum_{u,v} p_j(x,y|u,v) \, s(u,v).
\end{equation}
Assume the slit to be perfectly aligned with the \(y\) axis of the context imager at position \(x=x\slit\).  
The Zernike or KL modes can be scaled to the wavelength, yielding the same coefficients in the description of the PSF for all wavelength points (see Section \ref{sec:momfbd}).
Without loss of generality we can therefore assume the PSF to be the same for all wavelengths. For the mathematical model and the numerical feasibility we further assume that the width of the slit is negligible.\footnote{
In reality of course the slit is of a finite width and every wavelength point has contributions from various adjacent wavelength. But this contribution can be minimized with an optimized slit width and should be about the width of a pixel that would be needed to sample the image plane.
See also Chapter~\ref{ch:fiss_sp} for further details on optimal slit width selection.}
This allows us to neglect the \(\lambda\) component for the convolution function, and we can split
\begin{equation}  \label{eq:contribution}
    f_j(x\slit,y) \quad=\quad \sum_{\lambda} d_j(y,\lambda)  \quad=\quad \sum_{\lambda}  \sum_{u,v} p_j(x,y|u,v) \, s(u,v,\lambda),
\end{equation}
where \(d_j\) is the \(j\)-th spectral frame. Thus, we can evaluate the formula for each wavelength point in the spectrum independently.
Both, PSF \(p\) and intensity function \(s\), appear as discrete matrices. 
We can therefore define a mapping 
\[\nu: \NN_0^2 \ni (u,v) \mapsto n\in\NN_0,\] 
with \(\nu(0,0) = 0\). For each point on the slit in each frame we then get a vector
\[\mathbf{a_j} \quad:=\quad (a_{j,u,v})_{j,\nu(u,v)} \quad:=\quad \big(p_j(x\slit,y|u,v)\big)_{j,\nu(u,v)}. \]
such that we can describe each spectral point as a vector product  \(d_j(x\slit, y,\lambda) = \mathbf{a_j}^{\transpose} \cdot \mathbf{s}(\lambda)\).
Considering all points along the slit we get
\begin{equation}
    \mathbf{A}_j \cdot \mathbf{s} \quad=\quad \mathbf{d}_j \;:=\; f_j(x\slit,\cdot),
\end{equation}
with a row in \(\mathbf{A}_j\) for each slit-point in each frame \(f_j\). 

In general we cannot assume the contribution matrix \(\mathbf{A}_j\) to be invertible and even then, the numerical costs to invert such a matrix for every frame are not feasible. 
Thus we use the transposed matrix \(\mathbf{A}_j^\transpose\) as a pseudo-inverse to compute 
\begin{equation}
    \mathbf{A}_j^\transpose \cdot \mathbf{A}_j \cdot \mathbf{s} \quad=\quad \mathbf{A}_j^\transpose \cdot \mathbf{d}_j \quad=:\quad \mathbf{b}_j.
\end{equation}

Consider a CMOS sensor of a size of 2k \(\times\) 2k pixels. For a 1 meter class telescope we typically cover about 60 arc-seconds with such a sensor. Scanning with a scan speed of \(0.3 \nicefrac{\textnormal{"}}{\textnormal{s}}\) and with a frame-rate of 60 Hz, the number of variables in such system is in the order of \(10^7\). Modern CMOS sensors can easily exceed 4k \(\times\) 4k pixels, generating even bigger linear systems to solve. 

Observational data is usually only recorded under good to moderate seeing conditions. Under such conditions the influence of the PSF is in the range of a few arc-seconds.  
We can therefore again work in patches (with dimensions along the slit and in scanning directions) and reduce the number of variables to the numerically feasible order of e.g., \(10^4\) per patch.
Still, due to the sheer size of the linear system of equations, an iterative method needs to be applied. This also has the advantage that a noise-damping can be applied in every iteration step.

Since the original description in \cite{Noort17}, the algorithm to solve the linear system has evolved quite dramatically. We here give the algorithm in its current form. The algorithm follows the idea of a Lucy-Richardson (LR) deconvolution as independently described by \cite{Richardson72} and \cite{Lucy74}. The original LR deconvolution for a Poisson process with known PSF is given by
\begin{equation}
    \label{eq:lucy_richardson}
    \hat{s}^{(k+1)}\quad:=\quad 
    \hat{s}^{(k)} \cdot \left( \frac{f_j}{\hat{s}^{(k)} \circledast p_j} \circledast p_j^*\right),
\end{equation}
where \(k\in\NN_0\) is the current iteration step and \(p_j^*\) is the flipped point spread function. \cite{Shepp82} have later shown that this converges to the Maximum Likelihood (ML) estimation for \(\hat{s}\) if convergence is given.

\begin{figure}
  \centering
  \includegraphics[width=1\textwidth]{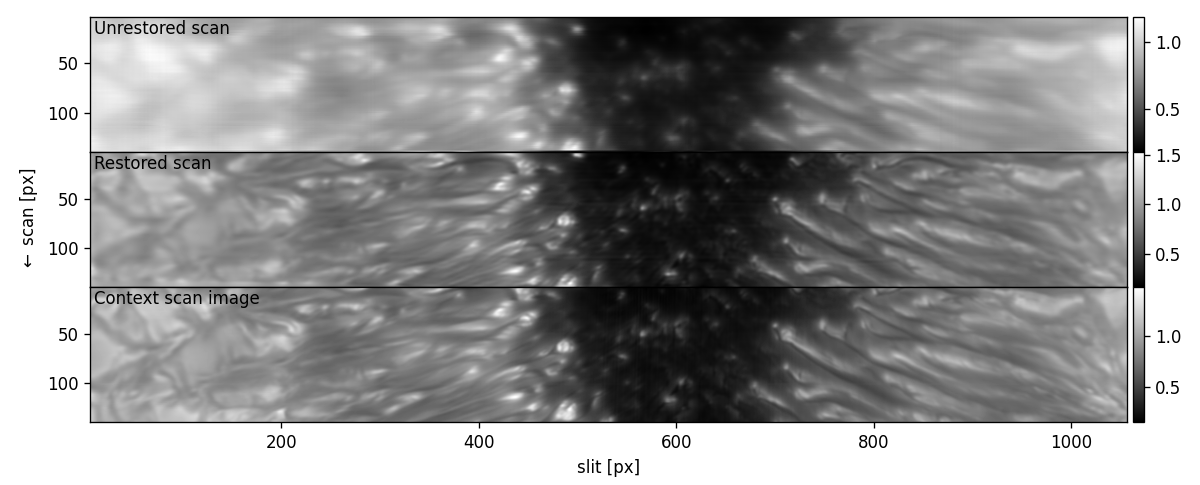} 
    \caption[Comparison of unrestored and restored scan with context scan image]{
        Comparison of intensity images integrated over the whole wavelength range and extracted from an unrestored scan (top panel) and a spatially restored scan (central panel) with a CSI (bottom panel). Intensity is scaled to average quiet Sun intensity and all tick marks are in pixels. The data were recorded with \fisssp{}.
    }
\label{fig:specrestore_result}
\end{figure}

In the case of spectral restoration we are confronted with an additional physical problem that we cannot neglect: Contributions from photons that are displaced by tip-tilt movements non-parallel to the slit and partial contributions from the wings of the PSF with an \(y\)-component. These contributions cannot simply be de-convoluted, as in the case of an 2D image, but have to be restored from spectral frames adjacent in time. Therefore Equation~\ref{eq:contribution} is not a finite-discrete convolution, but a contribution function as a weighted linear combination of PSF contributions. If that function would be uniform and symmetric, it would be a classical convolution function. In general, these assumptions do not hold. This means that, contrary to the estimation of the PSF in MOMFBD, the assumption of isoplanatism is not valid for the application to images of solar spectra. Fortunately, the spatial variation of the PSF is very slow within the size of a typical aperture of a solar telescope (\(\ll 5\) m) and is also smoothed by temporal averaging. Therefore, we neglect the spatial variations and perform a \textit{pseudo} Lucy-Richardson deconvolution with the contribution operator \(\mathbf{A_j}\):
\begin{equation}
    \label{eq:specrestore_alg}
    \mathbf{b}_j^{(k+1)}\quad:=\quad 
    \mathbf{b}_j^{(k)} \cdot \left( \mathbf{A}_j^{\transpose} \cdot \frac{\mathbf{d}_j}{\mathbf{A}_j \cdot \mathbf{b}_j^{(k)}}  \right),
\end{equation}
where the division is performed element-wise.
There is no mathematical proof that Equation~\ref{eq:specrestore_alg} still converges or makes sense in a physical way. However, in all cases this algorithm was applied to, the process did converge to a meaningful solution that matches the expectations. This can for instance be checked by comparing stacked row-wise cut-outs from MOMFBD restored 2D context images (context scan image, CSI) with an average intensity image extracted from the restored spectral scan, as shown in Figure~\ref{fig:specrestore_result}. The CSI can be seen as the best case scenario, since the deconvolution of each pixel was performed with the corresponding PSF. For the spectral camera the integration time is usually longer and thus an averaged PSF is used. Further, the spectral restoration uses a quasi deconvolution, as defined in Equation~\ref{eq:specrestore_alg}, instead of a normal deconvolution. Therefore, the quality of the spectral image restoration can never surpass the CSI. In the example shown in Figure~\ref{fig:specrestore_result}, a close match of CSI and restored scan can be found in the umbra and penumbra. In the granulation region, where the seeing effects were the strongest, the restored scan does not fully reach the quality of the CSI.
The comparison with the unrestored scan illustrates the improvements achieved by the spectral restoration.

\chapter{The \fisssp{} instrument}
\label{ch:fiss_sp}

\textit{The contents of this section correspond to the published article "\textbf{Large aperture diffraction limited spectro-polarimetry with \fisssp{}}" (M. van Noort, \textbf{J. Hölken} , H-P. Doerr, J. Chae, W.~Cao, N. Gorceix, J. Kang, K. Ahn, and S. K. Solanki) 2025, \AnA{}, 704, A215, \citep{Noort24}.}

\textit{
The work among the co-authors was distributed as follows: The extension to the \fiss{} instrument was designed and produced by M. van Noort and \mbox{H.P. Doerr}. The three of us installed and improved the setup at the Goode Solar Telescope (GST) during multiple technical and observational campaigns and conducted the observations. I calibrated, reduced, and demodulated the data, produced the statistics for the comparison with \hsp{} and wrote the first draft of this paper, which was then iterated with all co-authors. I further produced all plots but two.
}

\section*{Abstract}
\begin{footnotesize}
\textit{Context.} Small-scale events measuring only tens of kilometers can have significant implications for the overall energy state of various layers of the solar atmosphere. Current spectro-polarimetric observations lack either spatial or spectral resolution for a comprehensive study of these small scale events.

\noindent
\textit{Aims.} The slit-scanning spectro-polarimetric instrument described here is designed for spectral image reconstruction and in combination with the excellent optical performance of the 1.6 meter Goode Solar Telescope yields spectral hypercubes of the highest spatial and spectral resolution. Additionally, the instrument offers a huge spectral window of more than 30~\AA{}, allowing to observe many solar absorption lines simultaneously.

\noindent
\textit{Methods.} We extended the existing Fast Imaging Solar Spectrograph (\fiss{}) instrument with polarimetric capabilities, new customized cameras and a context imager.
We applied numerical methods to measure and correct for field-depended instrumental and atmospheric degradations, to obtain diffraction limited spectro-polarimetric scans.

\noindent
\textit{Results.} In this work we present the instrument design, the data reduction workflow and first light results. In comparison with a typical \hsp{} dataset we find a higher signal-to-noise ratio in our data within the resolution limits of the respective telescopes when utilizing the signal of all simultaneously observed spectral lines.

\noindent
\textit{Conclusions.} We have obtained the first diffraction limited full Stokes spectro-polarimetric data sets recorded with a slit-scanning spectrograph on a  telescope with an aperture exceeding 1.5 meter.

\end{footnotesize}

\newpage

\section{Introduction}
\label{sec:fisssp:introduction}
Magnetic fields play a crucial role in a variety of prominent solar phenomena, including sunspots, solar flares, and coronal mass ejections. Moreover, small-scale events, measuring only tens of kilometers, can significantly influence the overall energy state of various layers of the solar atmosphere. 
Phenomena such as magnetic reconnection or wave dissipation commonly occur in both active and quiet regions. 
Understanding the structure and dynamics of solar magnetic fields is essential for comprehending the solar activity cycle, predicting space weather events, and advancing the understanding of chromospheric and coronal heating. 
Hence, the deduction of magnetic fields in the photospheric and chromospheric layers is critical. Spectro-polarimetry allows for the investigation and study of these magnetic fields.

There are three principal remote sensing approaches for spectro-polarimetric observations: 
(1) Filtergraphs record images with high spatial resolution at different wavelengths and scan in the spectral dimension over time. 
These instruments typically sample only a few wavelength points in a given spectral line, leading to a relatively sparse spectral resolution. 
(2) Spectrographs record images in the spectral and one spatial dimension, scanning in the other spatial dimension over time. 
The spatial resolution of these instruments usually cannot match broad-band and filtergraph images recorded at the same time (with the same telescope). 
(3) Integral field, or snapshot instruments, record full spectral and spatial information at the same time. 
Unfortunately, these instruments are so far limited to a smaller field of view. 
A comprehensive overview of recent (instrumental) developments in this field can be found in the review by \cite{Iglesias19}.

Inversions of polarimetric observations from a spectrograph are generally more precise compared to filtergraphs, due to the better spectral sampling  \citep[see][]{Rubio06, ToroIniesta16}. 
This precision is enhanced further by the many-line inversion of spatially resolved solar observations, as suggested by \cite{Riethmueller19}. 
They found more precise inversion outcomes and a more complete coverage of atmospheric heights when employing many lines with a range of properties simultaneously. 
Their study effectively shows that by combining the signal of many spectral lines the total signal content can be increased and a higher polarimetric sensitivity reached. 
Modern large format detectors can observe broad spectral ranges spanning several nanometers, making them capable of registering tens to hundreds of spectral lines concurrently.

Restoration techniques for polarimetric filtergraphs have existed for a long time \citep{Keller92}. 
The concept of restoring spectroscopic data to remove the effects of variable instrumental or ambient optical aberrations is also not novel and was demonstrated by \cite{Keller95} and \cite{Suetterlin00}. 
However, their findings were not pursued further at that time. 
\cite{Noort17} presented a spatial reconstruction methodology for spectral observations based on the Multi Object Multi Frame Blind Deconvolution (MOMFBD) technique \citep{Noort05} that can be used to close the spatial resolution gap between filtergraphs and spectrographs. 
The capability to obtain spectral scans of very high spatial resolution across substantial fields of view was therein demonstrated using data acquired at the 1-meter Swedish Solar Telescope \citep[SST,][]{Scharmer02}. The restored spectral data were found to be nearly diffraction limited. Therefore, to reach an even higher spatial resolution by employing this technique, it must be applied to data from a telescope with a greater aperture.

In this work, we describe a modification of the Fast Imaging Solar Spectrograph (\fiss{}) instrument \citep{Chae13}, installed at the 1.6-meter clear aperture Goode Solar Telescope (GST) \citep{Goode10, cao2010b, Goode12} at the Big Bear Solar Observatory (BBSO) in California. 
The resulting \fiss{} Spectro-Polarimeter (\fisssp) prototype is a slit-scanning spectro-polarimeter with diffraction-limited\footnote{The Rayleigh criterion (\(1.22 \; \lambda / D\)) diffraction limit for the GST is 0.068" at 5250~\AA{}} imaging capability. 
It is specifically designed to allow for the spectral image reconstruction of an extended 2D field of view (FOV) using the method described in \cite{Noort17, Noort22}. 
State-of-the-art large format cameras are utilized, accommodating a spectral range exceeding 30~\AA{} at a central wavelength of 5\,241~\AA. 
This enables the simultaneous observation of more than 170 relevant solar absorption lines, potentially achieving a high polarimetric sensitivity by combining the information of many lines, as proposed by \citep{Riethmueller19}.


\section{Instrument}
\label{sec:instrumentation}
The Goode Solar Telescope \citep[GST,][]{Cao11}, is an off-axis solar telescope with a clear aperture of 1.6m. It is located on a small artificial island in Big Bear Lake as part of the Big Bear Solar Observatory (BBSO) and is operated by the New Jersey Institute of Technology (NJIT). The GST houses numerous instruments, including the Fast Imaging Solar Spectrograph \citep[FISS,][]{Chae13}. The FISS is a reflective high-resolution Littrow spectrograph furnished with a fast retro-reflector scanner, designed to perform scans of the solar surface in two chromospheric lines: the \ion{Ca}{II} IR line at 8542\AA, and the H\(\alpha\) line at 6563\AA. The scanner was specifically designed to facilitate rapid scans over a large FOV, with the specific intent to observe the propagation of waves in the solar chromosphere. However, it was not equipped with a polarimeter due to concerns that the Signal-to-Noise Ratio (\snr{}) achievable within the limited exposure times that a fast scan allows would be insufficient.

\begin{figure*}
    \centering
  \includegraphics[width=1\textwidth]{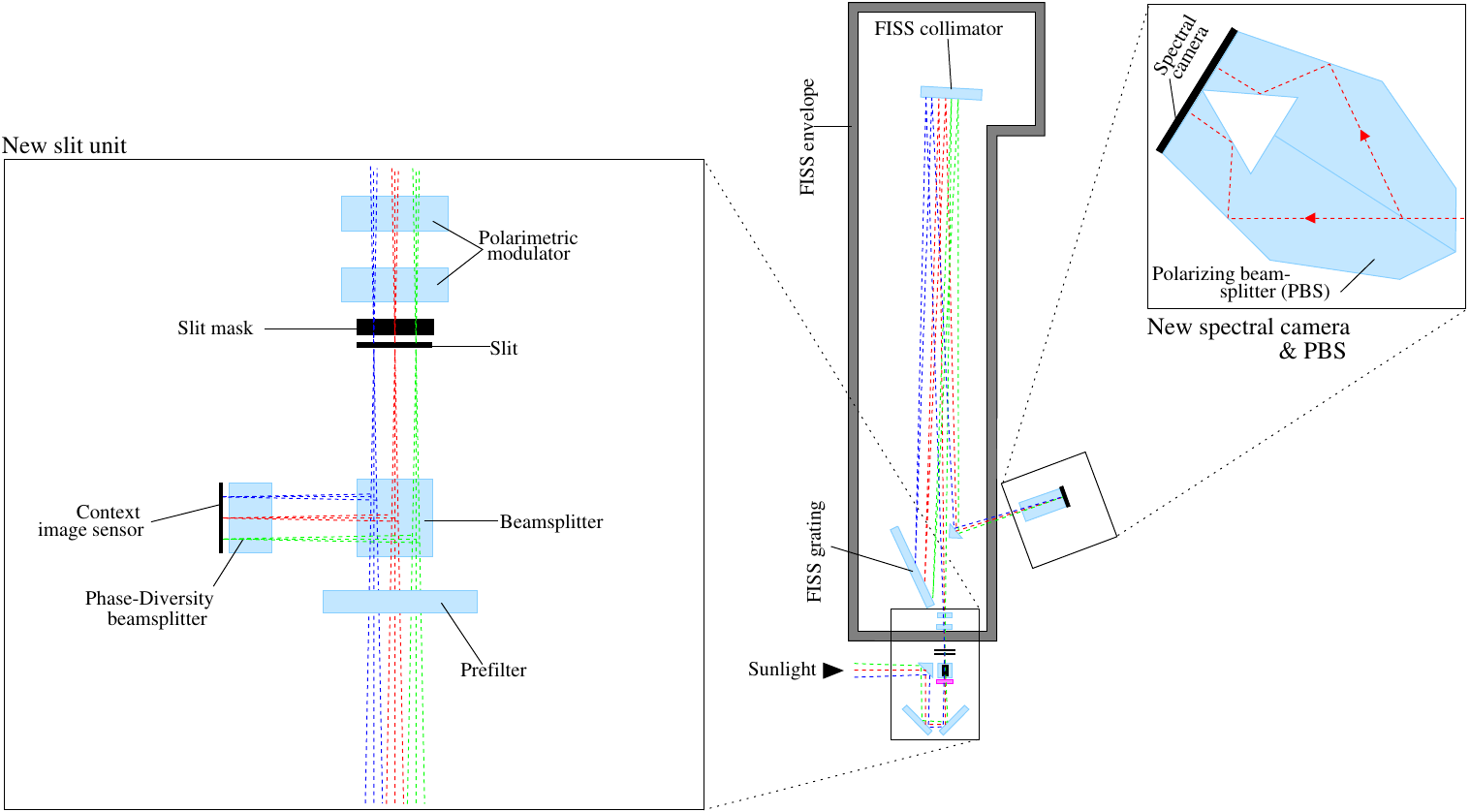}
  \caption[Schematic drawing of the FISS-SP optical layout]{
  Schematic drawing of the \fisssp{} assembly using the existing \fiss{} spectrograph (center), the new \fisssp{} slit unit (left) and the new spectral camera with polarizing beam splitter (PBS) (right). Rays with different colors correspond to different points in the field of view. In the interest of enhanced clarity, the projection of the detailed drawing of the slit unit was rotated by 90 degrees about the vertical, and is viewed from the right.  The spectral camera with PBS unit was rotated about the optical axis of the incoming beam and is viewed from the top-left. See the main text for more details.
  }
  \label{fig:cad}
\end{figure*}

The spectrograph could be applied to acquiring  high-resolution spectro-polarimetric scans suitable for MOMFBD reconstruction only after implementing several modifications. These could be consolidated into two separate modules, as shown in Figure~\ref{fig:cad}\footnote{Author's note: Compare with the CAD drawing of \fisssp{} (Figure~\ref{fig:fiss-sp_cad}) in the introduction of this thesis.}.

\subsection{Context Imager}
A context imager (sometimes also referred to as a slit-jaw camera, SJC) has a long tradition in solar spectroscopy as a means of qualitatively characterizing the feature that is sampled by a particular part of the slit in a particular spectrogram. Like traditional spectroscopy in general, the use of a context imager has fallen out of fashion in recent years, with many recent spectrographs lacking the ability to mount one altogether \citep[e.g.][]{deWijn20}. Since it was intended to be an imager itself, \fiss{} does not make use of the commonly used kind of slit that is etched into a reflecting metal coating, and also does not offer the possibility to mount a context camera anywhere. Instead it makes use of a mechanical slit, so that the width can be controlled and adjusted to the scan steps. 

In our target application, however, the use of a context camera is absolutely non-negotiable. It is needed not to characterize the features sampled by the slit, but rather to measure the instantaneous residual PSF that is degrading the solar image even after correction of the atmospheric seeing by the adaptive optics system.
This difference in the purpose of acquiring the data, allows for a different approach to its acquisition, since it is not essential to know where the slit is located with the highest possible precision, like it would be for a traditional slit-jaw camera. 

Although the atmospheric seeing varies substantially across the FOV, the characteristic scale of these changes is relatively large compared to the resolved spatial scales. As a consequence, a minor error in the assumed slit position does not result in a significant change in the PSF, and can thus be safely applied to the spectral data without introducing a significant error. To see the slit directly in the context image, however, we must pass the image through a re-imaging system, to re-image the slit plate onto the context camera. Since additional optical elements are required for this, there is a high potential for introducing additional optical aberrations, originating in these optics, into the beam, that are not present in the slit focus (a so-called non-common-path error). These non-common-path aberrations will subsequently be included and modeled from the context images, and wrongly applied to the spectral data, that have not been degraded by them.

Based on the above considerations, a simple solution in which the location of the slit could not be seen in the image, but could be calibrated for, was selected, which consisted of a 98:2 beam-splitter cube, placed directly in front of the spectrograph entrance slit. The reflected 2\% of the incoming light was split again using a 50:50 lateral displacement beam-splitter, after which the detector of the context camera was placed in the focal plane of the un-displaced transmitted beam. The laterally displaced reflected image was projected side-by-side to the transmitted image on the detector, but de-focused by some 3.366\,mm. This so-called phase-diversity (PD) channel can be used by image restoration algorithms to better separate image information from residual wavefront errors, so that the most accurate PSF can be obtained.

To recover the most accurate wavefront error, the filter that was used to limit the spectral range of the spectrograph was placed before the 98:2 beamsplitter, so that any wavefront errors introduced by that filter are also present in the context image. The width of this filter was more than sufficient to saturate the detector of the context camera, so that the best possible \snr{} in both the focused and the phase diversity channels could be achieved.

The context camera is a Jai Spark 12000, featuring a CMOSIS CMV12000 CMOS sensor, with a pixel size of 5.5$\mu$m, which oversamples the F/26 beam by approximately 25\%. The camera is capable of a pixel output rate of more than 1.6Gpx/s, allowing for two stripes of 65"x18"of the focal plane to be imaged at a rate of 360 frames per second. For scanning spectroscopy, a narrow stripe along the slit is sufficient for the determination of the PSF on the slit, and in no way limits the range over which a scan can be performed.

\subsection{Slit library}
\begin{figure}
\begin{minipage}[b]{0.62\textwidth}
  \includegraphics[width=1\textwidth]{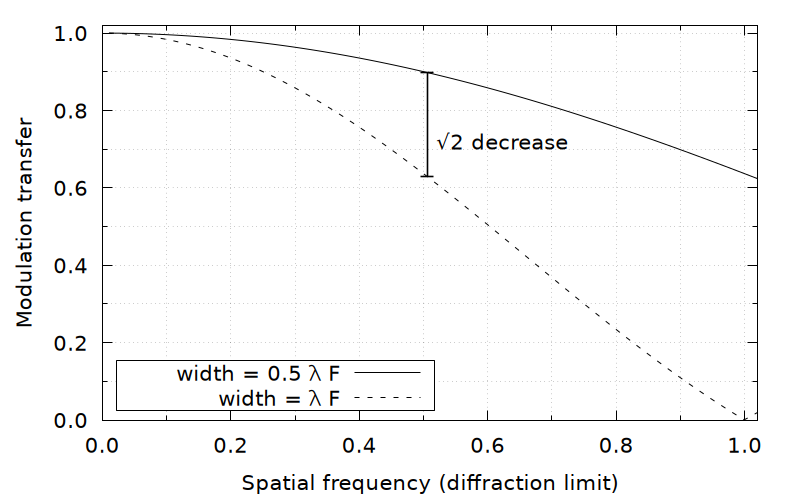}
\end{minipage}\quad\begin{minipage}[b]{0.35\textwidth}
    \caption[MTF for two slit widths]{
  MTF for two slit widths. A width of two Nyquist elements (dashed) significantly increases the damping compared to a strictly critical sampling slit (solid), resulting in a decrease in SNR for spatial frequencies higher than approx. 50\% of the diffraction limit.
  }
  \label{fig:mtf}
\end{minipage}
\end{figure}

The slit is an elongated entrance window that allows only a specific slice of light to enter the spectrograph. Since the selected slice represents a sample of the image plane in the direction perpendicular to the slit, it must have a width not much larger than the width of a critical sample of the focal plane, 
or image information is inevitably lost. Since diffraction on the edges of the slit causes the light entering the spectrograph to be spread over an angle that is inversely proportional to the width of the slit, it should not be made too narrow, or a significant fraction of the light that passes through the slit will fall outside the grated area of the spectrograph and is lost.

Since diffraction is wavelength dependent, so is the optimal choice of the slit width. To allow for flexibility, a slit plate containing a collection of slits of different widths was therefore used, covering the primary wavelength range of interest. A single slit was then selected from the collection by co-aligning it with a mask. 

For a strictly critical sampling of the focal plane, the full width at half maximum ($\text{FWHM} = \lambda F$, where $F$ denotes the focal ratio) of the PSF is to be sampled with at least two resolution elements. For a slit spectrograph, the dimension parallel to the slit is simply re-imaged onto the detector where it is sampled, but the dimension perpendicular to the direction of scanning is sampled by the slit itself.

Few solar slit spectrographs are designed for critical sampling (with  \hsp{} \citep{Lites01, Kosugi07, Tsuneta08} 
as the notable exception), probably because it has two undesired consequences: 1) the diffraction on the slit effectively halves the focal ratio in the spectral dimension, requiring larger optics and a more involved optical design, and 2) less light enters the spectrograph and reaches the detector, which is argued to reduce the SNR of the data unnecessarily. However, and perhaps somewhat counter-intuitively, a wider slit width actually \textit{decreases} the signal amplitude of high spatial frequencies.

This can be understood when taking into account the effect of the modulation transfer of an optical system: for two optical configurations with the same throughput but with different modulation transfer functions (MTFs), the SNR is no different, because it merely quantifies the statistical measurement error of the intensity due to the limited number of photons, but transmitted contrast (that is, signal amplitude) is reduced for the system with the lower MTF.

Diffraction is just one of the components that contribute to the total system MTF, another one is the sampling strategy \citep[see e.g.][]{boreman2001}. Wider pixels, or wider slits, lead to more averaging across a detector element, effectively reducing the transmitted contrast. Fig.~\ref{fig:mtf} displays the MTFs for two slit widths; one strictly Nyquist-critical slit ($\lambda F/2$, see above), and one with twice that width. Starting with spatial frequencies of 50\% of the diffraction limit, the wider slit reduces transmitted contrast by a factor of $\sqrt 2$ or more. This means that at 50\% of the diffraction limit, the expected increase in the SNR (a factor of $\sqrt 2$, because there is twice as much light) is canceled by a decrease of modulation transfer, and this is only getting worse with higher spatial frequencies.

This component of the detection MTF is sometimes referred to as the footprint MTF, because it is related to the size of the sampling elements. Another component, the sampling MTF, is related to the pitch of the detector elements. It however can be shown that the sampling MTF can be neglected if the slit is scanned across the focal plane in much finer intervals than the slit width \citep[dithering;][]{boreman2001}, as it is the case in the setup presented here.

The optimal slit width would appear to be the one yielding the highest detectibility of the highest resolved spatial frequencies, which cannot be determined easily based on fundamental considerations alone. It is in practice dependent on the SNR, the MTF, and on external factors. For the setup presented in this paper, for example, a compromise had to be found between the sampling of the focal plane and overfilling the grating, which would result in a loss of light due to truncation of the beam. Therefore, at 525\,nm, a slit width of 10\,\textmu{}m was typically used instead of the 6.8\,\textmu{}m (0.034") required for strictly critical sampling.

\subsection{Polarimetric modulator unit}
To accommodate the highest possible frame rate, the polarimetric modulator unit (PMU) was based on fast Ferroelectric Liquid Crystals (FLC). This type of Liquid Crystal (LC) unlike nematic LCs is characterized by a fixed retardence value, of which the fast axis can be aligned with two different directions, usually 0 and 45$^\circ$. An optimal modulation scheme can be constructed using two such FLCs, one with a retardence of half a wave, and one with a retardence of a quarter wave. 

Unfortunately, the retardence value of FLCs is fixed, so that as the wavelength changes, the retardence value in waves changes, which changes the efficiency of the modulation scheme. This can be partially compensated by changing the orientation angles of the FLCs relative to the analyzer polarizer and relative to each other. The FLCs were thus mounted on rotation stages, so that the modulation scheme can be optimized over a relatively wide wavelength range.

Many elements of the relay optics from the adaptive optics (AO) onward that feed the \fiss{} reflect light over considerable angles, leading to significant instrumental polarization. The loss of efficiency in specific Stokes parameters that is typically associated with such configurations can be distributed over all Stokes parameters evenly, if the modulator can be placed before the polarizing elements. Unfortunately, the beam properties are such that a large modulator would be required to accomplish this, leading to compromises in uniformity and speed. To keep the required diameter of the FLCs small, the modulator was located immediately behind the slit, the only location in which the required diameter is relatively small. The polarimetric efficiency is therefore not expected to be distributed very evenly over the Stokes parameters. 

\subsection{Spectral camera}
To use the spectral data for the restoration of high resolution spectro-polarimetric scans, the spectral camera (SPC) assembly must place the camera in the focal plane of the spectrograph, and simultaneously analyze the modulated light polarimetrically. In the interest of efficiency and cross-talk suppression, it is advantageous to split the beam into two orthogonally polarized channels, before detection. Unfortunately, most techniques for splitting orthogonally polarimetrically also split orthogonally geometrically, creating the need for two spectral cameras. 

To avoid the additional space required by a second camera,  we used a polarizing beamsplitter developed for the Fast Solar Polarimeter \citep[FSP,][]{Feller14}, which was designed to project both orthogonally polarized beams onto a single detector. Although this is a more compact solution, it is complicated by the off-axis design of the \fiss{}, which is responsible for a significant tilt of the focal plane. Although this is not a problem in itself, the thickness of the glass the FSP beamsplitter needs to re-project the beams side-by-side on a single detector is so large that mounting the camera parallel to the beamsplitter exit surface causes the beam to be vignetted by the beamsplitter entrance window. To reduce this vignetting, the camera therefore had to be mounted onto the beamsplitter at a significant angle. 

The spectral camera is a considerably modified IOI Flare 48 CXP camera, a machine vision camera that comprises an AMS CMOSIS CMV50000 CMOS detector and two separate electronics boards. To characterize these cameras with sufficient accuracy for our application, all components must be operated continuously and accurately thermally stabilized. To achieve this, the electronics boards of the camera were embedded in a custom built copper cooling block, allowing the detector temperature to be maintained at a constant set value to within approximately 65mK using a two-stage Peltier based thermal control loop.

The maximum pixel clock of this sensor exceeds 1.4Gpx/s, allowing this camera to achieve a maximum frame rate of 30 8k$\times$6k frames per second. The 4.6$\mu$m pixels are somewhat smaller than desired, and oversample the focal plane by a factor of approximately 1.5 in the spatial direction, but the huge dimensions of the detector still allow for the entire slit of approximately 60", projected side-by-side, to be accommodated.

\subsection{Polarimetric calibration unit}
Unlike many solar telescopes, the GST is pointed using a polar mount, which makes it particularly suitable for polarimetric observations. This is because the M\"uller matrix of the telescope on any given day consists of a nearly fixed matrix related to the main mirror and some relay mirrors, followed by a rotation over the hour angle. This simple form implies that an extensive telescope model from which this matrix can be calculated for any set of pointing angles is not necessary, the matrix can simply be measured directly once each day and corrected for the hour angle.

For this advantage to be exploited fully, the frame of reference of the polarimeter 
must be located directly after the polar axis, so that the response of all optical elements mounted between that point and the camera can be absorbed in a single modulation matrix. This advantage is so large that despite the limited accessibility of this location and the large clear aperture of at least 75mm it requires of the optical elements, the polarimetric calibration unit (PCU) was mounted here \citep[see][for details on instrumentation and mirror positions]{Cao2010a}.

The PCU consists of a high contrast linear polarizer (LP), followed by an achromatic quarter wave plate (QWP), both of which are mounted inside a motorized rotation mount. The high contrast of the LP ensures that the input light is fully polarized with a known orientation, regardless of the polarization state of the incoming solar light, whereas the QWP converts the linearly polarized light from the LP into a combination of Stokes parameters that can be calculated from theory based on the relative orientation of the LP and the QWP, see for instance \cite{2008A&A...489..429V} and references therein. 

\newpage
\section{Data reduction, calibration and restoration}
\label{sec:reduc}

\subsection{Flat fielding \& wavelength calibration}
During operation, flat measurements are frequently repeated, typically once per hour. In flat field mode the GST moves along a box pattern with a side length of 80" and a speed of approximately 5"/sec, following a path that is free from large scale solar magnetic features near, and optimally around, the center of the solar disk. In addition, the tip-tilt mirror of the AO system produces a random motion with a frequency of 50\,Hz and an amplitude of approximately 9". 
For two minutes during this movement frames are constantly recorded and later averaged to smear out any solar structure (yielding 3\,600 frames for the SP camera and 43\,200 frames for the context imager).
During the setup phase of the instrument we contrasted a two minutes flat  with a five minutes flat. Both where of the same quality and the two minutes flat was chosen as standard operation.

\begin{figure}
    \centering
    \includegraphics[width=1\textwidth]{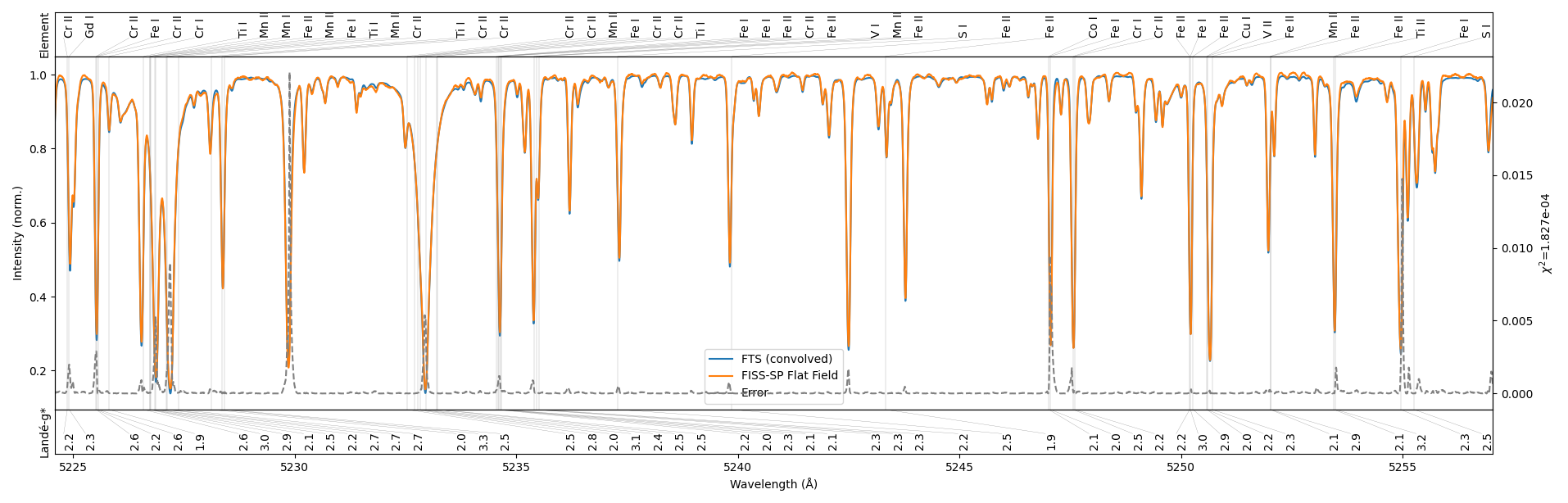}
  \caption[Comparison of FISS-SP and FTS measurements]{
   Comparison of a single spectrum from an temporally averaged and dark- and flatfield corrected \fisssp{} flat field measurement with the FTS after convolution with the \fisssp{} SPSF. 
   The error curve is the square difference between the convoluted FTS and the measured profile.
   Vertical lines in the plot indicate magnetically very sensitive absorption lines with an effective Landé-factor (\(g*\)) \citep{Shenstone29} greater than 1.9. 
  }
  \label{fig:SPSF_FTS_compare}
\end{figure}

These flat measurements are used for flat-field correction and wavelength calibration, to extract gain tables for sensor and slit features of the spectral-camera, and to characterize the spectrograph curvature effect we use the \spectroflat{} library \citep{Hoelken24}\footnote{Author's note: Reproduced as Chapter~\ref{ch:data_reduction} in this thesis.}.
The \texttt{atlas-fit} library from the same publication is used for the continuum correction and wavelength calibration, by contrasting the spectrum of an averaged and reduced flat field measurement with the solar atlas recorded with the Kitt Peak Fourier Transform Spectrometer \citep{Neckel99}, hereafter FTS. 
Figure~\ref{fig:SPSF_FTS_compare} presents a wavelength calibrated single row spectrum from a \fisssp{} flat field measurement.

\newpage
\subsection{Straylight and spectral PSF}
Ideally, the spectrograph's spectral PSF (SPSF) would be measured using a laser, but this possibility did not exist in the given setup. Unfortunately, the effect of gray stray light and an extended SPSF on the spectrum are very similar, and therefore they cannot be determined independently from a solar spectrum. 

\begin{figure}
\begin{minipage}[b]{0.59\textwidth}
    \includegraphics[width=1\textwidth]{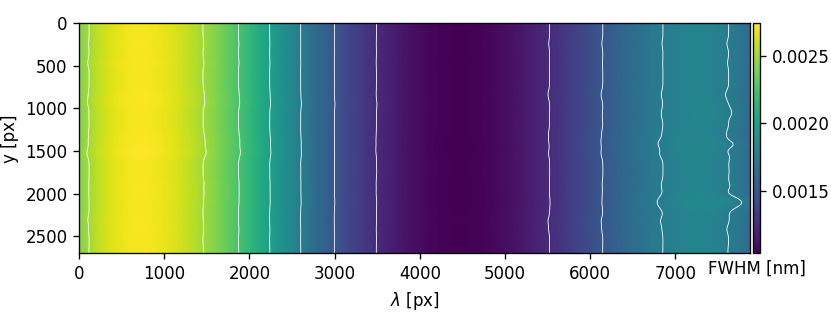}
\end{minipage}\quad%
\begin{minipage}[b]{0.38\textwidth}
  \caption[Spatial and spectral variation of the FISS-SP SPSF]{
  Variation of the SPSF core FWHM in spatial and spectral dimensions of the transmitted \fisssp{} beam lit area. Contours are added to highlight the shape of the degradation. 
  }
  \label{fig:SPSF_FOV}
\end{minipage}
\end{figure}

In a first step we determined the gray spectral stray-light contribution using the ratio of the mean intensity in the dark area on the spectrograph camera (which was shielded by a slit mask) and the non shielded area. This resulted in rather low mean stray-light measurements of 2.34~\% and 2.61~\% of the reflected and transmitted beam respectively. 
In a second step we estimated the asymmetry and contribution of the grating comparing a dark- and flatfield corrected and temporarily and spatially averaged flat field measurement (i.e., a spatially averaged quiet Sun spectrum) with the FTS atlas. For this we convolved the FTS with a Voigt function where we set the FWHM of the core to the theoretical value of \(9.36 \times 10^{-4}\)~nm given by the f-ratio, wavelength, pixel size and dispersion. We then fitted the FWHM of the wings and the asymmetry factor. The best match was achieved with an estimated wing FWHM of \(2.99 \times 10^{-4}\)~nm and an asymmetry factor of \(-1.91 \times 10^{-3}\).
In the last step we fixed the wing FWHM and the asymmetry factor and allowed the core FWHM to vary spatially and spectrally. 
In Figure~\ref{fig:SPSF_FTS_compare} we compare a single row spectrum from a \fisssp{} flat field measurement with the FTS atlas convolved with the corresponding estimated SPSF and
Figure~\ref{fig:SPSF_FOV} shows the variation of the core FWHM  of the field-dependent SPSF within FOV of the transmitted beam. The tilted oval structure of the imprint is in line with the expectations for an off-axis paraboloid imager mirror on a slightly rotated sensor.  

\subsection{Scaling and co-alignment}
After the wavelength calibration both channels are already on the same uniform linear scale in one dimension. 
For dual-beam recombination we extracted the keystone distortion in spatial direction at multiple wavelength positions and x,y-shift factors from a scan of a pinhole-grid target inserted in F3. After clipping to the nearest pixel a sub-pixel alignment  was performed using an FFT shift of the reflected beam to ensure beam alignment to a hundredth of a pixel in spatial direction. 

To prepare the data we applied binning in wavelength direction by a factor of two, yielding an effective linear dispersion of 8.38 m\AA{}/px, and a Fourier Transform resampling by a factor of \(\nicefrac{4.6}{5.5}\) in spatial direction. The latter is necessary to achieve a matching pixel scale of the SPC and SJC sensors. 
Once calibrated, the alignment and de-distortion can be applied to the data in a single processing step.

\subsection{Spatial Restoration of Spectra}
With the data dark-corrected, flatfield corrected, and mapped onto a calibrated orthogonal grid, we are ready to remove the effect of the residual seeing-induced image degradation 
present in the data. We follow the procedure outlined in \cite{Noort17} and obtain the PSF along the slit using the context images. 
With the use of the phase diversity information, 134 modes can be fitted reliably to each patch covering approximately 5"x5". To trace the spatial variability of the PSF, one such patch was placed every 10 pixels (0.27") along the slit, and each burst of images is restored with 50\% overlap in time with the previous one, so that the movement of the slit across the Sun can be accurately tracked.

To obtain accurate polarimetric results with a minimum of seeing induced cross-talk, each modulation state is restored separately using the spectral restoration technique described in \cite{Noort17}. For a given modulation state, only the subset of the PSFs that are co-temporal with that modulation state are used, but these are obtained using the context images of all modulation states in one restoration. Since the state of the MCU has no impact on the context images, they can all be combined, so that the MFBD algorithm determines the PSF of each image included in the restoration relative to the restored image. In this way the spectral data corresponding to each polarimetric state are automatically restored to a common frame of reference, much like in a multi-object MFBD restoration \citep[e.g.][]{Noort05}, so that all modulated restored images are accurately aligned. Since the PSFs used for the two polarimetric beams are identical, and the data are already aligned, no further alignment of the frames is required before demodulation and recombination of the modulated restored images.

\subsection{Polarimetric calibration}
For the polarimetric calibration we follow the strategy used by \cite{2008A&A...489..429V} and make measurements with the LP at four orientations with a 45 degree separation, and rotate the linear polarizer through the full 360 degrees, in steps of 5 degrees. The high density of the data points allows for an accurate fit to the data, which is not very sensitive to errors in the dark offset or in the input light level. 

The fit of the modulation matrix proceeds in two steps. In the first step the values of the QWP retardance and the offset of its fast axis with respect to the LP are fitted together with the modulation matrix elements to the spatial average of the data. Because this fit is highly non-linear, a start solution needs to be given for all parameters that is already fairly accurate, or the optimization algorithm will fail. Since the pupil location of the QWP and LP ensures that the properties of the LP and QWP must be very close to uniform across the FOV, in a second step they are fixed, and the modulation matrix elements are fitted for each pixel in the FOV.

\begin{figure}
    \centering
  \includegraphics[width=1\textwidth]{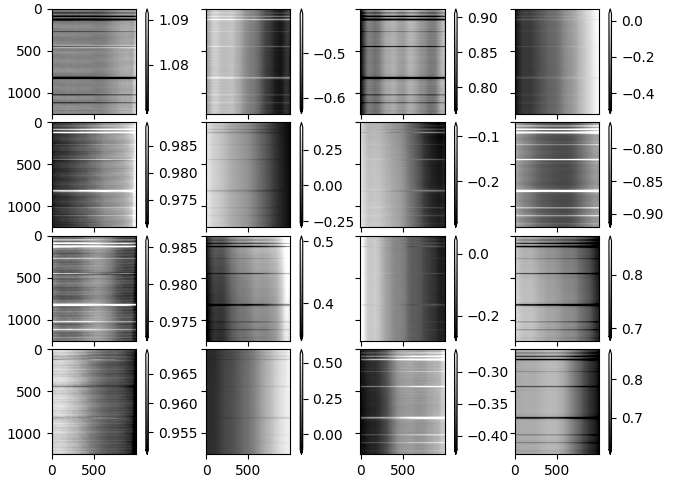}
  \caption[Elements of the FISS-SP modulation matrix]{
  Cutout from the elements of the reflected beam modulation matrix around the strongest spectral lines. Vertical is the spectral and horizontal the spatial dimension. The colour scale of each panel is clipped at \(2\sigma{}\) to highlight the imprints of the spectral lines and all axis tick marks are in pixels.
  }
  \label{fig:modmat}
\end{figure}

Unfortunately, the spatial and wavelength dependent calibration results were found to show strong imprints of the spectral lines (see Figure~\ref{fig:modmat}), implying a strong dependence of the modulation on the intensity level. Although such non-linear behavior can be the result of a strongly nonlinear response of the detector to the illumination level, the level of nonlinear behavior required to explain the observed behavior was found to exceed the expected level by an order of magnitude.

A more likely scenario is that the imprints are related to the placement of the PMU inside the spectrograph. When light is reflected between the different PMU elements, it can pass through one or both FLCs multiple times, resulting in a defocused ghost, with a modulation that differs from that of the main beam. 
If the PMU is placed before the entrance slit, this ghost is simply sampled as part of the image, and its modulation is included in the modulation matrix. If the PMU is placed inside the spectrogaph, however, the defocus of the ghost also blurs the spectral lines, and, if the defocus is severe enough, the ghost can mimic a uniform, gray, but modulated straylight contribution. 

Whereas in normal observations this contribution is usually only weakly polarized and therefore not modulated strongly, in the case of a polarimetric calibration the input light is fully polarized, and the modulation signal is strong. The measured modulation is thus a superposition of the modulated direct and ghost images, and as the relative weight between the two changes with the intensity, which is clearly the case for spectral lines, the modulation matrix changes accordingly. 

Since the modulation matrix of the direct and ghost images could not be determined separately, a surface was fitted to the the modulation matrix values in the highest intensity areas of the FOV, where the weight of the direct image is the highest and most appropriate to the normal observation. Since the flatfielding procedure estimated the gray straylight to be only 2.5\% of the continuum intensity, the error in the modulation matrix values fitted in this way is estimated to be limited to approximately that fraction. The field dependent matrix elements were calculated from this fit for the whole FOV.

To reduce the effect of the remaining calibration errors in the modulation matrix as much as possible, a final ad-hoc cross-talk removal was carried out on the data. 

\begin{figure}
    \centering
  \includegraphics[width=.95\textwidth]{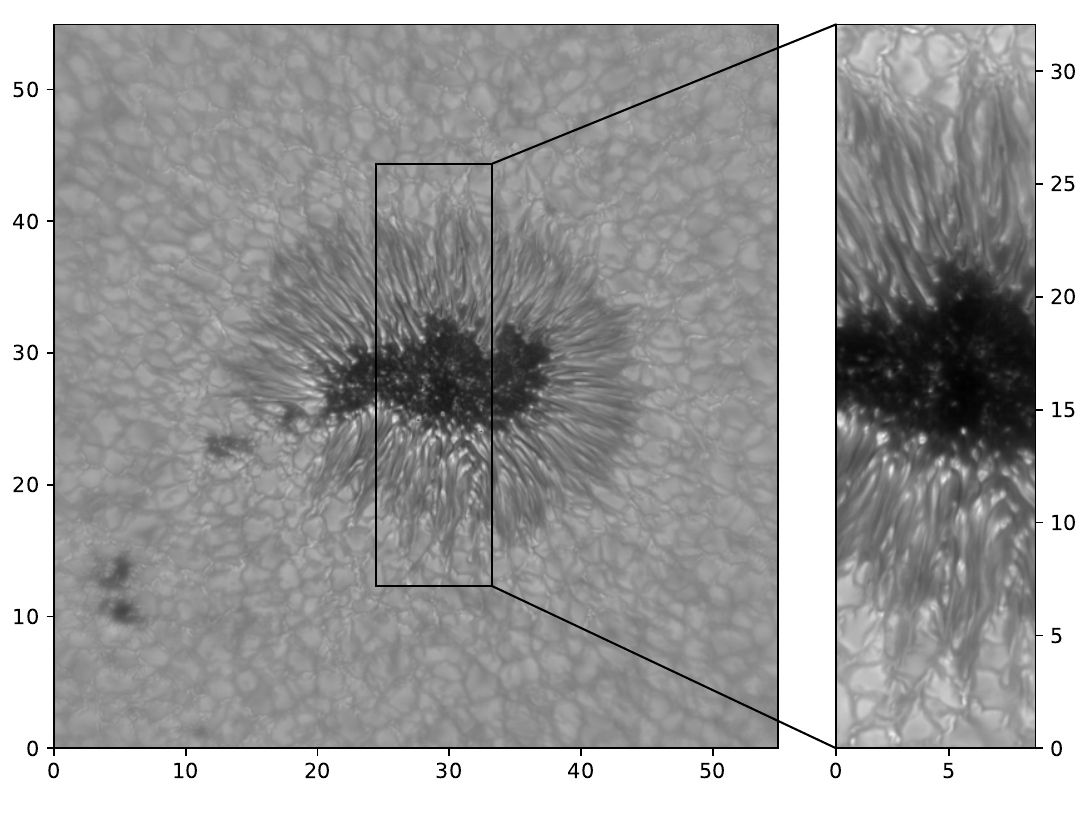}
  \caption[Restored context images]{
  Context image of the GST/TiO band camera (705.7~nm, speckle restored; left panel) and \fisssp{} context image (525.0~nm, MOMFB restored; right panel) of the scanned region. Placement of the \fisssp{} scanned region within the TiO image is indicated by a rectangle, all tick marks are in arc-seconds. Due to the hot-spot type seeing conditions, only a fraction of the possible FOV could be used.
  }
  \label{fig:context:2022-10-01/obs08}
\end{figure}

\section{Results}
\label{sec:fisssp:results}
Two modules were added to the \fiss{} in 2022. They extend the instrument with spectro-polarimetric capabilities and allow for the data to be restored. The extended \fisssp{} setup was used in a number of technical and observing campaigns in 2022 and 2023. After each \fisssp{} campaign the \fiss{} was restored to its original configuration. The data from two of these \fisssp{} campaigns were of sufficiently high quality that a good spectral restoration could be carried out. We will use two observations to discuss the performance of \fisssp{} below.

\subsection{Observations}
The first observation was a series of short scans of AR13111 recorded on the 1st of October 2022 at 19:42 UT at \(\mu=0.89\) (here $\mu$ is the cosine of the heliocentric angle). The length of each scan was approximately 6", which at a scanning speed of approximately 1.6\,s per pixel took about 6 minutes to complete. In contrast to a conventional stepped scan, we used a constantly moving slit with a scan speed of 0.04~"/sec as described in \cite{Noort17}. 

The adaptive optics system managed to correct the central 10"x10" of the FOV relatively well, allowing the spectral restoration to recover spatial structure with spatial scales that are close to the diffraction limit of the telescope.
Figure~\ref{fig:context:2022-10-01/obs08} provides an Overview of the sunspot and scanned region from GST/TiO and \fisssp{} context imagers.

\begin{figure}
    \centering
  \includegraphics[width=1\textwidth]{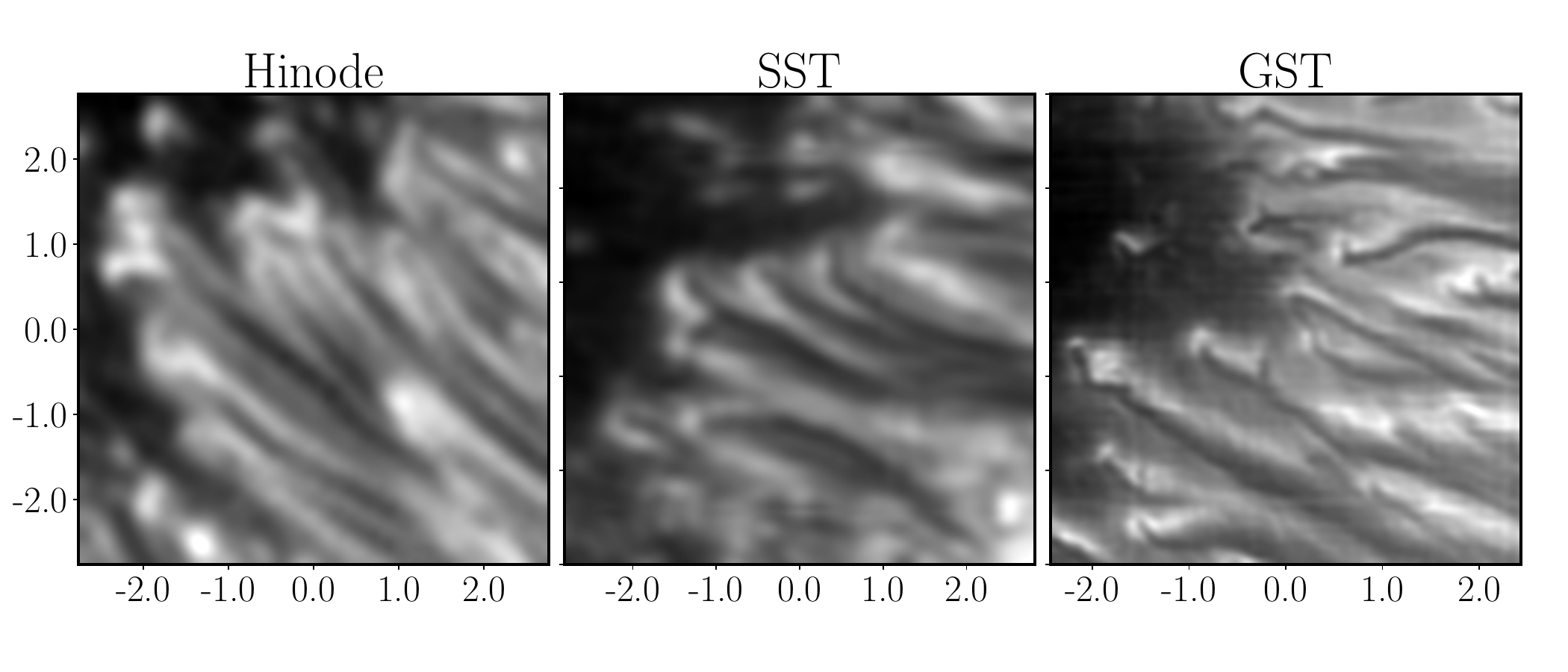}
  \caption[Comparison of scans from HINODE, SST, and GST]{
  Spectrograph scans of similar penumbral filaments, acquired with \hsp{} (0.5m, 6301.5\AA, deconvolved) TRIPPEL-SP@SST (0.98m, 6310.5\AA, restored) and \fisssp{}@GST (1.6m, 5250.6\AA{}, restored). All images show a spectral average over 0.15\AA{} of the respective hyperspectral cubes, tick marks are labeled in arc-seconds. 
  }
  \label{fig:specrestore_comparison}
\end{figure}

Focusing on the first and best scan of the series, covering the central part of the main spot of AR13111 together with an extended part of the penumbra, many structures appear to show considerable fine structure. Particularly the heads of the penumbral filaments appear to show signs of sub-structure, and many of them appear to be split. To place the spatial resolution in context, a small section of the umbra-penumbra border is shown in Figure~\ref{fig:specrestore_comparison}, alongside restored scans of similar targets from \hsp{} and TRIPPEL@SST. The increase in resolved structures when increasing the telescope aperture from 0.5m to 1m to 1.6m is notable, and there is no particular indication that the highest resolution scan is resolving all relevant structure.
Unfortunately, owing to technical difficulties, the polarimetric calibrations of the 2022 campaigns were found to be compromised, and fitting the Stokes spectra proved unsuccessful. 

\begin{figure}
    \centering
  \includegraphics[width=0.975\textwidth]{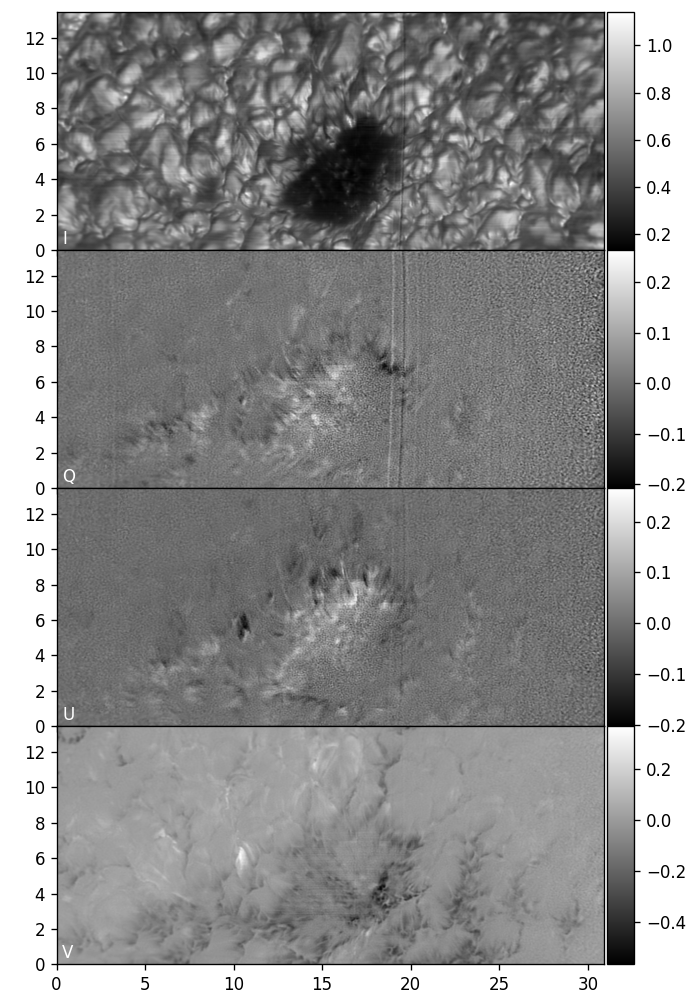}
  \caption[Narrow band scan images of all stokes parameters]{
  Narrow band images (0.06~\AA{} bandwidth) from a restored hyperspectral cube of the \mbox{May 11, 2023} observation, extracted in the red wing of the \ion{Fe}{I} line at 5250.2~\AA{}, the Stokes parameters are shown in individual panels. 
  Continuum intensity is normalized to one, $Q$, $U$ and $V$ maps show relative signal strength. The scan direction is vertical from bottom to top and tick marks are in arc seconds. See text for further discussion.
  }
  \label{fig:scan:2023-05-11/obs01}
\end{figure}

The best scan of the 2023 campaign was a large scan of an emerging active region (later assigned \mbox{AR~13304}).
The observation was recorded at \(\mu=0.81\) on May 11\(^{\textnormal{th}}\) 2023 starting at \mbox{17:01 (UTC)} until \mbox{18:48 (UTC)}. 
The dataset includes five scans of variable FOV around a large pore that was placed at the center to assist the AO. The largest scan covers 60".
Figure~\ref{fig:scan:2023-05-11/obs01} shows narrow band images of all Stokes parameters from the restored hyperspectral cube of the best scan in the red wing of the 5250.2~\AA{} \ion{Fe}{I} line.
Stokes $I$ shows various types of fine-structure as "canals" and "hairs" that have been reported on high resolution broad band images \citep[e.g.,][]{Scharmer02b, Schlichenmaier16}, but not yet on narrow band scan images. Clear linear polarization signals are detected around all pores in the FOV, especially at the umbra/granulation boundary of the main pore.
The vertical lines around 20" (predominantly visible in Stokes $Q$) are residual dust imprints from the slit, amplified by the restoration algorithm and deformed by the correction for the residual atmospheric seeing. 
The hot-spot type seeing condition becomes visible in the increasing noise towards the right border, again most notable in the linear polarization signals.

A more complete overview of and access to the reduced \fisssp{} data will be provided upon request to the authors. 

\subsection{Spatial resolution}
Solar images are altered during their journey from the Sun to a detector due to the Earth's atmospheric turbulence and telescope aperture diffraction. These cause distortion and blurring of small structures, reducing the amplitude of high-frequency components in the Fourier domain. The process of image restoration aims to undo these effects, and thus restore the image to its original form.
The detection of discrete events with imaging sensors, however, has an inherent uncertainty, known as photon or shot-noise, resulting in a measurement error in the Fourier phases and amplitudes. Image restoration can usually lessen Fourier phase errors significantly, while Fourier amplitude restoration is more sensitive to data noise. 

Due to diffraction on the edge of the telescope aperture, the amplitudes of the Fourier components are reduced monotonically as a function of Fourier frequency, and vanish completely at and beyond the highest frequency transmitted by the telescope, the so-called diffraction limit. The larger the telescope aperture, the less the amplitude reduction at all Fourier frequencies, and the higher the diffraction limit.
The noise, however, is not affected by these optical effects and maintains a constant power density independent of the Fourier frequency. As images cannot be completely free of photon noise, a resolution limit exists where the noise amplitude overwhelms the signal amplitude, making it impossible to restore meaningful image data.

The resolution limit of the \fisssp{} scans must be seen in that context.
Since the resolution limit of an image depends on the noise level, and the noise level depends on the width of the spectral band over which the spectra are integrated, the spatial resolution of the data is dependent on the spectral bandwidth \citep[see][for a more elaborate discussion of this issue.]{Noort22}. 

\begin{figure}
\begin{minipage}[b]{0.62\textwidth}
  \includegraphics[width=1\textwidth]{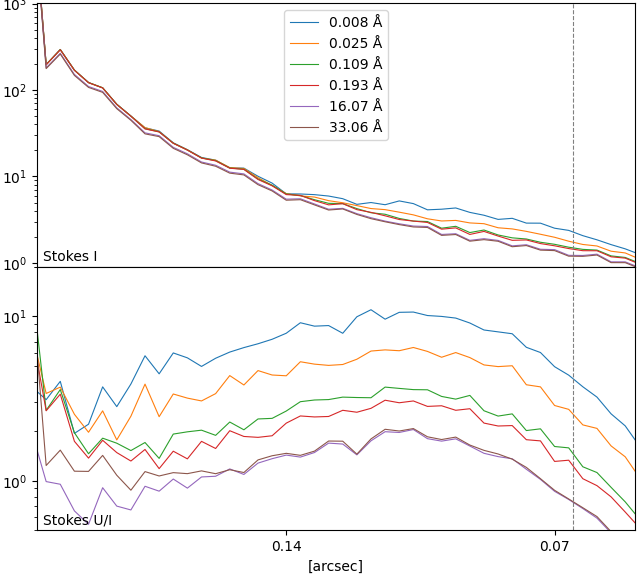}
\end{minipage}\quad\begin{minipage}[b]{0.35\textwidth}
  \caption[Power spectrum of integrated spectra]{
  Power spectrum of a quiet Sun continuum image extracted from Stokes $I$ and $U$/$I$ by integration  over one (0.008~\AA) spectral bin, 3 (0.025~\AA), 13 (0.109~\AA), 23 (0.193~\AA), half range of 1.9k (16.07~\AA) and the full range of 3.8k (33.06~\AA) spectral bins.
  The vertical dashed line marks the diffraction limit at 
  \mbox{\( 0.068" = 1.22 \, \lambda / D\)} at 5250~\AA{}.
  }
  \label{fig:power_spectrum_IU}
\end{minipage}
\end{figure}

Figure~\ref{fig:power_spectrum_IU} shows the azimuthally integrated power spectrum for Stokes $I$ and $U/I$ for a selection of bandwidths. From the small difference in the power near the diffraction limit between the integral over 16~\AA{} and 33~\AA{}, it can be concluded that the image information in Stokes I is diffraction limited when integrated over all wavelengths. For Stokes U, however, the signal level in the quiet Sun is so low, that even when accumulating all signal in the spectral range, the resolution falls well short of that, and does not appear to exceed 0.15".

\subsection{Signal to noise}

Besides the spatial resolution limit, the best metric of the quality of the \fisssp{} data is usually the polarimetric noise level. To compare our data to the de facto standard in this field, \hsp{}, it is important to make sure we compare like with like. First of all, the \fisssp{} spectral resolution is much higher than that of \hsp{}, we therefore bin the spectra down to 23m\AA\ per bin, since our spectral pixels are 3 times larger than those of \hsp{}. 
Secondly, the \hsp{} data are usually disseminated in their raw demodulated state, including the effect of telescope diffraction, whereas the \fisssp{} data are spatially restored. Since the photon flux for a critically sampling pixel is independent of the telescope aperture, we convolved the data with the diffraction limited PSF of a 1.6m telescope with a pupil function that is similar to that of the Hinode SOT, and scaled the pixel size such that it samples the focal plane critically relative to the Rayleigh limit, as it is for \hsp{} data.
We can now compare the noise levels in $Q$, $U$ and $V$ in the spectral continuum, as shown in Table~\ref{tab:fiss_hinode_noise}.

\begin{table}[ht]
    \centering
\caption[FISS-SP and HINODE/SP noise levels]{
RMS noise levels in the continuum for $I$, $Q$, $U$ and $V$ of re-degraded \fisssp{} and \hsp{}\ data. For convenience the equivalent \snr{} in the continuum (SNR$_c$=RMS$_c^{-1}$) is indicated for each.
}
\label{tab:fiss_hinode_noise}
\begin{tabular}{c | c | c | c | c} 
  & \multicolumn{2}{c}{FISS-SP} & \multicolumn{2}{c}{\hsp{}}\\
Stokes & RMS$_c$ (\(10^{-3}\)) & \snr{}$_c$ & RMS$_c$ (\(10^{-3}\)) & \snr{}$_c$\\
\hline 
$I/I_c$ & 4.0 & 250& 1.0 & 1000\\
Q/$I_c$ & 5.6 & 179& 1.0 & 1000\\
U/$I_c$ & 4.6 & 217& 1.0 & 1000\\
V/$I_c$ & 1.7 & 588& 1.0 & 1000 
\end{tabular}
\end{table}

As is clear from the values, the \fisssp{} noise is not too dissimilar from that of \hsp{} for Stokes $V$, but considerably worse for $Q$ and $U$, although the noise in $Q$ and $U$, relative to that in $V$, appears to be consistent with the polarimetric modulation efficiency as measured by the polarimetric calibration. 

The noise in Stokes $V$ is surprisingly low, considering that there are many elements in the beam that reduce the signal level of the \fisssp{} data as compared to the \hsp{} data. The most important losses can be estimated and are listed in Table~\ref{tab:signal_level_comparation}.
This indicates that the general noise level should be approximately 4 times higher than that of \hsp{}. Averaging the noise over the Stokes parameters, we get a value of 3.97$\cdot 10^{-3}$ which appears to be in excellent agreement with the value of 4$\cdot 10^{-3}$ expected from the transmission estimates in table \ref{tab:signal_level_comparation}.

\begin{figure*}
    \centering
  \includegraphics[width=1\textwidth]{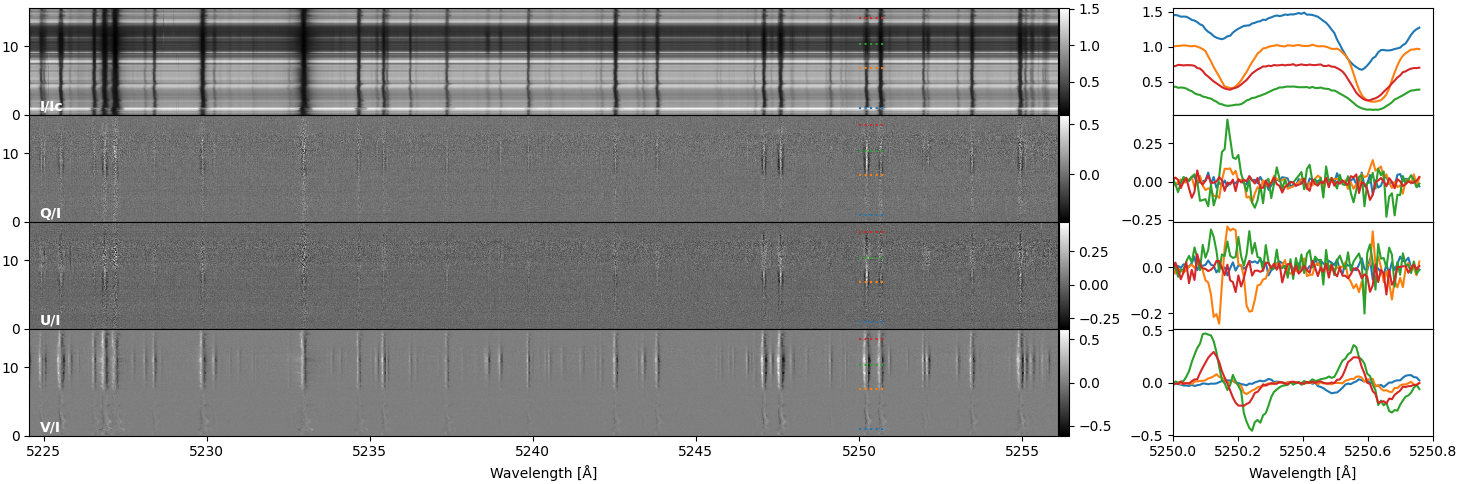}
  \caption[FISS-SP spectrum and profiles]{
  Restored Stokes spectrum and line profiles from a \fisssp{} observation of a pore in an active region. Y-axis of the left panel is in slit direction and corresponding tick-marks are in arc seconds. The right panel shows line profiles of the 5250~\AA{} \ion{Fe}{I} lines at positions marked in the left panel.
  }
  \label{fig:fiss_spectrum}
\end{figure*}

\begin{table}[ht]
    \centering
\caption[Estimated transmission of FISS-SP and HINODE/SP]{
Estimate of the transmission of the most important elements of \fisssp{} as compared to \hsp{}.
}
\label{tab:signal_level_comparation}
\begin{tabular}{l | c |c |c}  
&\hsp{} & \fisssp{} & ratio \\
\hline 
Atmosphere & 1.0 & 0.7 & 0.70\\
Relay optics & 0.9 & 0.43& 0.48\\
Prefilter & 0.9 & 0.85 & 0.94\\
Pixel area & 1.0 & 0.69 & 0.69\\
Quantum efficiency & 0.9 & 0.6 & 0.67 \\
Exposure & 3.2 & 1.6 & 0.50\\
\hline 
\textbf{Total} & && \textbf{0.07}
\end{tabular}
\end{table}

\begin{figure*}[!ht]
    \centering
  \includegraphics[width=0.46\textwidth]{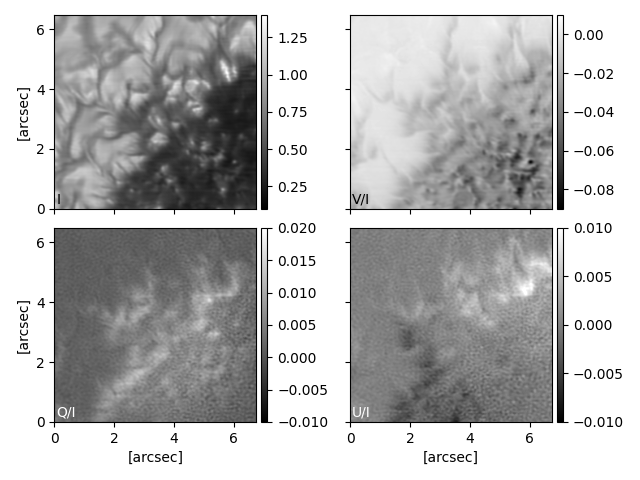}\;
  \includegraphics[width=0.51\textwidth]{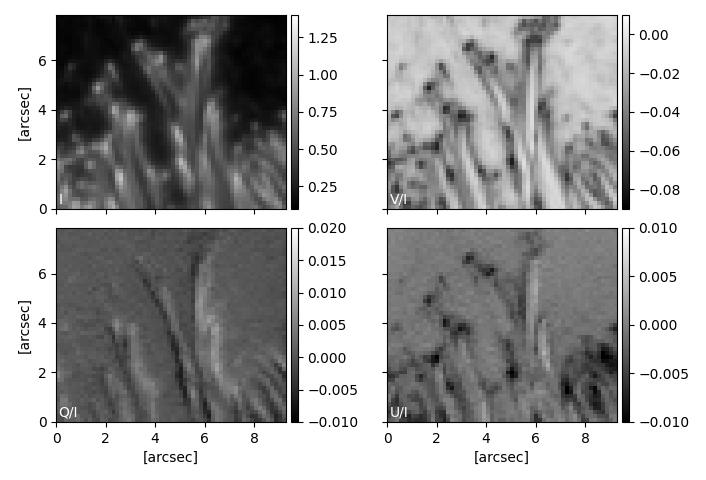}
  \caption[Comparison of total signal in FISS-SP and HINODE/SP scans]{
  Spectrally averaged signal in $I$, $Q$/$I$, $U$/$I$, $V$/$I$ of \fisssp{} (left) and \hsp{} (right).
 In both cases $V$ was multiplied with the sign of the first and $Q$, $U$ with the sign of the second derivative of $I$ before averaging. Unlike integrating the absolute value of the Stokes spectra, this procedure results in a predominantly unsigned polarimetric signal, but avoids the creation of single signed noise. This ensures that when integrating spectrally the signal accumulates, but the noise averages out.
  See text for further discussion.
  }
  \label{fig:context:2023-05-11/total_signal}
\end{figure*}

That the noise in Stokes $V$ is actually only a factor 1.7 worse than in \hsp\ can be understood when we consider the noise filtering that is applied during the spectral restoration procedure for \fisssp{}. 
The limit set on the number of iterations for the Richardson-Lucy deconvolution, which was not applied to the \hsp{} data, prevents the high frequency noise from being amplified, in effect filtering it to a very low level when the data are then re-degraded. The \hsp{} data, however, were compressed using JPEG compression, which amounts to a small amount of filtering of the \hsp{} data, that was not applied to the \fisssp{} data. 
Because the JPEG compression level can be varied and its effect on the data depends on the image content, it is not straightforward to work out the differences in filtering between the two data sets. We take the noise level of the \hsp{} data to be an upper limit of the real noise level, and assume it would probably be a bit lower when filtered in the same way as the \fisssp{} data. 
All things considered, we conclude that considering the differences in the instruments, the noise properties of the two data sets are consistent with each other. Since the modulation scheme was clearly biased towards Stokes $V$, we have a noise level that is less than a factor two worse than that of \hsp{}, whereas the noise in $Q$ and $U$ are about four to five times higher.

The total signal in the scan, however, is considerably higher in the \fisssp{} data than it is in the \hsp{} data, because \fisssp{} samples a much larger wavelength range than \hsp{}, centered on a wavelength where the spectral density of magnetically sensitive atomic lines is relatively high. Whereas \hsp{} has only 2 significant spectral lines in 112 spectral bins, the \fisssp{} data has more than 150 spectral lines in 3840 spectral bins. The benefits of utilizing information from many lines, even if the signal of the individual lines is buried in the noise, was already demonstrated for stellar spectra by \cite{Donati97}.

The increased signal content resulting from the large number of lines is immediately clear from the example spectrum, shown in Figure~\ref{fig:fiss_spectrum}, where signals from many lines can be discerned in all Stokes parameters. The impact of collecting all this signal is illustrated nicely in Figure~\ref{fig:context:2023-05-11/total_signal}, where the sign of the Stokes signals was flipped, based on their dependence on the spectral line profile, after which the spatio-spectral cube was integrated over all wavelengths.
It is clear from the noise in $Q$ and $U$ that although there is room for improvement, information on the horizontal magnetic field is present well above the noise, at least in the strong field regions. 

\section{Summary}
We have developed and built two modules for the \fiss{} spectrograph, currently installed at the GST. These modules add polarimetric capabilities and a context imager to the \fiss{}, allowing for the acquisition of data suitable for the restoration of spectro-polarimetric scans, and come with a significant increase of the spectral range. We refer to the combination of \fiss{} and the new modules as \fisssp{}. 

The new modules were installed and tested at a wavelength of 525~nm, during several experimental observing campaigns. Several data sets were recorded during episodes of good seeing, that were subsequently restored using the spectral restoration technique by \cite{Noort17}. 

At least two scans, and a corresponding sequence of broadband context images, could be produced with an image resolution that approaches the diffraction limit of 0.07" over a FOV of at least 6"$\times 20$" and 16"$\times 36$" and a spectral range of 33.06~\AA{}. All four Stokes parameters could be recorded.

An assessment of the noise revealed that although the restored data would appear to be much noisier than traditional scan spectra, they actually compare relatively favorably to traditional data when their states of reduction are properly harmonized. By making use of the accumulated signal provided by the large number of spectral lines that can be accommodated in the spectral range of \fisssp{}, the sensitivity to atmospheric properties should be greatly enhanced.

\chapter{Towards solar many-line inversions}
\label{ch:many_line}

\textit{The contents of this section correspond to the submitted version of the by now published article "\textbf{Towards solar many line inversion of high resolution spectro-polarimetric data}" (J. Hölken, M. van Noort, S.~K. Solanki, A.~Feller, \mbox{H.P. Doerr}, W. Cao, N. Gorceix, J. Kang, K. Ahn, and J. Chae), 2026, \AnA{}, 705, A220, \citep{Hoelken25}}

\textit{I ran the inversions, created and tested the line selection method, evaluated the results and wrote the majority of the text.}

\section*{Abstract}
\begin{footnotesize}    
\textit{Context.}  For the analysis of highly resolved solar spectra the simultaneous observation and interpretation (inversion) of only a few (often only one) spectral lines is still the norm. With modern instruments spatially resolved spectro-polarimetric data including many lines are available.

\noindent
\textit{Aims.} For the first time we combine the information from \nlines{} simultaneously observed absorption lines on spatially highly resolved data to test a proposed solar many-line inversion strategy.

\noindent
\textit{Methods.} We inverted full Stokes spectra recorded with the FISS spectro-polarimeter (\fisssp{}) at the 1.6-m Goode Solar Telescope in California, using the SPINOR code. We contrasted two different setups: one following the traditional approach of using a line doublet, and another innovative method inverting many-lines simultaneously.

\noindent
\textit{Results.} Compared to results from an inversion using the two lines of a line doublet, we discovered more fine-structure and better constrained values using the many-line technique.
    An average quiet Sun spectrum was successfully reproduced using a model atmosphere, but when inverting spatially resolved data, uncertainties in line parameters and blend configurations did not average out. Thus, a deliberate selection process of lines and line blends to be added to the model was required, in order to make the many-line case converge to a physically expected and coherent atmosphere. We successfully developed and tested such a selection method.
    Another finding is that, in contrast to the improvements in the inferred atmosphere, the fit quality decreases when many-lines are utilized.

\noindent
\textit{Conclusions.}Our results highlight that the many-line spectro-polarimetry method delivers more coherent results with superior line of sight (LOS) resolution of the atmospheric structure. Moreover, it effectively detects and utilizes even weak polarimetric signals in noisy data and thereby circumvents low noise requirements.
    It reveals uncertainties in atomic parameters of individual spectral lines and models, as the degree of freedom to compensate for those by compromising the inferred atmospheric parameters is considerably reduced.
    The study of these effects and their opportunities for enhancement of our models and tools has just begun. The innovative many-line method presents significant potential for solar physics and should be the preferred option for future observations with upcoming spectrographs.

\end{footnotesize}

\newpage
\section{Introduction}
\label{sec:ml:introduction}
To comprehend the solar lower atmosphere, it is essential to discern its physical features, most notably the temperature (\T), the line-of-sight velocity (\vlos{}), and the magnetic field defined by its strength (\B), azimuth (\(\phi)\), and inclination \((\gamma)\). The most effective way of determining these characteristics from observed polarization profiles is the application of Stokes inversions. Which leads to the creation of a model atmosphere that can accurately represent the observations, within the model's assumptions and constraints.

In contrast to stellar physics, for the analysis of solar spectra, the simultaneous observation and inversion of only a few spectral lines is still the norm. One reason for the limitation to a few lines is the preference for isolated, not blended, spectral lines embedded in a distinct continuum, which simplifies the calibration and inversion processes. 
In addition, in solar observations, much effort is invested in imaging and capturing the fast evolution of the solar scene, which necessitates fast data retrieval. This can be accomplished by reducing the number of pixels to be read out from a sensor. This can be done by restricting the field of view (FOV) or the wavelength range, with the choice typically being to limit the wavelength range. 
Hence, the use of many lines simultaneously has so far mostly been limited to the analysis of spatially low-resolution data from Fourier Transform Spectrometers \citep[FTS,][]{Brault1978}.

To achieve the required high polarimetric sensitivity, measurements need to possess a high resolution in the spatial, spectral, and temporal domains, paired with a low noise level. Ideally, the noise    should be between \(10^{-3}\) and \(10^{-5}\) of the continuum intensity \citep[See][and references therein]{Iglesias16}. Another way to reach the sensitivity goal is to use the signal from many spectral lines. In the era of fast large format imaging sensors and improved data handling capabilities, the need to observe only a few lines is no longer so strong. Consequently, \cite{Riethmueller19} suggested exploring the merits of using many absorption lines simultaneously. They showed that the outcomes from the many-line inversion were more precise and more robust against noise. 

The diffraction-limited data from \fisssp{} \citep[][here Chapter~\ref{ch:fiss_sp}]{Noort24}, the spectro-\-\mbox{polarimetric} extension of the Fast Imaging Solar Spectrograph \citep[FISS,][]{Chae13}, installed at the 1.6-meter Goode Solar Telescope \citep[GST,][]{Goode10, Cao10, Goode12} at the Big Bear Solar Observatory (BBSO) exceeds a spectral range of 30~\AA{} and covers \tlines{} relevant absorption lines. It is, therefore, particularly suited for such a many-line inversion technique.
In this study, we present the first application of the many-line inversion method to spatially highly resolved solar observations and compare it with a traditional inversion approach using a line doublet. 
We further discuss the necessity of, and a strategy for, the selection of lines for such a many-line inversion.

\section{Forward challenges and inversion problems}
\label{sec:background}
To appreciate the benefits and understand the challenges of solar many line inversions better, it is essential to remind ourselves on some inherent complications of the Stokes inversion process. The core of a Stokes inversion is the numerical solution of the differential radiative transfer equation (RTE). 
If the properties of the solar atmosphere, notably temperature, velocities, and magnetic field, are known, and local thermal equilibrium (LTE) can be assumed, the solution is relatively simple and results in synthetic \I,\Q,\U,\V-Stokes profiles for selected line(s). 
Unfortunately, for spectro-polarimetric observations the opposite is true: The Stokes profiles are known and the configuration of the solar atmosphere is the unknown parameter. 
The extraction of an descriptive atmospheric model is typically accomplished by iterative modifications to the assumed configuration, re-calculating the corresponding Stokes profiles, and minimizing the difference to the observed profiles. \citep{ToroIniesta16}.

Unfortunately, multiple atmospheric configurations can result in identical profiles. 
Moreover, as we can only observe the integral over the optical depth along the viewing angle, traces from different heights might cancel out or leave only a fuzzy footprint in the profiles. 
This degeneracy renders it impossible to derive the complete and turbulent stratification from observational data.
Nonetheless, by employing physical assumptions and using a micro turbulence factor to account for the fuzzy broadening, the extraction of stratified information is possible \citep[see, for instance,][]{Frutiger00, Danilovic16, Milic18, STIC, DeSIRe, MODEST24}
The inversion problem can usually be further constrained with the full set of Stokes parameters, but this constraint is still not perfect. For instance, the impact of even relatively strong horizontal fields on \Q, \U, or \V{} signals can be very small. 
The response of the line to a change in the atmospheric configuration is interconnected to the formation height of its core and wings, as outlined in \cite{Smitha20}. Using many lines, formed at different heights, hence adds constraints at those heights allowing to discern the atmospheric stratification better \citep[See, for instance,][]{Suetterlin96, Riethmueller19}.

A crucial ingredient for the success of a Stokes inversion is the knowledge of the atomic parameters of the observed spectral lines.
Factors such as the excitation potential of the transition's lower energy level, the central wavelength of the spectral line, and the weighted oscillator strength, typically given as \(\log(gf)\), significantly impact the line shape and depth. Not all of these parameters of all spectral lines of interest are known with the necessary accuracy from laboratory experiments \citep{Borrero03}.
The effects of a line being in LTE or not (non-LTE, NLTE) can influence the result of an inversion in a similar way \citep[see,][]{Shchukina01, Borrero02, Smitha20, Smitha21}. 
In addition, the influence of 3D NLTE and radiative transfer effects on line profiles becomes increasingly relevant, the higher resolution of the observation is \citep{Holzreuter12, Holzreuter13, Holzreuter15}.
Since atomic parameters and atmospheric configuration both influence the line profile, poorly determined parameters and incomplete models might lead to accurate fits of an observed profile, but the inferred parameters then might not represent the actual atmosphere.
From only a single spatial pixel and spectral line, it is comparatively difficult to affirm the accuracy of the line parameters.

Lastly, observational data is never ideal or perfect. 
Most notably the inevitable noise hides subtle signals in the observed profiles and increases the range of profiles considered to be a "good fit". This results in a further degradation of the already complex inversion problem \citep{Noort12, Danilovic16}.
Moreover, observational data is naturally limited in resolution and available photons and some spatial- and temporal averaging is inevitable.
Recently, \cite{Sinjan24} showed that this can lead to an underestimation of the magnetic flux everywhere on the solar disk.

All these factors affect each other in a complex and non-linear way, making the inversion of observational profiles a highly degenerate problem. 
The work by \cite{Riethmueller19} indicated that the consolidated signal of multiple lines can alleviate some of these issues. Also \cite{DiazBaso25} found it beneficial to use multiple lines to counteract spectral degradation and noise.
\cite{Vukadinovic24} devised a method to refine atomic parameters, like \(\log(gf)\), by combining information from multiple spectral lines and atmospheric configurations. 
However, all these studies used observables synthesized from numerical simulations, which circumvents some of the challenges when dealing with observations, highlighted above. In this study we will test the performance of many-line inversions on highly resolved solar observations.

\section{Observations and Data Calibration}
\begin{figure}
    \centering
  \includegraphics[width=1\textwidth]{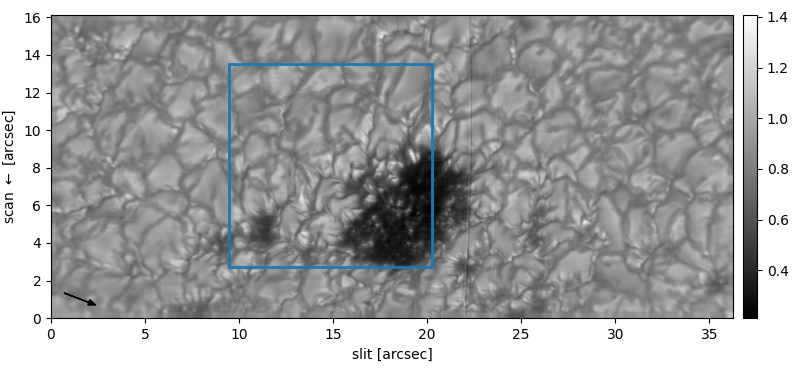}
  \caption[Continuum image of the observed active region extracted from the restored scan]{
  Continuum image of the observed active region extracted from the restored scan. The scan direction is from bottom to top along the \(y\)-axis, axis labels are in arc seconds, the intensity is normalized to \(I/I_C\). The blue box indicates the ROI used for this study, the black arrow indicates the direction to the disk center. 
  }
  \label{fig:context}
\end{figure}

\label{sec:data}
For this study we used a scan acquired with the slit-scanning spectrograph \fiss{} covering a spectral range of 33.06~\AA{}. By making use of a polarimetric modulation package, resulting in a setup referred to as \fisssp{}, all four Stokes parameters could be recorded.
For further instrumentation details, please refer to \cite{Noort24}\footnote{Author's note: Reproduced as Chapter~\ref{ch:fiss_sp} of this thesis.}.

The observation of an emerging active region was recorded at \(\mu=0.81\) (cosine of the heliocentric angle) on May 11\(^{\textnormal{th}}\), 2023, at \mbox{17:01 (UTC)}. 
The region was later assigned NOAA active region number \mbox{AR~13\,304}. For this study, we used a small region of interest (ROI) from the available data, which had a good representation of different types of solar features. 
Figure~\ref{fig:context} shows a continuum image extracted from the restored scan. The ROI used here is indicated with a blue box. Throughout the data reduction process, we followed the steps from \cite{Noort24}.
The \spectroflat{} algorithm \citep{Hoelken24}\footnote{Author's note: Reproduced as Chapter~\ref{ch:data_reduction} of this thesis.} allows for an accurate and field-dependent correction of the spectral curvature effect, which is crucial for our goal of using absorption lines from the full spectral field of view.
We used the Kitt Peak FTS solar atlas \citep{Neckel99} to perform the wavelength calibration and also to set up the accurate wavelength scale on the averaged flat field measurement that was closest in time, using the \texttt{atlas-fit} routine \citep{Hoelken24}. Based on a comparison with the FTS atlas we estimated and corrected for a gray spectral stray light level of 2.5\%.

The applied \specrestore{} algorithm \citep{Noort17} solves for the undegraded image using the seeing related point spread function (PSF) contribution to each pixel as determined from a simultaneously recorded set of context images. In doing so, it not only improves the contrast and resolution, but also reduces the spatial stray light contribution to each pixel \citep{Noort12, Noort17, Noort22}. 
After reconstruction the data show a noise level of about one percent in all parameters, which is comparable with other reconstructed or deconvolved datasets \citep[see][for a discussion]{Noort22}.
Unfiltered deconvolution is a linear operation that effects noise and signal in the same way. It therefore cannot change the signal to noise ratio (SNR). The Lucy-Richardson deconvolution algorithm used here only amplifies spatial frequencies below the diffraction limit and therefore acts as a filter to frequencies where there is only noise and no signal. Thus, the SNR of the restored data is enhanced \citep{Barnes04, Noort24}, compared to the original data.

After reconstruction we estimated the residual unpolarized wide-field spatial stray light, that cannot be removed by deconvolution. For this we spatially averaged over the quiet part of the scan and subtracted the resulting profile weighted by a number $<1$, from the umbral profiles. After subtracting 15\% of this averaged quiet-Sun line profile, the intensity in the cores of the strongest lines in the umbra becomes negative, producing a hard upper limit for the residual spatial stray light. As there are no means to estimate a lower limit in a similar way and the stray light must be below 15\%, we decided to not correct for the residual wide field stray light.

As described in \cite{Noort24}, there is a variation of the PSF along the wavelength dimension in the \fisssp{} data.
The \spinor{} inversion code \citep{Frutiger00phd, Frutiger00} assumes the same instrumental broadening for all lines. 
To allow for a simultaneous inversion of the complete spectral range, a homogenization of the spectral PSF was applied.
For this we convolved the recorded spectra with a Gaussian that had a spatially varying full width at half maximum (FWHM). 
The effective differential FWHM was determined for each wavelength position by the difference between the  FWHM of the widest part of the spectral PSF and the FWHM at the given wavelength position.

\newpage
\section{Inversion Setup}
\label{sec:setup}

The spectrum in the observed 5250~\AA{} region is dominated by iron, titanium, vanadium, and chromium. Table~\ref{tab:lines} provides an overview of the \tlines{} main contributors to the absorption spectrum and the used relative abundances of the respective elements. The line information was collected from the Kurucz line list \citep{Kurucz09, Kurucz18} and the Vienna Atomic Line Database \citep[VALD,][]{Ryabchikova15, Pakhomov19}. In the few cases of discrepancies where none of the parameters could sufficiently reproduce the lines we manually adjusted the \(\log(gf)\) until the discrepancy was minimal on multiple selected reference pixels \citep[Compare,][]{Vukadinovic24}. The abundances are based on \cite{Asplund21}.

\begin{table}
\begin{minipage}[b]{0.55\textwidth}
\begin{tabular}{l l r r r r} 
\hline\hline
Z & Atom  & Abundance & No. (I/II) \\
\hline 
12 & \io{Mg} & 7.55 & 6/0 \\
20 & \io{Ca} & 7.58 & 0/1 \\
21 & \io{Sc} & 3.10 & 0/1 \\
22 & \textbf{\io{Ti}} & 4.99 & 36/3  \\
23 & \textbf{\io{V}}  & 4.00 & 23/2 \\
24 & \textbf{\io{Cr}} & 5.67 & 18/5  \\
25 & \io{Mn} & 5.39 & 1/1 \\
26 & \textbf{\io{Fe}} & 7.46 & 40/9 \\
27 & \io{Co} & 4.92 & 8/3  \\
28 & \io{Ni} & 6.25 & 7/0  \\
39 & \io{Y}  & 2.21 & 1/0  \\
41 & \io{Nb} & 1.47 & 2/0  \\
57 & \io{La} & 1.11 & 1/0  \\
60 & \io{Nd} & 1.50 & 0/1  \\
62 & \io{Sm} & 0.95 & 2/0  \\
\hline 
\end{tabular}
\end{minipage}%
\begin{minipage}[!hb]{0.45\textwidth}
\vspace{-14em}
\caption[Number of identified solar lines per element]{
Relevant solar absorption lines per element and ionization state. 
The full line list consists of \tlines{} relevant lines.
}

\label{tab:lines}
\end{minipage}

\end{table}

\label{sec:2d_inversion}
\cite{Riethmueller19} have demonstrated the general suitability of \spinor{} for the inversion of spatially resolved many-line observations.
In a pixel-by-pixel inversion approach, information from neighboring pixels is not utilized. 
As discussed in Section~\ref{sec:background}, observed spectra can sometimes be reproduced by different atmospheric configurations without significantly affecting the \(\chi^2\) values of the pixel. 
This could lead to a large variance in the fitted parameters of adjacent spatial pixels, even though these contain information from essentially the same resolution element. 
To avoid this salt-and-pepper type noise and to converge into consistent maps of a spatially extended FOV, some form of spatial regularization is necessary. 
Intrinsically, \spinor{} inverts all pixels independently. 
To use the information from adjacent pixels, we applied Gaussian smoothing to the maps resulting from a given \spinor{} execution, and applied filters to the maps of the inclination and azimuth to keep them within the \(\pm90^\circ\) range. We then restarted the inversion process with the smoothed maps as improved initial estimates \citep{Noort12, noort13}.
In this article, we use the term "iterations" to describe the number of optimization steps \spinor{} uses to fit a single pixel profile, and "cycles" for the repeated runs of the \spinor{} code with improved initial estimates introduced between them. 

\begin{figure}
    \centering
  \includegraphics[width=1\textwidth]{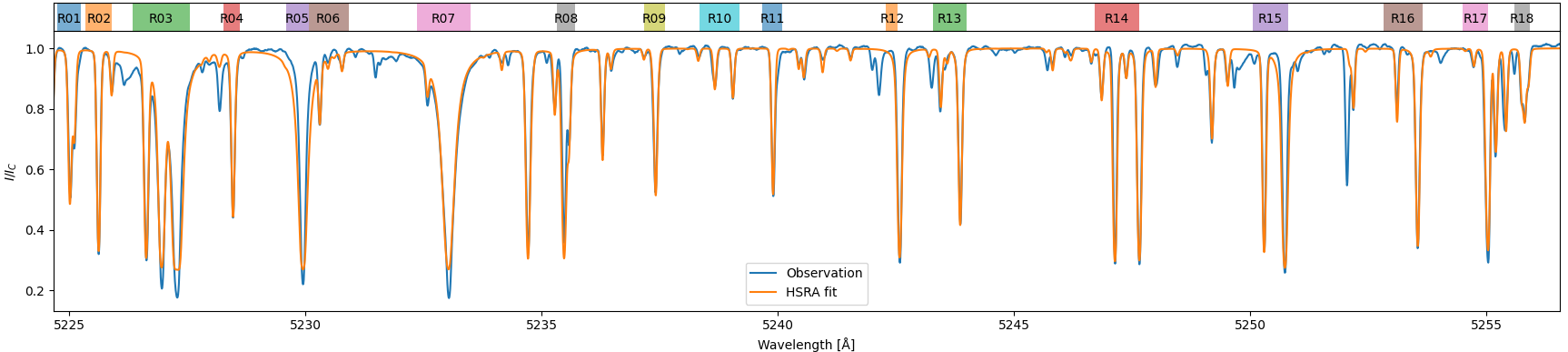}
  \caption[HSRA model vs. average quiet Sun profile from FISS-SP]{
  \blue{
  Comparison of the HSRA model atmosphere with a normalized Stokes \I{} profile from an averaged \fisssp{} flat field measurement. We allowed the code to adjust for micro-turbulence and temperature offset.
  The bar above the profile indicates the spectral windows R01 to R18 from Table~\ref{tab:wl?regions}. 
  }
  }
  \label{fig:hsra}
\end{figure}

To verify the setup of the inversion and the atomic line parameters in use, we replicated the spectrum of an averaged \fisssp{} flat-field measurement with the Harvard-Smithsonian reference atmosphere (HSRA) \citep{Gingerich71}, utilizing the strongest \tlines{} absorption lines within the spectral window of the data set (Table~\ref{tab:lines}). The average of the flat field measurement is particularly suitable for this exercise as it is recorded in the quiet Sun at disk centre and thus represents an averaged atmosphere, similar to the HSRA model. 
We adjusted the temperature offset to match the continuum levels of the synthetic spectrum with the observed one and further allowed the code to fit a height dependent micro-turbulence term, while all other values are taken from HSRA. The result is shown in Figure~\ref{fig:hsra}.

\begin{table}
\caption[Overview of the 18 spectral windows]{
Overview of the 18 spectral windows sorted by central wavelength (CWL) in \AA{}ngstrom.
The No. column notes the number of atmic lines in that window and an x in the last column marks the windows that have finally been excluded for the many-line setup.
An indication of the wavelength covered by each spectral window, R01--R18, is included in Figure~\ref{fig:hsra}.
}
\label{tab:wl?regions}
\centering 
\begin{tabular}{l l r r c} 
\hline\hline
Id & CWL & Atoms & No. & \\
\hline 
R01 & 5225.0 & 4\x\io{Cr}, 3\x\io{Ti}, \io{Nb}, \io{Ni}, \io{Y} & 10  & x \\
R02 & 5225.6 & \io{Cr}, \io{Fe} &  2 &  \\
R03 & 5226.9 & 7\x\io{Fe}, 4\x\io{Cr}, 3\x\io{Ti}, \io{Ca}, \io{V}, \io{Y} & 17 &  \\
R04 & 5228.4 & \io{Fe}, \io{Co}, \io{Ni}, \io{Y}&  4  &  x\\
R05 & 5229.8 & 3\x\io{Ni}, 2\x\io{Fe}, \io{Ca}&  6  &  x\\
R06 & 5230.3 & 3\x\io{Fe}, 2\x\io{Cr}, 2x\io{Co}, \io{Ca}, \io{Mn} & 9 &  \\
R07 & 5232.9 & \io{Cr}, \io{Fe}, \io{Ti}  &  3 &  \\
R08 & 5235.5 & 2\x\io{Fe}, \io{Cr}, \io{Ni} &  4  &  x\\
R09 & 5237.4 & 2\x\io{Cr}, \io{Co}, \io{Fe}, \io{Nb}, \io{Ti}, \io{Y} &  7 &  \\
R10 & 5238.8 & 3\x\io{Ti}, 2\x\io{Fe}, \io{Cr} &  6 &  \\
R11 & 5239.8 & \io{Cr}, \io{Sc}, \io{Ti} &  3 & x \\
R12 & 5242.4 & 2\x\io{Fe}, \io{Co}, \io{Ni}, \io{Sc}, \io{Y} &  6 &  \\
R13 & 5243.6 & 2\x\io{Cr}, 2\x\io{Ti}, \io{Ca}, \io{Fe}, \io{Mn} &  7 &  \\
R14 & 5247.2 & 2\x\ion{Cr}, \ion{Fe}, \ion{Ti}{}, \io{V} &  5 &  \\
R15 & 5250.4 & 2\x\ion{Fe}{} \io{Cr}, \io{Nd} &  4 &  \\
R16 & 5253.2 & 2x\io{Fe}, \io{La} &  3  &  \\
R17 & 5254.7 & 3\x\io{Fe}, 2\x\io{Co}, 2\x\io{Cr} &  7 &  \\
R18 & 5255.8 & 3\x\io{Fe}, \io{Ti} &  4  &   \\
\hline 
\end{tabular}
\end{table}

\subsection{Spatially resolved convergence challenges}
\label{sec:convergence_challenges}
We found it almost impossible to make the code converge to a consistent and physical solution when inverting the complete line list simultaneously on high spatial resolution spectra. For instance, we commonly found unrealistically high field values in the centers of granular cells, an unrealistic flow stratification, or strong inter-parameter cross-talk.
Even though we were able to reproduce the average spectrum with a slightly adjusted HSRA model, we were not able to extract a physically meaningful spatially resolved atmosphere from those lines. 
While in average quiet Sun profiles, magnetic field and velocities are negligibly small, this might not be the case in better resolved spectra. 

We tested different node setups, including number and position, to account for the higher complexity of the spatially better resolved data, but none of them showed a significantly better fit quality nor an atmosphere that would consistently match our expectations.
We finally adopted the configuration reported to be optimal by \cite{Danilovic16} and placed three height nodes at optical depth of 
\mbox{\(\log(\tau) := \log(\tau_{5000\AA})\in\{-2.0, -0.8, 0.0\}\)}. As atmospheric model we choose the established setup \citep{Riethmueller08, Noort12, MODEST24} with height dependent temperature, magnetic field strength, LOS velocity, and microturbulence, whereas field inclination and azimuth angle are assumed to be height-independent.

With this setup we could trace the issues with the resulting atmosphere to the selection of lines, by inverting the data using only the 5247/5250~\ion{Fe}{I} spectral lines. The result was noisy and imperfect, but matched our general expectations for the selected solar scene (details are given in Section~\ref{sec:doublet}).
Expanding the line list again soon led to the known issues of atmospheric configurations outside the physical expectations.
A careful examination of the lines, weights, and spectral windows and their influence on the inversion result was thus necessary. 
Facing similar challenges when analyzing stellar spectra, \cite{Bigot06} and \cite{Heiter21} compiled lists of "well-behaved" spectral lines that have well established atomic parameters and do not show anomalous behavior. 
\cite{laverick2019phd} compared a large number of line parameters from many different databases to remove systematic errors in the atomic data. They report that less than 40\% of the assessed \ion{Fe}{I} lines had sufficiently accurate \(\log(gf)\)-values.
However, all these lists focus on non-blended lines and are much too sparse for our endeavour. 
Therefore, we decided to try to identify problematic lines in our observed spectral region ourselves. 
Unfortunately, simply ignoring spectral lines that are present in the observation but which are not modelled by the inversion code, has a negative impact on the deduced atmosphere as well. 
To exclude such a line, we must exclude it from the fit, which is accomplished most conveniently by splitting the spectrum into discrete spectral windows.

We selected 18 such windows using the following three criteria: (1) The selected spectral window is free from lines that could not be reproduced by our code, (2) a section of continuum is present in the window, and (3) the lines within a spectral window need to provide enough information to constrain the target atmospheric stratification on their own.

In areas with large velocities the spectral profiles are strongly shifted. 
If this shifts a spectral line or an undescribed blend from a wavelength window with a strongly negative influence on the inversion result into (one of) the spectral window(s), the inversion result, again, suffers from the fact that this  part of the spectrum, that should be excluded, is now being inverted.
To mitigate this problem, the offending part of the spectrum must be either generously clipped or the border(s) of the spectral window(s) must be shifted on a per-spatial-pixel-basis according to the velocities (once known). We used the conservative first approach to avoid further complications in the comparison of the results.
For regions with strong lines that have a high effective Landé-\(g^*\)-factor \citep{Shenstone29} we have chosen higher weights on Stokes \V, as the determination of the magnetic field from \V{} is less affected by residual wide-field stray light, which is unlikely to be polarized.
We then individually inverted all 18 spectral windows independently from each other.
Table~\ref{tab:wl?regions} lists the spectral windows and included species that we used from our dataset. 
The full set of atomic line parameters and weights used can be found in Appendix~\ref{sec:atom_dat}. 

\subsection{Selection of spectral windows}
\label{sec:selection}
\begin{figure}
    \centering
  \includegraphics[width=1.01\textwidth]{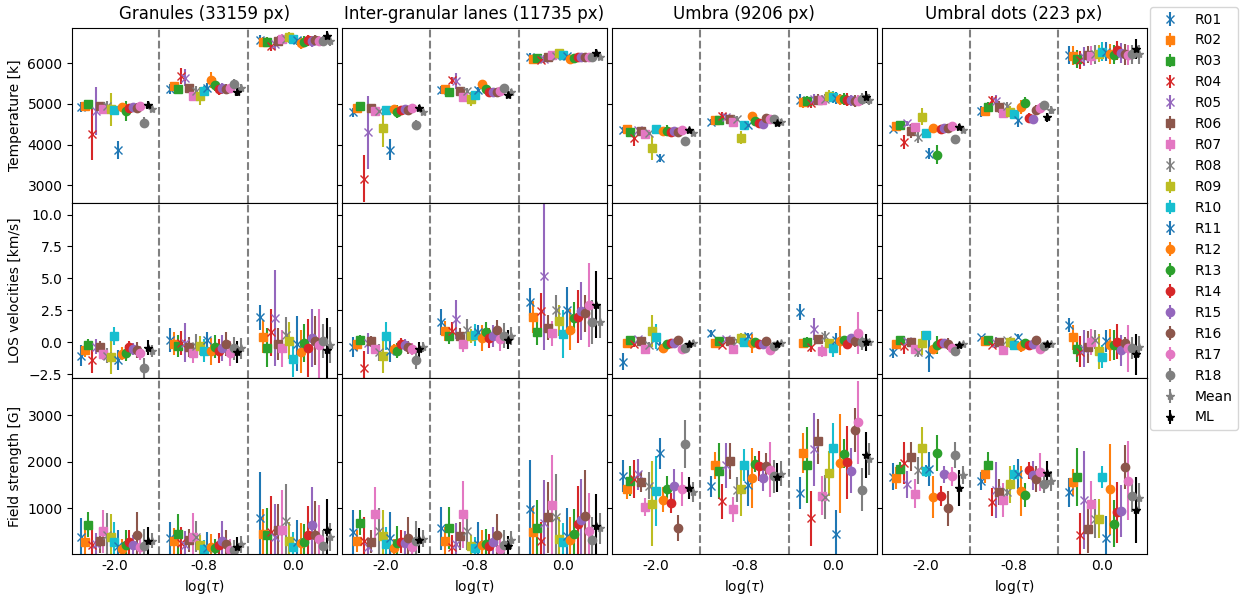}
  \caption[Mean values and standard deviation of atmospheric parameters from individual regions]{
 Comparison of atmospheric parameters  deduced from the 18 selected spectral windows (see Table~\ref{tab:wl?regions}), separated by type of solar feature. 
  Main columns show properties of granules, inter-granular lanes, bright points, umbra and umbral dots. Each column is sub-divided (separated by dashed lines) into the values of the parameters at the different optical depth nodes. Each row refers to one atmospheric parameter (temperature, LOS velocities and field strength). 
  Each spectral window is represented by a symbol, indicating the mean value of the retrieved parameter from all pixels corresponding to the given physical parameter, and a range bar, indicating the standard deviation. The excluded windows are marked with an \x{}-symbol. The gray entry indicates the mean of the finally used 14 spectral windows and the black entry the result from the simultaneous many line (ML) inversion. See text for detailed discussion. 
  }
  \label{fig:statistics}
\end{figure}

From the corresponding intensity image, we identified four atmospheric features, namely granules, intergranular lanes, umbra and umbral dots. 
Figure~\ref{fig:statistics} shows, for each of these four types of atmospheric features, the values of the atmospheric parameters temperature, line-of-sight velocity and magnetic field strength, given at the three height nodes used for the inversion. The individual symbols represent the mean values obtained from each of the 18 wavelength windows listed in Table~\ref{tab:wl?regions}. The error bars indicate standard deviation of the extracted atmospheric parameters per optical depth node.

While the results from most spectral windows align reasonably well, the variation in their sensitivity under different atmospheric conditions and at height nodes is apparent. 
R04, R05, and R11 all show strong deviations from the results obtained in the other windows in the \vlos{} and temperature, and we excluded these spectral regions from the final setup.
The temperature at \(\log(\tau)\!=\!-2\) of R09 also does not agree with the other spectral windows. 
However, since the temperature is generally so well constrained, we found that insensitivity to temperature in the higher nodes alone does not compromise the overall result when adding the corresponding lines to our many-line inversion setup.

\begin{figure}
\begin{minipage}[b]{0.5\textwidth}
  \includegraphics[width=1\textwidth]{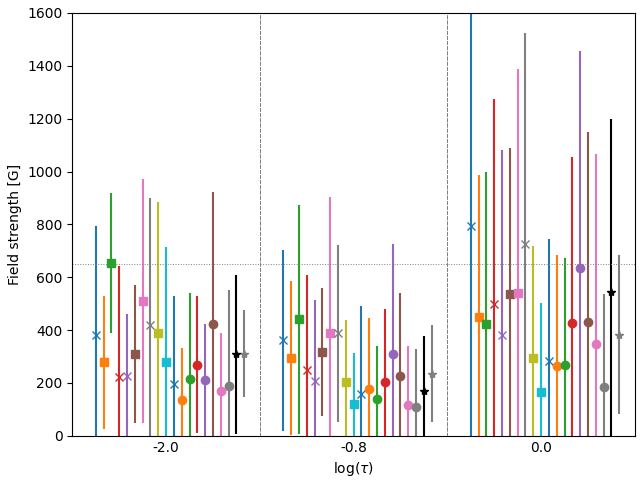}
\end{minipage}\quad
\begin{minipage}[b]{0.46\textwidth}
  \caption[Mean values and standard deviation of derived magnetic field values from individual regions]{
  Zoom into the lower left panel of Figure~\ref{fig:statistics} (please refer to it for the legend) showing the field strength statistics for granular pixels. 
  The threshold of (\(\approx\) 650 G) is highlighted with a dotted horizontal line.}
  \label{fig:field_statistics}
\end{minipage}
\end{figure}

The photospheric field strength needs a closer look, because we found that small deviations from the mean solution in this parameter have a strong influence on the combined solution.
This is especially apparent for the weak fields commonly found in granulation. 
As this is barely visible in Figure~\ref{fig:statistics}, we provide a zoomed-in view of the corresponding panel in Figure~\ref{fig:field_statistics}. 
We found that the addition of regions R01 and R08 drives the extracted atmosphere to implausible regimes. 
The regions R03 and R15 also show stronger field values in the highest and lowest node respectively, but the addition of the corresponding lines had no negative effect on the result of the many-line inversion.
Hence, we draw a line at approximately 650~G as the highest tolerable mean field value in granulation pixels.
In Figure~\ref{fig:field_statistics}, we indicate this with a dotted horizontal line. 

\begin{figure}
    \centering
  \includegraphics[width=1\textwidth]{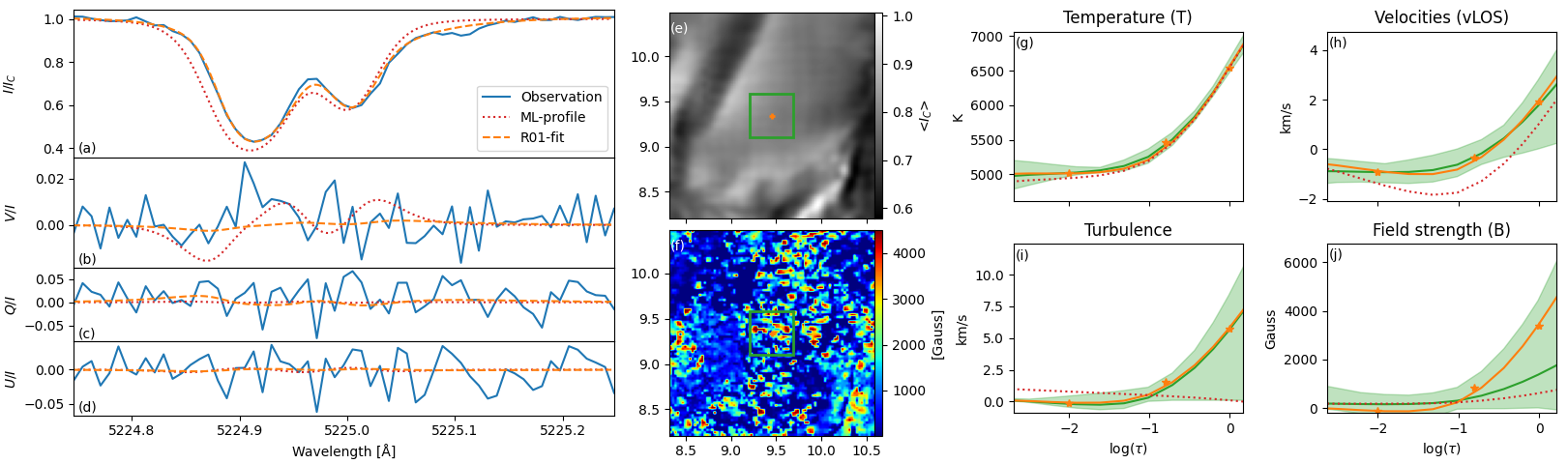}
  \caption[Stratification and fit of region R01]{
    Typical fit and stratification for the blended region R01 including ten spectral lines for a selected granulation pixel. 
    Panels (a-d):  Observation, and synthetic profiles in Stokes \I, \V, \Q, and \U{} in relative values of \I. 
    The R01-profile corresponds to the inversion result from the individual inversion of this region, while the ML-profile was synthesized using the atmospheric configuration extracted from the many-line inversion result (see Section~\ref{sec:ml:results}).
    Panel (e-f): Continuum image (e) of the granule with the selected pixel marked by the orange "+" and the corresponding map of the magnetic field strength (f) in the \(\log(\tau)=0\) node.
    Panels (g-j): Derived stratification for temperature (g), \vlos{} (h), turbulence (i) and field strength (j). The values at the \(\log(\tau)\) nodes are indicated with a symbol, the line represents the spline fit. Orange is the stratification for the selected pixel from the inversion result, green shows the mean and range of the values from the green box indicated in the middle panel, and red shows the stratification for the many-line atmosphere of the selected pixel.
  }
  \label{fig:r01_strat}
\end{figure}

Figure~\ref{fig:r01_strat} provides a typical example of a fit and corresponding stratification from the excluded spectral window R01.
Here, the code identifies a solution with a strong magnetic field in the lower photosphere without significant signals in Stokes \Q, \U, and \V. 
Given that this region contains two lines with an effective Landé-\(g^*\)-factor greater than 2.2, the catalyst for this solution could only be the near-continuum line broadening interpreted as a horizontal field. 
This solution is possible since the magnetic field is not sufficiently constraint from Stokes \I{} alone and the noise in Stokes \V{} allows this solution to satisfy the \(\chi^2\) metric.
The map of the magnetic field strength in the lowest optical depth node (panel f) shows many such field hot spots in the middle of a granule.
For reference the profile and stratification according to the final many-line result (see Section~\ref{sec:ml:results}) is included. 
The fit from the inversion of that region is a good representation of the observed line profile, while the profile synthesized from the atmosphere of the final result (ML-profile) diverges greatly.
Additionally, the displacement of the ML-profile indicates disagreement in the line wavelength position between the set of lines used in the final setup and R01.

The final line list consists of all regions from Table~\ref{tab:wl?regions} that were not disregarded. 
Additionally, we added three weak lines on the red side of R11 and three other weak lines between regions R14 and R15. 
These lines could not be tested with the same setup as the other lines, as they do not constrain the atmospheric stratification in the middle or upper photosphere. Unfortunately, they are also separated from adjacent windows.
However, these shallower lines help to constrain the parameters in the lower atmosphere, where the information can otherwise only be extracted from the wings of stronger lines.
We have carefully tested the additional lines with a reduced setup without height resolution of magnetic field and velocities. We further compared the result before and after incorporating them into the full setup to ensure their compatibility.
In total we identified \nlines{} solar absorption lines that do not show severe anomalous behaviour. This set includes most of the stronger ones, distributed over 15 wavelength windows.

\section{Results}
\label{sec:ml:results}
As stated in \cite{Noort24} the average \fisssp{} flat field measurement reproduces the Kitt Peak FTS atlas well. 
During the setup we recreated the flat field Stokes \I{} spectrum using HSRA model for the dominant \tlines{} absorption lines in our spectral window and achieved a good overall match with the temporally and spatially highly averaged profile. The result was presented in Figure~\ref{fig:hsra}.
The use of the average quiet Sun spectrum allows us to connect to previous studies using multiple solar lines from spatially heavily averaged profiles, for instance \cite{Balthasar84}, \cite{Solanki84, Solanki85}, \cite{Allende98}, or \cite{Borrero02}. 

Especially \ion{Fe}{I} lines are typically weakened by overionisation in non-LTE, and thus likely to have a somewhat too deep line-core when synthesized under the LTE assumption \citep{Lites72, Rutten88, Solanki88}. This effect is stronger the stronger the line.
On the other hand, the HSRA model includes the temperature rise in the chromosphere. When synthesized in LTE, this causes the rest intensities in the cores of the deepest lines to be too high, which likely explains why the depths of the deepest 4 computed lines are almost exactly the same.

\subsection{\io{Fe} line doublet inversions}
\label{sec:doublet}
\begin{figure*}[!hp]
    \centering
  \includegraphics[width=0.74\textwidth]{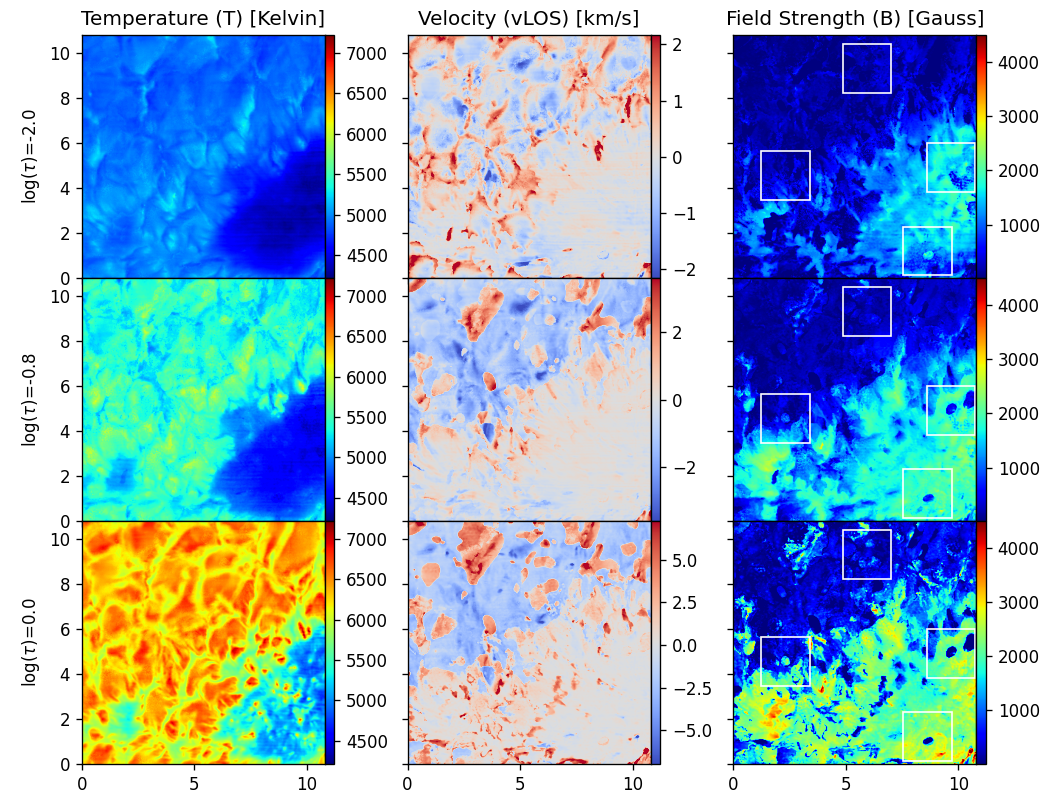}
  \includegraphics[width=0.246\textwidth]{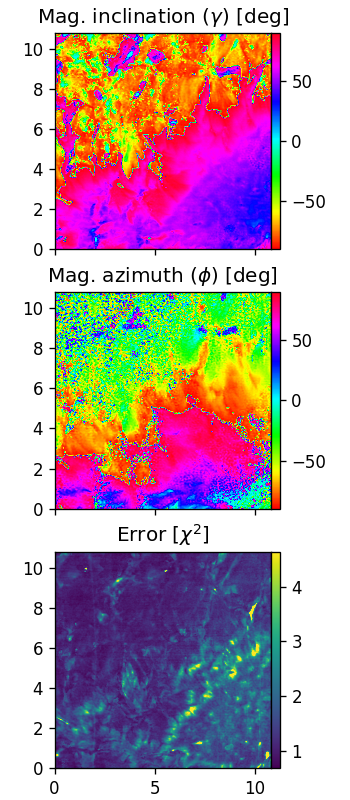}
  \caption[Atmospheric \(\log(\tau)\) maps from the Fe I doublet inversion]{
   Maps of the temperature, LOS velocities and magnetic field strength at \(\log(\tau) \in \{0.0, -0.8, -2.0\}\) and height independent inclination, azimuth, and \(\chi^2\). 
  This result is inferred with the traditional approach using the \ion{Fe}{I} doublet that consists of the 5247.05~\AA{} and the 5250.20~\AA{} lines. 
  We have used 4 cycles of 75 iterations each, the axis tick marks are in arc-seconds. See text for discussion.
  In all figures negative \vlos{} denote a flow towards the observer (i.e., up-flow) and positive LOS velocities denote a flow away from the observer (i.e., down-flow).
  }
  \label{fig:hinode_maps}
\end{figure*}

To establish a link to other spatially extended inversion results, we treated the data in a well-tested manner: 
The \ion{Fe}{I} doublet, consisting of the 5247.05~\AA{} and the 5250.20~\AA{} lines, is somewhat similar to the \ion{Fe}{I} doublet at 6301.5 and 6302.5~\AA{} observed by the Spectro-Polarimeter aboard the \mbox{HINODE} satellite \citep[HINODE/SP,][]{Lites01, Kosugi07, Tsuneta08}. 
Both lines of the 5250 \ion{Fe}{I} doublet are within the spectral range of our dataset. Hence, we used a setup that was already successfully used many times for the inversion of \mbox{HINODE/SP} data sets \citep[e.g.,][]{Riethmueller08, Noort12, noort13, MODEST24}.
With this, we inferred information on the physical parameters of the solar atmosphere including their height stratification. 
The result after 4 cycles of 75 iterations is shown in Figure~\ref{fig:hinode_maps}. 

The inferred maps feature salt-and-pepper type noise, especially in the higher temperature nodes and the azimuth and inclination of the magnetic field. 
Also immediately apparent is the missing granulation pattern in the low-photospheric \vlos{} maps. Instead, the typical imprint of the granular and inter-granular flows can be found in the highest node at \(\log(\tau)\!=\!-2\) only, which is higher than physically anticipated, but can be explained by the formation height of those lines.
The lower two maps are dominated by patches that resonate between the two height nodes and other parameters, most prominently with the magnetic field strength.
And also the maps of the magnetic field strength themselves prominently show "voids" (areas with apparently no field in a strong field regime, e.g. the umbra). We also observe the opposite: Areas with strong fields in a quiet region. 
These structures are almost always simultaneously present in multiple height nodes.
We will further refer to such cross-talk and void patches as areas with implausible values (AWIV).
The fits, however, seem of a constant quality, as the corresponding \(\chi^2\) map does not show the same structures. 
A typical example fit is presented in Figure~\ref{fig:fullfit_umbra_compare}.

\begin{figure*}[!hp]
    \centering
  \includegraphics[width=0.74\textwidth]{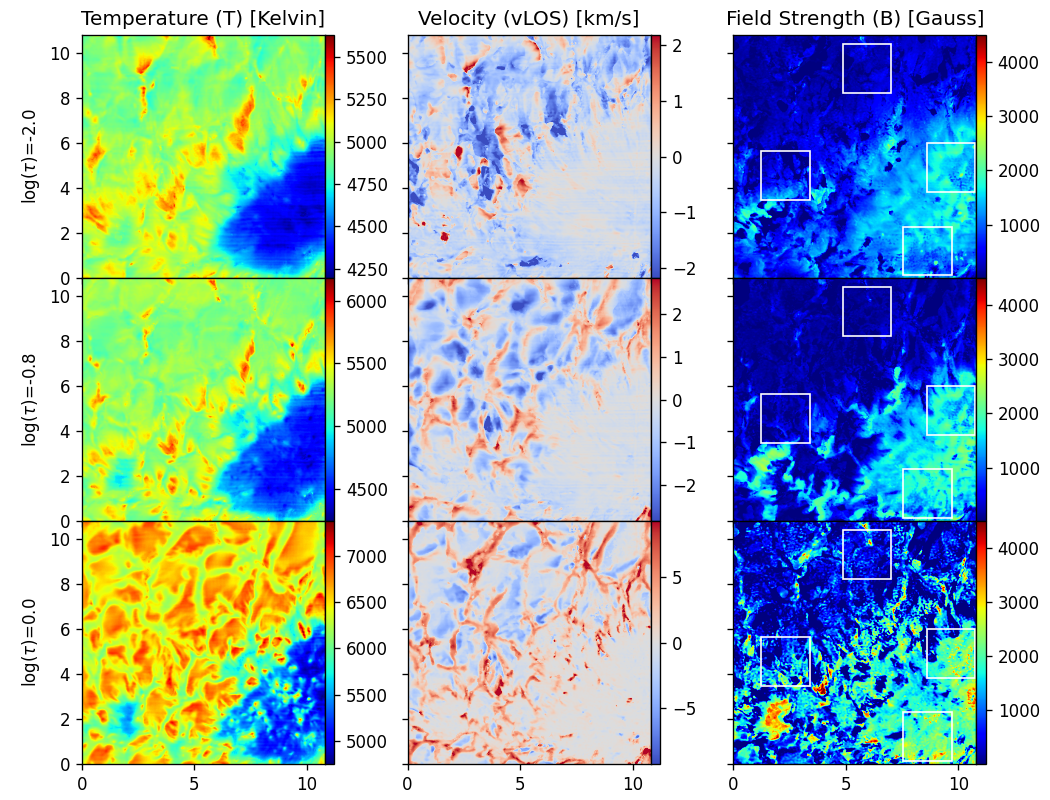}
  \includegraphics[width=0.246\textwidth]{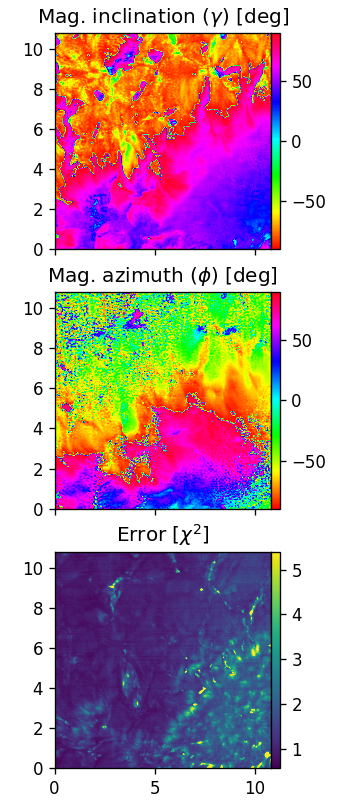}
  \caption[Atmospheric \(\log(\tau)\) maps from the many-line inversion]{
   Same as Figure~\ref{fig:hinode_maps}, but for the final many-line setup using 2 cycles a 75 iterations and \nlines{}~Lines. 
  }
  \label{fig:final_maps}
\end{figure*}

The limited information on height stratification encoded within the 5250/5247 \ion{Fe}{I}-line doublet is not unexpected. 
These lines are known to form at a very similar height and have therefore  been prominently used by \cite{Stenflo73} to determine magnetic field values at their formation height. Even though the lines sample the entire atmosphere where they are formed, the sensitivity in the line wings is not enough to constrain the magnetic stratification in multiple height nodes in many pixels in our dataset. 
Together with the one percent noise level of the observation the height resolution that can be extracted from these two lines is limited. This is especially true for gradients which are determined by contrasting information from the line wings and line cores.

\subsection{Many-line inversions}
For the many-line (ML) setup we simultaneously inverted all lines that could be identified as "well behaved" (see Section~\ref{sec:selection} and Appendix~\ref{sec:atom_dat}).
Figure~\ref{fig:final_maps} shows the inferred atmospheric configuration after two consecutive cycles of 75 iterations each on this setup. A third cycle was performed, but the differences to the previous result were negligible, so that we consider the inversion to be well converged.

In the direct comparison of the atmospheres extracted using the traditional two line approach and the new many-line approach we find that the utilization of multiple lines with different formation heights, magnetic sensitivities, and responses is able to constrain the inversion problem better.
While the general range of values in both setups match, the differences in the details are apparent.
Compared to the result from the \ion{Fe}{I} doublet, the new maps show the expected granular flows and, in all parameters, less salt-and-pepper type noise, even with only half as many cycles. 
In all maps, we notice a superior resolution of the fine structures. 
The maps of the azimuth and inclination of the magnetic field from the ML setup extend more coherent from the strong field into weak field regions. 

Also in the ML approach we find AWIVs. However, here we find them mostly in the form of voids restricted to the highest node of the magnetic field map, which can mostly be explained by too steep gradients along the optical depth axis caused by the three node setup.

\begin{figure}
\begin{minipage}[b]{0.64\textwidth}
\centering
  \includegraphics[width=0.45\linewidth]{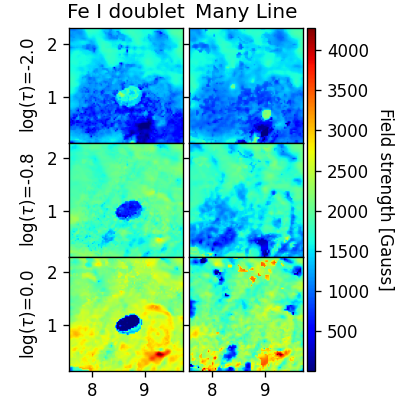}
  \includegraphics[width=0.45\linewidth]{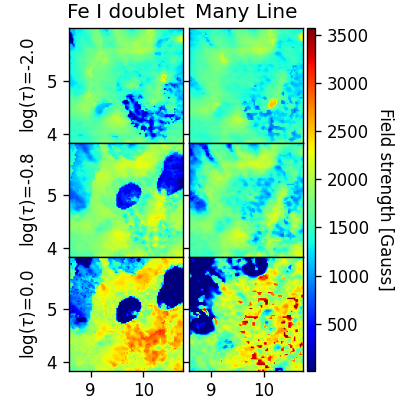}\\
  \includegraphics[width=0.6\linewidth]{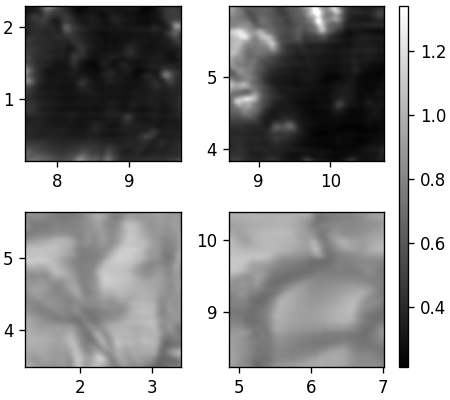}\\
  \includegraphics[width=0.45\linewidth]{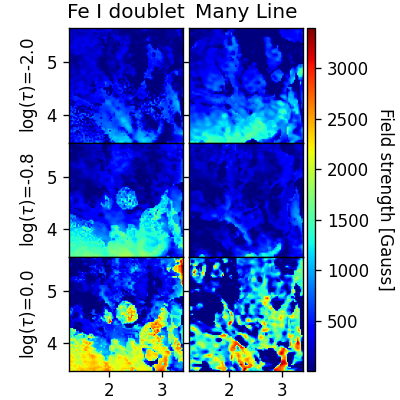}
  \includegraphics[width=0.45\linewidth]{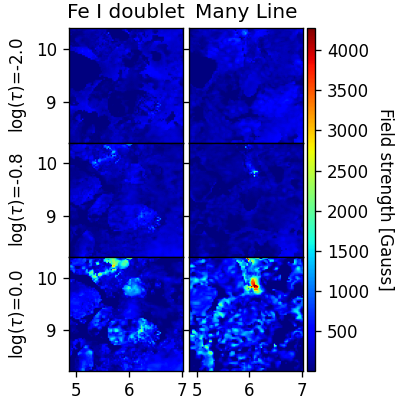}
\end{minipage}\quad%
\begin{minipage}[b]{0.32\textwidth}
  \caption[Comparison of AWIVs from \ion{Fe}{I}-doublet and many-line inversions]{
  Side-by-side comparison of magnetic field strength maps from both setups in four example areas. The continuum images in the middle panel correspond to the adjacent field maps panel.
  All tick marks are in arc seconds and within the reference frame of  Figures~\ref{fig:hinode_maps} and \ref{fig:final_maps} and are marked with white boxes therein.
  }
  \label{fig:map_comparison}
\end{minipage}
\end{figure}

Figure~\ref{fig:map_comparison} presents a side-by-side comparison of some AWIVs from both setups.
In the upper two panels we show typical examples of voids in the strong field regime of the umbra.
The field gradient within the AWIV in the top left panel is reversed, suggesting that the signal required to constrain it is obscured by noise.
In the corresponding ML columns not only the void is gone, but also more coherent fine-structure is visible.
The weak field regions in the umbra of the many-line result can be associated with flows, for instance from umbral dots, which is expected \citep[See for instance][]{Schuessler06, Riethmueller08}.

In the lower two panels we show typical examples from areas with weaker fields. In the left panel the (opposite) counterpart of a field void, an area with unexpected high field, is presented. In this case, the fitted profiles suggest that the code erroneously inferred a horizontal field from the noise present in the linear polarization profiles.
This artifact is also not present in the atmosphere extracted with the ML setup.
The lower right panel shows a granule with surrounding intergranular lanes. In the ML-case, we are presented with a well resolved small scale magnetic feature at the intersection of three intergranular lanes that can be associated with a bright point. In comparison only a fuzzy trace can be seen in the atmosphere extracted from two lines. 
In the ML case the combined information extracted simultaneously from many lines not only allows for a coherent height stratification and resolution of small scale events, it also eliminates most of the ambiguity in the profile which lead to these AWIVs.

\begin{figure*}
    \centering
  \includegraphics[width=1.01\textwidth]{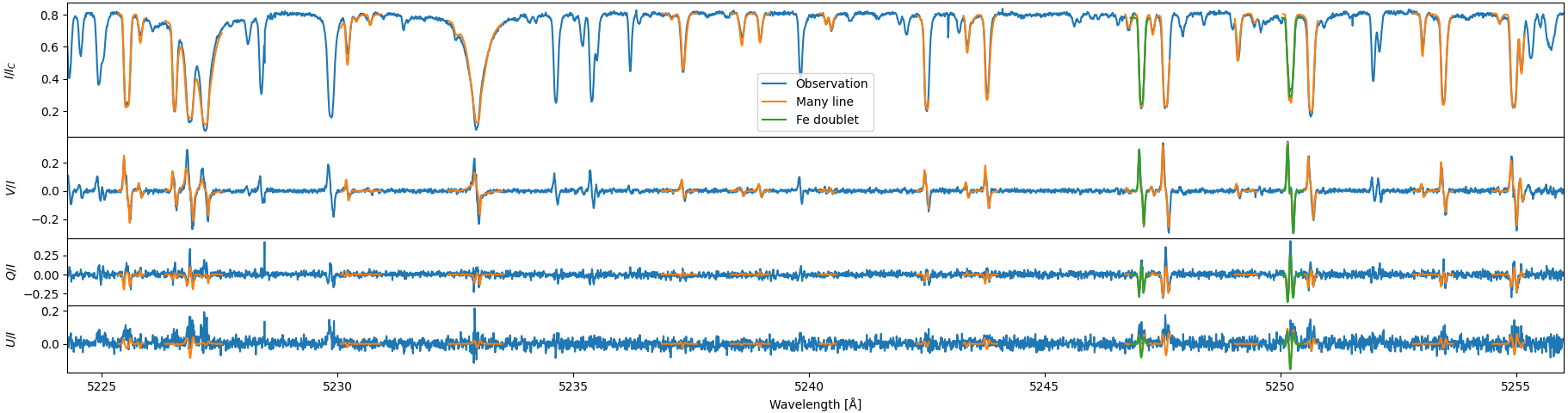}
  \caption[Full granular spectrum with fitted profiles from both inversions]{
   Typical example of the observed and fitted profiles for both setups. The topmost panel shows Stokes \(I/I_C\) while the other panels show relative signal strength of \V, \Q, and \U{} respectively.
  }
  \label{fig:fullfit}
\end{figure*}

\begin{figure}
\begin{minipage}[b]{0.6\textwidth}
  \includegraphics[width=1\textwidth]{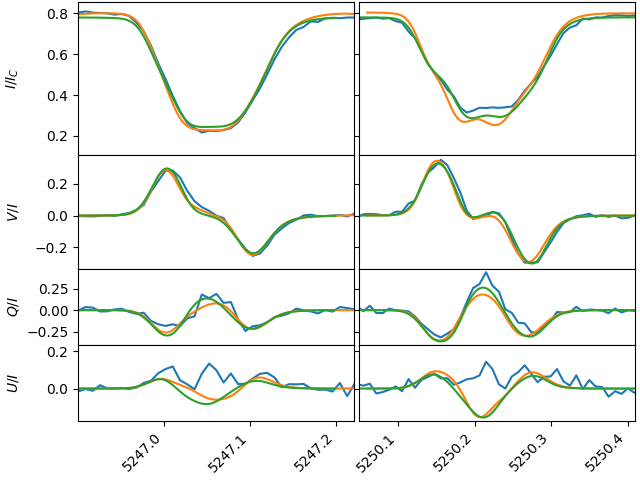}
\end{minipage}\quad 
\begin{minipage}[b]{0.37\textwidth}
  \caption[Observation and fitted profiles from both results on the 5247/50 lines]{
   Zoom in for Figure~\ref{fig:fullfit} on the two iron lines fitted in the traditional approach. Please refer there for legend.
  On this scale the difference in the fit quality, especially in \I{} and \Q{} becomes visible. We contribute the obvious discrepancy of both fits with Stokes \U{} to residual, field depended, cross-talk from Stokes \I{} in the data.
  }
  \label{fig:fullfit_umbra_compare}
\end{minipage}
\end{figure}

Comparing the normalized \(\chi^2\) maps we find that, in contrast to the more coherent atmosphere of the ML setup, the line doublet shows a better fit quality.
This also holds when contrasting the fitted profiles directly. Figure~\ref{fig:fullfit} presents a typical fit for Stokes \I, \Q, \U, and \V{} from both setups and Figure~\ref{fig:fullfit_umbra_compare} provides a zoom onto the two lines of the \ion{Fe}{I} doublet to allow for a more detailed comparison of the fits. 

\section{Discussion}
\label{sec:discussion}
We noticed that in our case the inversion of the 5247/5250~\ion{Fe}{I} line doublet does not readily lead to a coherent and physically meaningful model of the stratified solar atmosphere. This is possibly due to the relatively high noise level in the observations.  
When inverting our dataset with only those two spectral lines, spatial regularization, a reduced atmospheric stratification, or intense priming with expected values would have been necessary for the inversion code to converge further.
A simplified inversion setup (for instance, Milne-Eddington) might have produced better albeit more restricted result in the line-doublet case, but a too simple setup cannot account for the variety of lines used in the many-line inversion. 
On the other hand, the setup suggestion by \cite{Danilovic16} strives to optimize the atmospheric model for a few-line inversion. In our many-lines case, a more complex model might be better suited, as indicated by \cite{Riethmueller19}. However, accounting for the noise in our dataset and in order to allow for a direct comparison of the two setups we have chosen the same model for both inversions.

In the atmosphere inferred from the line doublet we find many areas where inter-node and inter-parameter crosstalk is apparent. 
Often these AWIVs feature sharp borders, where the atmospheric parameter is very different from one pixel to the other. These sharp features are stable over multiple inversion cycles and different initial guesses, indicating a converged solution.
The corresponding \(\chi^2\) map does not show the same structures and reports a similar fit quality inside and outside the AWIV.
Hence, inside an AWIV a non-physical atmosphere reproduces the spectral profile better than a physically expected one (compare Section~\ref{sec:background}). Another example for the conflict of matching fit and physically expected atmosphere was presented in Figure~\ref{fig:r01_strat}.

In the ML case the combined information extracted simultaneously from many lines facilitates the resolution of small scale structures and allows for a spatially coherent height stratification. 
Especially, the typical noise-related AWIVs from the two line case are not present in the ML case.
Combining many lines simultaneously allows signal that barely exceeds the noise floor to be reliably used (See, Figure~\ref{fig:fullfit_umbra_compare}), as the individual signals all constrain a common atmospheric property, which relaxes the low-noise requirement for precise polarimetry.
Utilizing lines that are diverse in formation height, Landé factor, excitation and ionization potential, and strength helps constrain all atmospheric parameters and their height stratification better, thus yielding a spatially smooth and physically more meaningful model of the solar atmosphere with a higher level of detail.

In the ML case we found boosted coherence of the inferred atmospheric configuration and improved noise robustness. In contradiction to this, we found the quality of the fits reduced. As noise and fit quality are important measures for acceptance of inversion results, both findings need to be better understood and will now be discussed further.

\subsection{Fit quality}
\label{sec:fit_quality}
\begin{figure*}
    \centering
  \includegraphics[width=1.01\textwidth]{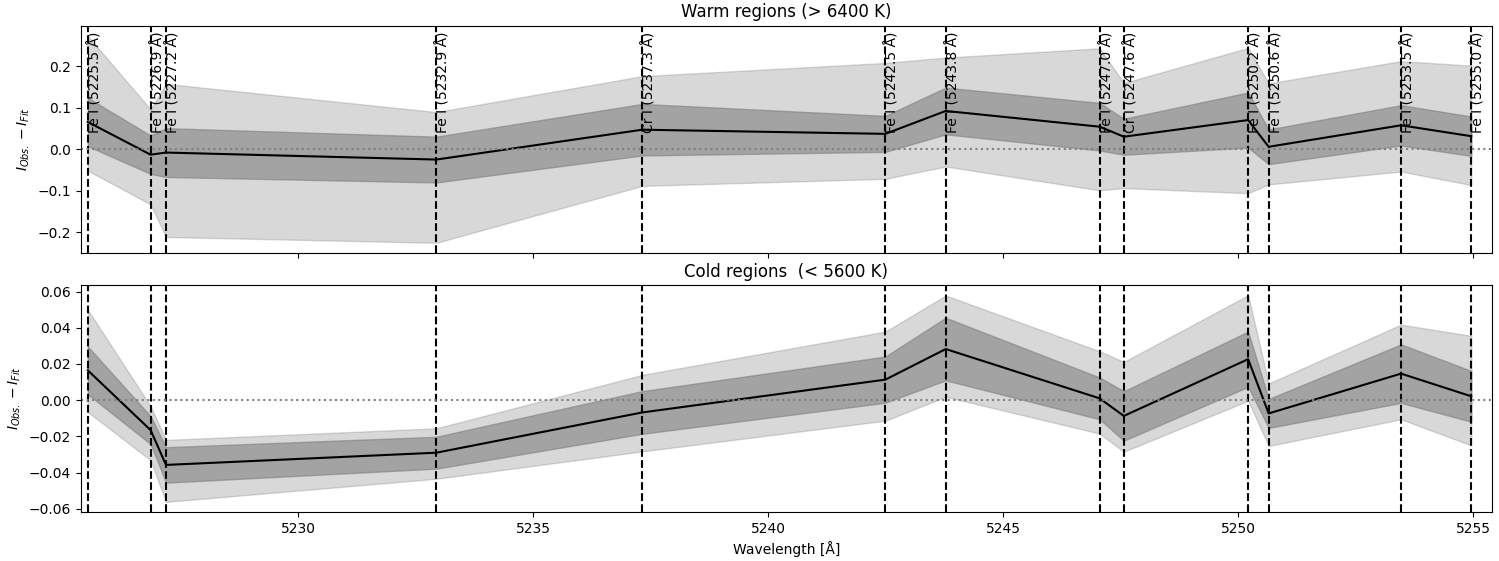}
  \caption[Typical fit residual for selected example lines]{
     Typical fit residual for some example absorption lines (indicated by dashed vertical lines).  The light area marks the full range of residuals measured, the darker area the standard deviation and the solid line the mean.
  }
  \label{fig:misfits}
\end{figure*}

Combining the individual results from the 13 selected spectral windows (see Table~\ref{tab:wl?regions}) into an average atmosphere closely resembles the many-line result, as can be seen from the "Mean" and "ML" markers in Figure~\ref{fig:statistics}. 
This similarity is a strong indication that the error for each spectral section is independent. Hence, the code has much less freedom in the ML case to compromise the atmosphere in order to keep the fit quality high (see Section~\ref{sec:background}).
The fit to the lines in each individual window is more accurate to the simultaneous fit of all windows, but the individual atmospheres extracted differ slightly from each other.
Hence, a synthesized profile based on the mean atmosphere cannot reproduce the "best fit" for any single window. 
Similarly, in the many-line approach the code needs to balance the responses of all lines to find the best solution for all used lines simultaneously. 
This is almost a fitting platitude: When the number of data points  increases while the number of free parameters remains fixed, the goodness of fit usually decreases and the credibility of the determined parameters increases.
Hence, with a high number of lines the usual global \(\chi^2\) measurement becomes less meaningful, as it does not reveal the line specific fit quality. 
To test for fixed patterns in the fit quality of the individual lines we assessed the over and under estimation of the line cores, typically referred to as fit residuals. We therefore computed the difference of the \(I/I_c\) intensity values in the core of all stronger lines in the observed and fitted profiles. Figure~\ref{fig:misfits} shows the result from various randomly selected spatial pixels from warmer and cooler regions for some selected prominent lines.
We find that for all atmospheric conditions the sum of the fit residuals over all lines is approximately zero in each spatial pixel, which is the expected effect of the necessary compromise to find the best global fit for all lines simultaneously.

At this stage we have already verified the reproducibility of an average quiet Sun profile and excluded the groups of lines that show severe anomalous behavior. But even in the selected line list that resulted in a coherent atmosphere with physically expected values, we still cannot be completely certain about the atomic line parameters in use. 
However, since all fitted lines show too shallow and two deep line core-fits in different pixels and we find lines that show trends in either direction, we assume the residual uncertainties in the \(\log(gf)\), excitation energy of the lower levels, and other line specific parameters to partly cancel out and hence play a minor role.
Another potential reason are errors in the relative abundances, but since we found over and under estimated line cores of each dominant species in most spatial pixels, we also found this unlikely to be the main driver.
It is nonetheless striking, that most lines show a clear and stable fit residual behavior relative to the other lines (see Figure~\ref{fig:misfits}).
One generic reason that we could not rule out is that, with just three height nodes, the code cannot create the steep gradients required in the temperature and velocity stratification to account for the different formation heights of the lines.
In the lowest atmospheric layers, steep temperature and velocity gradients are expected. However, these gradients are difficult to constrain with a few stronger absorption lines, since this information is encoded only in the near continuum line wings. When using many lines simultaneously each line forms at a different height and especially weaker lines provide information about the near-continuum region.
A further indication for this are the remaining AWIVs in the topmost magnetic field map of the ML result.
In our experiments, we found that seven nodes provided a well-constrained and reasonable temperature stratification. Unfortunately, due to the sensitivity and formation height of the observed lines, we were unable to constrain more nodes in the magnetic field and the Doppler velocity.
However, as it currently stands, SPINOR does not permit a different number of nodes for different atmospheric parameters, making the compromise of three height nodes necessary.

Another generic reason might be the use of an LTE code.
\cite{Lites72, Rutten88, Solanki88} and later also \cite{Shchukina01} have established that almost all neutral iron lines exhibit deviations from LTE. 
\cite{Asplund00} suggested that good fits to the observed lines could still be obtained by assuming LTE with an iron abundance of \(7.46\). 
Using lines with a wide range of atomic parameters can further help to minimize these effects from NLTE, because each transition will be influenced by the NLTE effects in different (and sometimes, opposing) ways. 
Indeed, \cite{Borrero02} were able to fit multiple \io{Fe} lines to an average quiet Sun spectrum from an FTS atlas better than anticipated. 
Also \cite{Holzreuter13} found that in such heavily temporally- and spatially-averaged line profiles NLTE effects are partially compensated.
This agrees with our well reproduced average quiet Sun spectrum (see Figure~\ref{fig:hsra}).
In our targeted observation we are presented with a spatially very highly resolved dataset from a single scan path, where the competition over atmospheric configuration between more LTE and more NLTE lines is no longer hidden in the average profile.
Systematic departures of up to 5\% (temperature) to 20\% (velocities and magnetic field) for using LTE inversions in NLTE conditions on spatially resolved data were demonstrated by \cite{Smitha20, Smitha21}.

The aim of this study was the application of the many-line technique, as suggested by \cite{Riethmueller19}, to highly resolved observations. For comparability with their work, and since \spinor{} was already successfully tested for many-line inversions, we used the same code here. We acknowledge that it is a LTE code and provides challenges in height node handling.
Unfortunately, an exploration of both issues is outside the scope of the current work. 

\subsection{Noise}
\label{sec:noise}
The undegraded spectra exhibit a relatively high noise level, which makes the deduced parameters less reliable \citep{Noort12}.
Especially in \Q{} and \U, the signal is almost buried in the \(10^{-2}\,I_c\) noise, and for multiple lines outside the 5247 and 5250 regions it barely exceeds the noise floor. 

In order to detect small scale features the scan speed cannot be arbitrary slow, as solar evolution would smear out the scene \citep[See,][Figure~1]{Iglesias19}. Therefore the number of frames accumulated per slit position cannot be higher. Further, deconvolution is needed to reach the diffraction limit and to resolve the desired small-scale features.
Consequently, a certain increased noise level is inevitable for this type of diffraction limited observation.

In \cite{Noort24}\footnote{Author's note: Reproduced as Chapter~\ref{ch:fiss_sp} of this thesis.} we compared the noise and signal levels of \fisssp{} with \hsp{}. Since \hsp{} observes two significant spectral lines over 112 spectral pixels while \fisssp{} observes more than 150 spectral lines over 3840 spectral pixels we concluded that the total signal in our observations is considerably higher. For this study we used \nlines{} lines distributed over 1281 spectral pixels. 
Considering the filtered deconvolution process applied, the SNR of \fisssp{} is boosted further. 
Furthermore, \cite{Noort22} argued that data taken with a bigger aperture improves the signal content across the whole frequency domain for spatial resolution elements (See Section~3.3 in their paper). 
Because of the enhanced SNR in the restored data, the detection of true positive signals, especially from small-scale features, is improved (see Figure~\ref{fig:map_comparison}). 
Unfortunately, the higher absolute noise level of the deconvolved data remains problematic. 
For instance, in regions where there is no detectable signal the noise allows fitting profiles with a relatively high background field \citep{Noort12, Danilovic16}.
A prominent example of such an unrealistic high field was shown in Figure~\ref{fig:r01_strat}. 
Since the field strength is an unsigned quantity, the corresponding extracted maps may look relatively coherent. In such a case the best indication for noise-driven erratic behaviour of the inferred field (as a vector) is given by the salt-and-pepper noise in the azimuth and inclination maps.

By combining the information from all \nlines{} lines, weak information that might not exceed the noise level anywhere can be reliably detected (see Figure~\ref{fig:fullfit}). Similar findings have been reported from stellar many-line approaches \citep{Donati97}. Consequently, the extracted stratification gradients from many-lines seem more realistic (see, Figure~\ref{fig:r01_strat}, Panels~g-j).
This agrees with the conclusion reached by \cite{Riethmueller19} and suggests that, when many lines can be inverted simultaneously, the low-spectral-noise requirement of \(10^{-3}\) to \(10^{-5}\) can be relaxed.
Also \cite{DiazBaso25} found a higher spectral sampling rate and undegraded data more beneficial than a lower noise level. The authors highlight that this effect is amplified when utilizing multiple lines.

\section{Conclusion}
We used the spatially reconstructed dataset that was obtained with \fisssp{} installed at the 1.6m GST and deconvolved using the method from \cite{Noort17}.
We applied the many-line inversion approach, proposed by \cite{Riethmueller19}, to this diffraction limited spectro-polarimetric dataset. 
In line with their findings the results we have obtained using manly-lines are more robust against noise compared with a line-doublet inversion.
The presented combination of spatial reconstruction and many-line inversion provides a way to mitigate losses from spatial degradation and shorter integration times, a conclusion recently also drawn by \cite{DiazBaso25}. Thus, the   approach discussed herein effectively enables us to use spectro-polarimetric data at the diffraction limit of large aperture telescopes while maintaining the reliability of the inferred atmospheric configuration.
Further, the combination of the information from many lines allows signal that barely exceeds the noise floor to be reliably used. This effect relaxes the low-noise requirement for precise polarimetry.

Our dataset included \tlines{} relevant absorption lines that were reasonably well reproduced with an HSRA based synthesis. 
It was therefore not expected that the inversion problem on spatially highly resolved data would not converge to a physically meaningful solution when incorporating all these lines.  
Instead we found that incorporating a line with anomalous behaviour (e.g., because of poorly known atomic parameters) can lead to non-physical results, potentially dominating the inversion outcome. In our tests we have seen that this can even drive an already converged solution to a non-physical one and hence a better initial guess of the atmospheric configuration does not suffice.
These problems are not new and have been predicted already, for instance by \cite{Shchukina01, Borrero02, Borrero03} and \cite{Smitha20, Smitha21}. 
To avoid the inclusion of such lines with anomalous behaviour we tested a selection strategy based on a statistical analysis of inversion results from individually inverted wavelength windows. In our case this resulted in the selection of \nlines{} lines for the many-line inversion setup. Such an analysis provides a relatively cost-effective and fast method for identifying incompatible contributions from not well understood lines.

While we had issues to extract a coherent and physically meaningful model of the stratified solar atmosphere from two prominent lines, we have shown that such issues can be alleviated by inverting many simultaneously observed spectral lines. Similar results are obtained if  atmospheres obtained individually from multiple simultaneously observed spectral windows are averaged. Here, the simultaneous inversion acts as a weighted average, using the knowledge on line response functions.
Although the result of averaging inverted atmospheres obtained from a large group of spectral windows, each with a small number of lines, somewhat resembles that obtained fitting all of the lines simultaneously, the quality of the fitted spectrum is poorer for the latter. 
This is an indication that fit quality alone is not a decisive measure for the quality of the atmosphere deduced from the inversion process, the accuracy of the atomic parameters, (non) LTE effects and the modelled complexity of the atmospheric stratification also have a significant influence. 
The adverse effect of using only a limited number of spectral lines on the spatial coherence of inversion results is at odds with the superior quality of the fit that is typically obtained in this way. 
As long as the "ground truth" is unknown, coherence and physicality of the extracted atmosphere must serve as our proxies to determine the (relative) reliability of our results.

We have just begun to explore and analyse highly resolved solar many-line data. Stokes inversions depend on reliable and exact atomic data, including \(\log(gf)\) values. Since the code can no longer so freely compensate for inaccuracies in the line parameters the here presented many-line technique is pointing to a need for improved values for many lines in the solar spectrum.
Further testing is required to determine if other regularization techniques and height node handling \citep[as in \texttt{SNAPI}, see][]{Milic18}, inversion codes that account for uncertainties in line parameters \citep[as in \texttt{globin}, see][]{Vukadinovic24}, or NLTE ready codes (as \texttt{SNAPI} or \texttt{NICOLE}, see \cite{SocasNavarro15}) can mitigate these problems.  

When HINODE/SP data first arrived, the proposed Milne-Edington inversions provided many challenges \citep[See for instance][]{Orozco07}. 
Improvements of the available tools helped overcome these challenges \citep[See for instance][]{Noort12}.
Since the decrease in fit quality we find when doing many-line inversions seems not to be erratic, we expect that yet again improved codes and atomic models are needed to reach a similar fit quality as we are used to from few-line inversions today.
However, even though the presented approach does not improve the quality of the fits, using many lines represents a noticeable step forward in terms of coherence, stability, and sensitivity. Finally, it allows for a higher spatial resolution, since it is more robust against noise.
Because of the availability of modern large format sensors, we, unlike \cite{DiazBaso25}, do not see a strong tension between wavelength range and spectral sampling, as long as data rates are not the limiting factor.
For upcoming spectrographs we therefore strongly recommend to aim for a larger simultaneous spectral coverage to allow for such many-line inversions. The spectro-polarimeters SUSI \citep{Feller25} and SCIP (Katsukawa et al. 2025 Solar Physics, submitted) on the Sunrise III observatory \citep{lagg25} are two further examples of instruments covering a broad wavelength range containing many spectral lines.
\chapter{New insights on small-scale features}
\label{ch:results}

\textit{The contents of this section correspond to the submitted version of the by now published article  "\textbf{On the magneto-convective nature of dark striations and moving bright grains at the edge of pores}" (J. Hölken, D.~Przybylski, M. van Noort, D. Cameron, S.-K. Solanki, \mbox{H.P. Doerr}, W. Cao, N. Gorceix, J.~Kang, K. Ahn, and J. Chae)  2026, \AnA{}, 708, A135, \citep{Hoelken26}.}

\textit{I ran the inversions, selected and analysed the features, performed the comparison with the simulation (which was provided by D. Przybylski), produced all figures, and wrote the majority of the text.}

\section*{Abstract}
\begin{footnotesize}    
\textit{Context.} 
High-resolution observations of the Sun reveal a multitude of small-scale striations, everywhere in the photosphere. While these features are well observed in broad-band intensity images, spectro-polarimetric observations remain rare.

\noindent
\textit{Aims.} 
In this study we want to characterize small dark striations on the pore-granulation boundary and bright grains moving along them. We aim to describe their magneto-convective nature.

\noindent
\textit{Methods.} 
We analysed restored context images and many-line Stokes inversions of a restored spectro-polarimetric scan from GST/\fisssp{} with a spatial resolution of 0.068". 
In the inversion we made simultaneous use of 85 solar absorption lines within a spectral window of 33~\AA{} in the 5250~\AA{} region. 
We compare the observations with a \muram{} simulation to discern the magneto-convective nature of striations and grains.

\noindent
\textit{Results.} 
We find multiple dark striations in the vicinity of pores or active region intergranular lanes with a typical width of 0.09" and moving bright grains that migrate along some of those striations towards the adjacent pore. 
Grains forming in a high-resolution \muram{} simulation of a pore show similar lifetimes of about 70~sec. A comparison of the atmospheric configuration of simulated and observed grains reveals a good qualitative agreement in structure, dynamics, as well as thermal and field stratification. The close agreement of observation and simulation supports the interpretation of a shared underlying process. 
The simulation show the dark striations are  formed at the top of a convective plume and their dark appearance is caused by plasma trapped in the field cusp at optical depth unity. The bright grains are composed of fresh upwards convecting material.

\noindent
\textit{Conclusions.} 
By utilizing high resolution spectro-polarimetry, many-line inversions, and comparison with a simulation we present the first analysis of the 3D fine-structure of small-scale striations and moving bright grains in the vicinity of a pore and describe their magneto-convective nature.

\end{footnotesize}    

\newpage
\section{Introduction}
\label{sec:grains:introduction}

Pores are the smaller siblings of the bigger and more thoroughly studied sunspots. 
Pores vary widely in size and flux, ranging from granular scales to small sunspots.
\cite{Peng24} recently presented a statistical comparison between pores and sunspots.
The defining difference between spots and pores is the presence of a penumbra. In the absence of a penumbra the body of a pore directly interfaces with granules. 
This boundary area of pores and surrounding granulation is dominated by small scale structures such as dark and bright striations. These features are of the order of or below 0.1 arc seconds in width. 
Observations have, for instance, been reported by, \cite{Scharmer02}, \cite{Lites04}, \cite{Schlichenmaier16} or \cite{kuridze25}. 
\cite{Schlichenmaier16} further reported some of those striations to form bright grains. 
A spectro-polarimetric observation of dark striations and bright grains was reported by \cite{Bharti16} using \hsp{} \citep{Lites01, Kosugi07, Tsuneta08}, but the resolution of \hsp{} is not sufficient to conclusively describe these features.
Recently, \cite{Zhao24} analysed striations found in an elongated granule.
To our knowledge, no other high resolution spectro-polarimetric description that allows to extract information on stratification has been published so far. 
\cite{kuridze25} contrasted high-resolution G-band images of striations forming at intergranular lanes with \muram{} \citep{muram, Rempel17} simulations and found a good match in the appearance of striations in the intensity images. They proposed the fluting instability as a driver for their formation. They however did not describe the formation of bright grains.

The data from the \fiss{} Spectro-Polarimeter \citep[\fisssp{},][]{Noort24}\footnote{Author's note: Reproduced as Chapter~\ref{ch:fiss_sp} in this thesis.} post focus setup at the 1.6 meter clear aperture Goode Solar Telescope \citep[GST,][]{Goode10, Cao10, Goode12} at the Big Bear Solar Observatory (BBSO) is so far the only slit-scanning spectrograph that can reach the diffraction limit in this telescope class and can hence resolve such small scale features.
In this study we use a time series from the context imager and a many-line inverted \citep{Riethmueller19, Hoelken25}\footnote{Author's note: The H\"olken et al. many-line paper is reproduced as Chapter~\ref{ch:many_line} in this thesis.} spectro-polarimetric scan to describe the magneto-convective nature of dark striations and moving bright grains found therein. We then compare the intensity images and the inferred atmosphere with a high-resolution \muram{} simulation.

\section{Observation \& Simulation}
\label{sec:observation}
\begin{figure}
    \centering
  \includegraphics[width=1\textwidth]{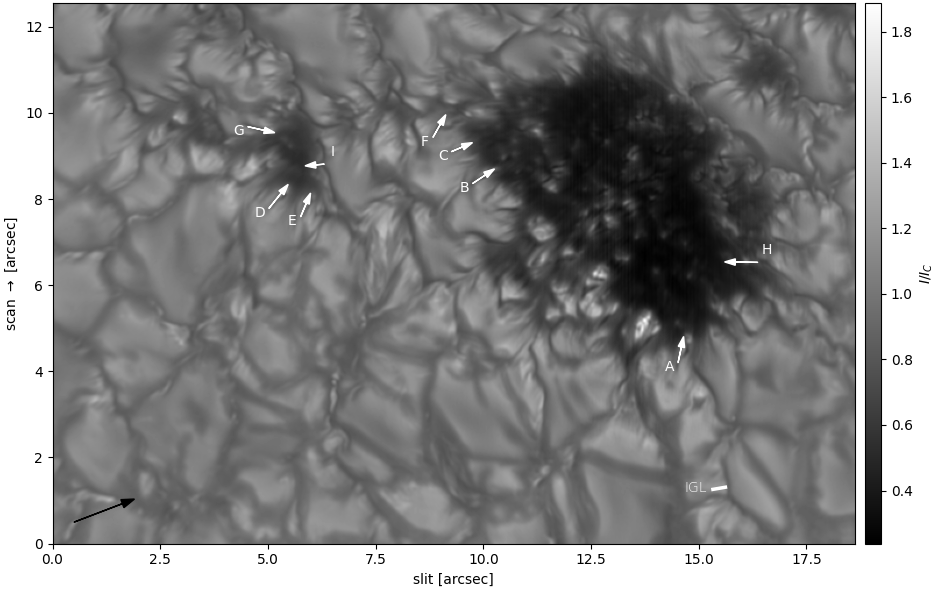}
  \caption[Continuum intensity scan image of the observed pore]{
      Continuum scan-image constructed from MOMFBD \citep{Noort05} reconstructed slit positions from the \fisssp{} context imager. The white arrows indicate striations which formed grains during the observation and indicate the movement direction of the grain. The labels correspond to panels in Figure~\ref{fig:grains}. The black arrow indicates the direction of the disk centre.
    }
  \label{fig:overview}
\end{figure}

For this study we used a scan with a diffraction-limited resolution of 0.068", obtained with the polarimetric extension of the slit-scanning Fast Imaging Solar Spectrograph \citep[\fiss{},][]{Chae13}, referred to as \fisssp{} \citep{Noort24}
An emerging active region was observed at \(\mu=0.81\) (cosine of the heliocentric angle) on May 11\(^{\textnormal{th}}\) 2023 starting from \mbox{17:01 (UTC)} until \mbox{18:48 (UTC)}.
The region was later assigned NOAA active region (AR) number \mbox{13\,304} and the central pore developed further into a sunspot in the following days \citep{Munjiba24}. 

The spatial reconstruction of the spectro-polarimetric dataset was performed following \cite{Noort17}.
The data reduction and inversion processes are described and discussed in detail in \cite{Hoelken25}. 
In the inversion we made simultaneous use of 85 solar absorption lines within a spectral window of 33~\AA{} in the 5250~\AA{} region. 
For this study, we use a slightly larger field of view (FOV), as shown in Figure~\ref{fig:overview}, from the same \spinor\ \citep{Frutiger00, Frutiger00phd} inversion. 
Simultaneously, the region was observed with the \fisssp{} context imager. From these intensity images a three~arc~seconds wide strip centred on the \fisssp{} slit was restored using the Multi Object Multi Frame Blind Deconvolution \citep[MOMFBD,][]{Noort05} method with a final cadence of \(2.667\)~seconds.

We will compare the \fisssp{} observation with a magnetohydrodynamic (MHD) simulation of an active region. We used the \muram{}-code \citep{muram, Rempel17} to produce a model of a small pore. We use a multi-group short-characteristics radiation transport scheme including 12 radiation bins, with the bin boundaries outlined in \cite{Beeck12}. The simulation uses the abundances of \cite{Asplund06}, the equation of state is generated using the FreeEoS code \citep{Irwin12} and the opacities are calculated using the MPS-ATLAS code \citep{Witzke21}. The horizontal boundary conditions are periodic. The upper boundary is open to outflows, closed to inflows and performs a potential field extrapolation into the ghost cells. The lower boundary is open, and allows passive advection of the magnetic field (OsB of \citep{Rempel14}). To avoid the stringent numerical timestep restriction from the strong magnetic fields we impose a limit on the Alfven speed of 250~km/s using the semi-relativistic 'Boris' correction.
The simulation domain consists of a box with \(x,y,z\)-edge length of roughly \(10 \times 10 \times 7\)  Mm, spanning from 4.9 Mm below and 2 Mm above the average $\tau_{500~\mathrm{nm}}=1$ layer. The simulation consists of \(1536\times 1536\times 1056\) gridpoints giving a regular cubiform voxel grid of ca. 6.5~km edge length. The simulation was initiated from a magnetoconvection simulation including only small-scale-dynamo (SSD) fields, similar to those of \citep{Rempel14,Przybylski2025}. 
An additional 400 Gauss uniform vertical field was added on top of the existing SSD fields, and the simulation was run until the RMS magnetic field in the box saturated. The resolution of this simulation was then increased and run for 15~minutes to allow the small scale structure to develop. Within those 15~minutes the evolution was saved every 14.3~seconds, defining the available cadence.
Our region of interest in this simulation is the formed small pore of about \(2\times 2\)~Mm that shows similar features as the \fisssp{} observation.

\section{Observational Results}
In our dataset we detect several dark striations bordering both the pores and strong network intergranular lanes with stronger flux concentrations. The striations are best visible on the limb side. They can have multiple shapes and may appear split or snake like. Some of the striations form bright grains moving along or with them. Such a grain is a small bright feature in (or directly attached to) a dark striation with an intensity that usually surpasses the intensity of the average quiet Sun (QS). 
Typical examples of striations and their grains are shown in Figure~\ref{fig:grains}, their apparent movement can be seen in the corresponding animated version.
The average full width at half maximum (FWHM) of grain hosting striations in our dataset is 0.09", with an average intensity of about 0.8~\(I_{QS}\) and all of them were found at the borders of the two dominant pores.


\subsection{Broad-band intensity images}
\begin{figure}[p]
\centering
  \includegraphics[width=1\textwidth]{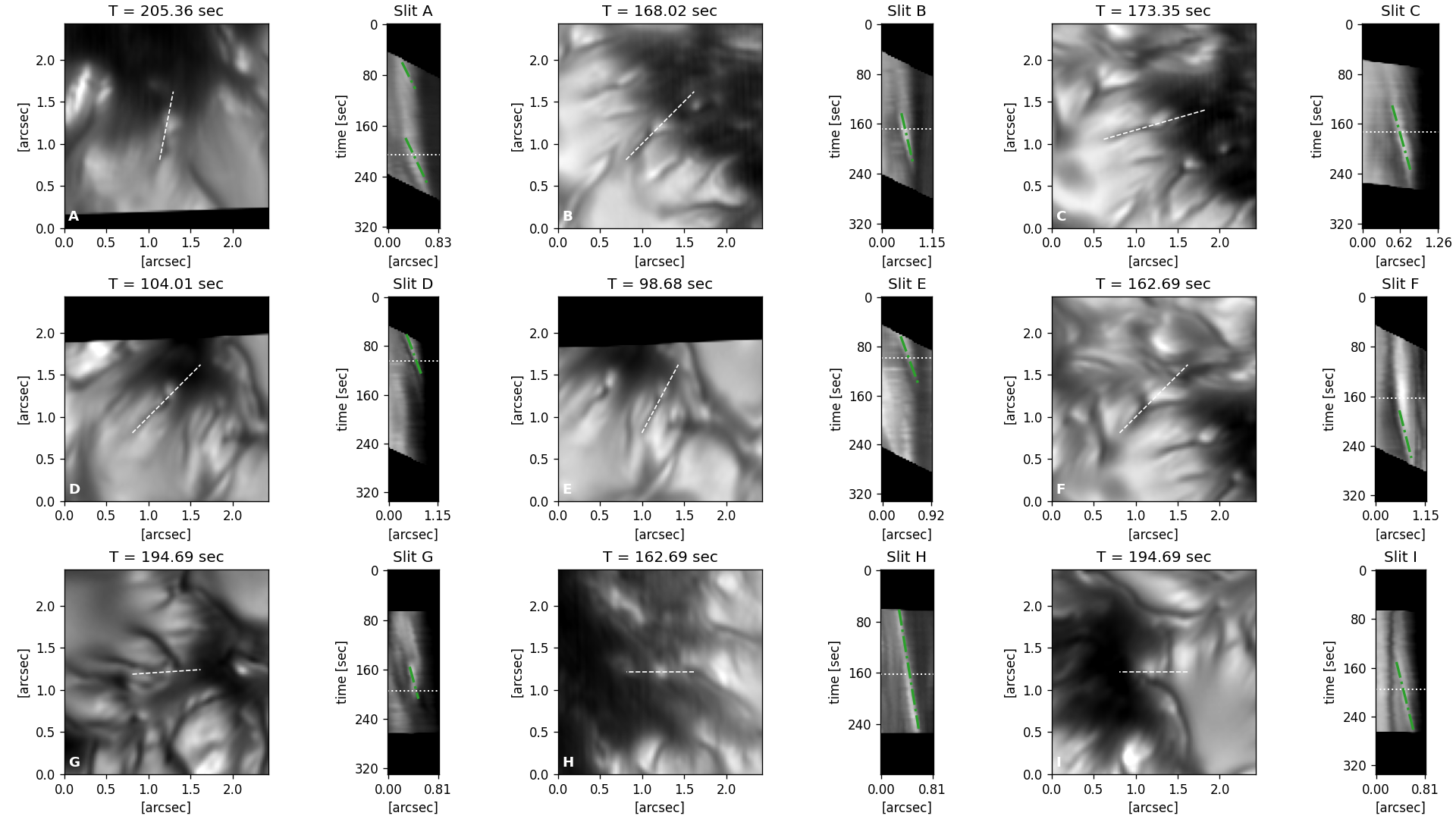}
  \caption[Context images and time-distance plots of striations and grains]{
       Overview of the nine striations hosting bright grains marked in Figure~\ref{fig:overview}. Each striation is represented by a continuum intensity image and a time-distance diagram (TDD). The latter was extracted along the dashed line indicated in the continuum image. The faint dotted line in the TDDs indicates the time of the adjacent context image, the dash-dotted green diagonal lines track the observed grains in time along the slit. 
    \href{https://www.aanda.org/10.1051/0004-6361/202659037/olm}{An animated version of this Figure is available with the online material of} \citep{Hoelken26}.
    }
  \label{fig:grains}
\end{figure}

In the broadband context images we frequently observe bright grains appearing  on the pore-side half of the dark striations. 
From the context images we have selected nine dark striations where the formation, evolution and decay of a grain was sufficiently sampled. Their location and orientation is indicated with white arrows in Figure~\ref{fig:overview}. 
We observe two distinct types of such grains. The grains of \textit{Type Y} are bright features between the legs of a Y-shaped striation and the pore. Similar to the grains described by \cite{Schlichenmaier16} and \cite{Bharti16}. The splitting point of the striation and the grain move simultaneous towards the pore. This bound movement is giving the impression of the grain being pushed by the striation. In our dataset the grains A.2, B, C, F, and G (compare, Figure~\ref{fig:grains} and Table~\ref{tab:grains}) are of this type. 
In comparison, the grains of \textit{Type S} seem to form as a small bright dot in a strongly curved section at the side of a snake-like striation, which seems to be transported towards the pore by the swaying motion of the striation. Grains A.1, D and I are of this type. Grain H cannot conclusively characterized.

\begin{table}[p]
\begin{minipage}[b]{0.55\textwidth}
\begin{tabular}{l|r|r|r }
         Id & orientation & speed & lifetime  \\
         \hline
         A.1  &  57.79\supcirc{}DC & 3.6 km/s & 42.13 s \\
         A.2  &  57.79\supcirc{}DC & 3.5 km/s & 71.74 s \\
         B  & 24.10\supcirc{}DC & 2.3 km/s & 80.01 s \\
         C   & 4.79\supcirc{}DC & 2.1 km/s & 103.20 s \\
         D  & 24.10\supcirc{}DC & 3.3 km/s & 64.81 s \\
         E  & 41.03\supcirc{}DC & 3.0 km/s & 80.81 s \\
         F  & 24.10\supcirc{}DC & 2.5 km/s & 77.08 s \\
         G  & -17.08\supcirc{}DC & 2.0 km/s & 51.74 s \\
         \hline 
          \multicolumn{2}{l|}{Limb-side grains} &  2.97 km/s & 71,44 s \\
         \hline 
         H  & 159.10\supcirc{}DC & 1.2 km/s & 186.40 s \\
         I  & 159.10\supcirc{}DC & 1.8 km/s & 111.48 s \\
         \hline 
          \multicolumn{2}{l|}{Disc centre-side grains} &  1.47 km/s & 148,94 s \\         
    \end{tabular}
\end{minipage}\qquad%
\begin{minipage}[!hb]{0.42\textwidth}
\vspace{-11em}
    \caption[Speed, orientation, and lifetime of grains]{
    Apparent horizontal speed, orientation with respect to disk centre, and lifetime of bright grains.
    }
    \label{tab:grains}
\end{minipage}
\end{table}

Since we only restored a three~arc~seconds wide strip from the context images, which is centred on the \fisssp{} slit and moves along with it, a scan pass with the restored context images allows for a high cadence broad-band observation of a striation of about 200~s. 
Figure~\ref{fig:grains} presents a region of interest (ROI) from a single context image for each grain forming striation and related time distance diagrams (TDD) that track their temporal evolution along a slit placed parallel to the movement direction. 
All of the observed grains moved towards their adjacent pore, as can be seen from the animated version of this figure. In the TDDs we fitted a slope to each grain, indicated as dashed green lines, to estimate lifetime and apparent horizontal speed of the grains. 
Table~\ref{tab:grains} lists the corresponding measured lifetimes and apparent speeds. 

We define the apparent rotation of a grain as deviation of its apparent movement direction from an idealized straight line towards disk centre. With this the limb-side grains A to G show rotations of -20\supcirc{} to 60\supcirc{}, where 0\supcirc{} represents the idealized disk-centre line. The grains H and I are formed on the disk-centre side of the pore and are moving towards the nearest limb with a relative angle of about 159\supcirc{}. The rotation of all grains is given in Table~\ref{tab:grains}.
In terms of rotation, Grain~F shows an anomaly in the sense that it is not directed towards the biggest pore, but to the chain of micro-pores that connect the two bigger pores in the dataset.

For grains A1, D, and E we did not fully catch their formation, thus the mean lifetime of 71.44~s is a lower limit. The mean apparent speed of all those bright grains is about 2.97~km/s.
From the animated version of Figure~\ref{fig:grains} it is clear that most striations that form a bright grain form multiple grains in short succession.

For the two grains on the disk-centre side we find lower speeds and longer lifetimes. The sample sizes of seven and two are too small to allow for a statistical analysis. We speculate that the discrepancy between disk-centre and limb-side grains, if not purely statistical, may be due to the difference in viewing angles: When we are looking along the feature, following the apparent movement direction of the grain into the pore, we are also looking downwards along the slope of the Wilson depression. This effect can decrease the apparent horizontal speed. We can also not exclude multiple consecutive grains lining up in the line of sight (LOS), which would increase the observed lifetime.

\subsection{Stratification inferred from spectro-polarimetric scan}
Unfortunately, not all grains that can be seen in the context camera, are observed with the spectrograph,
as the slit covers only one row in the centre of the context camera and the average lifetime of a grain is less than half of the coverage with the context imager. However, the grains B, C, and F were observed with the slit. A cut through the stratified atmosphere, as obtained from the many-line inversions of the polarised \fisssp{} spectra, of striation and grain B is presented in Figure~\ref{fig:cut_grain}, the other two in Appendix~\ref{sec:supplementary_obs}. 

The observed temperature stratification aligns well with the expectations: We observe a cooler striation and a hotter grain in the lower photosphere, while the upper photospheric temperature seems to be partly determined by the expanding field of the pore.

\begin{figure}
    \centering
  \includegraphics[width=1\textwidth]{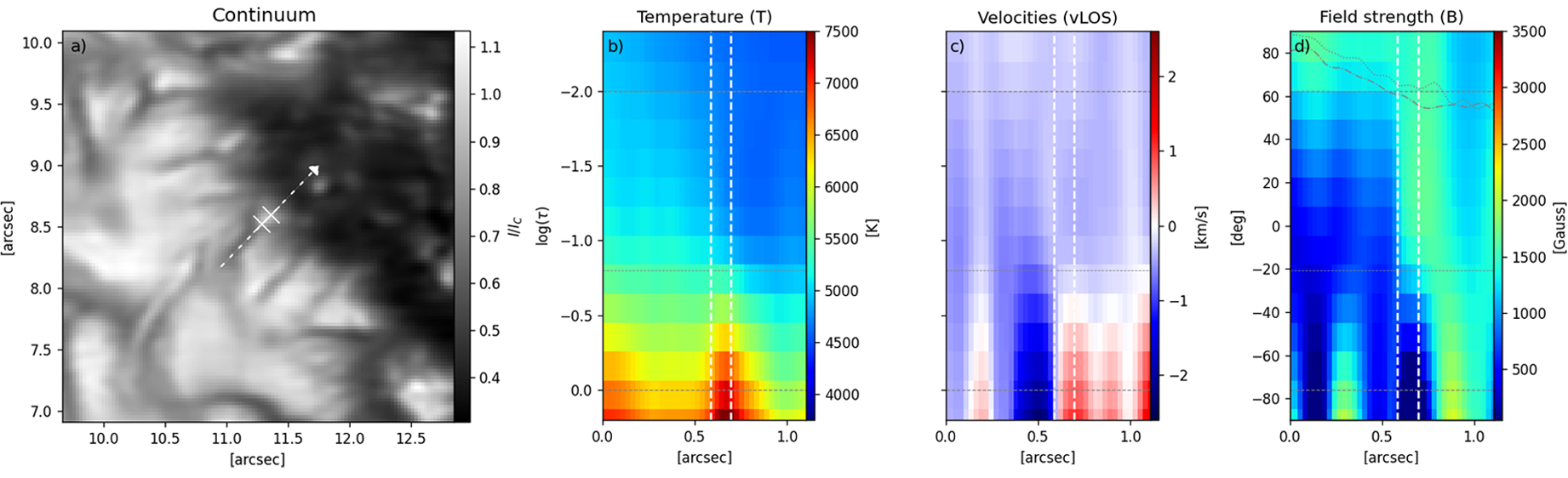}
  \caption[Cut through the 3D atmosphere of Grain B]{
    Stratification and context of bright Grain B.  
    Panel (a) presents an \fisssp{} integrated intensity image from the restored scan. 
    The white dashed line indicates the area shown in the other panels.
    The remaining three panels show cuts through: (b) the temperature (\T), (c) line of sight velocities (\vlos), and (d) field strength (\B) stratification. The vertical axis of all three panels represents height in \(\log(\tau)\), indicated on the left of panel (b).
    The \spinor\ nodes are indicated with dashed gray horizontal lines, the beginning and end of the grain, based on the intensity image, is marked with white dashed vertical lines in panels b, c, and d, and with a white | in panel (a).
    In panel (d) the dotted line indicates azimuth angle and dash-dotted line indicates inclination of the magnetic field corresponding to the y axis of panel (d). Here, 0\supcirc{} is parallel to the LOS, the value range is \(\pm90\)\supcirc{}.
    In all figures negative \vlos{} denote a flow towards the observer (i.e., up-flow) and positive LOS velocities denote a flow away from the observer (i.e., down-flow).
    }
  \label{fig:cut_grain}
\end{figure}

\begin{figure}
    \centering
  \includegraphics[width=1\textwidth]{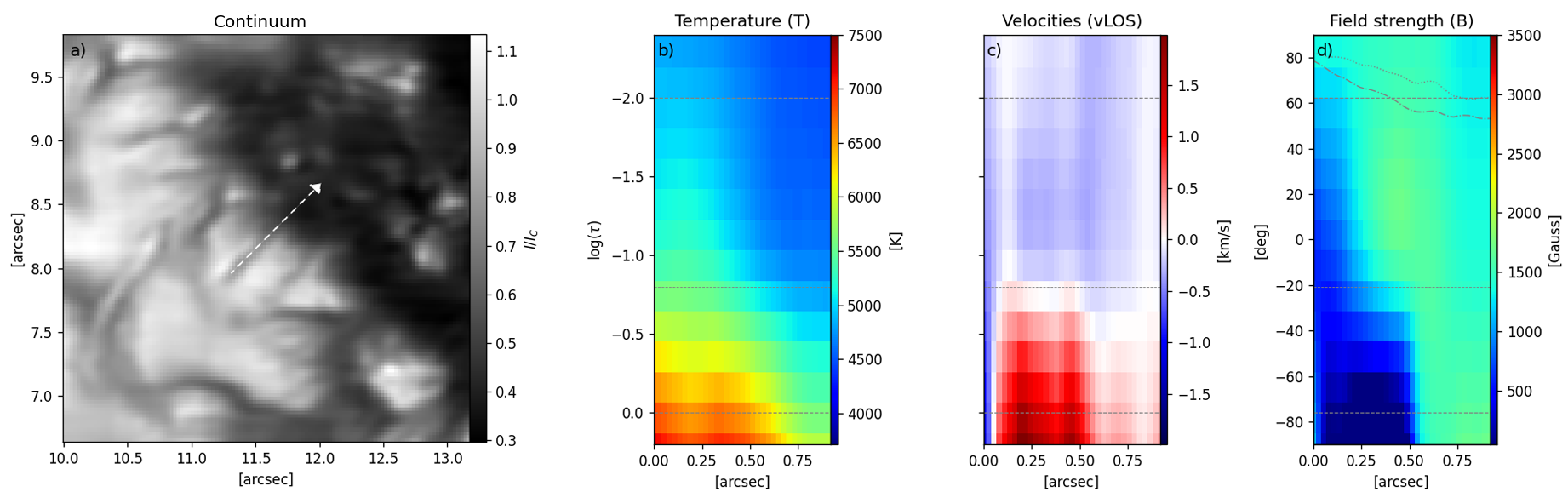}
  \caption[Cut through the 3D atmosphere of a striation without a grain]{
     Same as Figure~\ref{fig:cut_grain} but through an adjacent striation that did not form a grain during observation.
    }
  \label{fig:cut_no_grain}
\end{figure}

We find a magnetic canopy with a strongly inclined field spanning from the pore body over the granule (see, Figure~\ref{fig:cut_grain} panel d). The figure shows the structure as a function of optical depth in \(\log(\tau)\) and distance along the slit. The canopy like configuration of the magnetic field, therefore, not necessarily corresponds to an actual inclination of the field vector. From Stokes~\Q\ and \U\ we inferred an height independent LOS inclination angle. The extracted mean inclination along this funnel is 72.3\supcirc{} (with 0\supcirc{} parallel to LOS) and after correction for the angle of incidence (AOI) we get \(\approx\)53\supcirc{} (with 0\supcirc{} being parallel to surface normal).
We further observe a secondary magnetic enhancement in the lower \(\log(\tau)\)-layers in the wake of the spectroscopically observed grains B, C and F (see also, Appendix~\ref{sec:supplementary_obs}). We defer its discussion to Section~\ref{dsg:sec:muram_compare}, where we compare our observations with a simulation. 

The canopy like field configuration is also observed along striations that did not form a grain during our observation. A typical example is shown in Figure~\ref{fig:cut_no_grain}.
In such "quiet" striations we further observe a consistent \vlos{} downflow signature, typically in the order of 1.5~km/s, over the full length of the feature.
In contrast we observe a structured \vlos{} pattern in the wake of the grains. All three grains show signatures of a strong upflow following the temperature enhancement in the lower photosphere and a strong downflow in front of it and show additional small‑scale alternations between upward and downward flows. We find secondary magnetic enhancements and structured \vlos{} pattern only in striations that form grains, but in none of the others. 

\begin{figure}
    \centering
  \includegraphics[width=1\textwidth]{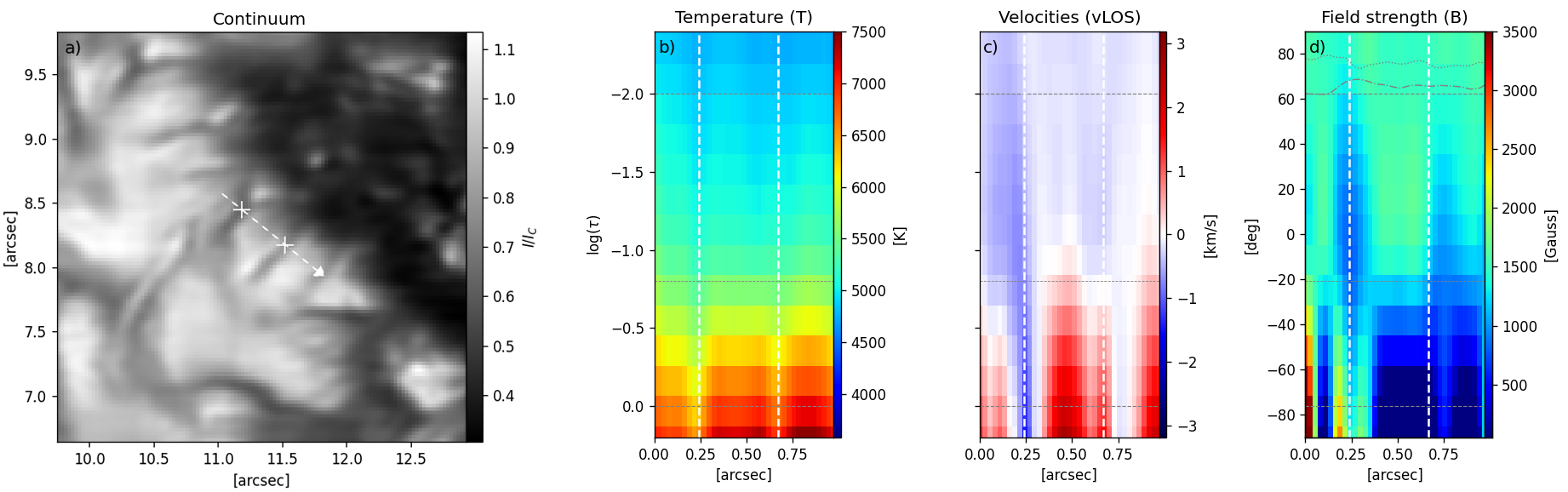}
  \caption[Cut through the 3D atmosphere perpendicular to striations]{
    Same as Figure~\ref{fig:cut_grain} but perpendicular to the striations. Here, the +-markings in panel (a) and the corresponding white dashed vertical lines in panels (b-d) mark the centre of the striations, based on the intensity image.
    }
  \label{fig:cut_perp}
\end{figure}

Cuts perpendicular to the striations, as shown in Figure~\ref{fig:cut_perp}, reveal a clear correlation of the striations with lower temperature, especially in the lower atmospheric layers, but visible even in the upper part of the inverted atmosphere. The \vlos{} is dominated by a hot downflow, even in the middle of the cut granule. A ring of such hot downflows can be clearly seen in the low photosphere \vlos{} maps around all pores in our dataset \citep[see,][Figure~7 or Appendix~\ref{sec:supplementary_obs}]{Hoelken25}, which is expected \citep{Stein00, Bharti16}. In Figure~\ref{fig:cut_perp} this downflow is interrupted with the upflow associated with the grain forming striation on the left (Grain B, shown in Figure~\ref{fig:cut_grain}) and two regions with weak (up-)flows left and right to the striation on the right (shown in  Figure~\ref{fig:cut_no_grain}).

The magnetic field configuration presented in Figure~\ref{fig:cut_perp} shows roughly the same inclination inside and outside a striation. However, in terms of optical depth, the expanding field from the pore seems to be elevated above the striations. Outside the striations it is detected from \(\log(\tau)\!=\!-0.8\) upwards, while above the striation the canopy field is detected more around \(\log(\tau)\!=\!-1.8\). If this is caused by an increased opacity gradient or a real elevation of the field cannot be conclusively decided from the 3D atmospheric model based on inversions. In the lower nodes we find the secondary magnetic enhancement in the left striation, as already seen in Figure~\ref{fig:cut_grain}. In the right striation we find a weak magnetic enhancement, co-aligned with the weak \vlos{} upflow just to the right of the striation.

\begin{figure}
    \centering
  \includegraphics[width=1\textwidth]{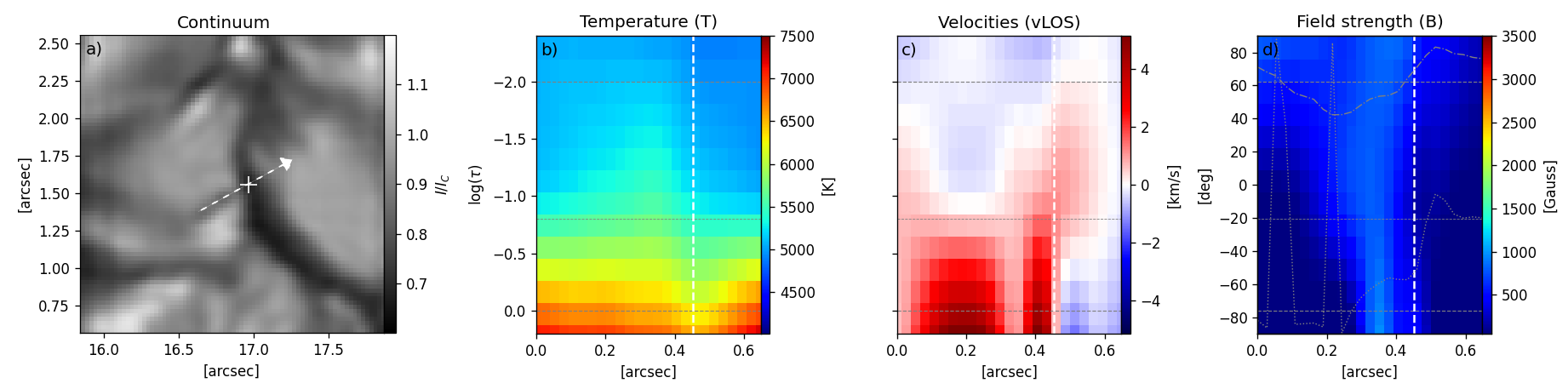}
  \caption[Cut through the 3D atmosphere of a striation formed at an IGL]{
    Same as Figure~\ref{fig:cut_grain} but for a dark striation perpendicular to a dark intergranular lane.
    }
  \label{fig:cut_igl_striation}
\end{figure}

\newpage
As mentioned earlier, dark striations are not only observed near pores. In our dataset we also find them next to strong intergranular lanes (IGL). Figure~\ref{fig:cut_igl_striation} shows a cut through the stratified atmosphere along such an IGL striation. The location of this example is indicated with a white line marked 'IGL' in Figure~\ref{fig:overview}. As in all other cases we find an expanding magnetic element, forming in the IGL. 
The mean AOI-corrected field inclination angle is about 54\supcirc{}, which is similar to the canopy we found at the pore boundary.  We did not resolve any grains forming at IGL striations. We find a consistent downflow pattern in the IGL striations, similar to the pattern we find in the pore-striations that did not form a grain during our observation (compare Figure~\ref{fig:cut_no_grain}). 
Given these similarities we conclude that striations forming on strong IGLs and on the border of pores are of the same type.

\section{Comparison with a \muram{} simulation}
\label{dsg:sec:muram_compare}
In the presented observations, we find the grains enclosed in the legs of a split dark striation (Y-Type) or adjacent to a strongly curved striation (S-Type) to be associated with a secondary magnetic enhancement in the lower photosphere. These enhancements, which are observed in the wake of the apparent movement direction of the grains, could not be explained from the inferred atmospheric snapshot.
Further, the typical swaying, snake like movement of the dark striations (See animation of  Figure~\ref{fig:grains}) hints towards the existence of an instability, for instance fluting and Rayleigh–Taylor instabilities, which could both lead to these kind of finger like structures. 

\begin{figure}
    \centering
  \includegraphics[width=1\textwidth]{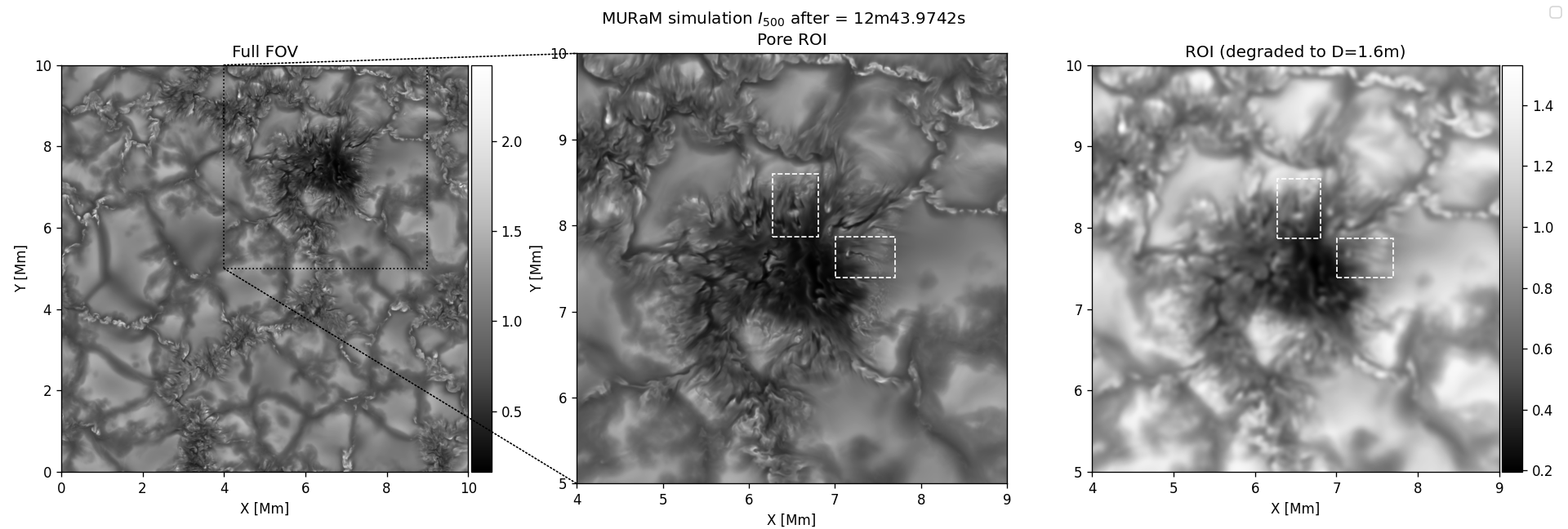}
  \caption[Intensity snapshot of the MuRAM simulation]{
    Continuum intensity snapshot of the \muram{} simulation at the optical surface \(\tau_{500}=1\). The left panel shows the complete simulation domain. The middle panel shows a blow-up of the pore and the right most panel shows the blow-up convolved with the point spread function (PSF) of an ideal 1.6~m telescope.
    The white rectangles mark the locations of the striations and grains shown in Figures~\ref{fig:sim_grain_1_still} and \ref{fig:sim_grain_2_still}.
    }
  \label{fig:sim_overview}
\end{figure}

\begin{figure}
    \centering
  \includegraphics[width=1\textwidth]{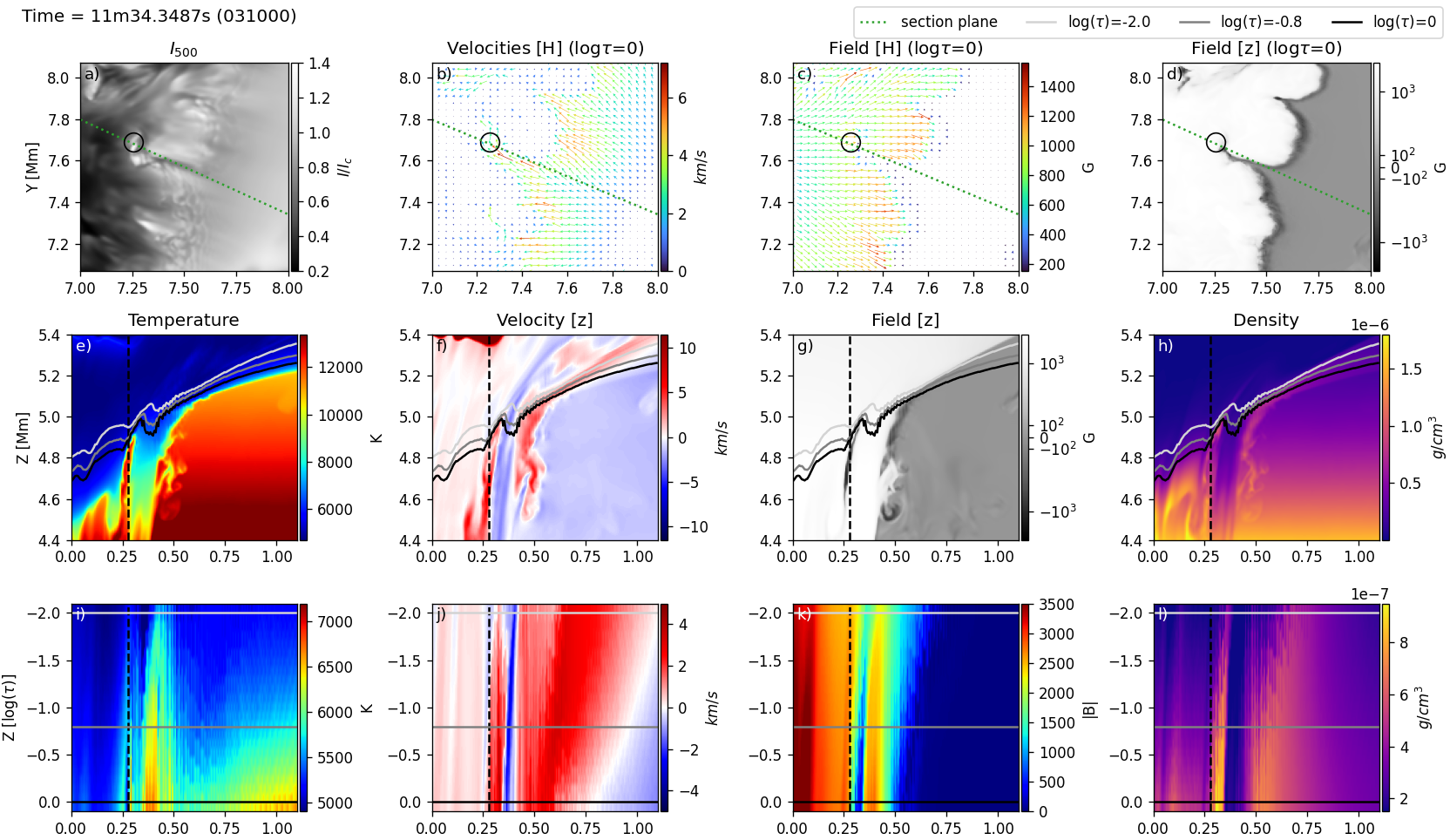}
  \caption[Cut through a grain from a \muram{} simulation]{
 Cut through a grain in a \muram{} simulation. Top row shows (a) continuum intensity at 500~nm and horizontal (b) velocities and (c) field as vector maps, and (d) vertical field strength, all at \(\log(\tau)=0\). All tick marks are in Mm.
    Middle row shows cuts along the green dotted line 
    indicated in the top row. Plotted are (e) temperature, (f) vertical velocity, vertical field (g) and density (h). All tick marks are in Mm.
    Bottom row shows vertical cuts through the same quantities as in panels (e)-(h), again along the green dotted line, but with the vertical axis now given in optical depth between \(\log(\tau)=0\) and \(\log(\tau)=-2\), which is similar to the optical depth range covered by the atmospheres obtained from the observations. Tick marks on the x axis are in Mm, tick marks on y axis in \(\log(\tau)\). 
    We have adjusted the sign of the vertical flows in all panels to match inversion results, thus negative values along the \(z\)-axis denote an up-flow and positive \mbox{\(z\)-velocities} denote a down-flow.
    The simulated grain is marked with a black circle in panels (a-d) and a vertical dashed line in panels (e-l).
    \href{https://www.aanda.org/10.1051/0004-6361/202659037/olm}{An animated version of this Figure is available with the online material of} \citep{Hoelken26}. 
    }
  \label{fig:sim_grain_1_still}
\end{figure}

To deepen our understanding of the underlying processes we compared the atmosphere extracted from an observation with a \muram{} simulation, shown in Figure~\ref{fig:sim_overview}. Our region of interest in this simulation is the formed small pore of about \(2 \times 2\)~Mm that shows similar features to the \fisssp{} observation: On the pore umbra/granulation boundary in the simulation we found numerous  dark striations forming bright grains. The simulated intensity images are comparable with the observed context images. Figure~\ref{fig:sim_grain_1_still} shows a cut through the simulated atmosphere of such a grain.

Comparing the simulated \(\log(\tau)\)-stratification (last row of the Figure~\ref{fig:sim_grain_1_still}) with the stratification we extracted from our inverted atmosphere (e.g., Figure~\ref{fig:cut_grain}) we find a good qualitative match of the observed signature with the simulation. 
Simulated and observed grains also have comparable lifetimes. At \mbox{\(\log(\tau)\!\in\![0,\!-2]\)}, simulated and inversion based atmospheres show an increased temperature at the bright feature, a structured up- and downflow pattern, a canopy-like field structure and a secondary magnetic enhancement in the low photosphere in the wake of the grain.

This similarity holds for all observed and all simulated events that we investigated on. From the simulation we have analysed four striations forming a total of 14 grains, whereof ten were still visible after convolution with a point spread function (PSF) of an ideal 1.6~m aperture. Figure~\ref{fig:sim_grain_1_still} shows an S-Type striation that formed a sequence of grains. A Y-Type striation with grain is presented in Appendix~\ref{sec:supplementary_sim}. In the simulation we find similar processes acting in both types.
We hence conclude that we see the same process in simulations and observations. Having established the similarity of the features observed and simulated, we will now use the simulation to discern their formation and evolution.

\begin{figure}
    \centering
  \includegraphics[width=0.9\textwidth]{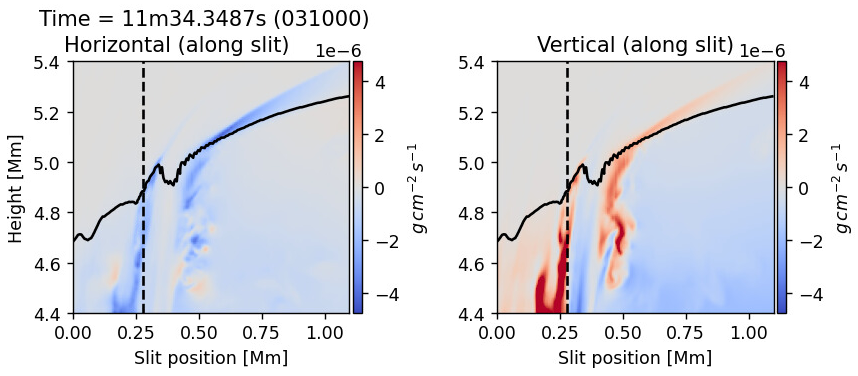}
    \caption[Momentum within a grain forming striation (\muram{} simulation)]{
    Momentum along the slit indicated in Figure~\ref{fig:sim_grain_1_still} panel (a). The vertical dashed line indicates the location of the grain and the black line indicates the \(\log(\tau)\!=\!0\) surface. Left panel shows horizontal momentum along the slit, right panel shows vertical momentum. 
   The negative horizontal momentum is pointing towards the pore on the left and negative vertical momentum is pointing upwards.
    }
  \label{fig:sim_grain_1_momentum}
\end{figure}

\begin{figure}
    \centering
  \includegraphics[width=0.9\textwidth]{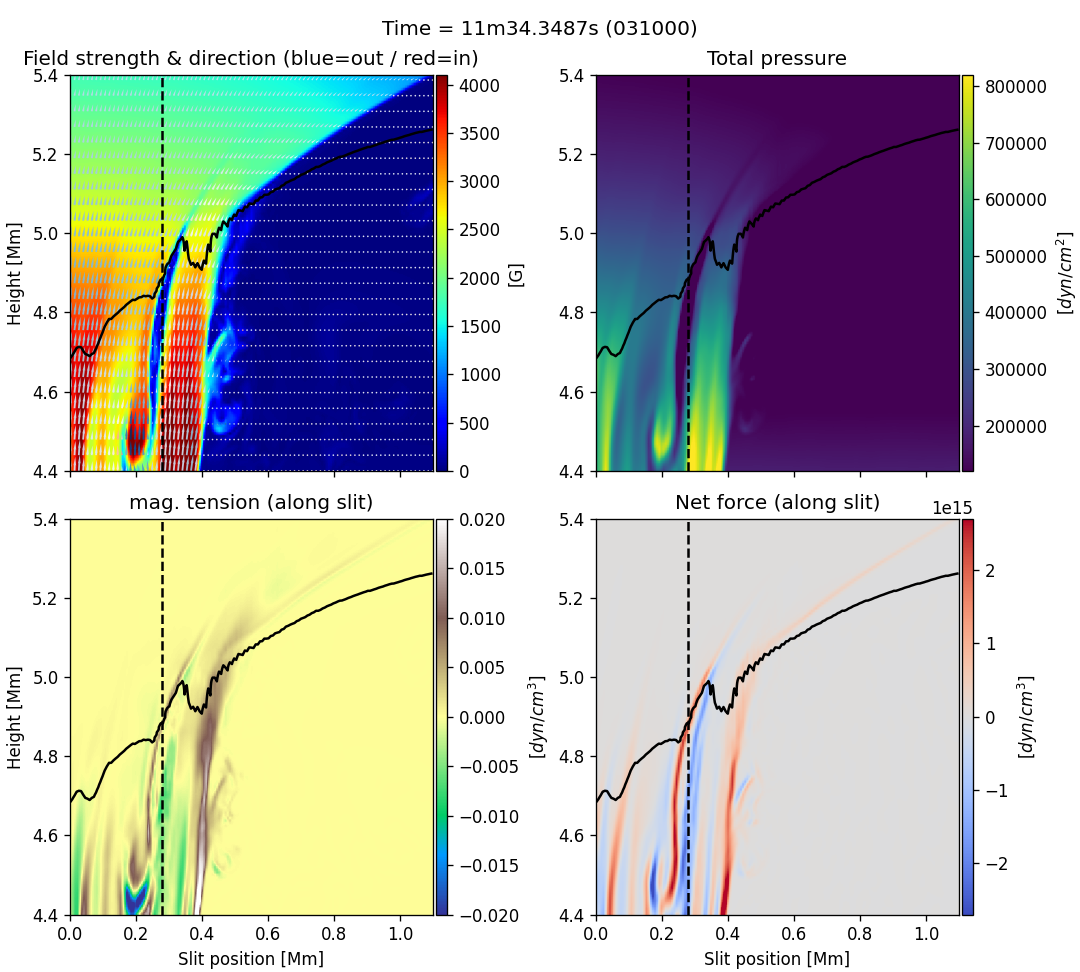}
  \caption[Magnetic field within a grain forming striation (\muram{} simulation)]{
      Magnetic field vector, magnetic tension, total pressure and horizontal net force along the slit indicated in Figure~\ref{fig:sim_grain_1_still} panel (a). The vertical dashed line indicates the location of the grain and the black line indicates the \(\log(\tau)\!=\!0\) surface.
    The negative horizontal horizontal net force is pointing towards the pore
    \href{https://www.aanda.org/10.1051/0004-6361/202659037/olm}{An animated version of this Figure is available with the online material of} \citep{Hoelken26}.
    }
  \label{fig:sim_grain_1_field}
\end{figure}

\begin{figure*}
    \centering
  \includegraphics[width=1\textwidth]{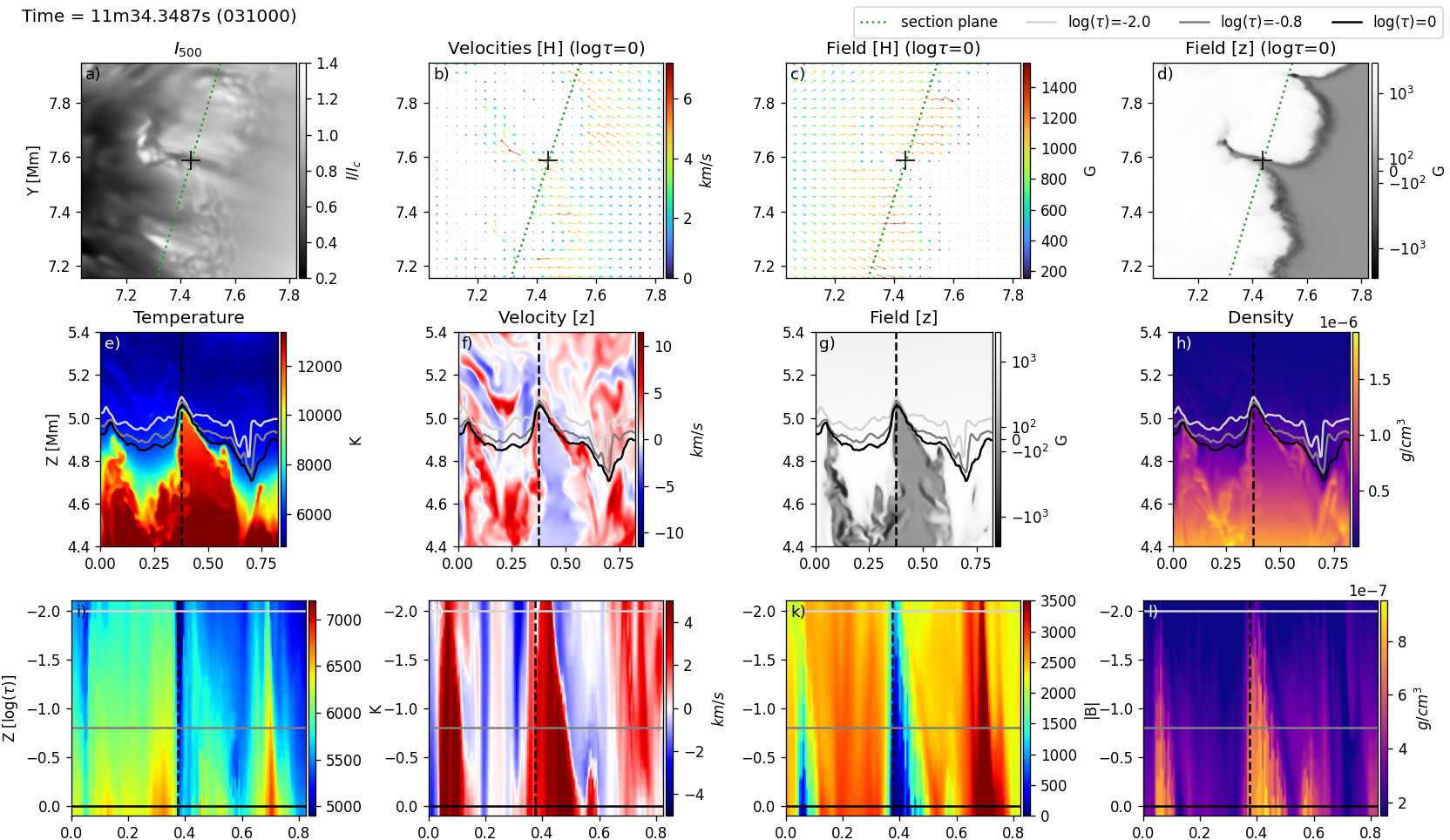}
  \caption[Cut perpendicular to a striation (\muram{} simulation)]{
     Same as Figure~\ref{fig:sim_grain_1_still} but perpendicular through the grain shown in Figure~\ref{fig:sim_grain_1_still}. In this Figure the black + in panels (a-d) and the vertical dashed lines in panels (e-l) mark the centre of the striation (based on the intensity).
    }
  \label{fig:sim_grain_1_perp}
\end{figure*}

The physical processes that lead to the appearance of dark striations at the interface between pores and granulation is composed of multiple aspects. From the momentum presented in Figure~\ref{fig:sim_grain_1_momentum} we see that material is moving from the forming granule on one side towards the pore on the other side, were it hits the magnetic field which acts as a boundary. 
Because of the field's curvature the fluting instability, as suggested by \cite{kuridze25}, is a possible driver for the appearance of the finger-like weak field structure that coincides with the dark striations. 
However, the pore is almost evacuated and its strong field blocks the denser material flowing towards it from the granule. This behaviour might also indicate a convection‑driven Rayleigh–Taylor instability (RTI), in which the rapid deceleration of the plasma as it approaches the flux‑tube boundary plays the role normally attributed to gravitational force.
The field geometry is favourable for  an interpretation in terms of fluting instability, but suppression by the strong vertical field (Figure~\ref{fig:sim_grain_1_field}) could favour RTI. 
Cuts perpendicular to the striation, as shown in Figure~\ref{fig:sim_grain_1_perp}, reveal a different picture. The plume of hot material upwelling underneath the striation indicates a predominantly convective nature of the striations.
In such a complex situation it is almost impossible disentangle the driving processes, as all of them presumably play a role.

\newpage
While the dominant driver behind the presence of the striations remains unsolved, their darker appearance can be understood when we analyse the plasma flows. Cuts perpendicular to the striation (Figure~\ref{fig:sim_grain_1_perp}), show a plume of hot material convecting upwards in the weak field passage, elevating the \(\log(\tau)=0\) surface.
Also in this case the simulated observable signatures show similarities to the signatures that we find in the observations (Compare Figure~\ref{fig:cut_perp}).
At the top, at the \(\log(\tau)=0\) surface, the field forms a cusp where the up-flowing material gets trapped and then cools due to radiation. This explains the lower temperature and hence the darkness of the striations.
Similar processes of material trapped in a field cusp have been reported for the dark lanes atop umbral dots \citep{Schuessler06} or for dark lanes in penumbral filaments \citep{Tiwari13}.
When viewed in geometric height, the cut through the vertical velocities shows the convective nature of the striation, with material piling up from below, slowing down and becoming suspended on the top, and is then downflowing on the sides.

To understand the formation, evolution, and decay of the bright grains it is best to study the animations of Figures~\ref{fig:sim_grain_1_still} and \ref{fig:sim_grain_1_field}. 
Along the dark striation convection is pushing non-magnetic material against the flux tube boundary from the side and from below. The deceleration by the magnetic field is well visible in the net force panel of Figure~\ref{fig:sim_grain_1_field}. This process forms fast downflows, in the order of 5-7~km/s, in front of the grain (Figure~\ref{fig:sim_grain_1_still}, panel (f)). This evacuating process prevents material from getting trapped atop the bright grain and is exposing the hot upflowing material.
This evacuation process also leads to a steep density gradient, which leads to a sharp increase of the optical depth surface. In Figure~\ref{fig:sim_grain_1_still} the steep angle of the \(\log(\tau)\)-surfaces illustrate this nicely. This causes the grains to appear bright and hot, when observed from a viewing angle that is at near-normal incidence to the slope of the optical depth surface. This explains why we predominantly see these features at the limb side in observations. The flow direction from the granule centre to its boundary with the umbra, which causes the grains, further explains why all observed and simulated grains move inwards towards the umbra.

\section{Discussion}
The atmospheric stratification inferred from a \fisssp{} scan matches very well with the observable signature extracted from a \muram{} simulation. On first glance, there appears to be one discrepancy that must be discussed further: While in the observations only some dark striations formed bright grains, in the simulation all dark striations in the vicinity of the pore form bright grains, sooner or later. 
Two mechanisms explain this apparent discrepancy: (i) The observation with the \fisssp{} context camera covers 200~s, while the simulation spans roughly 15 minutes, of which we analysed the last ten. In the simulation grains are constantly forming, but their appearance is scattered over the analysed ten minutes. (ii) The diffraction limited \fisssp{} scans have a resolution of  0.068" which corresponds roughly to 49~km. The simulation has a pixel size of 6~km. Degradation of the simulation to resemble our observation shows that not all simulated grains could be resolved with a 1.6~m telescope. 
In a randomly selected 200~s sample of degraded simulated intensity images, as shown in the rightmost panel of Figure~\ref{fig:sim_overview}, we found a total of seven grains. The apparent discrepancy can hence be accounted to missed (in terms of timing) and unresolved formation of grains in some of the striations in the observations. Latest broad-band observations\footnote{See DKIST fastcam dataset (\url{https://dkist.virtualsolar.org/vanNoortfastcam/}) acquired with a diagnostic setup at NSO/DKIST, built in collaboration with M. van Noort.} from the Daniel K. Inouye Solar Telescope \citep[DKIST,][]{Rimmele20} seem to further support the assumption that also on the Sun grains form at even smaller scales.

The similarity of our observations (see Figure~\ref{fig:overview}, especially panels B and F) with the described Y-shaped striations in the smaller granular light bridges reported by \cite{Schlichenmaier16} and \cite{Zhang18} is striking. As in our pore scenario, granular light bridges have granulation next to a strong magnetic field, which indicates a similarity of these structures. We further report a new S-shaped striation type with a grain forming at a curved section of a striation.

Other typical dark lanes features, as above umbral dots \citep{Schuessler06}, penumbral filaments \citep{Zakharov08, Tiwari13}, or above filamentary light bridges \citep{Schlichenmaier16} are formed just under the cusp where two fields merge. 
Indeed, already \cite{Bharti16} reported such a split field above a striation at a pore boundary based on \hsp{} observations and our data agrees well with these results.
Our simulation indicates that elevated trapped material in a field cusp atop a convective pillar causes the appearance, also in the case of the dark striations discussed here. 

\cite{Bharti16} further connected the grain forming \mbox{Y-shaped} striations with the penumbra formation process. 
At first glance the moving bright grains seem to have a strong similarity with penumbral filament grains \citep{Muller73, Zhang13, Tiwari13}. On closer inspection we find them to be much smaller, faster, and not as long living. 
A further difference is that we only find grains moving 'inwards' towards the pore and none moving 'outwards' towards the surrounding granulation. Also the general flow direction is only inwards (compare, Figure~\ref{fig:sim_grain_1_still}, panel (b)) and does therefore not agree with more complex flow patterns of a developed penumbra \citep{Tiwari13}.
\cite{Zhao24} observed striations in an elongated granule associated with a moving magnetic feature \citep[MMF,][]{Harvey73} and found a rather stable magnetic field configuration over the evolution of different optical appearances. They found a similar magnetic field configuration and strength around and above their striations as we found at the boundary of a pore.
Also \cite{Zhao24} connected the striations they studied to formation processes of penumbral filaments.
\cite{Munjiba24} reported multiple penumbra formations and losses for the AR observed here. However, due to the lack of extended temporal high-resolution coverage, our dataset does not allow for speculations about penumbra formation or similarities between penumbral grains and the grains we observe here.  

In our observations we also find striations at IGLs hosting a magnetic element. One example positioned a bit away from the pore was presented in  Figure~\ref{fig:cut_igl_striation}.
Considering the similarity of IGL striations and pore striations (compare  Figures~\ref{fig:cut_no_grain} and \ref{fig:cut_igl_striation}) in our dataset, we assume that at least the stronger IGL striations are formed by the same processes as the striations observed at a pore boundary. 
\cite{kuridze25} report a similar formation process for striations observed in a plague region. The size of the IGL-striations from our dataset is at the upper end of the striations studied in their work. This connects our findings to striations of even smaller scales.

\section{Conclusion}
In this study we analysed a spectrally and spatially highly resolved spectro-polarimetric dataset from GST/\fisssp{} \citep{Noort24}\footnote{Author's note: Reproduced as Chapter~\ref{ch:fiss_sp} in this thesis.}, inverted with the many-line inversions technique \citep{Riethmueller19, Hoelken25}\footnote{Author's note: The H\"olken et al. many-line paper is reproduced as Chapter~\ref{ch:many_line} in this thesis.} using 85 lines. This allowed us to investigate small-scale features in the presence of a pore. 
Due to the close match of atmospheric signatures, inferred from the inverted observations, with the atmospheric configuration of a pore in a \muram{} simulation, we were able to describe the formation and evolution of bright grains by a plasma flow penetrating a magnetic field tube, visible as dark striation. 
Further, we offered a physical explanation for the viewing angle based selection bias.

Based on the configuration of density, flows, and magnetic field in the simulation, we identified the dark striations as convective features. Multiple instabilities could account for their appearance. Especially, fluting and RTI seem to be likely. But also the upwelling of hot convection-driven gas maybe producing the cusp of the field. The dark appearance of such striations can be explained by trapped plasma in the cusp of the closing magnetic field, similar to dark lanes on umbral dots, penumbral filaments, or filamentary light bridges.
The bright grains are identified as fresh material being pushed up- and sidewards. Since this material did not have time to cool, it appears bright. During the grain evolution fast downflows  evacuate cooling material in front of the grain. 

Our description matches well with findings from intensity images \citep[e.g.,][]{Schlichenmaier16, kuridze25} and spectro-polarimetric observations \citep[e.g.,][]{Zhao24, Bharti16}. 
The good qualitative match with the \muram{}-based MHD simulation in intensity and atmospheric signature enabled us to provide a description of the underlying magneto-convective processes. Our results further establish the good agreement of \muram{} simulations with observed solar features.
From the observation of nine bright grains, where of three were also observed with the spectrograph, some statistical relevance could be established. 
However, further high-resolution spectro-polarimetric observations are desirable to confirm the generality of our findings.

\chapter{Summarizing remarks and outlook}
\label{ch:conclusion}
In this thesis I have tackled the challenges that arise when we want to explore the full resolution power of modern large-aperture solar telescopes for spectro-polarimetric observations. In Chapter~\ref{ch:introduction} I have introduced why image restoration of solar spectra is needed to explore small-scale features near the diffraction limit. I have further detailed why the unavoidably increased noise floor, that comes with this process, provides a challenge for the interpretation of the restored data using the Stokes inversion technique.
As a solution I proposed the solar many-line inversions technique for high spatial resolution spectra, based on a theoretical pre-study by \cite{Riethmueller19}.
The chapters that form the main body of the thesis describe the building blocks needed to put the solar many-line inversion technique into practice. They draw an arc from data calibration and numerical methods over instrumental requirements to the application of many-line inversions on solar observations. The previous chapter can then be seen as a proof of concept, that the proposed solution indeed improves our possibilities to describe and understand solar small-scale features.

As most chapters correspond to standalone articles with their own conclusion sections, I here wish to bring all these results together under the umbrella of their common contribution towards solar many-line inversions, discuss open questions, and put the results in the context of latest developments in this field.

\section{High resolution spectro-polarimetry}
To exploit the potential of spectra with a high spatial resolution, we need to deal with the noise. Using the combined signal of many simultaneously observed lines, enhances coherence and physicality of the inferred solar atmosphere. This allows us to detect signals near the (increased) noise floor of the restored spectra and hence lessens the dependence of precise polarimetry on low noise.
Unfortunately, I cannot provide quantitative estimates or speculate about the degree of relaxation, as the tolerable noise level strongly depends on the specific scientific use case, and so will the possible relaxation. 
As a summary of the benefits found in  Chapter~\ref{ch:many_line} I can state that diverse lines, spanning formation heights and various atomic properties, better constrain atmospheric parameters, leading to a comprehensive, detailed, and meaningful representation of an observed solar atmospheric configuration.

However, the many line approach turned out to be not as straight forward as one might think. Similar to challenges faced by stellar astronomers, we now must carefully select "well-behaved" lines. Fitting lines with missing blends and not sufficiently well described parameters require large adaptations in the atmospheric model to reach a reasonable fit. This can drive the inferred atmosphere to physically implausible regimes (see, Figure~\ref{fig:statistics}), since the degree of freedom to adapt line shapes by altering the atmospheric composition based on a \(\chi^2\)-metric is quite high. The proposed selection process circumvents this issue, by selecting only lines that agree well and produce results within the physically meaningful regime. This allows us to extract information from a multitude of lines and improves the coherence and noise-robustness of the inversion results. However, this is not completely satisfactory, as we only select lines with somewhat similar behaviour and thereby neglect information from other lines. In the case discussed in Chapter~\ref{ch:many_line} I was able to use \nlines{} out of \tlines, which is almost exactly 50\% of the lines. 

Any Stokes inversion relies on the accuracy of the atomic line parameters, including \(\log(gf)\) values. When inverting many-lines simultaneously the inversion-code can no longer compensate so freely for inaccuracies in those line parameters. Therefore, the many-line technique reveals those uncertainties more clearly than previous approaches and is thereby pointing to a need for improved parameters of many of the lines in the solar spectrum. In addition to poorly understood lines, we can speculate that lines, which are not formed under LTE conditions, do not match the behaviour of other lines in an LTE inversion approach, but would contribute valuable information when treated in NLTE. Both tasks, setting up a multi-atom, multi-state atomic model for a many-line NLTE inversions and improving the atomic line parameters are complex tasks, worth of a thesis on their own \citep[See, for instance, ][]{laverick2019phd, vukadinovic2024phd}. It was therefore not feasible to explore this realm in the scope of this work.

A further problem lies in the transferability of the results to other wavelength regions. As I have shown in Chapter~\ref{ch:many_line}, the possibility to reproduce an average quiet Sun spectrum does not automatically imply that the atomic parameters of the lines included are well enough known to use them for the inversion of a high spatial resolution spectrum. Thus, the need for a similar line selection process can be expected for observations with large-spectral windows in other wavelength regimes. The observations from SUSI (nUV) and  SCIP (IR), as well as for the \ion{Mg}{I} observations around 5180~\AA{} conducted during the 2025 \fisssp{}-campaign are of this kind. 
Especially the SUSI observations will provide new challenges for the line selection process described in this work, as in the highly blended nUV regime no section of clear continuum that separates the individual spectral windows to test (see, Section~\ref{sec:convergence_challenges}) can be expected to exist anywhere in the spectra observed by SUSI.

Despite all these remaining challenges, with the atmospheric configuration, inferred from a \fisssp{} observation using the many-line inversions approach, it was possible to discern the magneto-convective nature
of dark striations and bright grains. These are small-scale features in the vicinity of magnetic elements in intergranular lanes and pores, which are similar to those on granular light bridges in sunspots. Such features have been observed and reported in broad-band images previously, but due to the lack of spectro-polarimetric information with a high spatial resolution, no conclusive match with numerical simulations and no satisfactory description of the underlying magneto-convective processes existed so far. In Chapter~\ref{ch:results} I provided an interpretation of the dark lanes as convective features where the dark lane forms in the cusp of closing magnetic fields, similar to the dark lanes on penumbral filaments, umbral dots, or filamentary light bridges. I further also discerned the process causing the bright grains as a plasma flow evacuating the topmost layer of the upwelling material in front of the striations. This description was only possible because of the good agreement of the atmospheric signatures inferred from the inverted high-resolution observations utilising many-lines simultaneously with the atmospheric configuration of similar features in a \muram{} simulation.

In summary, the presented results showcase the potential but also the limits of the herein developed many-line approach. 
An obvious advantage of the traditional approach using only a very limited number of lines lies in the natural limitation to lines that have well-studied and relatively well-known atomic parameters. But with many-line inversions we were able to use the reconstructed and noisy observations, which spatially resolve small-scale features spectro-polarimetrically, to describe the corresponding composition of the solar atmosphere. The method has thereby been proven as a possible way to mitigate the noise dilemma in spectro-polarimetry. It can hence be regarded as a success in terms of the goals set in Chapter~\ref{ch:introduction}, which justifies the additional effort. However, as all Stokes inversions, also the many-line method depends on the input model parameters, assumptions, and available atomic data. Moreover, it reveals this dependency much stronger than the currently common approach of using a few non-blended lines only. Further work is therefore needed to improve atomic line parameters, to extend the line selection approach to other wavelength regimes, and to test the performance of other codes.

\section{Outlook on future studies}
\label{sec:summary:outlook}
With the availability of the many-line method the question if it is worth the increased effort arises. In Chapter~\ref{ch:results} I have shown one result, but dark striations and moving bright grains are not the only small-scale features that are worth (re-)visiting. As a further example I want to sketch a possible future study with some preliminary findings. 

The dataset analysed in Chapters~\ref{ch:many_line} and \ref{ch:results} includes a large pore with a fully developed umbra that shows a typical umbral feature: UDs. Recent UD studies  \citep{Yadav18, Kilcik20, Calisir23} focus on the lifetime and motion of umbral dots within an umbra. Most studies of UDs were so far focused on umbrae of sunspots, while the analysis of UDs in pores is under-represented.
In general, high-resolution spectro-polarimetric observations of pores remain very rare. 

I have introduced the current state of agreement between simulations \citep[i.e.,][]{Schuessler06} and observations \citep[e.g.,][]{Socas-Navarro04, Riethmueller08, Watanabe12, Riethmueller13} in Chapter~\ref{ch:introduction}. In summary, the main open points are the amount of the field strength reduction in and the details of the predicted plasma flows in and around UDs. 

\begin{figure}[thp]
    \centering
  \includegraphics[width=1\textwidth]{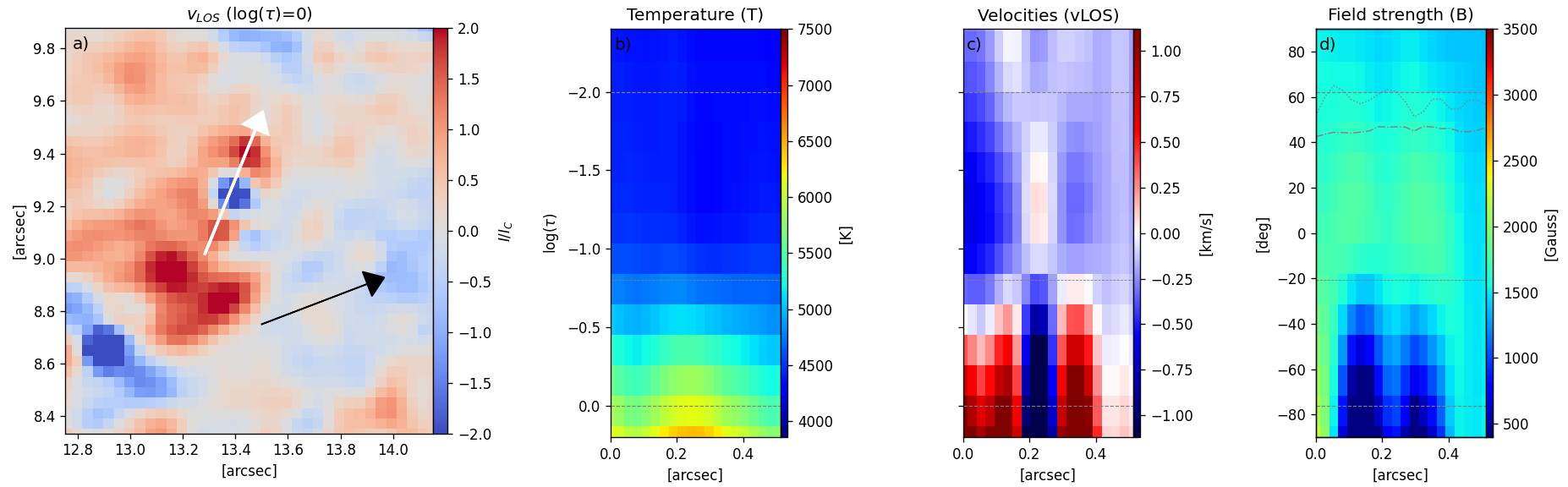}
  \caption[Fast downflow channels of an umbral dot]{
   Fast downflow channels of an umbral dot. Panel (a) shows the \vlos{} at the optical surface.
   The white arrow indicates the length and direction of the cut shown on the right while the black arrow points towards disk centre. 
    The remaining three panels show cuts through: (b) the \mbox{temperature (\(T\))}, (c) line of sight velocities (\vlos{}), and (d) field strength (\(B\)) stratification. The vertical axis of all three panels represents height in \(\log(\tau)\), indicated on the left of panel (b). The SPINOR nodes are located at \mbox{\(\log(\tau) \!\in\!\{0, -0.8, -2\}\)}. In panel (d) the dotted line indicates azimuth angle and dash-dotted line indicates inclination of the magnetic field corresponding to the y axis of panel (d). Here, 0\supcirc{} is parallel to the LOS, the value range is \textpm{}90\supcirc{}. In all panels negative \vlos{} denote a flow towards the observer (i.e., up-flow) and positive LOS velocities denote a flow away from the observer (i.e., down-flow).
    }
  \label{fig:UD_cuts}
\end{figure}

In the dataset presented in Chapters~\ref{ch:results}, all upflowing UDs show a clear region of descending plasma in their perimeters, which is in agreement with \cite{Schuessler06} and \cite{Riethmueller13}.
Figure~\ref{fig:UD_cuts} presents cuts through the atmospheric stratification of a typical \(\log(\tau)\)-stratification of an UD as extracted from our inversion. In Panel (a) the two downflow channels (dark red) are visible at the top-right and bottom-left of the central upflow plume (dark blue)
We usually find such narrow fast downflow-channels in the slowly downflowing perimeter surrounding an upflowing UD, which is in line with the predictions of \cite{Schuessler06} but was to date not confirmed observationally. In contradiction to \cite{Schuessler06} we do not find such channels in the perimeters of all upflowing UDs.

\begin{figure}[hp]
\begin{minipage}[b]{0.65\textwidth}
  \includegraphics[width=1\textwidth]{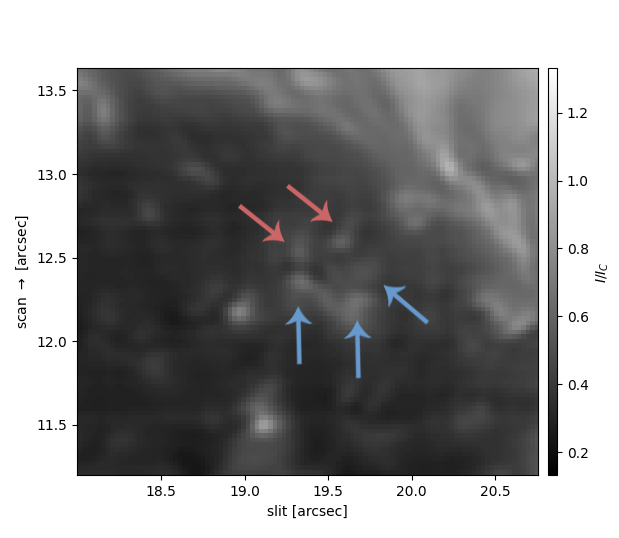}
\end{minipage}\quad 
\begin{minipage}[b]{0.32\textwidth}
  \caption[Intensity image of an UD cluster]{
   A group of UDs forming a cluster. The figure shows normalized intensity of Stokes \I{} from the restored \fisssp{} scan, integrated over the full wavelength range. The arrows point to the UDs forming the cluster; blue for upflowing, red for downflowing.
  }
  \label{fig:ud_cluster}
\end{minipage}
\end{figure}

\newpage
Five UDs (three upflowing, two downflowing) in the upper right corner of the umbra form a cluster, shown in Figure~\ref{fig:ud_cluster}. Between the upflowing UDs and the downflowing UDs is a separating dark lane. This feature resembles the impression of a short faint light bridge, similar to the description of \cite{Schlichenmaier16}. The three upflowing UDs in this cluster are among the fastest in the dataset with speeds above 6~km/s.

As expected we find a reduction of the magnetic field in most UDs, but also counter examples, as also reported by \cite{Ortiz10} and \cite{Watanabe12}. Out of the 44 detected UDs six seem to be associated with hot downflows, the others with upflows. The mean flow speeds are about 2.1~km/s upflow and 1.6~km/s downflow.
The inverted field maps indicate a field strength reduction of about 600~Gauss with respect to the measured average field strength in the umbral background of about 2.1k~Gauss. 

Why we do not detect the downflow channels on every upflowing UD and also the UD cluster is yet to be studied in more detail. 
But the good agreement of the inverted atmosphere with the literature in terms of field reduction in the UDs, average flow speeds, and height stratification allows us to trust the new observation of the fast downflow channels already. However, border cases, outliers, and the reliability of the results are still to be analysed. Connections to the position of the UDs within the umbra or to the size and brightness of the umbra are still to be drawn (or not).

\section{Follow up and extension of this work}
As we continue our exploration and analysis of solar many-line data, we can hope to advance our understanding of the solar atmosphere beyond what pure LTE codes can achieve. Although LTE codes fall short in fully reproducing many-line spectra with mixed N/LTE populations, I was able to show that they are still effective in extracting coherent configurations of the solar atmosphere from noisy data. Future work should hence include the testing of various regularization techniques and a less strict height node handling, as exemplified by \texttt{SNAPI} or \texttt{globin}, as well as considering NLTE-ready codes like \texttt{SNAPI} or \texttt{NICOLE}, to mitigate described challenges.

Surely, focusing on codes and stratification models does not address the fundamental issue of insufficiently well known atomic parameters. But from a pragmatic point of view, this approach still appears promising in the short-term: Improving atomic parameters for highly blended lines is significantly more complex than for isolated ones, and already there the currently available data is incomplete. In contrast, examining whether the results can be improved using more flexible atmospheric models seems less demanding. Via the \texttt{globin} method it is plausible that a better understanding of the observed spectra and features may facilitate the determination of the atomic parameters as well. 
Ultimately, however, also the improvement of the many-line method remains limited by the availability and correctness of the atomic data. This is especially the case for NLTE inversions. Hence, in the long term, improvements in the accuracy of the atomic line parameters are needed.

The data analysis on the other hand suggests that the pragmatic boundary, set by feature evolution time, is almost reached for spatially scanning instruments. The time scales seen from the \fisssp{} context imager that I discussed in Chapter~\ref{ch:results} are in the order of 70 seconds, including formation, evolution, and decay. Other, yet not fully resolved features, seem to evolve on even shorter timescales. The scan speed for high resolution spectro-polarimetry is limited, as sufficient photons of the highly dispersed spectrum need to be collected. With increasing aperture, and therefore resolution, the evolution time scales of the smallest resolvable features do not leave much room for a scan any more, especially, if a multi-path observation is targeted that allows to discern the time evolution of such features. This is also nicely presented in  Figure 1 of \cite{Iglesias19}. 
Filtergraphs, and other spectrally scanning instruments, currently do not allow to sample a wide spectral range in a quasi-simultaneous way. They are, therefore, not suitable for the application of a many-line inversion technique.
One possible way out of this new dilemma would be a many-line ready integral field unit. However, an instrument like the Microlensed Hyperspectral Imager \citep[MiHI,][]{MiHI} with a spectral coverage of 30~\AA{}, installed at a four~meter class telescope like DIKIST or the European Solar Telescope \citep[EST,][]{EST}, will create new instrumentation, data handling, and data processing challenges. 

\begin{figure}
\centering
  \includegraphics[width=.75\textwidth]{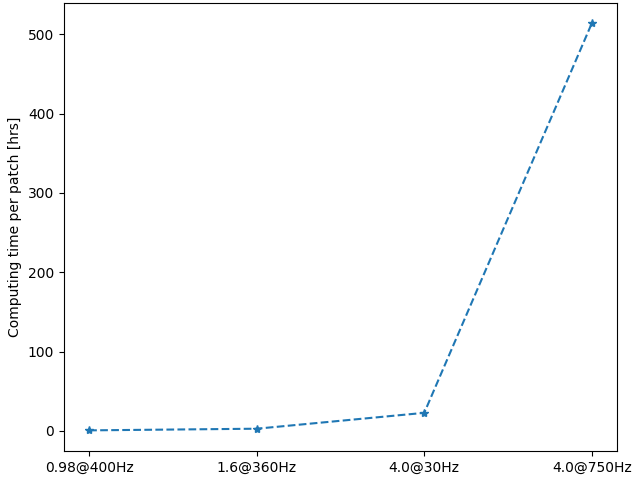}  
\caption[Estimation of numerical image reconstruction costs]{Estimated costs of image reconstruction per patch as CPU time (y-axis) versus aperture size [m] and frame rate [Hz] (x-axis). The Plot is based on estimates by Michiel van Noort (private communication).}
  \label{fig:momf_costs}
\end{figure}

With approximately 40 minutes of CPU time\footnote{CPU time is a standard measure of the computational work done by a computer's processor, calculated by multiplying the number of active CPU cores by the time they are active. In High Performance Computing (HPC) this is a useful unit to compare computational effort.} needed, the estimation of the PSF and reconstruction of the scene on a single sub-patch, obtained at a one-meter-class telescope, is relatively affordable. 
One main assumption for the blind reconstruction approach, as in MOMFBD, is that the solar scene remains unchanged for the whole batch of images and the seeing must not change within each single exposure. 
The larger the aperture and, thus, the diffraction-limited resolution, the shorter both time periods become. 
Hence, the number of individual exposures must be adjusted accordingly to maximise the camera duty-cycle in order to collect all available photons. Since the isoplanatic patch also depends on the aperture size, the number of needed frame-sub-patches might increase as well.
At the same time, the Fourier frequency, up until which the data can be reasonably restored, increases.
Consequently, the CPU time per reconstructed patch already increases to about three hours for the 1.6-meter aperture of GST. 
However, for the four-meter aperture of DKIST or EST and with the estimated needed frame-rate of 750~Hz the CPU time for a single patch skyrockets to over 500 hours, as depicted in Figure~\ref{fig:momf_costs}.
This is no longer a sustainable use of (computing) resources and for a timely release of observational data not tolerable. 
Hence, new approaches to speed up the estimation of the residual wavefront error and extracting a PSF are needed. 

What comes directly to mind is the use of a Machine Learning algorithm, and indeed, success is reported in the restoration of intensity images \citep[see, for example,][]{Schirninger25}. However, extensive testing has so far only partially been successful in estimating the wavefront error \citep[see, among others,][]{AsensioRamos23, Ge24}, which is needed to reconstruct also spectrograph data. One reason for the PSF estimation being more problematic is that the problem seems to be too degenerate and more constraints must be added to improve the convergence. New prototypes that explore the merits of more objects from other wavelengths or the addition of an instrument specific Shack-Hartman type wavefront sensor could help constraining this problem. 
Using a parallel implementation on graphical processor units (GPUs) of MFBD \cite{Li25} recently reported an acceleration factor of up to 266, while maintaining image reconstruction and PSF estimation quality. However, the solar part of their study was based on only eight frames from a one-meter class telescope. It will be interesting to see if this scales reasonably with the huge number of frames needed for a large aperture and a large FOV.
Testing these new approaches of PSF estimation on data from a very large aperture telescope and to combine it with the spectral image reconstruction could provide the next advancements is solar data acquisition.

A similar issue arises when interpreting the data by using inversion codes. In a comparison of artificial intelligence (AI) enhanced inversion codes for high resolution spectro-polarimetry \cite{Socas-Navarro21} report some progress on this topic. But given the findings form this thesis, further consideration of the inaccuracy of our training and validation data sets, for instance because of incomplete line parameters, need to be incorporated in such approaches. The concurrent information from multiple simultaneously observed lines might also be beneficial in this case. AI benchmark studies with diffraction-limited SP data from large aperture telescopes using multiple lines simultaneously have not been published so far. However, one foreseeable problem is the need of a new training for every change in the atomic model, the inclusion or removal of lines, or the change of any other line specific parameter. If this is feasible - or in the end costs more then we gain - is yet to be clarified.

The vast amount of data recorded with modern instruments is required to further enhance our understanding of our host star. While introducing AI into the calibration and interpretation of solar data seems to provide a possibility to cope with those data volumes, it also introduces new challenges in terms of validation \citep[See, for instance,][]{Oneto24} and training. With the high precision and reliability needed (compare, Chapters \ref{ch:data_reduction} and \ref{ch:many_line}) and the huge set of unknowns and inter-dependencies (compare, Chapters~\ref{ch:introduction}, \ref{ch:reconstruction}, and \ref{ch:many_line}) the question on how to build physically educated models is still an open field.

\section{Conclusion}
This thesis presents a step in improving our interpretation methodology and providing a path for future developments in this field. It uses advancements in inversion and image reconstruction codes, as well as in observational and computational hardware to make it possible to use a many-line inversion approach on solar spectra with a high spatial resolution. It also underlines open questions as further advancements in the description of atomic line parameters and also new strategies to timely reduce and interpret the vast amount of observational data are needed.

The results presented in Chapter~\ref{ch:results} show that, with the application of the herein developed many-line inversions on high spatial resolution spectra, a new realm of solar small-scale features can be made accessible to interpretation based on SP data. The brief outlook presented in Section~\ref{sec:summary:outlook} of this summarizing chapter shows further potential to decipher known unknowns. However, taking into account its higher demands and complexity it is not a one-size-fits-all tool for the interpretation of solar data and its application needs to be decided depending on the specific science case. In conclusion, the here presented solar many-line inversion approach is a powerful extension of our toolbox to decipher solar observations.

\appendix
\cleardoublepage
\thispagestyle{empty}
\vspace*{2.2cm}
{\Huge\bf Appendix}

\chapter{An update from \spectroflat{}}\label{app:spectroflat}
The original description of the \spectroflat{} library (Chapter~\ref{ch:data_reduction}) was published in spring 2024. Now, in late summer 2025, almost 1.5 years have passed and it seems appropriate to re-evaluate the performance and impact of the library. 
Already in 2024 we where able to show that it surpasses the precision that can be achieved with the traditional approach of extracting the smile-distortion from a single line only. The fully field dependent calibration is a necessary step to prepare data from grating based spectro-polarimeters for restoration and many-line inversions alike. 

Since the original publication, \spectroflat{} has been extensively used  in the calibration of \fisssp{} data (see, Chapters~\ref{ch:fiss_sp} and \ref{ch:many_line}). After the successful re-flight of Sunrise-III in Summer 2024 it is now also an integral part of the \mbox{Sunrise-III/SUSI} data reduction pipeline \citep{Feller25}. With that it has now prooven its utility in practice. 

Successful testing has been performed with other instruments as well. Among these instruments are the Fibre Resolved OpticAl and Near-Ultraviolet Czerny–Turner Imaging Spectropolarimeter \citep[FRANCIS,][]{Jess23} (Glen Chambers, private communication), the Zurich Imaging Stokes Polarimeters \citep[ZIMPOL,][]{Keller94, Povel95, Ramelli10} (Franziska Zeuner, private communication), and \mbox{GREGOR/GRIS}.
It has also been successfully used with amateur instruments, for instance with the spectrograph at the historic Hainberg-Observatory in Göttingen\footnote{Amateurastronomische Vereinigung Göttingen e.V. (AVG), \url{https://www.avgoe.de/}} and the spectrograph of Alessandro Ravagnin (private communication).
This adds to the wide range of applications already shown in the original publication and underlines the general applicability and usefulness of a shared, commonly used library. 

\spectroflat{} is still under constant development and maintenance to provide a service to the solar community. 
A major improvement was achieved when changing the smile determination process to an iterative one. The first iteration is exactly the same as in the previous version, but we now get a relative correction from the line residuals in the dust flat (Figure~\ref{fig:slitsensor-flat}) and small higher order line offsets in consecutive iterations. Depending on the number of iterations, a smile correction down to residuals in the \(10^{-5}\)~px regime is now possible.
Other recent developments include further improvements of outlier masking in the pre-flat preparation, a global post-factum alignment of different modulation states in the generated offset-maps, and some improvements for smile-detection in cases with very strong smile for larger slits. In the process I have improved the documentation and user friendliness. Since October 2025, \spectroflat{} is an affiliated package\footnote{Overview of SunPy affiliated packages: \url{https://sunpy.org/affiliated/}} of the SunPy project \citep{sunpy_community2020}.

\chapter{Supplementary figures to Chapter~\ref{ch:results}}
\label{sec:supplementary_obs}
\begin{figure}[!ht]
    \centering
  \includegraphics[width=1\textwidth]{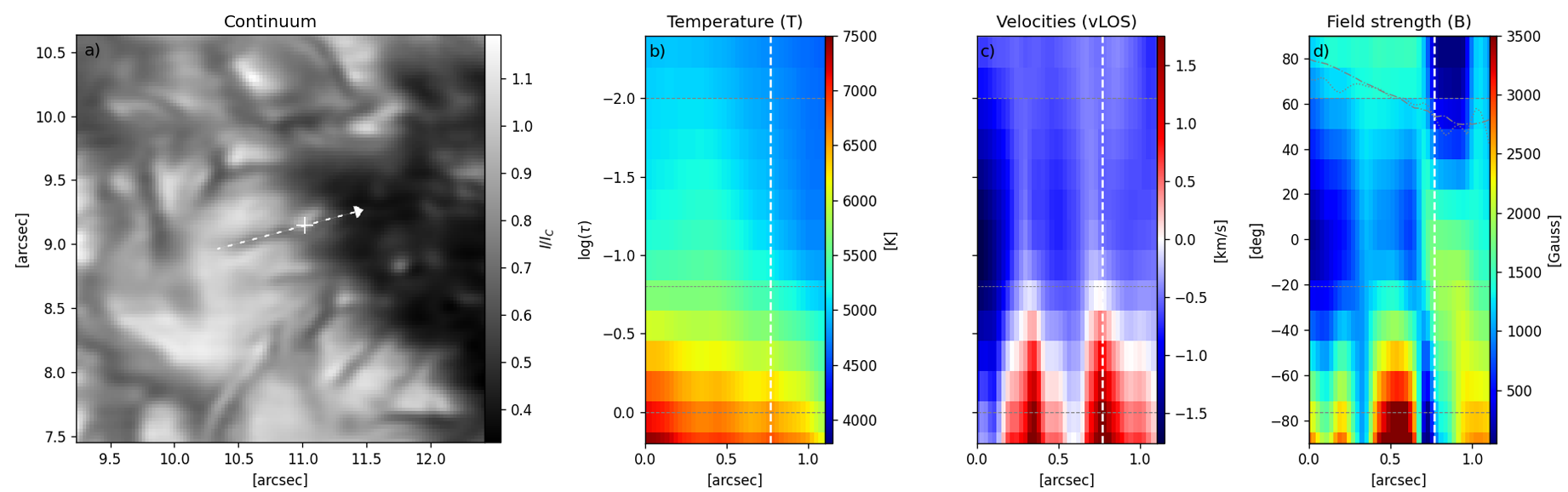}
  \caption[Cut through the 3D atmosphere of Grain C]{
    Same as Figure~\ref{fig:cut_grain}, but for Grain C. The temperature increase in this grain is much more localized as the ones shown in Figures~\ref{fig:cut_grain} and \ref{fig:cut_grainF}. The peak is marked with a + in panel (a) and a vertical dashed line in panels (b) to (d).
    }
    \label{fig:cut_grainC}
\end{figure}
\begin{figure}[!h]
    \centering
  \includegraphics[width=1\textwidth]{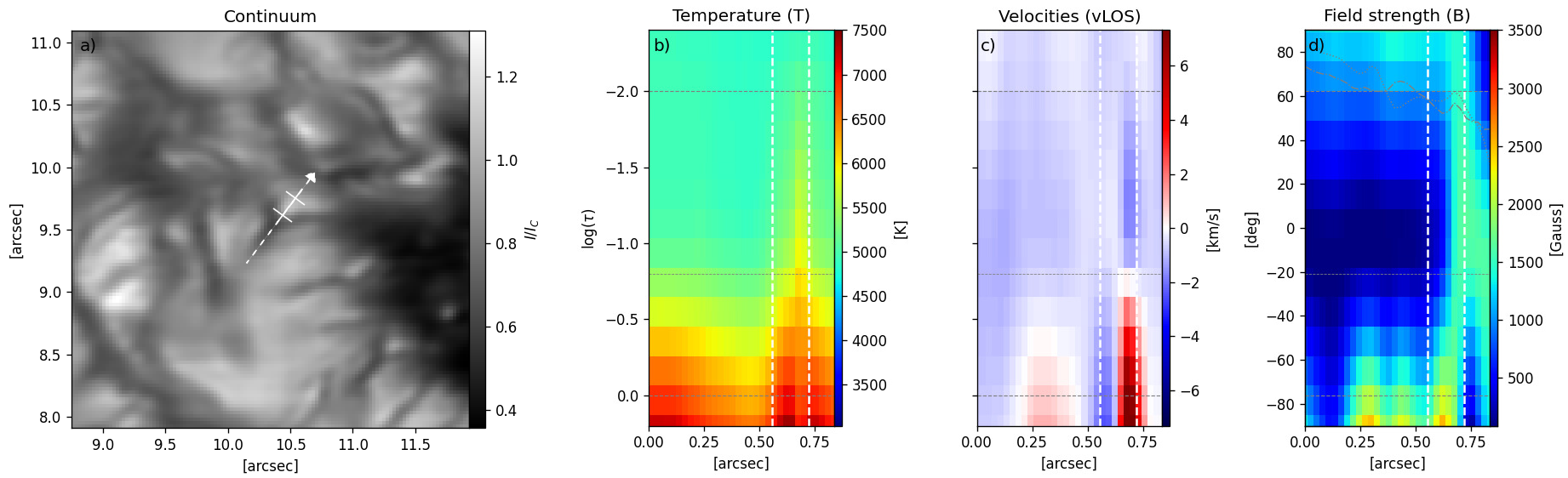}
  \caption[Cut through the 3D atmosphere of Grain F]{
    Same as Figure~\ref{fig:cut_grain}, but for Grain F.
    }
    \label{fig:cut_grainF}
\end{figure}

\begin{figure*}[!h]
    \centering
  \includegraphics[width=0.74\textwidth]{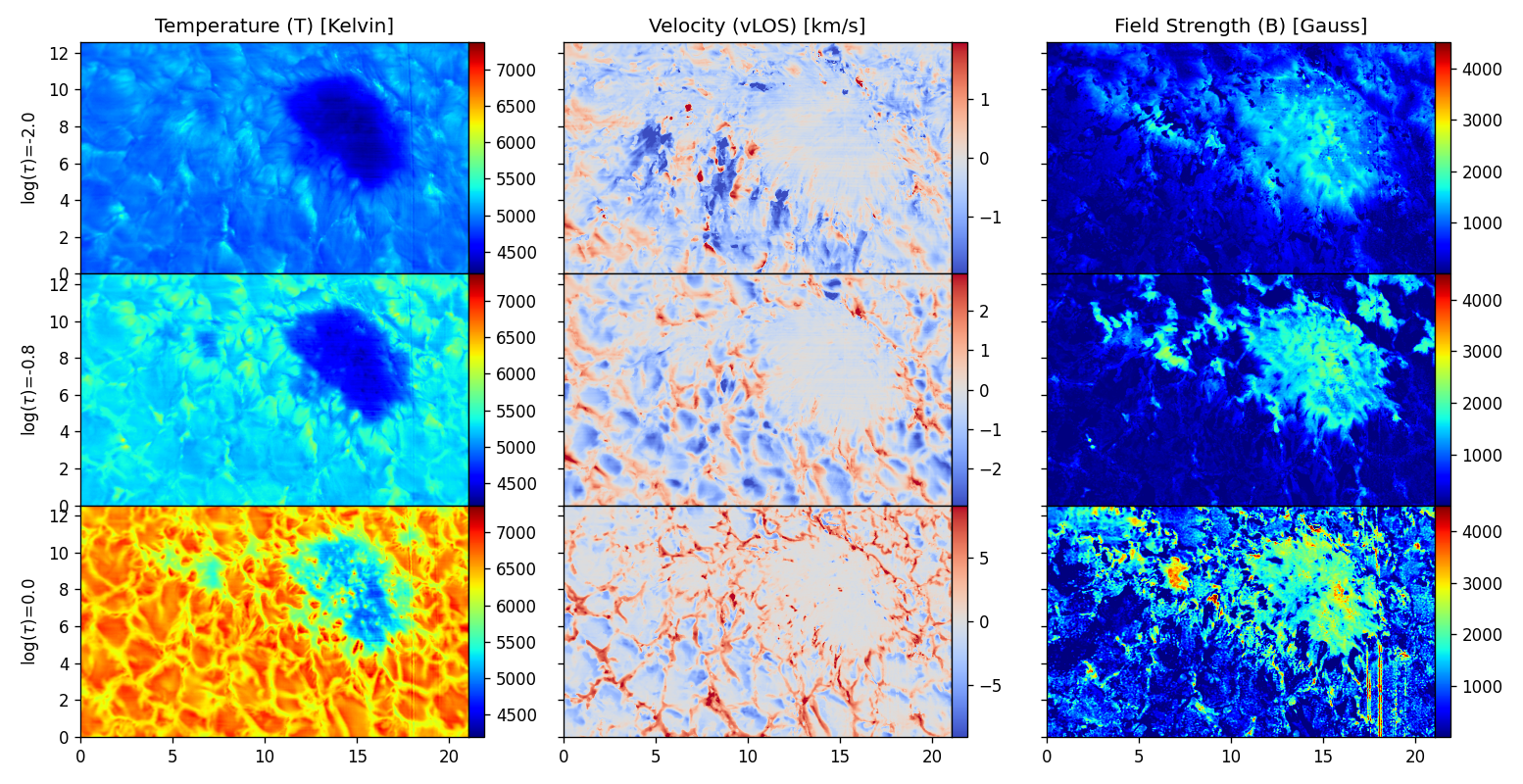}
  \includegraphics[width=0.248\textwidth]{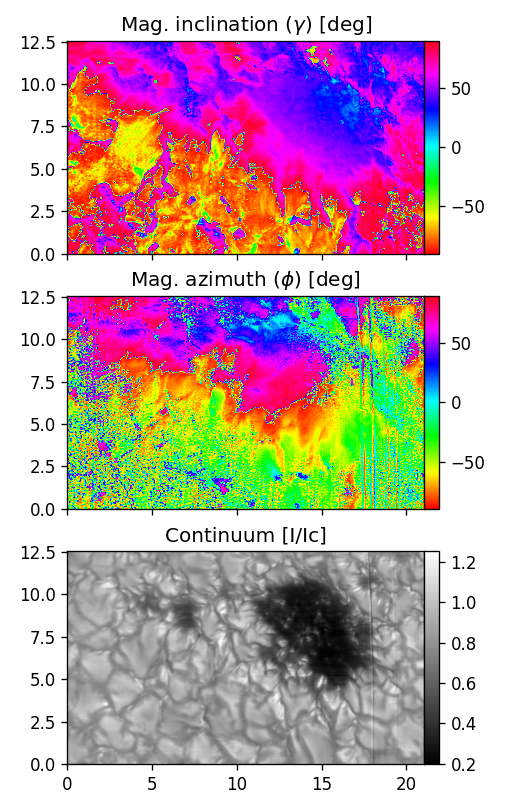}
  \caption[Full inverted FOV]{
    Maps of the inverted atmosphere similar to Figure~\ref{fig:final_maps} but for the complete inverted FOV and flipped y-axis. 
    Here, the lower right panel shows a continuum image. 
    In all figures negative \vlos{}  denote a flow towards the observer (i.e., up-flow) and positive LOS velocities denote a flow away from the observer (i.e., down-flow). Disk-centre is towards the upper right.
    }
  \label{fig:inversion_overview}
\end{figure*}

\label{sec:supplementary_sim}
\begin{figure}[!h]
    \centering
  \includegraphics[width=1\textwidth]{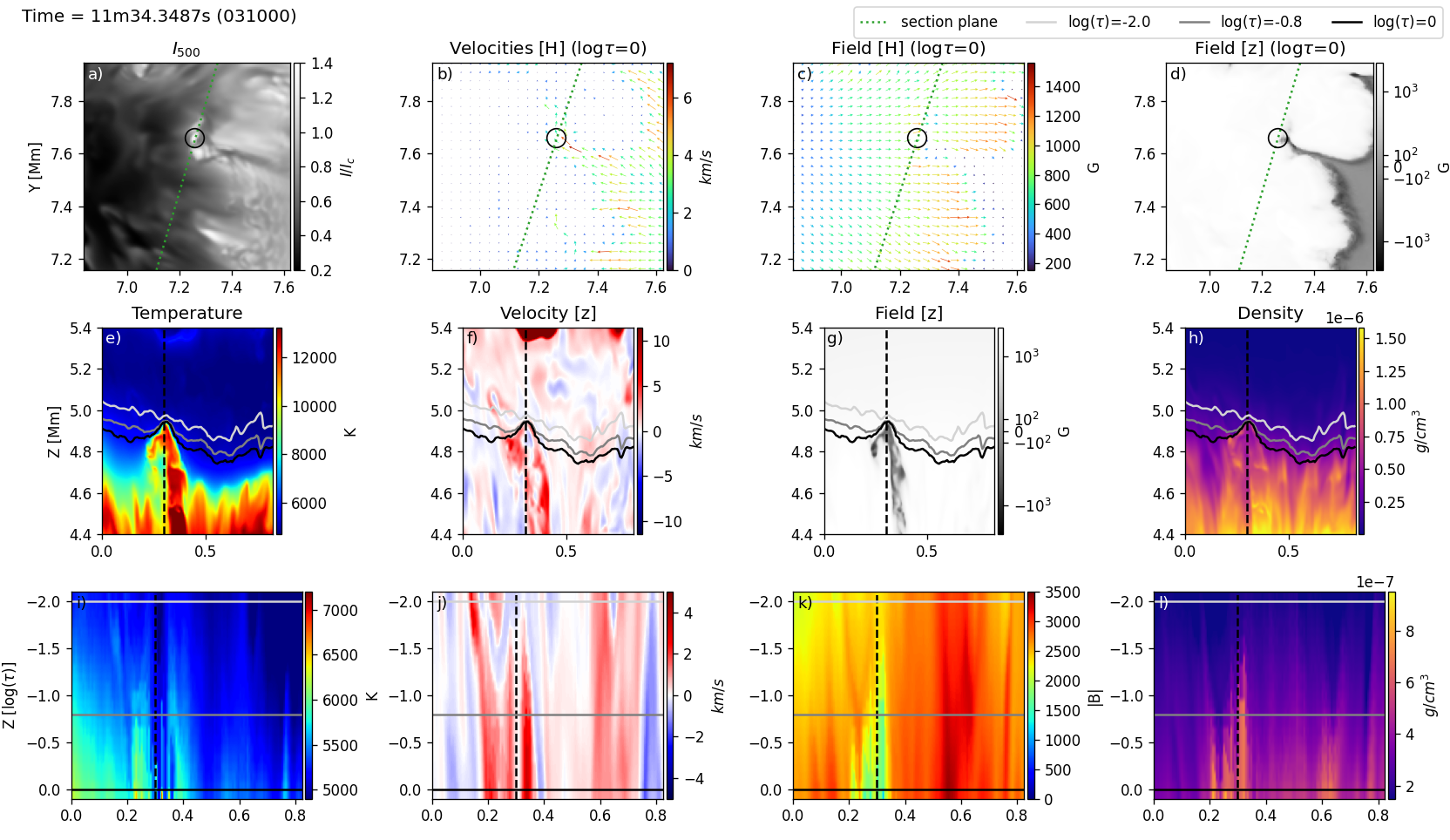}
  \caption[Cut perpendicular to a striation, through the grain (\muram{} simulation)]{
   Same as Figure~\ref{fig:sim_grain_1_perp} but cut through the grain shown in Figure~\ref{fig:sim_grain_1_still}. The circle in panels (a-d) marks again the grain.
    }
  \label{fig:sim_grain_2_still}
\end{figure}

\begin{figure}[!h]
    \centering
  \includegraphics[width=1\textwidth]{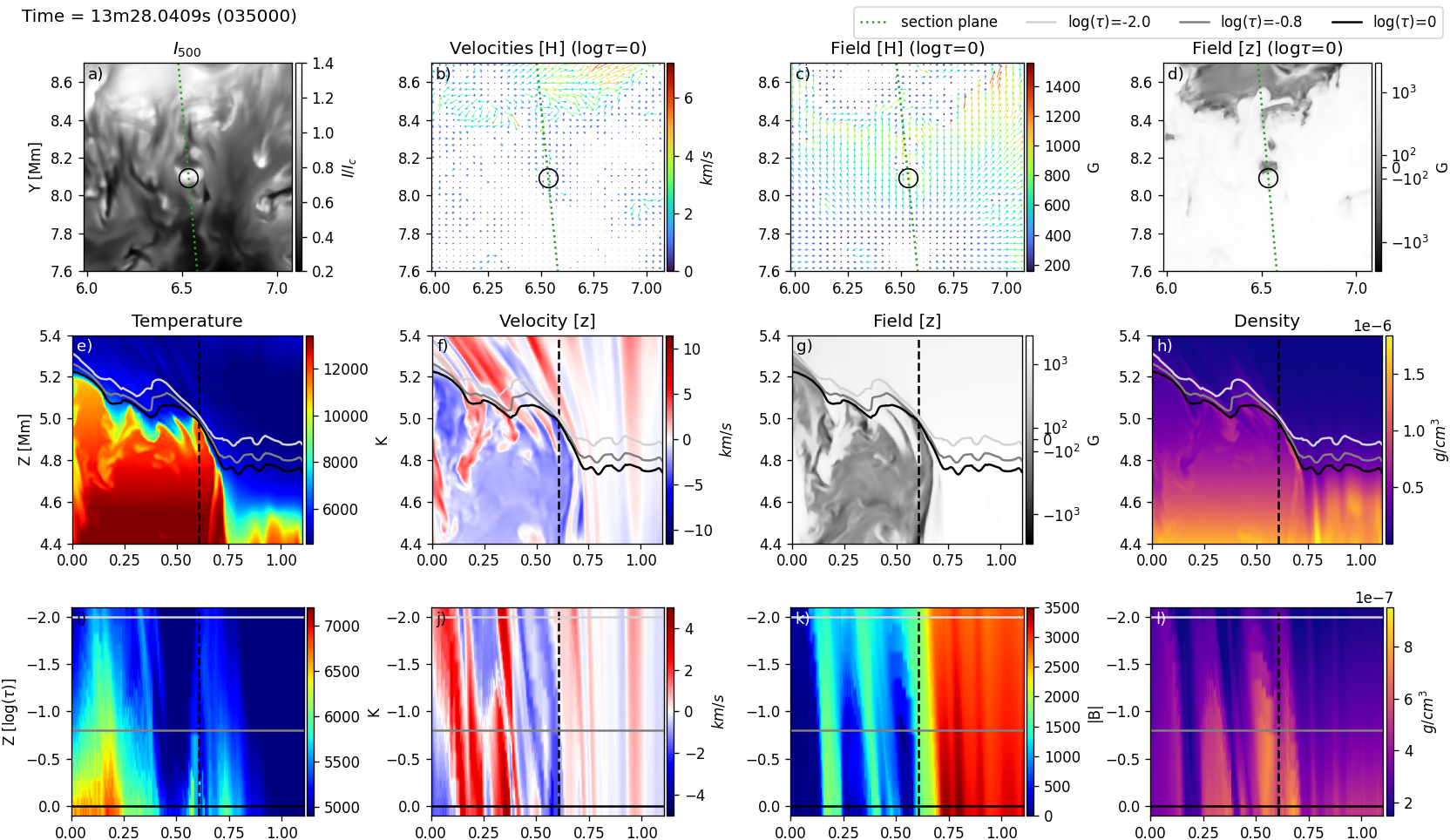}
  \caption[Cut through another grain forming striation (\muram{} simulation)]{
   Same as Figure~\ref{fig:sim_grain_1_still}, but for a Y-Type grain. 
    \href{https://owncloud.gwdg.de/index.php/s/KnNAXc1kRAOwufR}{An animated version of this Figure is available with the online material.}
    }
  \label{fig:sim_grain_2_still}
\end{figure}

\begin{figure}[!h]
    \centering
  \includegraphics[width=1\textwidth]{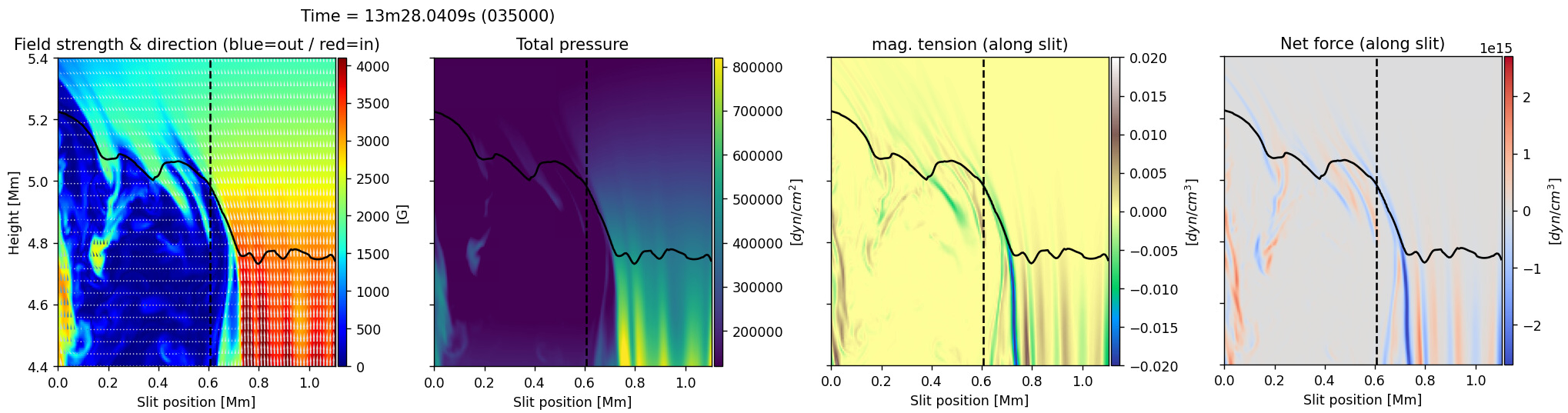}
  \caption[Field within a grain forming striation (\muram{} simulation)]{
   Same as Figure~\ref{fig:sim_grain_1_field}, but for the grain shown in Figure~\ref{fig:sim_grain_2_still}. Please note that the pore is on the right in this plot and hence the positive horizontal horizontal net force is pointing towards the pore.
    }
  \label{fig:sim_grain_2_field}
\end{figure}
\begin{figure}[!ht]
    \centering
  \includegraphics[width=1\textwidth]{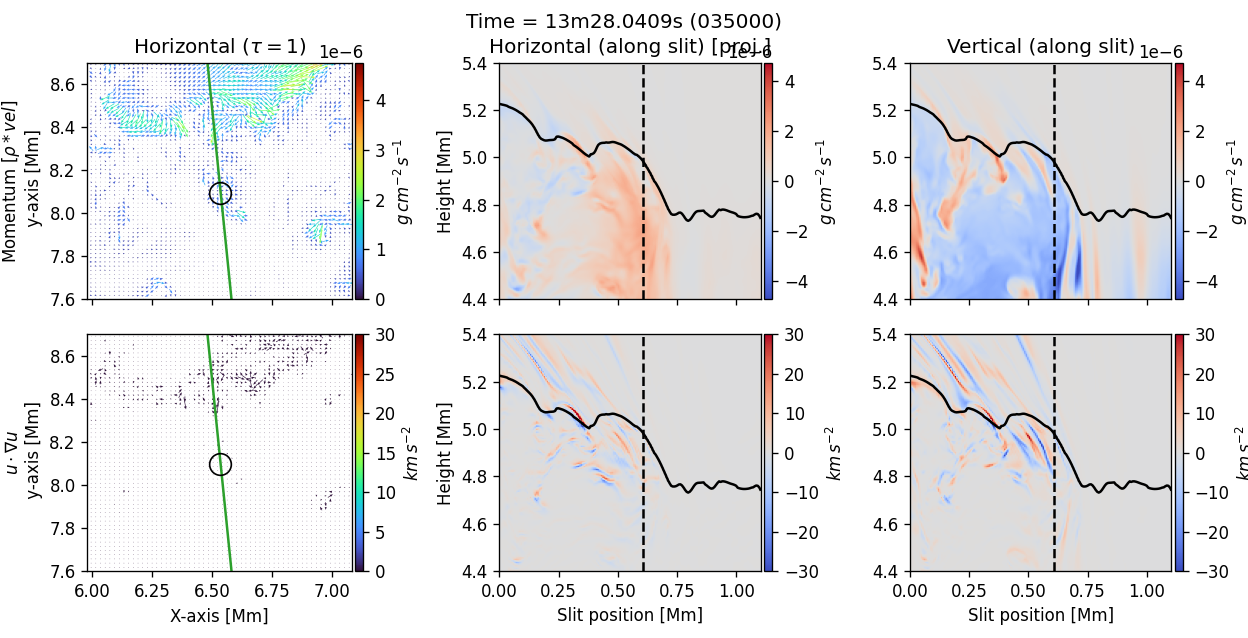}
  \caption[Momentum within a grain forming striation (\muram{} simulation)]{
    Momentum along the slit for the grain shown in Figure~\ref{fig:sim_grain_2_still}. Upper row shows the momentum, lower row the non-linear part. 
    Left column shows horizontal parts within the \(\log(\tau)\!=\!0\)-surface, the green line indicates the slit plane with the grain marked by the black circle. Middle column shows horizontal part, projected onto the slit plane, right column the vertical part. The black solid line indicates the \(\log(\tau)=0\)-surface and the vertical black dashed lines marks the grain position. The positive horizontal momentum is pointing towards
    the pore on the right and negative vertical momentum is pointing upwards.
    }
  \label{fig:sim_grain_2_momentum}
\end{figure}

\onecolumn

\chapter{Atomic data listings}
\label{sec:atom_dat}

This Appendix presents the atomic line parameters used in the inversion process of the \fisssp{} data, as detailed in Chapter~\ref{ch:many_line}. The configurations presented in Table~\ref{tab:atom_dat} were then also used for the inversion results presented and discussed in Chapter~\ref{ch:results}.

\begin{landscape}
\captionsetup{width=23cm, margin={0cm,0cm}}
\begin{small}
\begin{longtable}{l|l|l|r|rrr|rr|rrrr}
\caption{\label{tab:atom_dat} Atomic line data used for the many-line SPINOR inversion sorted by central wavelength (CWL). spectral windows are separated by horizontal lines. The region ID (RID) refers to Table~\ref{tab:wl?regions}. 
Please note that the significance concept in SPINOR is inspired by a noise level approximation and translates to the inverse of the used weight. Thus, a high significance value leads to a low weight on the respective profile. 
}\\
\hline\hline
\multirow{2}{*}{RID}&\multirow{2}{*}{CWL [Å]}&\multirow{2}{*}{Element}&\multirow{2}{*}{\(\log(gf)\)}&\multicolumn{3}{c}{Lower}&\multicolumn{2}{|c|}{Upper}&\multicolumn{4}{c}{Significance}\\
&&&&E [eV]&J &Config&J&Config&I&Q&U&V\\
\hline
\endfirsthead
\caption{Atomic line data used for the many-line inversion (continued).}\\
\hline\hline
\multirow{2}{*}{RID}&\multirow{2}{*}{CWL [Å]}&\multirow{2}{*}{Element}&\multirow{2}{*}{\(\log(gf)\)}&\multicolumn{3}{c}{Lower}&\multicolumn{2}{|c|}{Upper}&\multicolumn{4}{c}{Significance}\\
&&&&E [eV]&J &Config&J&Config&I&Q&U&V\\
\hline
\endhead
\hline
\multirow{2}{*}{R02}& 5225.53&Fe I&-4.889&0.110121&1.0&s2 a$^5$D&1.0&(5D)sp z$^7$D&\rdelim\}{2}{*}[3.0e-03]&\multirow{2}{*}{2.0e-03}&\multirow{2}{*}{2.0e-03}&\multirow{2}{*}{1.0e-03}\\
& 5225.82&Cr I&-1.420&2.707900&3.0 & 3d$^5$(4P)4s a$^5$P & 2.0 & 4s4p($^1$P) y$^5$F & & & & \\
\hline
\multirow{17}{*}{R03}& 5226.38&Cr I&-2.724&4.116400&5.0& 3d$^5$(2G$_2$)4s c$^3$G & 5.0 & 3d$^5$(4F)4p v$^5$G &\rdelim\}{17}{*}[1.0e-02]&\multirow{17}{*}{9.9e-01}&\multirow{17}{*}{9.9e-01}&\multirow{17}{*}{1.0e-03}\\
&5226.54&Ti II&-1.260&1.565800&1.5&3d$^3$ b$^2$D & 1.5 & 3d$^2$(3F)4p z$^2$D & & & & \\
&5226.59&Ti I&-3.460&3.323200&2.0& 4s4p(3P) x$^3$F & 3.0 & 3d$^2$4s($^4$S)5s $^5$P& & & & \\
&5226.86&Fe I&-0.555&3.038500&2.0&4s4p(3P) z$^7$P & 2.0 & 4s($^6$D)5s e$^7$D& & & & \\
&5226.90&Fe I&-1.906&4.294400&2.0& 3d$^6$4s$^2$ b$^1$D$_2$ & 3.0 & 3d$^7$(2D$_2$)4p x$^1$F& & & & \\
&5226.90&Cr I&-0.932&3.375200&2.0&3d$^5$(4P)4s b$^3$P & 2.0 & 4s4p(3P) w$^5$D& & & & \\
&5226.92&Ca I&-3.407&3.728400&2.0&3p$^6$4s4f $^3$F & 1.0 & 3p$^6$3d8g $^3$D& & & & \\
&5226.98&Cr II&-3.416&10.761500&3.5&4s4p(3P) u$^4$G & 2.5 & 3d$^4$(a$^3$F)4d $^4$G& & & & \\
&5227.03&Y II&-2.272&7.257200&3.0&4d5d e$^1$F & 2.0 & 4d7p $^3$F & & & & \\
&5227.07&Cr I&-1.720&2.709900&1.0&3d$^5$(4P)4s a$^5$P & 2.0 & 4s4p($^1$P) x$^5$P & & & & \\
&5227.10&Fe II&-2.452&6.703000&0.5& 3d$^6$(3P$_1$)4s c$^2$P & 1.5 & 3d$^6$(3D)4p y$^2$P & & & & \\
&5227.12&Fe II&-3.353&9.575400&4.5&3d$^5$4s$^2$ $^2$H & 3.5 & 4s4p(3P) t$^2$F & & & & \\
&5227.15&Fe I&-1.352&2.424200&1.0&3d$^6$4s$^2$ a$^3$P$_2$ & 2.0 & 3d$^7$(4F)4p y$^3$D& & & & \\
&5227.19&Fe I&-1.328&1.557400&3.0&3d$^7$(4F)4s a$^3$F & 2.0 & 4s4p(3P) z$^3$D& & & & \\
&5227.52&Fe I&-2.511&5.351600&4.0&4s($^6$D)5s e$^7$D & 3.0 & 3d$^7$(4F)7p $^5$G & & & & \\
&5227.55&Ti I&-2.017&3.708700&3.0&4s4p($^1$P) w$^3$D & 4.0 & 3d$^2$4p$^2$ h$^5$D& & & & \\
&5227.56&V I&-1.323&2.256300&2.5&4s4p(3P) z$^6$D & 3.5 & 3d$^4$(5D)5s e$^6$D& & & & \\
\pagebreak
\multirow{9}{*}{R06} &5229.85 & Fe I & -1.057 & 3.283246 & 1.0 & (5D)sp z$^5$D & 0.0 & s$^6$D)5s e$^5$D & \rdelim\}{9}{*}[3.0e-03] & \multirow{9}{*}{2.0e-03} & \multirow{9}{*}{2.0e-03} & \multirow{9}{*}{1.0e-03} \\
 & 5229.87 & Fe I & -0.201 & 4.220646 & 4.0 & (4F)4p y$^5$F & 4.0 & (4F)4d h$^5$D & & & & \\
 & 5230.19 & Ca I & -3.229 & 3.599700 & 2.0 & 3p$^6$4s5d $^3$D & 1.0 & 3p$^6$3d$^{10}$f $^3$P & & & & \\
 & 5230.21 & Co I & -1.940 & 1.740300 & 1.5 & 3p$^6$3d$^7$4s$^2$ a$^4$P & 1.5 & 3p$^6$3d$^8$(3F)4p y$^4$D & & & & \\
 & 5230.21 & Cr II & -3.486 & 8.727800 & 4.5 & 3d$^4$(3G)4p y$^2$H & 5.5 & 3d$^4$(5D)4d e$^4$G & & & & \\
 & 5230.22 & Cr I & -1.830 & 2.709900 & 1.0 & 3d$^5$(4P)4s a$^5$P & 2.0 & 4s4p($^1$P) y$^5$F & & & & \\
 & 5230.27 & Mn II & -2.599 & 10.214800 & 3.0 & 3d$^5$(2F)4p y$^3$G & 3.0 & 3d$^5$(4G)5s e$^5$G & & & & \\
 & 5230.40 & Co I & -0.489 & 4.110438 & 1.5 & (3F)4p y$^4$D & 1.5 & (3F)4d e$^4$D & & & & \\
 & 5230.69 & Fe I & -2.490 & 4.294703 & 2.0 & s$^2$ b$^1$D$_2$ & 2.0 & (4F)5p $^5$F & & & & \\
\hline
\multirow{3}{*}{R07} & 5232.50 & Cr II & -2.340 & 4.071504 & 3.5 & d$^5$ b$^4$F & 4.5 & (5D)4p z$^4$F & \rdelim\}{3}{*}[1.0e-02] & \multirow{3}{*}{9.9e-01} & \multirow{3}{*}{9.9e-01} & \multirow{3}{*}{5.0e-04} \\
    & 5232.95 & Fe I & -0.200 & 2.940044 & 4.0 & (5D)sp z$^7$P & 5.0 & s$^6$D)5s e$^7$D & & & & \\
    & 5233.21 & Ti I & -3.357 & 1.733633 & 1.0 & (4P)4s a$^5$P & 0.0 & (4P)4p x$^3$P & & & & \\
\multirow{7}{*}{R09} & 5237.08 & Co I & -0.355 & 4.175373 & 2.5 & (3F)4p y$^4$G & 3.5 & 3F)s5s g$^2$F & \rdelim\}{7}{*}[2.3e-03] & \multirow{7}{*}{3.5e-03} & \multirow{7}{*}{2.6e-03} & \multirow{7}{*}{1.0e-03} \\
 & 5237.24 & Y II & -1.560 & 7.945200 & 2.0 & 4d6p $^3$D & 3.0 & 4d8s $^3$D & & & & \\
 & 5237.26 & Fe I & -3.451 & 3.546500 & 2.0 & 3d$^7$(2D$_2$)4s a$^1$D & 3.0 & 4s4p(3P) w$^5$G & & & & \\
 & 5237.33 & Cr II & -1.380 & 4.073400 & 4.5 & 3d$^5$ b$^4$F & 4.5 & 3d$^4$(5D)4p z$^4$F & & & & \\
 & 5237.35 & Ti I & -3.000 & 1.502591 & 4.0 & s$^2$ a$^1$G & 3.0 & (3P)sp v$^3$D & & & & \\
 & 5237.42 & Cr I & -1.569 & 4.261500 & 5.0 & 3d$^5$(4G)4p y$^5$H & 4.0 & 3d$^4$4s5s e$^5$H & & & & \\
 & 5237.43 & Nb I & -1.390 & 1.042800 & 0.5 & 4d$^4$(a$^5$D)5s a$^4$D & 0.5 & 4d$^3$5s(c$^3$P)5p y$^4$P & & & &\\
\hline
\multirow{6}{*}{R10} & 5238.54 & Ti I & -0.996 & 2.092200 & 2.0 & 4s4p(3P) z$^5$F & 1.0 & 3d$^2$4s($^4$F)5s e$^5$F & \rdelim\}{6}{*}[2.0e-04] & \multirow{6}{*}{2.0e-03} & \multirow{6}{*}{2.0e-03} & \multirow{6}{*}{1.0e-04} \\
 & 5238.57 & Ti I & -1.091 & 2.092398 & 2.0 & (3F)sp z$^5$F & 1.0 & s$^4$F)5s e$^5$F & & & & \\
 & 5238.59 & Ti I & -2.074 & 0.848400 & 5.0 & 3d$^3$(4F)4s a$^5$F & 4.0 & (3P)sp y$^5$D & & & & \\
 & 5238.62 & Fe II & -5.295 & 2.891213 & 2.5 & d$^5$ 4s$^2$ a$^6$S & 1.5 & (5D)4p z$^6$F & & & & \\
 & 5238.95 & Fe II & -2.788 & 10.857600 & 2.5 & 4s4p(3P) y$^6$F & 2.5 & 4p$^2$(1D) $^6$D & & & & \\
 & 5238.96 & Cr I & -1.450 & 2.708800 & 2.0 & 3d$^5$(4P)4s a$^5$P & 1.0 & 4s4p(1P) x$^5$P & & & & \\
\hline
& 5240.36 & Fe I & -3.537 & 3.267333 & 5.0 & (2H)4s b$^3$H & 4.0 & (3H)sp y$^3$G & \rdelim\}{2}{*}[2.0e-03] & \multirow{2}{*}{2.0e-02} & \multirow{2}{*}{2.0e-02} & \multirow{2}{*}{1.0e-02} \\
& 5240.48 & Cr I & -0.804 & 3.668273 & 2.0 & (5D)sp y$^5$P & 3.0 & 4s5s f$^5$D & & & & \\
\pagebreak
\multirow{6}{*}{R12} & 5242.40 & Fe I & -3.248 & 4.638200 & 1.0 & 4s4p(1P) y$^5$P & 2.0 & 3d$^6$4s($^6$D)5d $^5$D & \rdelim\}{6}{*}[3.0e-03] & \multirow{6}{*}{1.0e-02} & \multirow{6}{*}{1.0e-02} & \multirow{6}{*}{1.0e-03} \\
 & 5242.42 & Ni II & -1.631 & 12.552900 & 0.5 & 3p$^6$3d$^8$(3F)4d $^2$P & 0.5 & 3p$^6$3d$^8$(3P)5p $^4$P & & & & \\
 & 5242.42 & Sc I & -1.542 & 2.319900 & 0.5 & 4s$^2$4p $^2$P & 1.5 & 3d4s($^1$D)5s $^2$D & & & & \\
 & 5242.45 & Y II & -1.717 & 7.882100 & 1.0 & 4d6p $^3$D & 1.0 & 4d8s $^3$D & & & & \\
 & 5242.50 & Fe I & -0.967 & 3.634300 & 6.0 & 3d$^6$4s$^2$ a$^1$I & 5.0 & 3d$^7$(2G)4p z$^1$H & & & & \\
 & 5242.50 & Co II & -1.669 & 12.034300 & 4.0 & 3p$^6$3d$^7$(4F)5p $^3$F & 3.0 & 3p$^6$3d$^7$(2H)4d $^3$F & & & & \\
\hline
\multirow{7}{*}{R13} & 5243.32 & Mn II & -2.611 & 10.661300 & 2.0 & 3d$^5$(4F)4p $^3$F & 2.0 & 3d$^5$(4P)5s e$^3$P & \rdelim\}{7}{*}[3.0e-03] & \multirow{7}{*}{1.0e-02} & \multirow{7}{*}{1.0e-02} & \multirow{7}{*}{1.0e-03} \\
 & 5243.35 & Cr I & -0.580 & 3.395000 & 2.0 & 4s4p(3P) z$^7$D & 1.0 & 3d$^4$4s5s f$^7$D & & & & \\
 & 5243.45 & Cr II & -2.764 & 4.042300 & 2.5 & 3d$^5$ a$^2$F & 2.5 & 3d$^4$(5D)4p z$^4$F & & & & \\
 & 5243.46 & Ti I & -2.658 & 2.305400 & 3.0 & 4s4p(3P) z$^5$D & 3.0 & 3d$^2$4s($^4$F)5s e$^3$F & & & & \\
 & 5243.49 & Ca I & -3.442 & 3.599700 & 2.0 & 3p$^6$4s5d $^3$D & 1.0 & 3p$^6$3d$^{10}$f $^3$D & & & & \\
 & 5243.70 & Ti I & -1.450 & 2.344714 & 2.0 & (2P)4s c$^3$P & 2.0 & (3P)sp u$^3$D & & & & \\
 & 5243.78 & Fe I & -0.950 & 4.256509 & 3.0 & (4F)4p y$^5$F & 4.0 & (4F)4d g$^5$F & & & & \\
\hline
\multirow{5}{*}{R14} & 5246.77 & Cr II & -2.480 & 3.713810 & 0.5 & (3P)4s b$^4$P & 1.5 & (5D)4p z$^4$P & \rdelim\}{5}{*}[4.0e-04] & \multirow{5}{*}{2.0e-04} & \multirow{5}{*}{2.0e-04} & \multirow{5}{*}{1.0e-04} \\
 & 5246.96 & V I & -1.683 & 2.253448 & 1.5 & (4F)sp z$^6$F & 2.5 & (5D)5s e$^6$D & & & & \\
 & 5247.05 & Fe I & -4.946 & 0.087291 & 2.0 & s$^2$ a$^5$D & 3.0 & (5D)sp z$^7$D & & & & \\
 & 5247.29 & Ti I & -0.727 & 2.103090 & 3.0 & (3F)sp z$^5$F & 2.0 & s$^4$F)5s e$^5$F & & & & \\
 & 5247.57 & Cr I & -1.640 & 0.961039 & 0.0 & s$^2$ a$^5$D & 1.0 & (6S)4p z$^5$P & & & & \\
\hline
 & 5249.09 & Fe I & -1.480 & 4.473573 & 3.0 & (4F)4p z$^3$G & 3.0 & (4F)4d f$^3$F & \rdelim\}{3}{*}[2.3e-03] & \multirow{3}{*}{9.9e-01} & \multirow{3}{*}{9.9e-01} & \multirow{3}{*}{1.0e-03} \\
 & 5249.27 & Ti I & -4.653 & 0.818196 & 2.0 & (4F)4s a$^5$F & 3.0 & (4F)4p y$^3$D & & & & \\
 & 5249.44 & Cr II & -2.620 & 3.757897 & 1.5 & (3P)4s b$^4$P & 2.5 & (5D)4p z$^6$D & & & & \\
\hline
\multirow{4}{*}{R15} & 5250.21 & Fe I & -4.938 & 0.121274 & 0.0 & s$^2$ a$^5$D & 1.0 & (5D)sp z$^7$D & \rdelim\}{4}{*}[4.0e-04] & \multirow{4}{*}{2.0e-04} & \multirow{4}{*}{2.0e-04} & \multirow{4}{*}{1.0e-04} \\
 & 5250.65 & Fe I & -2.281 & 2.198014 & 2.0 & (4P)4s a$^5$P & 3.0 & (5D)sp y$^5$P & & & & \\
 & 5250.81 & Nd II & -2.720 & 0.744600 & 4.5 & 4f$^4$(5I)5d $^6$K & 3.5 & 4f$^4$(5I)6p $^6$I & & & & \\
 & 5250.82 & Cr I & -3.456 & 3.987700 & 3.0 & 4s4p(3P) x$^5$D & 2.0 & 3d$^4$4s5s e$^5$P & & & & \\
\pagebreak
\multirow{3}{*}{R16} & 5253.02 & Fe I & -3.840 & 2.278758 & 2.0 & s$^2$ a$^3$P$_2$ & 1.0 & (5D)sp y$^5$P & \rdelim\}{3}{*}[2.3e-03] & \multirow{3}{*}{3.5e-03} & \multirow{3}{*}{5.0e-01} & \multirow{3}{*}{1.0e-03} \\
 & 5253.46 & La I & -0.820 & 0.130589 & 2.5 & 6s$^2$ a$^2$D & 1.5 & 6p  y$^4$F & & & & \\
 & 5253.47 & Fe I & -1.670 & 3.283246 & 1.0 & (5D)sp z$^5$D & 1.0 & s$^6$D)5s e$^5$D & & & & \\
\hline
\multirow{7}{*}{R17} & 5254.65 & Co I & -0.389 & 3.970900 & 3.5 & 3p$^6$3d$^8$(3F)4p y$^4$D & 3.5 & 3p$^6$3d$^8$(3F)4d e$^4$D & \rdelim\}{7}{*}[2.0e-04] & \multirow{7}{*}{2.0e-03} & \multirow{7}{*}{2.0e-03} & \multirow{7}{*}{1.0e-04} \\
 & 5254.85 & Co I & -2.805 & 4.208700 & 2.5 & 3p$^6$3d$^8$(3F)4p $^6$S & 2.5 & 3p$^6$3d$^8$(3F)4d f$^2$F & & & & \\
 & 5254.90 & Cr I & -0.610 & 3.409600 & 3.0 & 4s4p(3P) z$^7$D & 2.0 & 3d$^4$4s5s f$^7$D & & & & \\
 & 5254.92 & Fe II & -3.336 & 3.230500 & 2.5 & 3d$^6$(3G)4s a$^4$G & 2.5 & 3d$^6$(5D)4p z$^4$F & & & & \\
 & 5254.95 & Fe I & -4.834 & 0.110121 & 1.0 & s$^2$ a$^5$D & 2.0 & (5D)sp z$^7$D & & & & \\
 & 5254.98 & Fe I & -1.742 & 4.294400 & 2.0 & 3d$^6$4s$^2$ b$^1$D$_2$ & 3.0 & 3d$^7$(4F)5p $^5$F & & & & \\
 & 5255.11 & Cr I & -0.200 & 3.463529 & 4.0 & (5D)sp y$^7$P & 5.0 & 4s5s f$^7$D & & & & \\
\hline
\multirow{4}{*}{R18} & 5255.66 & Fe I & -2.050 & 4.220400 & 4.0 & 3d$^7$(4F)4p y$^5$F & 5.0 & 3d$^7$(4F)4d g$^5$F & \rdelim\}{4}{*}[3.0e-03] & \multirow{4}{*}{1.0e-02} & \multirow{4}{*}{1.0e-02} & \multirow{4}{*}{1.0e-03} \\
 & 5255.73 & Fe I & -1.801 & 4.283300 & 2.0 & 3d$^7$(4F)4p y$^5$F & 2.0 & 3d$^7$(4F)4d f$^5$P & & & & \\
 & 5255.81 & Ti I & -0.650 & 2.117100 & 4.0 & 4s4p(3P) z$^5$F & 3.0 & 3d$^2$4s($^4$F)5s e$^5$F & & & & \\
 & 5255.82 & Fe II & -3.075 & 10.857600 & 2.5 & 4s4p(3P) y$^6$F & 2.5 & 4s($^7$S)5s $^6$S$_2$ & & & & \\
\end{longtable}
\end{small}
\newpage

\begin{small}
\begin{longtable}{l|l|l|r|rrr|rr|rrrr}
\caption{\label{tab:atom_dat_unused} Same as \ref{tab:atom_dat} but for lines that where disregarded during the selection process.}\\
\hline\hline
\multirow{2}{*}{RID}&\multirow{2}{*}{CWL [Å]}&\multirow{2}{*}{Element}&\multirow{2}{*}{\(\log(gf)\)}&\multicolumn{3}{c}{Lower}&\multicolumn{2}{|c|}{Upper}&\multicolumn{4}{c}{Significance}\\
&&&&E [eV]&J &Config&J&Config&I&Q&U&V\\
\hline
\endfirsthead
\caption{Atomic line data excluded from the many-line inversion (continued).}\\
\hline\hline
\multirow{2}{*}{RID}&\multirow{2}{*}{CWL [Å]}&\multirow{2}{*}{Element}&\multirow{2}{*}{\(\log(gf)\)}&\multicolumn{3}{c}{Lower}&\multicolumn{2}{|c|}{Upper}&\multicolumn{4}{c}{Significance}\\
&&&&E [eV]&J &Config&J&Config&I&Q&U&V\\
\hline
\endhead
\hline
\multirow{10}{*}{R01} & 5224.87 & Cr I & -3.470 & 4.355100 & 4.0 & 3d$^5$(2G$_2$)4s b$^1$G & 4.0 & 3d$^5$(4F)4p r$^5$D & \rdelim\}{10}{*}[3.0e-03] & \multirow{10}{*}{2.0e-03} & \multirow{10}{*}{2.0e-03} & \multirow{10}{*}{1.0e-03} \\
    & 5224.91 & Ni I & -1.208 & 4.418989 & 3.0 & (3F)sp y$^1$F & 3.0 & (2D)5d h$^3$F & & & & \\
    & 5224.92 & Ti I & -0.196 & 3.449900 & 5.0 & (5D)sp z$^7$D & 5.0 & (3F)sp z$^5$F & & & & \\
    & 5224.92 & Cr I & -0.196 & 3.450149 & 5.0 & (5D)sp z$^7$D & 5.0 & 4s5s f$^7$D & & & & \\
    & 5224.93 & Ti I & -0.312 & 2.117062 & 4.0 & (3F)sp z$^5$F & 4.0 & s$^4$F)5s e$^5$F & & & & \\
    & 5224.95 & Ti I & -0.252 & 2.117211 & 4.0 & (3F)sp z$^5$F & 4.0 & s$^4$F)5s e$^5$F & & & & \\
    & 5225.02 & Cr I & -0.360 & 3.428100 & 4.0 & 4s4p(3P) z$^7$D & 4.0 & 3d$^4$4s5s f$^7$D & & & & \\
    & 5225.02 & Cr II & -2.875 & 10.635200 & 2.5 & 4s4p(3P) u$^4$D & 2.5 & 3d$^4$(a$^3$P)4d $^4$P & & & & \\
    & 5225.07 & Y I & -2.633 & 3.068200 & 2.5 & 4d5s($^1$D)5p y$^2$D & 2.5 & 4d5s($^1$D)5d $^2$F & & & & \\
    & 5225.13 & Nb I & -1.530 & 1.428900 & 6.5 & 4d$^4$(a$^3$H)5s a$^4$H & 5.5 & 4d$^4$(a$^3$H)5p y$^4$G & & & & \\
\hline
\multirow{4}{*}{R04} & 5228.31 & Y II & -3.270 & 7.398200 & 2.0 & 5p$^2$ e$^3$P & 1.0 & 4d$^7$p $^1$P & \rdelim\}{4}{*}[2.0e-03] & \multirow{4}{*}{1.0e-02} & \multirow{4}{*}{1.0e-02} & \multirow{4}{*}{1.0e-03} \\
    & 5228.38 & Fe I & -1.000 & 4.220400 & 4.0 & 3d$^7$(4F)4p y$^5$F & 3.0 & 3d$^7$(4F)4d f$^5$P & & & & \\
    & 5228.41 & Co II & -2.181 & 11.397700 & 2.0 & 3p$^6$3d$^7$(4F)4d e$^5$G & 2.0 & 3p$^6$3d$^7$(4P)5p $^5$D & & & & \\
    & 5228.47 & Ni II & -1.303 & 14.414600 & 0.5 & 3p$^6$3d$^8$(3P)4d $^4$P & 1.5 & 4s4p($^1$P) $^2$P & & & & \\
\hline
\multirow{6}{*}{R05} & 5229.68 & Ni II & -2.661 & 14.400200 & 3.5 & 3p$^6$3d$^8$(3P)4d $^2$F & 4.5 & 3p$^6$3d$^8$(3F)8p $^4$G & \rdelim\}{6}{*}[3.0e-04] & \multirow{6}{*}{1.0e-03} & \multirow{6}{*}{1.0e-03} & \multirow{6}{*}{1.0e-04} \\
    & 5229.70 & Ca I & -2.427 & 3.699500 & 1.0 & 3p$^6$4s5d $^3$D & 1.0 & 3p$^6$3d$^{10}$f $^3$P & & & & \\
    & 5229.71 & Ni II & -2.852 & 12.290900 & 5.5 & 3p$^6$3d$^8$(3F)4d $^4$G & 5.5 & 4s4p(3P) $^2$I & & & & \\
    & 5229.84 & Fe I & -0.962 & 3.283000 & 1.0 & 4s4p(3P) z$^5$D & 0.0 & 4s($^6$D)5s e$^5$D & & & & \\
    & 5229.87 & Fe I & -0.310 & 4.220400 & 4.0 & 3d$^7$(4F)4p y$^5$F & 4.0 & 3d$^7$(4F)4d h$^5$D & & & & \\
    & 5229.99 & Ni II & -2.163 & 13.975500 & 3.5 & 3p$^6$3d$^8$(1D)4d $^2$F & 2.5 & 4s4p(1P) $^4$P & & & & \\
\multirow{4}{*}{R08} & 5235.39 & Fe I & -0.980 & 4.076084 & 4.0 & d$^8$ c$^3$F & 3.0 & (2P)4p u$^3$D & \rdelim\}{4}{*}[2.0e-03] & \multirow{4}{*}{2.0e-03} & \multirow{4}{*}{2.0e-03} & \multirow{4}{*}{1.0e-03} \\
    & 5235.42 & Fe I & -4.201 & 2.588281 & 3.0 & s$^2$ b$^3$F$_2$ & 3.0 & (5D)sp x$^5$D & & & & \\
    & 5235.50 & Ni I & -1.037 & 3.898533 & 2.0 & (2D)4p z$^1$D & 1.0 & (2D)4d e$^1$P & & & & \\
    & 5235.56 & Cr I & -1.586 & 4.171200 & 7.0 & 3d$^5$(4G)4p y$^5$H & 6.0 & 3d$^4$4s5s e$^5$H & & & & \\
\hline
\multirow{3}{*}{R11} & 5239.75 & Cr II & -3.250 & 11.318800 & 5.5 & 3d$^4$(b$^3$F)4p w$^4$G & 4.5 & 3d$^4$(5D)5d $^4$G & \rdelim\}{3}{*}[2.0e-04] & \multirow{3}{*}{2.0e-03} & \multirow{3}{*}{2.0e-03} & \multirow{3}{*}{1.0e-04} \\
    & 5239.81 & Sc II & -0.850 & 1.455100 & 0.0 & 3p$^6$4s$^2$ a$^1$S & 1.0 & 3p$^6$3d4p $^1$P & & & & \\
    & 5239.94 & Ti I & -3.270 & 0.812900 & 1.0 & 3d$^3$(4F)4s a$^5$F & 1.0 & 4s4p(3P) y$^5$D & & & & \\
\end{longtable}
\end{small}

\end{landscape}
\clearpage

\chapter{Notation}\label{app:notation}
\section{Symbols}
\textit{This listing holds the typical definition/use of the symbols. Some definitions can be locally overwritten.
Symbols with local/temporal definition only are not listed.}
\begin{tabbing}
    \(I, Q, U, V\)\;\, \= Description \kill
    \(\EE[X]\) \> Expected value of a random variable \(X\)\\[5pt]
    \(\mathbb{C}\) \> Complex numbers\\[5pt]
    \(I_c\), \(I_{QS}\) \> Quiet Sun continuum intensity\\[5pt]
    \(\mu\)\> Cosine of the heliocentric angle\\[5pt]
    \(\FC\) \> Fourier transform operator\\[5pt]
    \(r_0\) \> Fried parameter \\[5pt]
    \(\one_{x\in X}\) \> Indicator function (\(1\) if \(x\in X\), \(0\) otherwise) \\[5pt]
    \(\delta_{ij}\) \> Kronecker Delta\\[5pt]
    \(B\) \> Magnetic field strength [G]\\[5pt]
    \(\mathbf{L}\) \> Maximum Likelihood Estimator\\[5pt]
    \(\NN_0\) \> Natural numbers including \(0\) (\(\{0, 1, 2, ...\}\))\\[5pt]
    \(\tau, \tau_{x}\) \> Optical depth (at wavelength \(x\))\\[5pt]
    \(\mathbb{R}\) \> Real numbers\\[5pt]
    \(I, Q, U, V\) \> Stokes parameters\\[5pt]
    \(L^2(X)\) \> Set of square integrable functions on \(X\)\\[5pt]
    \(T\) \> Temperature [K]\\[5pt]
    \(\sun\) \> The Sun\\[5pt]
    \vlos \> (Line of sight) velocities [km/s]\\[2pt]
    \(\chi^2\) \> Error metric, typically computed as \(\sum_{\lambda} \frac{(x_{\lambda}^\text{obs} - x_{\lambda}^\text{fit})^2}{x_{\lambda}^\text{obs}}\) for all wavelength bins \(\lambda\)\\[2pt]
    \(\Psi_n\) \> \(n\)-th Zernike polynomial according to the Noll index scheme
\end{tabbing}

\newpage
\section{Acronyms}
\begin{tabbing}
    ACRONYM... \= Description \kill
    ADU \> Analogue-to-digital units (photon counts)\\
    AO \> Adaptive Optics\\
    AOI \> Angle of Incidence\\
    AR \> Active Region \\
    AU \> Astronomical Unit (\(\sim 1.5\times 10^8\) km)\\
    AWGN \> Additive White Gaussian Noise \\
    AWIV \> Area with implausible value \\
    BBSO \> Big Bear Solar Observatory\\
    CAD \> Computer Aided Design\\
    CCD \> Charge-coupled device\\
    CMOS \> Complementary metal-oxide-semiconductor\\
    CWL \> Central Wavelength\\
    DKIST \> Daniel K. Inouye Solar Telescope\\
    EM \> Expectation Maximization (algorithm) \\
    EST \> European Solar Telescope\\
    EUV \> Extreme Ultra Violet\\
    FFT \> Fast Fourier Transform \\
    \fiss{} \> Fast Imaging Solar Spectrograph \\
    \fisssp{} \> \fiss{} spectro-polarimeter\\
    FLC \> Ferro-electric Liquid Crystals\\
    FOV \> Field of View \\
    FTS \> Fourier Transform Spectrograph\\
    FWHM \> Full Width at Half Maximum\\
    GRIS \> GREGOR Infrared Spectrograph\\
    GS \> Gerchberg-Saxton (algorithm) \\
    GST \> Goode Solar Telescope \\
    HSRA \> Harvard-Smithsonian reference atmosphere \\
    IGL \> Intergranular lane\\
    IR \> Infrared part of the spectrum\\
    KL \> Karhunen-Loève\\
    LB \> Light bridge\\
    LOS \> Line Of Sight\\
    LP \> Linear Polarizer\\
    LR \> Lucy-Richardson (algorithm)\\
    LTE \> Local Thermal Equilibrium \\
    ML \> Maximum Likelihood / Many Line\\
    (MO)MFBD \> (Multi Object) Multi Frame Blind Deconvolution \\
    MSE \> Mean Square Error \\
    MTF \> Modulation Transfer Function\\
    NL \> (Camera) Non-Linearity\\
    NLTE \> Non-LTE \\
    NOAA \> National Oceanic and Atmospheric Administration\\
    nUV \> Near Ultraviolet part of the spectrum\\
    OTF \> Optical Transfer Function\\
    PBS \> Polarizing Beam Splitter\\
    PCU \> Polarimetric Calibration Unit\\
    PD \> Phase Diversity\\
    PMU \> Polarimetric Modulation Unit\\
    PSF \> Point Spread Function \\
    RMS \> Root Mean Square (\(\sqrt{\nicefrac{\sum_{n=1}^N x^2_n}{N}}\))\\
    ROI \> Region of Interest\\
    RTE \> Radiative Transfer Equation\\
    RTI \> Rayleigh–Taylor instability\\
    SJC \> Slit-Jaw Camera \\
    \snr{} \> Signal to noise ratio\\
    SST \> Swedish Solar Telescope\\
    SP \> Spectro-Polarimeter \\
    SPC \> Spectrograph Camera\\
    SPSF \> Spectrograph PSF \\
    SUSI \> Sunrise UV Spectrograph and Imager\\
    UD \> Umbral dot\\
    UTC \> Coordinated Universal Time\\
    UV \> Ultraviolet part of the spectrum\\
    VALD \>  Vienna Atomic Line Database\\
    VIS \> Visible part of the spectrum\\
    WL \> Wavelength\\
    QWP \> Quarter Wave Plate 
\end{tabbing}

\addcontentsline{toc}{chapter}{List Of Figures}
\listoffigures

\addcontentsline{toc}{chapter}{List Of Tables}
\listoftables

\bibliography{thesis}

@article{Dialetis86,
       author = {{Dialetis}, D. and {Macris}, C. and {Prokakis}, T. and {Sarris}, E.},
        title = "{The lifetime and evolution of solar granules.}",
      journal = {\aap},
         year = 1986,
        month = nov,
       volume = {168},
       number = {1-2},
        pages = {330-334},
}

@book{Harlim18, 
    place={Cambridge}, 
    title={Data-Driven Computational Methods: Parameter and Operator Estimations},
    DOI={10.1017/9781108562461}, 
    URL={https://doi.org/10.1017/9781108562461},
    publisher={Cambridge University Press}, 
    author={Harlim, John}, 
    year={2018}
}

@ARTICLE{Hoelken24,
       author = {{H{\"o}lken}, J. and {Doerr}, H. -P. and {Feller}, A. and {Iglesias}, F.~A.},
        title = "{Spectroflat: A generic spectrum and flat-field calibration library for spectro-polarimetric data}",
      journal = {\aap},
         year = 2024,
        month = jul,
       volume = {687},
          eid = {A22},
        pages = {A22},
          doi = {10.1051/0004-6361/202348877},
       adsurl = {https://ui.adsabs.harvard.edu/abs/2024A&A...687A..22H}
}

@article{Hoelken25,
       author = {{H{\"o}lken}, J. and {van Noort}, M. and {Solanki}, S.~K. and {Feller}, A. and {Doerr}, H.-P. and {Cao}, W. and {Gorceix}, N. and {Kang}, J. and {Ahn}, K. and {Chae}, J.},
        title = "{Toward solar many-line inversions of high-resolution spectropolarimetric data}",
      journal = {\aap},
         year = 2026,
        month = jan,
       volume = {705},
          eid = {A220},
        pages = {A220},
          doi = {10.1051/0004-6361/202556890},
       adsurl = {https://ui.adsabs.harvard.edu/abs/2026A&A...705A.220H}
}

@article{Hoelken26,
       author = {{H{\"o}lken}, J. and {Przybylski}, D. and {van Noort}, M. and {Cameraon}, R. and {Solanki}, S.~K. and {Doerr}, H.-P. and {Cao}, W. and {Gorceix}, N. and {Kang}, J. and {Ahn}, K. and {Chae}, J.},
        title = "{The magneto-convective nature of dark striations and moving bright grains at the edge of pores}",
      journal = {\aap},
         year = 2026,
       volume = {708},
          eid = {A135},
        pages = {A135},
          doi = {10.1051/0004-6361/202659037 },
}

@article{Hohage2016,
  author    = {Thorsten Hohage and Frank Werner},
  title     = {Inverse problems with Poisson data: statistical regularization theory, applications and algorithms},
  journal   = {Inverse Problems},
  year      = {2016},
  volume    = {32},
  number    = {9},
  pages     = {093001},
  doi       = {10.1088/0266-5611/32/9/093001},
  url       = {https://iopscience.iop.org/article/10.1088/0266-5611/32/9/093001}
}

@article{Gerchberg72,
  author = {{Gerchberg}, R. W. and {Saxton}, W. O.},
  journal = {OPTIK},
  pages = {237-256},
  title = {A practical algorithm for the determination of phase from image and diffraction plane pictures},
  type = {Article},
  volume = {35},
  year = {1972}
}

@ARTICLE{STIC,
       author = {{de la Cruz Rodr{\'\i}guez}, J. and {Leenaarts}, J. and {Danilovic}, S. and {Uitenbroek}, H.},
        title = "{STiC: A multiatom non-LTE PRD inversion code for full-Stokes solar observations}",
      journal = {\aap},
         year = 2019,
        month = mar,
       volume = {623},
          eid = {A74},
        pages = {A74},
          doi = {10.1051/0004-6361/201834464},
archivePrefix = {arXiv},
       eprint = {1810.08441},
 primaryClass = {astro-ph.SR},
       adsurl = {https://ui.adsabs.harvard.edu/abs/2019A&A...623A..74D}
}

@ARTICLE{DeSIRe,
       author = {{Ruiz Cobo}, B. and {Quintero Noda}, C. and {Gafeira}, R. and {Uitenbroek}, H. and {Orozco Su{\'a}rez}, D. and {P{\'a}ez Ma{\~n}{\'a}}, E.},
        title = "{DeSIRe: Departure coefficient aided Stokes Inversion based on Response functions}",
      journal = {\aap},
         year = 2022,
        month = apr,
       volume = {660},
          eid = {A37},
        pages = {A37},
          doi = {10.1051/0004-6361/202140877},
archivePrefix = {arXiv},
       eprint = {2202.02226},
 primaryClass = {astro-ph.SR},
       adsurl = {https://ui.adsabs.harvard.edu/abs/2022A&A...660A..37R}
}

@book{Gonzales03,
  author    = "{Gonzalez}, R. and {Woods}, R and {Eddins}, S.",
  title     = "Digital Image Processing Using Matlab",
  publisher = "Prentice Hall",
  address   = "Hoboken, New Jersey",
  year      = 2003
}

@article{Meynadier99,
    author = {Meynadier, Ludovic and Michau, Vincent and Velluet, Marie-Th\'{e}r\`{e}se and Conan, Jean-Marc and Mugnier, Laurent M. and Rousset, G\'{e}rard},
    journal = {Appl. Opt.},
    number = {23},
    pages = {4967--4979},
    publisher = {Optica Publishing Group},
    title = {Noise propagation in wave-front sensing with phase diversity},
    volume = {38},
    month = {Aug},
    year = {1999},
    url = {https://opg.optica.org/ao/abstract.cfm?URI=ao-38-23-4967},
    doi = {10.1364/AO.38.004967}
}

@ARTICLE{Noort05,
       author = {{Van Noort}, Michiel and {Der Voort}, Luc Rouppe Van and {L{\"o}fdahl}, Mats G.},
        title = "{Solar Image Restoration By Use Of Multi-frame Blind De-convolution With Multiple Objects And Phase Diversity}",
      journal = {\solphys},
         year = 2005,
        month = may,
       volume = {228},
       number = {1-2},
        pages = {191-215},
          doi = {10.1007/s11207-005-5782-z},
       adsurl = {https://ui.adsabs.harvard.edu/abs/2005SoPh..228..191V}
}

@ARTICLE{Noort17,
       author = {{Van Noort}, M.},
        title = "{Image restoration of solar spectra}",
      journal = {\aap},
         year = 2017,
        month = dec,
       volume = {608},
          eid = {A76},
        pages = {A76},
          doi = {10.1051/0004-6361/201731339},
archivePrefix = {arXiv},
       eprint = {1711.09629},
 primaryClass = {astro-ph.IM},
       adsurl = {https://ui.adsabs.harvard.edu/abs/2017A&A...608A..76V}
}

@article{Noll76,
    author = {Noll, Robert J.},
    journal = {J. Opt. Soc. Am.},
    number = {3},
    pages = {207--211},
    publisher = {Optica Publishing Group},
    title = {Zernike polynomials and atmospheric turbulence$\ast$},
    volume = {66},
    month = {Mar},
    year = {1976},
    url = {https://opg.optica.org/abstract.cfm?URI=josa-66-3-207},
    doi = {10.1364/JOSA.66.000207}
}

@ARTICLE{Paxman96,
       author = {{Paxman}, Richard G. and {Seldin}, John H. and {Loefdahl}, Mats G. and {Scharmer}, Goeran B. and {Keller}, Christoph U.},
        title = "{Evaluation of Phase-Diversity Techniques for Solar-Image Restoration}",
      journal = {\apj},
         year = 1996,
        month = aug,
       volume = {466},
        pages = {1087},
          doi = {10.1086/177578},
       adsurl = {https://ui.adsabs.harvard.edu/abs/1996ApJ...466.1087P}
}

@ARTICLE{Miura94,
       author = {{Miura}, Noriaki and {Ohsawa}, Kenro and {Baba}, Naoshi},
        title = "{Parallel blind deconvolution applied to solar image}",
      journal = {Optical Review},
         year = 1994,
        month = jan,
       volume = {1},
       number = {2},
        pages = {208-210},
          doi = {10.1007/BF03254864},
       adsurl = {https://ui.adsabs.harvard.edu/abs/1994OptRv...1..208M}
}

@ARTICLE{Roddier90,
       author = {{Roddier}, Nicolas},
        title = "{Atmospheric wavefront simulation using Zernike polynomials.}",
      journal = {Optical Engineering},
         year = 1990,
        month = oct,
       volume = {29},
        pages = {1174-1180},
          doi = {10.1117/12.55712},
       adsurl = {https://ui.adsabs.harvard.edu/abs/1990OptEn..29.1174R}
}

@article{Schulz93,
    author = {Schulz, Timothy J.},
    journal = {J. Opt. Soc. Am. A},
    number = {5},
    pages = {1064--1073},
    publisher = {Optica Publishing Group},
    title = {Multiframe blind deconvolution of astronomical images},
    volume = {10},
    month = {May},
    year = {1993},
    url = {https://opg.optica.org/josaa/abstract.cfm?URI=josaa-10-5-1064},
    doi = {10.1364/JOSAA.10.001064}
}

@ARTICLE{Shepp82,
  author={Shepp, L. A. and Vardi, Y.},
  journal={IEEE Transactions on Medical Imaging}, 
  title={Maximum Likelihood Reconstruction for Emission Tomography}, 
  year={1982},
  volume={1},
  number={2},
  pages={113-122},
  url={https://doi.org/10.1109/TMI.1982.4307558},
  doi={10.1109/TMI.1982.4307558}
}

@book{Snyder91,
  author    = "{Snyder}, Donald L. and {Miller}, Michael I.",
  title     = "Random Point Processes in Time and Space",
  publisher = "Springer",
  address   = "New York, NY",
  year      = 1991,
  edition = {2},
  doi={10.1007/978-1-4612-3166-0},
  url={https://doi.org/10.1007/978-1-4612-3166-0}
}

@book{Basalla06,
    author = "{Basalla}, George",
    title = "Civilized life in the universe : scientists on intelligent extraterrestrials",
    publisher = "Oxford University Press",
    address   = "New York, NY",
    url = {https://archive.org/details/civilizedlifeinu0000basa/page/52/mode/2up},
    year = 2006,
    pages = {52,53},
    isbn = "978-0195171815"
}

@ARTICLE{Kawaler81,
       author = {{Kawaler}, S. and {Veverka}, J.},
        title = "{The Habitable Sun - One of Herschel, William's Stranger Ideas}",
      journal = {\jrasc},
         year = 1981,
        month = feb,
       volume = {75},
        pages = {46},
       adsurl = {https://ui.adsabs.harvard.edu/abs/1981JRASC..75...46K}
}

@article{Zhivanovich20,
  title = {Relationship of the Magnetic-Field Strength and the Brightness of the Sunspot Umbra and the Center of a Facular Knot},
  volume = {60},
  ISSN = {1555-645X},
  url = {http://dx.doi.org/10.1134/S0016793220070324},
  DOI = {10.1134/s0016793220070324},
  number = {7},
  journal = {Geomagnetism and Aeronomy},
  publisher = {Pleiades Publishing Ltd},
  author = {Zhivanovich,  I. and Solov’ev,  A. A. and Efremov,  V. I. and Miller,  N. O.},
  year = {2020},
  month = dec,
  pages = {865–871}
}

@article{Peng24,
  title = {Statistical Comparison between Pores and Sunspots during the Time Interval 2010–2023},
  volume = {975},
  ISSN = {1538-4357},
  url = {http://dx.doi.org/10.3847/1538-4357/ad7858},
  DOI = {10.3847/1538-4357/ad7858},
  number = {1},
  journal = {The Astrophysical Journal},
  publisher = {American Astronomical Society},
  author = {Peng,  Yang and Fei,  Yu and Xiang,  Nan-bin and Deng,  Lin-hua and Xu,  Ting-ting and Zheng,  Sheng and Zeng,  Shu-guang and Zhang,  Hai-yang and Liu,  Shi-hu},
  year = {2024},
  month = oct,
  pages = {23}
}

@book{French2023Sun,
  author    = {French, Ryan},
  title     = {The {Sun}: Beginner's Guide to Our Closest Star},
  year      = {2023},
  publisher = {Collins},
  isbn      = {9780008580230},
  note      = {ISBN-10: 0008580235}
}

@ARTICLE{Wilson74,
       author = {{Wilson}, Alexander and {Maskelyne}, Nevil},
        title = "{Observations on the Solar Spots. By Alexander Wilson, M. D. Professor of Practical Astronomy in the University of Glasgow. Communicated by the Rev. Nevil Maskelyne, Astronomer Royal}",
      journal = {Philosophical Transactions of the Royal Society of London Series I},
         year = 1774,
        month = jan,
       volume = {64},
        pages = {1-30},
       adsurl = {https://ui.adsabs.harvard.edu/abs/1774RSPT...64....1W}
}

@INPROCEEDINGS{Harra21,
       author = {{Harra}, Louise K.},
        title = "{Coronal 'camp-fires' in the quiet Sun as observed by Solar Orbiter EUV Imagers}",
    booktitle = {43rd COSPAR Scientific Assembly. Held 28 January - 4 February},
         year = 2021,
       volume = {43},
        month = jan,
          eid = {950},
        pages = {950},
       adsurl = {https://ui.adsabs.harvard.edu/abs/2021cosp...43E.950H}
}

@ARTICLE{muram,
       author = {{V{\"o}gler}, A. and {Shelyag}, S. and {Sch{\"u}ssler}, M. and {Cattaneo}, F. and {Emonet}, T. and {Linde}, T.},
        title = "{Simulations of magneto-convection in the solar photosphere.  Equations, methods, and results of the MURaM code}",
      journal = {\aap},
         year = 2005,
        month = jan,
       volume = {429},
        pages = {335-351},
          doi = {10.1051/0004-6361:20041507},
       adsurl = {https://ui.adsabs.harvard.edu/abs/2005A&A...429..335V}
}

@ARTICLE{Bhatia25,
       author = {{Bhatia}, Tanayveer Singh and {Panja}, Mayukh and {Cameron}, Robert H. and {Solanki}, Sami K.},
        title = "{3D radiative MHD simulations of starspots: II. Large-scale structure}",
      journal = {\aap},
         year = 2025,
        month = jan,
       volume = {693},
          eid = {A264},
        pages = {A264},
          doi = {10.1051/0004-6361/202452747},
archivePrefix = {arXiv},
       eprint = {2412.16921},
 primaryClass = {astro-ph.SR},
       adsurl = {https://ui.adsabs.harvard.edu/abs/2025A&A...693A.264B}
}

@ARTICLE{Schmassmann21,
       author = {{Schmassmann}, M. and {Rempel}, M. and {Bello Gonz{\'a}lez}, N. and {Schlichenmaier}, R. and {Jur{\v{c}}{\'a}k}, J.},
        title = "{Characterization of magneto-convection in sunspots. The Gough-Tayler stability criterion in MURaM sunspot simulations}",
      journal = {\aap},
         year = 2021,
        month = dec,
       volume = {656},
          eid = {A92},
        pages = {A92},
          doi = {10.1051/0004-6361/202141607},
archivePrefix = {arXiv},
       eprint = {2210.17245},
 primaryClass = {astro-ph.SR},
       adsurl = {https://ui.adsabs.harvard.edu/abs/2021A&A...656A..92S}
}

@ARTICLE{Rempel12,
       author = {{Rempel}, M.},
        title = "{Numerical Sunspot Models: Robustness of Photospheric Velocity and Magnetic Field Structure}",
      journal = {\apj},
         year = 2012,
        month = may,
       volume = {750},
       number = {1},
          eid = {62},
        pages = {62},
          doi = {10.1088/0004-637X/750/1/62},
archivePrefix = {arXiv},
       eprint = {1203.0534},
 primaryClass = {astro-ph.SR},
       adsurl = {https://ui.adsabs.harvard.edu/abs/2012ApJ...750...62R}
}

@ARTICLE{Cowling85,
       author = {{Cowling}, T.~G.},
        title = "{Astronomer by accident.}",
      journal = {\araa},
         year = 1985,
        month = jan,
       volume = {23},
        pages = {1-18},
          doi = {10.1146/annurev.aa.23.090185.000245},
       adsurl = {https://ui.adsabs.harvard.edu/abs/1985ARA&A..23....1C}
}

@ARTICLE{Hale08,
       author = {{Hale}, George E.},
        title = "{The Zeeman Effect in the Sun}",
      journal = {\pasp},
         year = 1908,
        month = dec,
       volume = {20},
       number = {123},
        pages = {287},
          doi = {10.1086/121847},
       adsurl = {https://ui.adsabs.harvard.edu/abs/1908PASP...20..287H}
}

@INPROCEEDINGS{Spruit81,
       author = {{Spruit}, H.~C.},
        title = "{A cluster model for sunspots}",
    booktitle = {The Physics of Sunspots},
         year = 1981,
       editor = {{Cram}, L.~E. and {Thomas}, J.~H.},
        month = jan,
        pages = {98-103},
       adsurl = {https://ui.adsabs.harvard.edu/abs/1981phss.conf...98S}
}

@ARTICLE{Tatarski61,
       author = {{Tatarski}, V.~I. and {Silverman}, R.~A. and {Chako}, Nicholas},
        title = "{Wave Propagation in a Turbulent Medium}",
      journal = {Physics Today},
         year = 1961,
        month = jan,
       volume = {14},
       number = {12},
        pages = {46},
          doi = {10.1063/1.3057286},
       adsurl = {https://ui.adsabs.harvard.edu/abs/1961PhT....14l..46T}
}

@ARTICLE{Voitsekhovich95,
       author = {{Voitsekhovich}, V.~V. and {Cuevas}, S.},
        title = "{Adaptive optics and the outer scale of turbulence}",
      journal = {Journal of the Optical Society of America A},
         year = 1995,
        month = nov,
       volume = {12},
       number = {11},
        pages = {2523-2531},
          doi = {10.1364/JOSAA.12.002523},
       adsurl = {https://ui.adsabs.harvard.edu/abs/1995JOSAA..12.2523V}
}

@article{Wang80,
    author = {J. Y. Wang and D. E. Silva},
    journal = {Appl. Opt.},
    number = {9},
    pages = {1510--1518},
    publisher = {Optica Publishing Group},
    title = {Wave-front interpretation with Zernike polynomials},
    volume = {19},
    month = {May},
    year = {1980},
    url = {https://opg.optica.org/ao/abstract.cfm?URI=ao-19-9-1510},
    doi = {10.1364/AO.19.001510}
}

@article{Thimons18, 
    title={Investigating the Gerchberg-Saxton Phase Retrieval Algorithm}, 
    volume={11}, 
    ISSN={2327-7807}, 
    url={http://dx.doi.org/10.1137/17s016610}, 
    DOI={10.1137/17s016610}, 
    journal={SIAM Undergraduate Research Online}, 
    publisher={Society for Industrial & Applied Mathematics (SIAM)}, 
    author={{Thimons}, Theresa and {Wittle}, Lily}, year={2018} 
}

@ARTICLE{Orlov97,
       author = {{Orlov}, V.~G. and {Voitsekhovich}, V.~V. and {Cuevas}, S.},
        title = "{Karhunen-Loeve Functions in Simulations of Atmospheric Distortions}",
      journal = {\rmxaa},
         year = 1997,
        month = oct,
       volume = {33},
        pages = {187-195},
       adsurl = {https://ui.adsabs.harvard.edu/abs/1997RMxAA..33..187O}
}

@misc{lagg25,
      author = {Korpi-Lagg, Andreas and Gandorfer, Achim and Solanki, Sami K. and del Toro Iniesta, Jose Carlos and Katsukawa, Yukio and Bernasconi, Pietro and Berkefeld, Thomas and Feller, Alex and Riethmüller, Tino L. and Álvarez-Herrero, Alberto and Kubo, Masahito and Martínez Pillet, Valentín and Smitha, H. N. and Orozco Suárez, David and Grauf, Bianca and Carpenter, Michael and Bell, Alexander and Álvarez-Alonso, María-Teresa and Álvarez García, Daniel and Aparicio del Moral, Beatriz and Atiénzar, Julia and Ayoub, Daniel and Bailén, Francisco Javier and Bailón-Martínez, Eduardo and Balaguer Jiménez, Maria and Barthol, Peter and Bayon Laguna, Montserrat and Bellot-Rubio, Luis R. and Bergmann, Melani and Blanco Rodríguez, Julian and Bochmann, Jan and Borrero, Juan Manuel and Campos-Jara, Antonio and Castellanos Durán, Juan Sebastián and Cebollero, María and Conde Rodríguez, Aitor and Deutsch, Werner and Eaton, Harry and Fernández-Medina, Ana Belen and Fernandez-Rico, German and Ferreres, Agustin and García, Andrés and García Alarcia, Ramón María and García Parejo, Pilar and Garranzo-García, Daniel and Gasent Blesa, José Luis and Gerber, Karin and Germerott, Dietmar and Gilabert Palmer, David and Gizon, Laurent and Gómez Sánchez-Tirado, Miguel Angel and González-Bárcena, David and Gonzalo Melchor, Alejandro and Goodyear, Sam and Hara, Hirohisa and Harnes, Edvarda and Heerlein, Klaus and Heidecke, Frank and Heinrichs, Jan and Hernández Expósito, David and Hirzberger, Johann and Hölken, Johannes and Hyun, Sangwon and Iglesias, Francisco A. and Ishikawa, Ryohtaroh T. and Jeon, Minwoo and Kawabata, Yusuke and Kolleck, Martin and Laguna, Hugo and Lomas, Julian and López Jiménez, Antonio C. and Manzano, Paula and Matsumoto, Takuma and Mayo Turrado, David and Meierdierks, Thimo and Meining, Stefan and Monecke, Markus and Morales-Fernández, José Miguel and Moreno Mantas, Antonio Jesús and Moreno Vacas, Alejandro and Müller, Marc Ferenc and Müller, Reinhard and Naito, Yoshihiro and Nakai, Eiji and Núñez Peral, Armonía and Oba, Takayoshi and Palo, Geoffrey and Pérez-Grande, Isabel and Piqueras Carreño, Javier and Preis, Tobias and Przybylski, Damien and Quintero Noda, Carlos and Ramanath, Sandeep and Ramos Más, Jose Luis and Raouafi, Nour and Rivas-Martínez, María-Jesús and Rodríguez-Martínez, Pedro and Rodríguez Valido, Manuel and Ruiz Cobo, Basilio and Sánchez Rodríguez, Antonio and Sanchez Toledo, Mariano and Sánchez Gómez, Antonio and Sanchis Kilders, Esteban and Sant, Kamal and Santamarina Guerrero, Pablo and Schulze, Erich and Shimizu, Toshifumi and Silva-López, Manuel and Singh, Kunal and Siu-Tapia, Azaymi L. and Sonner, Thomas and Staub, Jan and Strecker, Hanna and Tobaruela, Angel and Torralbo, Ignacio and Tritschler, Alexandra and Tsuzuki, Toshihiro and Uraguchi, Fumihiro and Volkmer, Reiner and Vourlidas, Angelos and Vukadinović, Dušan and Werner, Stephan and Zerr, Andreas},
  title = {Sunrise iii: Overview of Observatory and Instruments},
  journal = {Solar Physics},
  volume = {300},
  number = {5},
  pages = {75},
  year = {2025},
  month = {May},
  day = {30},
  doi = {10.1007/s11207-025-02485-1},
  url = {https://doi.org/10.1007/s11207-025-02485-1},
  issn = {1573-093X}
}

@article{Feller25,
  title = {The Sunrise Ultraviolet Spectropolarimeter and Imager: Instrument Description},
  volume = {300},
  ISSN = {1573-093X},
  url = {http://dx.doi.org/10.1007/s11207-025-02471-7},
  DOI = {10.1007/s11207-025-02471-7},
  number = {5},
  journal = {Solar Physics},
  publisher = {Springer Science and Business Media LLC},
  author = {Feller,  Alex and Gandorfer,  Achim and Grauf,  Bianca and H\"{o}lken,  Johannes and Iglesias,  Francisco A. and Korpi-Lagg,  Andreas and Riethm\"{u}ller,  Tino L. and Staub,  Jan and Fernandez-Rico,  German and Castellanos Durán,  Juan Sebastián and Solanki,  Sami K. and Smitha,  H. N. and Sant,  Kamal and Barthol,  Peter and Bayon Laguna,  Montserrat and Bergmann,  Melani and Bischoff,  J\"{o}rg and Bochmann,  Jan and Bruns,  Stefan and Deutsch,  Werner and Eberhardt,  Michel and Enge,  Rainer and Goodyear,  Sam and Heerlein,  Klaus and Heinrichs,  Jan and Hirche,  Dennis and Meining,  Stefan and Mende,  Roland and Meyer,  Sabrina and M\"{u}hlhaus,  Maria and M\"{u}ller,  Marc Ferenc and Monecke,  Markus and Oberdorfer,  Dietmar and Papagiannaki,  Ioanna and Ramanath,  Sandeep and Verg\"{o}hl,  Michael and Vukadinović,  Dušan and Werner,  Stephan and Zerr,  Andreas and Berkefeld,  Thomas and Bernasconi,  Pietro and Katsukawa,  Yukio and del Toro Iniesta,  Jose Carlos and Bell,  Alexander and Carpenter,  Michael and Álvarez Herrero,  Alberto and Kubo,  Masahito and Martínez Pillet,  Valentín and Orozco Suárez,  David},
  year = {2025},
  month = may 
}

@ARTICLE{Lande23,
       author = {{Land{\'e}}, Alfred},
        title = "{Fortschritte beim Zeemaneffekt}",
      journal = {Ergebnisse der exakten Naturwissenschaften},
         year = 1923,
        month = jan,
       volume = {2},
        pages = {147},
          doi = {10.1007/BFb0111702},
       adsurl = {https://ui.adsabs.harvard.edu/abs/1923ErNW....2..147L}
}

@article{Fried66,
    author = {D. L. Fried},
    journal = {J. Opt. Soc. Am.},
    number = {10},
    pages = {1372--1379},
    publisher = {Optica Publishing Group},
    title = {Optical Resolution Through a Randomly Inhomogeneous Medium for Very Long and Very Short Exposures},
    volume = {56},
    month = {Oct},
    year = {1966},
    url = {https://opg.optica.org/abstract.cfm?URI=josa-56-10-1372},
    doi = {10.1364/JOSA.56.001372},
}

@article{Lucy74,
       author = {{Lucy}, L.~B.},
        title = "{An iterative technique for the rectification of observed distributions}",
      journal = {\aj},
         year = 1974,
        month = jun,
       volume = {79},
        pages = {745},
          doi = {10.1086/111605},
          url = {https://doi.org/10.1086/111605}
}

@article{Richardson72,
    author = {Richardson, William Hadley},
    journal = {J. Opt. Soc. Am.},
    number = {1},
    pages = {55--59},
    publisher = {Optica Publishing Group},
    title = {Bayesian-Based Iterative Method of Image Restoration$\ast$},
    volume = {62},
    month = {Jan},
    year = {1972},
    url = {https://opg.optica.org/abstract.cfm?URI=josa-62-1-55},
    doi = {10.1364/JOSA.62.000055}
}

@ARTICLE{Schlichenmaier16,
       author = {{Schlichenmaier}, R. and {von der L{\"u}he}, O. and {Hoch}, S. and {Soltau}, D. and {Berkefeld}, T. and {Schmidt}, D. and {Schmidt}, W. and {Denker}, C. and {Balthasar}, H. and {Hofmann}, A. and {Strassmeier}, K.~G. and {Staude}, J. and {Feller}, A. and {Lagg}, A. and {Solanki}, S.~K. and {Collados}, M. and {Sigwarth}, M. and {Volkmer}, R. and {Waldmann}, T. and {Kneer}, F. and {Nicklas}, H. and {Sobotka}, M.},
        title = "{Active region fine structure observed at 0.08 arcsec resolution}",
      journal = {\aap},
         year = 2016,
        month = nov,
       volume = {596},
          eid = {A7},
        pages = {A7},
          doi = {10.1051/0004-6361/201628561},
archivePrefix = {arXiv},
       eprint = {1607.07094},
 primaryClass = {astro-ph.SR},
       adsurl = {https://ui.adsabs.harvard.edu/abs/2016A&A...596A...7S}
}

@ARTICLE{Scharmer02,
       author = {{Scharmer}, G{\"o}ran B. and {Gudiksen}, Boris V. and {Kiselman}, Dan and {L{\"o}fdahl}, Mats G. and {Rouppe van der Voort}, Luc H.~M.},
        title = "{Dark cores in sunspot penumbral filaments}",
      journal = {\nat},
         year = 2002,
        month = nov,
       volume = {420},
       number = {6912},
        pages = {151-153},
          doi = {10.1038/nature01173},
       adsurl = {https://ui.adsabs.harvard.edu/abs/2002Natur.420..151S}
}

@misc{spectroflat,
  title        = "spectroflat v1.0.0",
  author       = "{H\"olken}, J.",
  howpublished = "\url{https://gitlab.gwdg.de/hoelken/spectroflat}",
  year         = 2023,
  note         = "Gitlab repository of the spectroflat python library."
}

@misc{atlas-fit,
  title        = "atlas-fit v1.0.0",
  author       = "{H\"olken}, J.",
  howpublished = "\url{https://gitlab.gwdg.de/hoelken/atlas-fit}",
  year         = 2023,
  note         = "Gitlab repository of the atlas-fit python library."
}

@ARTICLE{Woehl02,
       author = {{W{\"o}hl}, H. and {Ku{\v{c}}era}, A. and {Ryb{\'a}k}, J. and {Hanslmeier}, A.},
        title = "{Precise reduction of solar spectra obtained with large CCD arrays}",
      journal = {\aap},
         year = 2002,
        month = nov,
       volume = {394},
        pages = {1077-1091},
          doi = {10.1051/0004-6361:20021203},
       adsurl = {https://ui.adsabs.harvard.edu/abs/2002A&A...394.1077W}
}

@ARTICLE{Qiu22,
       author = {{Qiu}, Ye and {Rao}, ShiHao and {Li}, Chuan and {Fang}, Cheng and {Ding}, MingDe and {Li}, Zhen and {Ni}, YiWei and {Wang}, WenBo and {Hong}, Jie and {Hao}, Qi and {Dai}, Yu and {Chen}, PengFei and {Wan}, XiaoSheng and {Xu}, Zhi and {You}, Wei and {Yuan}, Yuan and {Tao}, HongJiang and {Li}, XianSheng and {He}, YuKun and {Liu}, Qiang},
        title = "{Calibration procedures for the CHASE/HIS science data}",
      journal = {Science China Physics, Mechanics, and Astronomy},
         year = 2022,
        month = aug,
       volume = {65},
       number = {8},
          eid = {289603},
        pages = {289603},
          doi = {10.1007/s11433-022-1900-5},
archivePrefix = {arXiv},
       eprint = {2205.06075},
 primaryClass = {astro-ph.SR},
       adsurl = {https://ui.adsabs.harvard.edu/abs/2022SCPMA..6589603Q}
}

@ARTICLE{Bommier02,
       author = {{Bommier}, V. and {Rayrole}, J.},
        title = "{Search for polarimetric sensitivity in the first observations with THEMIS spectropolarimetric mode MTR (August 1998 campaign)}",
      journal = {\aap},
         year = 2002,
        month = jan,
       volume = {381},
        pages = {227-240},
          doi = {10.1051/0004-6361:20011420},
       adsurl = {https://ui.adsabs.harvard.edu/abs/2002A&A...381..227B}
}

@ARTICLE{Jones03J,
       author = {{Jones}, Harrison P.},
        title = "{Data Calibration and Analysis for He I 1083 nm Imaging Spectroscopy}",
      journal = {\solphys},
         year = 2003,
        month = dec,
       volume = {218},
       number = {1},
        pages = {1-16},
          doi = {10.1023/B:SOLA.0000013031.13781.66},
       adsurl = {https://ui.adsabs.harvard.edu/abs/2003SoPh..218....1J}
}

@ARTICLE{Chae13,
       author = {{Chae}, Jongchul and {Park}, Hyung-Min and {Ahn}, Kwangsu and {Yang}, Heesu and {Park}, Young-Deuk and {Nah}, Jakyoung and {Jang}, Bi Ho and {Cho}, Kyung-Suk and {Cao}, Wenda and {Goode}, Philip R.},
        title = "{Fast Imaging Solar Spectrograph of the 1.6 Meter New Solar Telescope at Big Bear Solar Observatory}",
      journal = {\solphys},
         year = 2013,
        month = nov,
       volume = {288},
       number = {1},
        pages = {1-22},
          doi = {10.1007/s11207-012-0147-x},
       adsurl = {https://ui.adsabs.harvard.edu/abs/2013SoPh..288....1C}
}

@ARTICLE{Cai17,
       author = {{Cai}, Yunfang and {Xu}, Zhi and {Li}, Zhenggang and {Xiang}, Yongyuan and {Chen}, Yuchao and {Fu}, Yu and {Ji}, Kaifan},
        title = "{Precise Reduction of Solar Spectra Observed by the One-Meter New Vacuum Solar Telescope}",
      journal = {\solphys},
         year = 2017,
        month = oct,
       volume = {292},
       number = {10},
          eid = {150},
        pages = {150},
          doi = {10.1007/s11207-017-1178-0},
archivePrefix = {arXiv},
       eprint = {1709.09067},
 primaryClass = {astro-ph.SR},
       adsurl = {https://ui.adsabs.harvard.edu/abs/2017SoPh..292..150C}
}

@ARTICLE{Wueser18,
       author = {{W{\"u}lser}, J. -P. and {Jaeggli}, S. and {De Pontieu}, B. and {Tarbell}, T. and {Boerner}, P. and {Freeland}, S. and {Liu}, W. and {Timmons}, R. and {Brannon}, S. and {Kankelborg}, C. and {Madsen}, C. and {McKillop}, S. and {Prchlik}, J. and {Saar}, S. and {Schanche}, N. and {Testa}, P. and {Bryans}, P. and {Wiesmann}, M.},
        title = "{Instrument Calibration of the Interface Region Imaging Spectrograph (IRIS) Mission}",
      journal = {\solphys},
         year = 2018,
        month = nov,
       volume = {293},
       number = {11},
          eid = {149},
        pages = {149},
          doi = {10.1007/s11207-018-1364-8},
       adsurl = {https://ui.adsabs.harvard.edu/abs/2018SoPh..293..149W}
}

@misc{spice.3.0,
  doi = {10.48326/IDOC.MEDOC.SPICE.3.0},
  url = {https://spice.osups.universite-paris-saclay.fr/spice-data/release-3.0/release-notes.html},
  author = {{SPICE Consortium} and Auchère, Frédéric and Aznar Cuadrado, Regina and Buchlin, Eric and Caminade, Stéphane and Carlsson, Mats and Fludra, Andrzej and Fredvik, Terje and Giunta, Alessandra S. and Grundy, Timothy W. and Guest, Steve and Hassler, Donald M. and Haugan, Stein V. H. and Janvier, Miho and Kucera, Therese A. and Kouliche, Dimitri and Leeks, Sarah and Mercier, Claude and Parenti, Susanna and Pelouze, Gabriel and Peter, Hardi and Picard, David and Sidher, Sunil and Teriaca, Luca and Thompson, William T. and Williams, David R. and Müller, Daniel},
  title = {Data issued from SPICE instrument on Solar Orbiter: 3rd data release},
  publisher = {IDOC},
  year = {2023}
}

@ARTICLE{SPICE,
       author = {{SPICE Consortium} and {Anderson}, M. and {Appourchaux}, T. and {Auch{\`e}re}, F. and {Aznar Cuadrado}, R. and {Barbay}, J. and {Baudin}, F. and {Beardsley}, S. and {Bocchialini}, K. and {Borgo}, B. and {Bruzzi}, D. and {Buchlin}, E. and {Burton}, G. and {B{\"u}chel}, V. and {Caldwell}, M. and {Caminade}, S. and {Carlsson}, M. and {Curdt}, W. and {Davenne}, J. and {Davila}, J. and {Deforest}, C.~E. and {Del Zanna}, G. and {Drummond}, D. and {Dubau}, J. and {Dumesnil}, C. and {Dunn}, G. and {Eccleston}, P. and {Fludra}, A. and {Fredvik}, T. and {Gabriel}, A. and {Giunta}, A. and {Gottwald}, A. and {Griffin}, D. and {Grundy}, T. and {Guest}, S. and {Gyo}, M. and {Haberreiter}, M. and {Hansteen}, V. and {Harrison}, R. and {Hassler}, D.~M. and {Haugan}, S.~V.~H. and {Howe}, C. and {Janvier}, M. and {Klein}, R. and {Koller}, S. and {Kucera}, T.~A. and {Kouliche}, D. and {Marsch}, E. and {Marshall}, A. and {Marshall}, G. and {Matthews}, S.~A. and {McQuirk}, C. and {Meining}, S. and {Mercier}, C. and {Morris}, N. and {Morse}, T. and {Munro}, G. and {Parenti}, S. and {Pastor-Santos}, C. and {Peter}, H. and {Pfiffner}, D. and {Phelan}, P. and {Philippon}, A. and {Richards}, A. and {Rogers}, K. and {Sawyer}, C. and {Schlatter}, P. and {Schmutz}, W. and {Sch{\"u}hle}, U. and {Shaughnessy}, B. and {Sidher}, S. and {Solanki}, S.~K. and {Speight}, R. and {Spescha}, M. and {Szwec}, N. and {Tamiatto}, C. and {Teriaca}, L. and {Thompson}, W. and {Tosh}, I. and {Tustain}, S. and {Vial}, J. -C. and {Walls}, B. and {Waltham}, N. and {Wimmer-Schweingruber}, R. and {Woodward}, S. and {Young}, P. and {de Groof}, A. and {Pacros}, A. and {Williams}, D. and {M{\"u}ller}, D.},
        title = "{The Solar Orbiter SPICE instrument. An extreme UV imaging spectrometer}",
      journal = {\aap},
         year = 2020,
        month = oct,
       volume = {642},
          eid = {A14},
        pages = {A14},
          doi = {10.1051/0004-6361/201935574},
archivePrefix = {arXiv},
       eprint = {1909.01183},
 primaryClass = {astro-ph.IM},
       adsurl = {https://ui.adsabs.harvard.edu/abs/2020A&A...642A..14S}
}

@ARTICLE{EST,
       author = {{Quintero Noda}, C. and {Schlichenmaier}, R. and {Bellot Rubio}, L.~R. and {L{\"o}fdahl}, M.~G. and {Khomenko}, E. and {Jur{\v{c}}{\'a}k}, J. and {Leenaarts}, J. and {Kuckein}, C. and {Gonz{\'a}lez Manrique}, S.~J. and {Gun{\'a}r}, S. and {Nelson}, C.~J. and {de la Cruz Rodr{\'\i}guez}, J. and {Tziotziou}, K. and {Tsiropoula}, G. and {Aulanier}, G. and {Aboudarham}, J. and {Allegri}, D. and {Alsina Ballester}, E. and {Amans}, J.~P. and {Asensio Ramos}, A. and {Bail{\'e}n}, F.~J. and {Balaguer}, M. and {Baldini}, V. and {Balthasar}, H. and {Barata}, T. and {Barczynski}, K. and {Barreto Cabrera}, M. and {Baur}, A. and {B{\'e}chet}, C. and {Beck}, C. and {Bel{\'\i}o-As{\'\i}n}, M. and {Bello-Gonz{\'a}lez}, N. and {Belluzzi}, L. and {Bentley}, R.~D. and {Berdyugina}, S.~V. and {Berghmans}, D. and {Berlicki}, A. and {Berrilli}, F. and {Berkefeld}, T. and {Bettonvil}, F. and {Bianda}, M. and {Bienes P{\'e}rez}, J. and {Bonaque-Gonz{\'a}lez}, S. and {Braj{\v{s}}a}, R. and {Bommier}, V. and {Bourdin}, P. -A. and {Burgos Mart{\'\i}n}, J. and {Calchetti}, D. and {Calcines}, A. and {Calvo Tovar}, J. and {Campbell}, R.~J. and {Carballo-Mart{\'\i}n}, Y. and {Carbone}, V. and {Carlin}, E.~S. and {Carlsson}, M. and {Castro L{\'o}pez}, J. and {Cavaller}, L. and {Cavallini}, F. and {Cauzzi}, G. and {Cecconi}, M. and {Chulani}, H.~M. and {Cirami}, R. and {Consolini}, G. and {Coretti}, I. and {Cosentino}, R. and {C{\'o}zar-Castellano}, J. and {Dalmasse}, K. and {Danilovic}, S. and {De Juan Ovelar}, M. and {Del Moro}, D. and {del Pino Alem{\'a}n}, T. and {del Toro Iniesta}, J.~C. and {Denker}, C. and {Dhara}, S.~K. and {Di Marcantonio}, P. and {D{\'\i}az Baso}, C.~J. and {Diercke}, A. and {Dineva}, E. and {D{\'\i}az-Garc{\'\i}a}, J.~J. and {Doerr}, H. -P. and {Doyle}, G. and {Erdelyi}, R. and {Ermolli}, I. and {Escobar Rodr{\'\i}guez}, A. and {Esteban Pozuelo}, S. and {Faurobert}, M. and {Felipe}, T. and {Feller}, A. and {Feijoo Amoedo}, N. and {Femen{\'\i}a Castell{\'a}}, B. and {Fernandes}, J. and {Ferro Rodr{\'\i}guez}, I. and {Figueroa}, I. and {Fletcher}, L. and {Franco Ordovas}, A. and {Gafeira}, R. and {Gardenghi}, R. and {Gelly}, B. and {Giorgi}, F. and {Gisler}, D. and {Giovannelli}, L. and {Gonz{\'a}lez}, F. and {Gonz{\'a}lez}, J.~B. and {Gonz{\'a}lez-Cava}, J.~M. and {Gonz{\'a}lez Garc{\'\i}a}, M. and {G{\"o}m{\"o}ry}, P. and {Gracia}, F. and {Grauf}, B. and {Greco}, V. and {Grivel}, C. and {Guerreiro}, N. and {Guglielmino}, S.~L. and {Hammerschlag}, R. and {Hanslmeier}, A. and {Hansteen}, V. and {Heinzel}, P. and {Hern{\'a}ndez-Delgado}, A. and {Hern{\'a}ndez Su{\'a}rez}, E. and {Hidalgo}, S.~L. and {Hill}, F. and {Hizberger}, J. and {Hofmeister}, S. and {J{\"a}gers}, A. and {Janett}, G. and {Jarolim}, R. and {Jess}, D. and {Jim{\'e}nez Mej{\'\i}as}, D. and {Jolissaint}, L. and {Kamlah}, R. and {Kapit{\'a}n}, J. and {Ka{\v{s}}parov{\'a}}, J. and {Keller}, C.~U. and {Kentischer}, T. and {Kiselman}, D. and {Kleint}, L. and {Klvana}, M. and {Kontogiannis}, I. and {Krishnappa}, N. and {Ku{\v{c}}era}, A. and {Labrosse}, N. and {Lagg}, A. and {Landi Degl'Innocenti}, E. and {Langlois}, M. and {Lafon}, M. and {Laforgue}, D. and {Le Men}, C. and {Lepori}, B. and {Lepreti}, F. and {Lindberg}, B. and {Lilje}, P.~B. and {L{\'o}pez Ariste}, A. and {L{\'o}pez Fern{\'a}ndez}, V.~A. and {L{\'o}pez Jim{\'e}nez}, A.~C. and {L{\'o}pez L{\'o}pez}, R. and {Manso Sainz}, R. and {Marassi}, A. and {Marco de la Rosa}, J. and {Marino}, J. and {Marrero}, J. and {Mart{\'\i}n}, A. and {Mart{\'\i}n G{\'a}lvez}, A. and {Mart{\'\i}n Hernando}, Y. and {Masciadri}, E. and {Mart{\'\i}nez Gonz{\'a}lez}, M. and {Matta-G{\'o}mez}, A. and {Mato}, A. and {Mathioudakis}, M. and {Matthews}, S. and {Mein}, P. and {Merlos Garc{\'\i}a}, F. and {Moity}, J. and {Montilla}, I. and {Molinaro}, M. and {Molodij}, G. and {Montoya}, L.~M. and {Munari}, M. and {Murabito}, M. and {N{\'u}{\~n}ez Cagigal}, M. and {Oliviero}, M. and {Orozco Su{\'a}rez}, D. and {Ortiz}, A. and {Padilla-Hern{\'a}ndez}, C. and {Pa{\'e}z Ma{\~n}{\'a}}, E. and {Paletou}, F. and {Pancorbo}, J. and {Pastor Ca{\~n}edo}, A. and {Pastor Yabar}, A. and {Peat}, A.~W. and {Pedichini}, F. and {Peixinho}, N. and {Pe{\~n}ate}, J. and {P{\'e}rez de Taoro}, A. and {Peter}, H. and {Petrovay}, K. and {Piazzesi}, R. and {Pietropaolo}, E. and {Pleier}, O. and {Poedts}, S. and {P{\"o}tzi}, W. and {Podladchikova}, T.},
        title = "{The European Solar Telescope}",
      journal = {\aap},
         year = 2022,
        month = oct,
       volume = {666},
          eid = {A21},
        pages = {A21},
          doi = {10.1051/0004-6361/202243867},
archivePrefix = {arXiv},
       eprint = {2207.10905},
 primaryClass = {astro-ph.SR},
       adsurl = {https://ui.adsabs.harvard.edu/abs/2022A&A...666A..21Q}
}

@ARTICLE{AsensioRamos23,
       author = {{Asensio Ramos}, Andr{\'e}s and {Esteban Pozuelo}, Sara and {Kuckein}, Christoph},
        title = "{Accelerating Multiframe Blind Deconvolution via Deep Learning}",
      journal = {\solphys},
         year = 2023,
        month = jul,
       volume = {298},
       number = {7},
          eid = {91},
        pages = {91},
          doi = {10.1007/s11207-023-02185-8},
archivePrefix = {arXiv},
       eprint = {2306.12078},
 primaryClass = {astro-ph.IM},
       adsurl = {https://ui.adsabs.harvard.edu/abs/2023SoPh..298...91A}
}

@ARTICLE{Rempel17,
       author = {{Rempel}, M.},
        title = "{Extension of the MURaM Radiative MHD Code for Coronal Simulations}",
      journal = {\apj},
         year = 2017,
        month = jan,
       volume = {834},
       number = {1},
          eid = {10},
        pages = {10},
          doi = {10.3847/1538-4357/834/1/10},
archivePrefix = {arXiv},
       eprint = {1609.09818},
 primaryClass = {astro-ph.SR},
       adsurl = {https://ui.adsabs.harvard.edu/abs/2017ApJ...834...10R}
}

@ARTICLE{Witzke21,
       author = {{Witzke}, V. and {Shapiro}, A.~I. and {Cernetic}, M. and {Tagirov}, R.~V. and {Kostogryz}, N.~M. and {Anusha}, L.~S. and {Unruh}, Y.~C. and {Solanki}, S.~K. and {Kurucz}, R.~L.},
        title = "{MPS-ATLAS: A fast all-in-one code for synthesising stellar spectra}",
      journal = {\aap},
         year = 2021,
        month = sep,
       volume = {653},
          eid = {A65},
        pages = {A65},
          doi = {10.1051/0004-6361/202140275},
archivePrefix = {arXiv},
       eprint = {2105.13611},
 primaryClass = {astro-ph.SR},
       adsurl = {https://ui.adsabs.harvard.edu/abs/2021A&A...653A..65W}
}

@misc{Irwin12,
       author = {{Irwin}, Alan W.},
        title = "{FreeEOS: Equation of State for stellar interiors calculations}",
 howpublished = {Astrophysics Source Code Library, record ascl:1211.002},
         year = 2012,
        month = nov,
          eid = {ascl:1211.002},
       adsurl = {https://ui.adsabs.harvard.edu/abs/2012ascl.soft11002I}
}

@article{Przybylski2025,
       author = {{Przybylski}, D. and {Cameron}, R. and {Solanki}, S.~K. and {Rempel}, M. and {Danilovic}, S. and {Leenaarts}, J.},
        title = "{Structure and dynamics of the internetwork solar chromosphere: results of a small-scale dynamo simulation}",
      journal = {arXiv e-prints},
         year = 2025,
        month = aug,
          eid = {arXiv:2508.19809},
        pages = {arXiv:2508.19809},
          doi = {10.48550/arXiv.2508.19809},
archivePrefix = {arXiv},
       eprint = {2508.19809},
 primaryClass = {astro-ph.SR},
       adsurl = {https://ui.adsabs.harvard.edu/abs/2025arXiv250819809P}
}

@ARTICLE{Cho21,
       author = {{Cho}, Kyuhyoun and {Chae}, Jongchul and {Madjarska}, Maria S.},
        title = "{Investigation of the subsurface structure of a sunspot based on the spatial distribution of oscillation centers inferred from umbral flashes}",
      journal = {\aap},
         year = 2021,
        month = dec,
       volume = {656},
          eid = {A86},
        pages = {A86},
          doi = {10.1051/0004-6361/202141500},
archivePrefix = {arXiv},
       eprint = {2109.11185},
 primaryClass = {astro-ph.SR},
       adsurl = {https://ui.adsabs.harvard.edu/abs/2021A&A...656A..86C}
}

@ARTICLE{Asplund06,
       author = {{Asplund}, Martin and {Grevesse}, Nicolas and {Jacques Sauval}, A.},
        title = "{The solar chemical composition}",
      journal = {\nphysa},
         year = 2006,
        month = oct,
       volume = {777},
        pages = {1-4},
          doi = {10.1016/j.nuclphysa.2005.06.010},
archivePrefix = {arXiv},
       eprint = {astro-ph/0410214},
 primaryClass = {astro-ph},
       adsurl = {https://ui.adsabs.harvard.edu/abs/2006NuPhA.777....1A}
}

@ARTICLE{Beeck12,
       author = {{Beeck}, B. and {Collet}, R. and {Steffen}, M. and {Asplund}, M. and {Cameron}, R.~H. and {Freytag}, B. and {Hayek}, W. and {Ludwig}, H. -G. and {Sch{\"u}ssler}, M.},
        title = "{Simulations of the solar near-surface layers with the CO5BOLD, MURaM, and Stagger codes}",
      journal = {\aap},
         year = 2012,
        month = mar,
       volume = {539},
          eid = {A121},
        pages = {A121},
          doi = {10.1051/0004-6361/201118252},
archivePrefix = {arXiv},
       eprint = {1201.1103},
 primaryClass = {astro-ph.SR},
       adsurl = {https://ui.adsabs.harvard.edu/abs/2012A&A...539A.121B}
}

@phdthesis{vukadinovic2024phd,
  author    = {Dušan Vukadinović},
  title     = {Inferring atomic line parameters from solar spectra},
  school    = {Georg-August-Universität Göttingen},
  year      = {2024},
  address   = {Göttingen},
  url       = {https://ediss.uni-goettingen.de/handle/11858/15495},
  doi       = {10.53846/goediss-11947},
  note      = {Doctoral thesis, Faculty of Physics; date of oral examination: 2024-01-29; supervisors: Prof. Dr. Sami K. Solanki, Prof. Dr. Lucia Kleint},
  language  = {en}
}

@article{Oneto24,
title = {Towards algorithms and models that we can trust: A theoretical perspective},
journal = {Neurocomputing},
volume = {592},
pages = {127798},
year = {2024},
issn = {0925-2312},
doi = {https://doi.org/10.1016/j.neucom.2024.127798},
url = {https://www.sciencedirect.com/science/article/pii/S0925231224005691},
author = {Luca Oneto and Sandro Ridella and Davide Anguita}
}

@ARTICLE{Ge24,
       author = {{Ge}, Xinlan and {Zhu}, Licheng and {Gao}, Zeyu and {Ma}, Shiqing and {Li}, Ao and {Wang}, Shuai and {Yang}, Ping},
        title = "{Breaking Boundaries: A Universal Wavefront Reconstruction Approach for High-resolution Solar Imaging}",
      journal = {\apjl},
         year = 2024,
        month = jul,
       volume = {970},
       number = {1},
          eid = {L1},
        pages = {L1},
          doi = {10.3847/2041-8213/ad5b53},
       adsurl = {https://ui.adsabs.harvard.edu/abs/2024ApJ...970L...1G}
}

@ARTICLE{Schirninger25,
       author = {{Schirninger}, C. and {Jarolim}, R. and {Veronig}, A.~M. and {Kuckein}, C.},
        title = "{Deep learning image burst stacking to reconstruct high-resolution ground-based solar observations}",
      journal = {\aap},
         year = 2025,
        month = jan,
       volume = {693},
          eid = {A6},
        pages = {A6},
          doi = {10.1051/0004-6361/202451850},
archivePrefix = {arXiv},
       eprint = {2506.04781},
 primaryClass = {astro-ph.SR},
       adsurl = {https://ui.adsabs.harvard.edu/abs/2025A&A...693A...6S}
}

@ARTICLE{Li25,
       author = {{Li}, Wentao and {Liang}, Yonghui and {Liu}, Xuewen and {Yang}, Huizhe and {Liu}, Jin},
        title = "{Real-time multi-frame blind deconvolution with Nesterov-like optimizer based on GPU}",
      journal = {Optics Laser Technology},
         year = 2025,
        month = nov,
       volume = {189},
          eid = {113042},
        pages = {113042},
          doi = {10.1016/j.optlastec.2025.113042},
       adsurl = {https://ui.adsabs.harvard.edu/abs/2025OptLT.18913042L}
}

@INPROCEEDINGS{keller94,
       author = {{Keller}, Christoph U. and {Povel}, Hans-Peter and {Stenflo}, Jan O.},
        title = "{Zurich Imaging Stokes polarimeters I and II}",
    booktitle = {Polarization Analysis and Measurement II},
         year = 1994,
       editor = {{Goldstein}, Dennis H. and {Chenault}, David B.},
       series = {Society of Photo-Optical Instrumentation Engineers (SPIE) Conference Series},
       volume = {2265},
        month = sep,
        pages = {222-230},
          doi = {10.1117/12.186673},
       adsurl = {https://ui.adsabs.harvard.edu/abs/1994SPIE.2265..222K}
}

@ARTICLE{Povel95,
       author = {{Povel}, Hans-Peter},
        title = "{Imaging Stokes polarimetry with piezoelastic modulators and charge-coupled-device image sensors}",
      journal = {Optical Engineering},
         year = 1995,
        month = jul,
       volume = {34},
        pages = {1870-1878},
          doi = {10.1117/12.200596},
       adsurl = {https://ui.adsabs.harvard.edu/abs/1995OptEn..34.1870P}
}

@ARTICLE{Tiwari13,
       author = {{Tiwari}, Sanjiv Kumar and {van Noort}, Michiel and {Lagg}, Andreas and {Solanki}, Sami K.},
        title = "{Structure of sunspot penumbral filaments: a remarkable uniformity of properties}",
      journal = {\aap},
         year = 2013,
        month = sep,
       volume = {557},
          eid = {A25},
        pages = {A25},
          doi = {10.1051/0004-6361/201321391},
archivePrefix = {arXiv},
       eprint = {1307.3668},
 primaryClass = {astro-ph.SR},
       adsurl = {https://ui.adsabs.harvard.edu/abs/2013A&A...557A..25T}
}

@ARTICLE{Muller73,
       author = {{Muller}, R.},
        title = "{{\'E}tude morphologique et cin{\'e}matique des structures fines d'une tache solaire}",
      journal = {\solphys},
         year = 1973,
        month = mar,
       volume = {29},
       number = {1},
        pages = {55-73},
          doi = {10.1007/BF00153440},
       adsurl = {https://ui.adsabs.harvard.edu/abs/1973SoPh...29...55M}
}

@ARTICLE{Zakharov08,
       author = {{Zakharov}, V. and {Hirzberger}, J. and {Riethm{\"u}ller}, T.~L. and {Solanki}, S.~K. and {Kobel}, P.},
        title = "{Evidence of convective rolls in a sunspot penumbra}",
      journal = {\aap},
         year = 2008,
        month = sep,
       volume = {488},
       number = {2},
        pages = {L17-L20},
          doi = {10.1051/0004-6361:200810266},
archivePrefix = {arXiv},
       eprint = {0808.2317},
 primaryClass = {astro-ph},
       adsurl = {https://ui.adsabs.harvard.edu/abs/2008A&A...488L..17Z}
}

@INPROCEEDINGS{Ramelli10,
       author = {{Ramelli}, Renzo and {Balemi}, Silvano and {Bianda}, Michele and {Defilippis}, Ivan and {Gamma}, Luca and {Hagenbuch}, Stephan and {Rogantini}, Marco and {Steiner}, Peter and {Stenflo}, Jan O.},
        title = "{ZIMPOL-3: a powerful solar polarimeter}",
    booktitle = {Ground-based and Airborne Instrumentation for Astronomy III},
         year = 2010,
       editor = {{McLean}, Ian S. and {Ramsay}, Suzanne K. and {Takami}, Hideki},
       series = {Society of Photo-Optical Instrumentation Engineers (SPIE) Conference Series},
       volume = {7735},
        month = jul,
          eid = {77351Y},
        pages = {77351Y},
          doi = {10.1117/12.857120},
       adsurl = {https://ui.adsabs.harvard.edu/abs/2010SPIE.7735E..1YR}
}

@ARTICLE{Saranathan21,
       author = {{Saranathan}, S. and {van Noort}, M. and {Solanki}, S.~K.},
        title = "{Correction of atmospheric stray light in restored slit spectra}",
      journal = {\aap},
         year = 2021,
        month = sep,
       volume = {653},
          eid = {A17},
        pages = {A17},
          doi = {10.1051/0004-6361/201937100},
archivePrefix = {arXiv},
       eprint = {2112.02648},
 primaryClass = {astro-ph.SR},
       adsurl = {https://ui.adsabs.harvard.edu/abs/2021A&A...653A..17S}
}

@ARTICLE{Neckel99,
       author = {{Neckel}, H.},
        title = "{Announcement}",
      journal = {\solphys},
         year = 1999,
        month = jan,
       volume = {184},
        pages = {421-422},
          doi = {10.1023/A:1017165208013},
       adsurl = {https://ui.adsabs.harvard.edu/abs/1999SoPh..184..421N}
}

@misc{stenflo14,
    title = "{Atlas of the Second Solar Spectrum (SS2)}",
    author = "{{Stenflo}, J.O.} and {{Gandorfer}, A.}",
    year = 2014,
    howpublished = "\url{https://www.irsol.usi.ch/data-archive/second-solar-spectrum-ss2-atlas/}"
}

@inproceedings{Feller20,
author = {Alex Feller and Achim Gandorfer and Francisco A. Iglesias and Andreas Lagg and Tino L. Riethm{\"u}ller and Sami K. Solanki and Yukio Katsukawa and Masahito Kubo},
title = {{The SUNRISE UV Spectropolarimeter and imager for SUNRISE III}},
volume = {11447},
booktitle = {Ground-based and Airborne Instrumentation for Astronomy VIII},
editor = {Christopher J. Evans and Julia J. Bryant and Kentaro Motohara},
organization = {International Society for Optics and Photonics},
publisher = {SPIE},
pages = {11447AK},
year = {2020},
doi = {10.1117/12.2562666},
URL = {https://doi.org/10.1117/12.2562666}
}

@article{Zhao03,
    author = {Jun Zhao},
    journal = {Appl. Spectrosc.},
    number = {11},
    pages = {1368--1375},
    publisher = {Optica Publishing Group},
    title = {Image Curvature Correction and Cosmic Removal for High-Throughput Dispersive Raman Spectroscopy},
    volume = {57},
    month = {Nov},
    year = {2003},
    url = {https://opg.optica.org/as/abstract.cfm?URI=as-57-11-1368},
}

@inproceedings{Reimers15,
author = {Jacob Reimers and Eric M. Schiesser and Kevin P. Thompson and Kevin L. Whiteaker and Dennis Yates and Jannick P. Rolland},
booktitle = {Imaging and Applied Optics 2015},
journal = {Imaging and Applied Optics 2015},
pages = {FM3B.3},
publisher = {Optica Publishing Group},
title = {Comparison of Freeform Imaging Spectrometer Design Forms Using Spectral Full-Field Displays},
year = {2015},
url = {https://opg.optica.org/abstract.cfm?URI=Freeform-2015-FM3B.3},
doi = {10.1364/FREEFORM.2015.FM3B.3},
}

@inproceedings{Kaiser08,
author = {Stefan Kaiser and Bernhard Sang and Josef Schubert and Stefan Hofer and Timo Stuffler},
title = {{Compact prism spectrometer of pushbroom type for hyperspectral imaging}},
volume = {7100},
booktitle = {Optical Design and Engineering III},
editor = {Laurent Mazuray and Rolf Wartmann and Andrew Wood and Jean-Luc Tissot and Jeffrey M. Raynor},
organization = {International Society for Optics and Photonics},
publisher = {SPIE},
pages = {710014},
year = {2008},
doi = {10.1117/12.797177},
URL = {https://doi.org/10.1117/12.797177}
}

@inproceedings{Wei16,
author = {Lidong Wei and Lei Feng and Jinsong Zhou and Juanjuan Jing and Yacan Li},
title = {{Optical design of Offner-Chrisp imaging spectrometer with freeform surfaces}},
volume = {10021},
booktitle = {Optical Design and Testing VII},
editor = {Yongtian Wang and Tina E. Kidger and Kimio Tatsuno},
organization = {International Society for Optics and Photonics},
publisher = {SPIE},
pages = {100211P},
year = {2016},
doi = {10.1117/12.2246355},
URL = {https://doi.org/10.1117/12.2246355}
}

@ARTICLE{Wang19,
       author = {{Wang}, Yiran and {Bai}, Xianyong and {Liu}, Siqing and {Deng}, Yuanyong and {Zhang}, Zhiyong and {Sun}, Yueqiang},
        title = "{Flat-fielding of Full-disk Solar Images with a Gaussian-type Diffuser}",
      journal = {\solphys},
         year = 2019,
        month = sep,
       volume = {294},
       number = {9},
          eid = {127},
        pages = {127},
          doi = {10.1007/s11207-019-1527-2},
       adsurl = {https://ui.adsabs.harvard.edu/abs/2019SoPh..294..127W}
}

@ARTICLE{Denker99,
       author = {{Denker}, C. and {Johannesson}, A. and {Marquette}, W. and {Goode}, P.~R. and {Wang}, H. and {Zirin}, H.},
        title = "{Synoptic H{\ensuremath{\alpha}} Full-Disk Observations of the Sun from Big Bear Solar Observatory - I. Instrumentation, Image Processing, Data Products, and First Results}",
      journal = {\solphys},
         year = 1999,
        month = jan,
       volume = {184},
       number = {1},
        pages = {87-102},
          doi = {10.1023/A:1005047906097},
       adsurl = {https://ui.adsabs.harvard.edu/abs/1999SoPh..184...87D}
}

@ARTICLE{Song22,
       author = {{Song}, Qiao and {Bai}, Xianyong and {Chen}, Bo and {Hu}, Xiuqing and {Chen}, Yajie and {Hou}, Zhenyong and {Zhang}, Xiaofan and {He}, Lingping and {Song}, Kefei and {Zhang}, Peng and {Wang}, Jing-Song and {Zhang}, Xiaoxin and {Zong}, Weiguo and {Dun}, Jinping and {Tian}, Hui and {Deng}, Yuanyong},
        title = "{A New Post-hoc Flat Field Measurement Method for the Solar X-Ray and Extreme Ultraviolet Imager Onboard the FengYun-3E Satellite}",
      journal = {Research in Astronomy and Astrophysics},
         year = 2022,
        month = oct,
       volume = {22},
       number = {10},
          eid = {105001},
        pages = {105001},
          doi = {10.1088/1674-4527/ac7f0e},
archivePrefix = {arXiv},
       eprint = {2207.01829},
 primaryClass = {astro-ph.SR},
       adsurl = {https://ui.adsabs.harvard.edu/abs/2022RAA....22j5001S}
}

@ARTICLE{Kotrc16,
       author = {{Kotr{\v{c}}}, P. and {Proch{\'a}zka}, O. and {Heinzel}, P.},
        title = "{New Observations of Balmer Continuum Flux in Solar Flares. Instrument Description and First Results}",
      journal = {\solphys},
         year = 2016,
        month = mar,
       volume = {291},
       number = {3},
        pages = {779-789},
          doi = {10.1007/s11207-016-0860-y},
archivePrefix = {arXiv},
       eprint = {1601.04610},
 primaryClass = {astro-ph.SR},
       adsurl = {https://ui.adsabs.harvard.edu/abs/2016SoPh..291..779K}
}

@INPROCEEDINGS{Keller03,
       author = {{Keller}, Christoph U. and {Harvey}, John W. and {Giampapa}, Mark S.},
        title = "{SOLIS: an innovative suite of synoptic instruments}",
    booktitle = {Innovative Telescopes and Instrumentation for Solar Astrophysics},
         year = 2003,
       editor = {{Keil}, Stephen L. and {Avakyan}, Sergey V.},
       series = {Society of Photo-Optical Instrumentation Engineers (SPIE) Conference Series},
       volume = {4853},
        month = feb,
        pages = {194-204},
          doi = {10.1117/12.460373},
       adsurl = {https://ui.adsabs.harvard.edu/abs/2003SPIE.4853..194K}
}

@ARTICLE{Judge04,
       author = {{Judge}, Philip G. and {Elmore}, David F. and {Lites}, Bruce W. and {Keller}, Christoph U. and {Rimmele}, Thomas},
        title = "{Evaluation of Seeing-Induced Cross Talk in Tip-Tilt-Corrected Solar Polarimetry}",
      journal = {\ao},
         year = 2004,
        month = jul,
       volume = {43},
       number = {19},
        pages = {3817-3828},
          doi = {10.1364/AO.43.003817},
       adsurl = {https://ui.adsabs.harvard.edu/abs/2004ApOpt..43.3817J}
}

@ARTICLE{MartinezPillet92,
       author = {{Martinez Pillet}, V.},
        title = "{Stray Light Effects on the Solar Intensity Distribution}",
      journal = {\solphys},
         year = 1992,
        month = aug,
       volume = {140},
       number = {2},
        pages = {207-237},
          doi = {10.1007/BF00146310},
       adsurl = {https://ui.adsabs.harvard.edu/abs/1992SoPh..140..207M}
}

@ARTICLE{Rubio03,
       author = {{Bellot Rubio}, L.~R. and {Collados}, M.},
        title = "{Understanding internetwork magnetic fields as determined  from visible and infrared spectral lines}",
      journal = {\aap},
         year = 2003,
        month = jul,
       volume = {406},
        pages = {357-362},
          doi = {10.1051/0004-6361:20030704},
       adsurl = {https://ui.adsabs.harvard.edu/abs/2003A&A...406..357B}
}

@ARTICLE{Longcope09,
       author = {{Longcope}, D.~W. and {Parnell}, C.~E.},
        title = "{The Number of Magnetic Null Points in the Quiet Sun Corona}",
      journal = {\solphys},
         year = 2009,
        month = jan,
       volume = {254},
       number = {1},
        pages = {51-75},
          doi = {10.1007/s11207-008-9281-x},
archivePrefix = {arXiv},
       eprint = {0811.0097},
 primaryClass = {astro-ph},
       adsurl = {https://ui.adsabs.harvard.edu/abs/2009SoPh..254...51L}
}

@ARTICLE{Lagg04,
       author = {{Lagg}, A. and {Woch}, J. and {Krupp}, N. and {Solanki}, S.~K.},
        title = "{Retrieval of the full magnetic vector with the He I multiplet at 1083 nm. Maps of an emerging flux region}",
      journal = {\aap},
         year = 2004,
        month = feb,
       volume = {414},
        pages = {1109-1120},
          doi = {10.1051/0004-6361:20031643},
       adsurl = {https://ui.adsabs.harvard.edu/abs/2004A&A...414.1109L}
}

@INPROCEEDINGS{Lagg09,
       author = {{Lagg}, A. and {Ishikawa}, R. and {Merenda}, L. and {Wiegelmann}, T. and {Tsuneta}, S. and {Solanki}, S.~K.},
        title = "{Internetwork Horizontal Magnetic Fields in the Quiet Sun Chromosphere: Results from a Joint Hinode/VTT Study}",
    booktitle = {The Second Hinode Science Meeting: Beyond Discovery-Toward Understanding},
         year = 2009,
       editor = {{Lites}, B. and {Cheung}, M. and {Magara}, T. and {Mariska}, J. and {Reeves}, K.},
       series = {Astronomical Society of the Pacific Conference Series},
       volume = {415},
        month = dec,
        pages = {327},
       adsurl = {https://ui.adsabs.harvard.edu/abs/2009ASPC..415..327L}
}

@ARTICLE{Tyson86,
       author = {{Tyson}, J. Anthony},
        title = "{Low-light-level charge-coupled device imaging in astronomy.}",
      journal = {Journal of the Optical Society of America A},
         year = 1986,
        month = dec,
       volume = {3},
        pages = {2131-2138},
          doi = {10.1364/JOSAA.3.002131},
       adsurl = {https://ui.adsabs.harvard.edu/abs/1986JOSAA...3.2131T}
}

@ARTICLE{KLL91,
       author = {{Kuhn}, J.~R. and {Lin}, H. and {Loranz}, D.},
        title = "{Gain calibrating NonUniform Image-Array Data Using Only the Image Data}",
      journal = {\pasp},
         year = 1991,
        month = oct,
       volume = {103},
        pages = {1097},
          doi = {10.1086/132932},
       adsurl = {https://ui.adsabs.harvard.edu/abs/1991PASP..103.1097K}
}

@ARTICLE{Riethmueller19,
       author = {{Riethm{\"u}ller}, T.~L. and {Solanki}, S.~K.},
        title = "{The potential of many-line inversions of photospheric spectropolarimetric data in the visible and near UV}",
      journal = {\aap},
         year = 2019,
        month = feb,
       volume = {622},
          eid = {A36},
        pages = {A36},
          doi = {10.1051/0004-6361/201833379},
archivePrefix = {arXiv},
       eprint = {1812.03757},
 primaryClass = {astro-ph.IM},
       adsurl = {https://ui.adsabs.harvard.edu/abs/2019A&A...622A..36R}
}

@ARTICLE{Iglesias16,
       author = {{Iglesias}, F.~A. and {Feller}, A. and {Nagaraju}, K. and {Solanki}, S.~K.},
        title = "{High-resolution, high-sensitivity, ground-based solar spectropolarimetry with a new fast imaging polarimeter. I. Prototype characterization}",
      journal = {\aap},
         year = 2016,
        month = may,
       volume = {590},
          eid = {A89},
        pages = {A89},
          doi = {10.1051/0004-6361/201628376},
archivePrefix = {arXiv},
       eprint = {1604.01521},
 primaryClass = {astro-ph.IM},
       adsurl = {https://ui.adsabs.harvard.edu/abs/2016A&A...590A..89I}
}

@ARTICLE{Kiselman11,
       author = {{Kiselman}, D. and {Pereira}, T.~M.~D. and {Gustafsson}, B. and {Asplund}, M. and {Mel{\'e}ndez}, J. and {Langhans}, K.},
        title = "{Is the solar spectrum latitude-dependent?. An investigation with SST/TRIPPEL}",
      journal = {\aap},
         year = 2011,
        month = nov,
       volume = {535},
          eid = {A14},
        pages = {A14},
          doi = {10.1051/0004-6361/201117553},
archivePrefix = {arXiv},
       eprint = {1108.4527},
 primaryClass = {astro-ph.SR},
       adsurl = {https://ui.adsabs.harvard.edu/abs/2011A&A...535A..14K}
}

@ARTICLE{Collados12,
       author = {{Collados}, M. and {L{\'o}pez}, R. and {P{\'a}ez}, E. and {Hern{\'a}ndez}, E. and {Reyes}, M. and {Calcines}, A. and {Ballesteros}, E. and {D{\'\i}az}, J.~J. and {Denker}, C. and {Lagg}, A. and {Schlichenmaier}, R. and {Schmidt}, W. and {Solanki}, S.~K. and {Strassmeier}, K.~G. and {von der L{\"u}he}, O. and {Volkmer}, R.},
        title = "{GRIS: The GREGOR Infrared Spectrograph}",
      journal = {Astronomische Nachrichten},
         year = 2012,
        month = nov,
       volume = {333},
       number = {9},
        pages = {872},
          doi = {10.1002/asna.201211738},
       adsurl = {https://ui.adsabs.harvard.edu/abs/2012AN....333..872C}
}

@ARTICLE{Solanki03A,
       author = {{Solanki}, Sami K.},
        title = "{Sunspots: An overview}",
      journal = {\aapr},
         year = 2003,
        month = jan,
       volume = {11},
       number = {2-3},
        pages = {153-286},
          doi = {10.1007/s00159-003-0018-4},
       adsurl = {https://ui.adsabs.harvard.edu/abs/2003A&ARv..11..153S}
}

@ARTICLE{Zhang18,
       author = {{Zhang}, Jingwen and {Tian}, Hui and {Solanki}, Sami K. and {Wang}, Haimin and {Peter}, Hardi and {Ahn}, Kwangsu and {Xu}, Yan and {Zhu}, Yingjie and {Cao}, Wenda and {He}, Jiansen and {Wang}, Linghua},
        title = "{Dark Structures in Sunspot Light Bridges}",
      journal = {\apj},
         year = 2018,
        month = sep,
       volume = {865},
       number = {1},
          eid = {29},
        pages = {29},
          doi = {10.3847/1538-4357/aada0a},
archivePrefix = {arXiv},
       eprint = {1809.00146},
 primaryClass = {astro-ph.SR},
       adsurl = {https://ui.adsabs.harvard.edu/abs/2018ApJ...865...29Z}
}

@ARTICLE{Berghmans21,
       author = {{Berghmans}, D. and {Auch{\`e}re}, F. and {Long}, D.~M. and {Soubri{\'e}}, E. and {Mierla}, M. and {Zhukov}, A.~N. and {Sch{\"u}hle}, U. and {Antolin}, P. and {Harra}, L. and {Parenti}, S. and {Podladchikova}, O. and {Aznar Cuadrado}, R. and {Buchlin}, {\'E}. and {Dolla}, L. and {Verbeeck}, C. and {Gissot}, S. and {Teriaca}, L. and {Haberreiter}, M. and {Katsiyannis}, A.~C. and {Rodriguez}, L. and {Kraaikamp}, E. and {Smith}, P.~J. and {Stegen}, K. and {Rochus}, P. and {Halain}, J.~P. and {Jacques}, L. and {Thompson}, W.~T. and {Inhester}, B.},
        title = "{Extreme-UV quiet Sun brightenings observed by the Solar Orbiter/EUI}",
      journal = {\aap},
         year = 2021,
        month = dec,
       volume = {656},
          eid = {L4},
        pages = {L4},
          doi = {10.1051/0004-6361/202140380},
archivePrefix = {arXiv},
       eprint = {2104.03382},
 primaryClass = {astro-ph.SR},
       adsurl = {https://ui.adsabs.harvard.edu/abs/2021A&A...656L...4B}
}

@article{DELTOROINIESTA96,
title = {On the discovery of the zeeman effect on the sun and in the laboratory},
journal = {Vistas in Astronomy},
volume = {40},
number = {2},
pages = {241-256},
year = {1996},
issn = {0083-6656},
doi = {https://doi.org/10.1016/0083-6656(96)00005-0},
url = {https://www.sciencedirect.com/science/article/pii/0083665696000050},
author = {Jose Carlos {del Toro Iniesta}}
}

@ARTICLE{sunpy_community2020,
  doi = {10.3847/1538-4357/ab4f7a},
  url = {https://iopscience.iop.org/article/10.3847/1538-4357/ab4f7a},
  author = {{The SunPy Community} and Barnes, Will T. and Bobra, Monica G. and Christe, Steven D. and Freij, Nabil and Hayes, Laura A. and Ireland, Jack and Mumford, Stuart and Perez-Suarez, David and Ryan, Daniel F. and Shih, Albert Y. and Chanda, Prateek and Glogowski, Kolja and Hewett, Russell and Hughitt, V. Keith and Hill, Andrew and Hiware, Kaustubh and Inglis, Andrew and Kirk, Michael S. F. and Konge, Sudarshan and Mason, James Paul and Maloney, Shane Anthony and Murray, Sophie A. and Panda, Asish and Park, Jongyeob and Pereira, Tiago M. D. and Reardon, Kevin and Savage, Sabrina and Sipőcz, Brigitta M. and Stansby, David and Jain, Yash and Taylor, Garrison and Yadav, Tannmay and Rajul and Dang, Trung Kien},
  title = {The SunPy Project: Open Source Development and Status of the Version 1.0 Core Package},
  journal = {The Astrophysical Journal},
  volume = {890},
  issue = {1},
  pages = {68-},
  publisher = {American Astronomical Society},
  year = {2020}
}

@ARTICLE{Solanki10,
       author = {{Solanki}, S.~K. and {Barthol}, P. and {Danilovic}, S. and {Feller}, A. and {Gandorfer}, A. and {Hirzberger}, J. and {Riethm{\"u}ller}, T.~L. and {Sch{\"u}ssler}, M. and {Bonet}, J.~A. and {Mart{\'\i}nez Pillet}, V. and {del Toro Iniesta}, J.~C. and {Domingo}, V. and {Palacios}, J. and {Kn{\"o}lker}, M. and {Bello Gonz{\'a}lez}, N. and {Berkefeld}, T. and {Franz}, M. and {Schmidt}, W. and {Title}, A.~M.},
        title = "{SUNRISE: Instrument, Mission, Data, and First Results}",
      journal = {\apjl},
         year = 2010,
        month = nov,
       volume = {723},
       number = {2},
        pages = {L127-L133},
          doi = {10.1088/2041-8205/723/2/L127},
archivePrefix = {arXiv},
       eprint = {1008.3460},
 primaryClass = {astro-ph.SR},
       adsurl = {https://ui.adsabs.harvard.edu/abs/2010ApJ...723L.127S}
}

@ARTICLE{Solanki17,
       author = {{Solanki}, S.~K. and {Riethm{\"u}ller}, T.~L. and {Barthol}, P. and {Danilovic}, S. and {Deutsch}, W. and {Doerr}, H. -P. and {Feller}, A. and {Gandorfer}, A. and {Germerott}, D. and {Gizon}, L. and {Grauf}, B. and {Heerlein}, K. and {Hirzberger}, J. and {Kolleck}, M. and {Lagg}, A. and {Meller}, R. and {Tomasch}, G. and {van Noort}, M. and {Blanco Rodr{\'\i}guez}, J. and {Gasent Blesa}, J.~L. and {Balaguer Jim{\'e}nez}, M. and {Del Toro Iniesta}, J.~C. and {L{\'o}pez Jim{\'e}nez}, A.~C. and {Orozco Suarez}, D. and {Berkefeld}, T. and {Halbgewachs}, C. and {Schmidt}, W. and {{\'A}lvarez-Herrero}, A. and {Sabau-Graziati}, L. and {P{\'e}rez Grande}, I. and {Mart{\'\i}nez Pillet}, V. and {Card}, G. and {Centeno}, R. and {Kn{\"o}lker}, M. and {Lecinski}, A.},
        title = "{The Second Flight of the Sunrise Balloon-borne Solar Observatory: Overview of Instrument Updates, the Flight, the Data, and First Results}",
      journal = {\apjs},
         year = 2017,
        month = mar,
       volume = {229},
       number = {1},
          eid = {2},
        pages = {2},
          doi = {10.3847/1538-4365/229/1/2},
archivePrefix = {arXiv},
       eprint = {1701.01555},
 primaryClass = {astro-ph.SR},
       adsurl = {https://ui.adsabs.harvard.edu/abs/2017ApJS..229....2S}
}

@INPROCEEDINGS{cao2010b,
       author = {{Cao}, Wenda and {Gorceix}, Nicolas and {Coulter}, Roy and {Coulter}, Aaron and {Goode}, Philip R.},
        title = "{First light of the 1.6 meter off-axis New Solar Telescope at Big Bear Solar Observatory}",
    booktitle = {Ground-based and Airborne Telescopes III},
         year = 2010,
       editor = {{Stepp}, Larry M. and {Gilmozzi}, Roberto and {Hall}, Helen J.},
       series = {Society of Photo-Optical Instrumentation Engineers (SPIE) Conference Series},
       volume = {7733},
        month = jul,
          eid = {773330},
        pages = {773330},
          doi = {10.1117/12.856611},
       adsurl = {https://ui.adsabs.harvard.edu/abs/2010SPIE.7733E..30C}
}

@ARTICLE{Cao2010a,
       author = {{Cao}, W. and {Gorceix}, N. and {Coulter}, R. and {Ahn}, K. and {Rimmele}, T.~R. and {Goode}, P.~R.},
        title = "{Scientific instrumentation for the 1.6 m New Solar Telescope in Big Bear}",
      journal = {Astronomische Nachrichten},
         year = 2010,
        month = jun,
       volume = {331},
       number = {6},
        pages = {636},
          doi = {10.1002/asna.201011390},
       adsurl = {https://ui.adsabs.harvard.edu/abs/2010AN....331..636C}
}

@INPROCEEDINGS{Cao11,
       author = {{Cao}, W. and {Ahn}, K. and {Goode}, P.~R. and {Shumko}, S. and {Gorceix}, N. and {Coulter}, R.},
        title = "{The New Solar Telescope in Big Bear: Polarimetry II}",
    booktitle = {Solar Polarization 6},
         year = 2011,
       editor = {{Kuhn}, J.~R. and {Harrington}, D.~M. and {Lin}, H. and {Berdyugina}, S.~V. and {Trujillo-Bueno}, J. and {Keil}, S.~L. and {Rimmele}, T.},
       series = {Astronomical Society of the Pacific Conference Series},
       volume = {437},
        month = apr,
        pages = {345},
       adsurl = {https://ui.adsabs.harvard.edu/abs/2011ASPC..437..345C}
}

@INPROCEEDINGS{Feller14,
       author = {{Feller}, A. and {Iglesias}, F.~A. and {Nagaraju}, K. and {Solanki}, S.~K. and {Ihle}, S.},
        title = "{Fast Solar Polarimeter: Description and First Results}",
    booktitle = {Solar Polarization 7},
         year = 2014,
       editor = {{Nagendra}, K.~N. and {Stenflo}, J.~O. and {Qu}, Z.~Q. and {Sampoorna}, M.},
       series = {Astronomical Society of the Pacific Conference Series},
       volume = {489},
        month = oct,
        pages = {271},
       adsurl = {https://ui.adsabs.harvard.edu/abs/2014ASPC..489..271F}
}

@ARTICLE{Goode10,
       author = {{Goode}, P.~R. and {Coulter}, R. and {Gorceix}, N. and {Yurchyshyn}, V. and {Cao}, W.},
        title = "{The NST: First results and some lessons for ATST and EST}",
      journal = {Astronomische Nachrichten},
         year = 2010,
        month = jun,
       volume = {331},
       number = {6},
        pages = {620},
          doi = {10.1002/asna.201011387},
       adsurl = {https://ui.adsabs.harvard.edu/abs/2010AN....331..620G}
}

@INPROCEEDINGS{Goode12,
       author = {{Goode}, Philip R. and {Cao}, Wenda},
        title = "{The 1.6 m off-axis New Solar Telescope (NST) in Big Bear}",
    booktitle = {Ground-based and Airborne Telescopes IV},
         year = 2012,
       editor = {{Stepp}, Larry M. and {Gilmozzi}, Roberto and {Hall}, Helen J.},
       series = {Society of Photo-Optical Instrumentation Engineers (SPIE) Conference Series},
       volume = {8444},
        month = sep,
          eid = {844403},
        pages = {844403},
          doi = {10.1117/12.925494},
       adsurl = {https://ui.adsabs.harvard.edu/abs/2012SPIE.8444E..03G}
}

@ARTICLE{Lites04,
       author = {{Lites}, B.~W. and {Scharmer}, G.~B. and {Berger}, T.~E. and {Title}, A.~M.},
        title = "{Three-Dimensional Structure of the Active Region Photosphere as Revealed by High Angular Resolution}",
      journal = {\solphys},
         year = 2004,
        month = may,
       volume = {221},
       number = {1},
        pages = {65-84},
          doi = {10.1023/B:SOLA.0000033355.24845.5a},
       adsurl = {https://ui.adsabs.harvard.edu/abs/2004SoPh..221...65L}
}

@ARTICLE{Rempel14,
       author = {{Rempel}, M.},
        title = "{Numerical Simulations of Quiet Sun Magnetism: On the Contribution from a Small-scale Dynamo}",
      journal = {\apj},
         year = 2014,
        month = jul,
       volume = {789},
       number = {2},
          eid = {132},
        pages = {132},
          doi = {10.1088/0004-637X/789/2/132},
archivePrefix = {arXiv},
       eprint = {1405.6814},
 primaryClass = {astro-ph.SR},
       adsurl = {https://ui.adsabs.harvard.edu/abs/2014ApJ...789..132R}
}

@article{Iglesias19,
  author        = {{Iglesias}, Francisco A. and {Feller}, Alex},
  title         = {{Instrumentation for solar spectropolarimetry: state of the art and prospects}},
  journal       = {Optical Engineering},
  year          = 2019,
  month         = aug,
  volume        = {58},
  eid           = {082417},
  pages         = {082417},
  doi           = {10.1117/1.OE.58.8.082417},
  archiveprefix = {arXiv},
  eprint        = {1911.01368},
  primaryclass  = {astro-ph.IM},
  adsurl        = {https://ui.adsabs.harvard.edu/abs/2019OptEn..58h2417I}
}

@ARTICLE{Donati97,
       author = {{Donati}, J. -F. and {Semel}, M. and {Carter}, B.~D. and {Rees}, D.~E. and {Collier Cameron}, A.},
        title = "{Spectropolarimetric observations of active stars}",
      journal = {\mnras},
         year = 1997,
        month = nov,
       volume = {291},
       number = {4},
        pages = {658-682},
          doi = {10.1093/mnras/291.4.658},
       adsurl = {https://ui.adsabs.harvard.edu/abs/1997MNRAS.291..658D}
}

@ARTICLE{Stenflo83,
       author = {{Stenflo}, J.~O. and {Twerenbold}, D. and {Harvey}, J.~W.},
        title = "{Coherent scattering in the solar spectrum - Survey of linear polarization in the range 3165-4230 A}",
      journal = {\aaps},
         year = 1983,
        month = apr,
       volume = {52},
        pages = {161-180},
       adsurl = {https://ui.adsabs.harvard.edu/abs/1983A&AS...52..161S}
}

@ARTICLE{Gough66,
       author = {{Gough}, D.~O. and {Tayler}, R.~J.},
        title = "{The influence of a magnetic field on Schwarzschild's criterion for convective instability in an ideally conducting fluid}",
      journal = {\mnras},
         year = 1966,
        month = jan,
       volume = {133},
        pages = {85},
          doi = {10.1093/mnras/133.1.85},
       adsurl = {https://ui.adsabs.harvard.edu/abs/1966MNRAS.133...85G}
}

@ARTICLE{Solanki84,
       author = {{Solanki}, S.~K. and {Stenflo}, J.~O.},
        title = "{Properties of solar magnetic fluxtubes as revealed by Fe I lines}",
      journal = {\aap},
         year = 1984,
        month = nov,
       volume = {140},
       number = {1},
        pages = {185-198},
       adsurl = {https://ui.adsabs.harvard.edu/abs/1984A&A...140..185S}
}

@ARTICLE{Keller92,
       author = {{Keller}, C.~U. and {von der Luehe}, O.},
        title = "{Solar speckle polarimetry}",
      journal = {\aap},
         year = 1992,
        month = jul,
       volume = {261},
       number = {1},
        pages = {321-328},
       adsurl = {https://ui.adsabs.harvard.edu/abs/1992A&A...261..321K}
}

@ARTICLE{Keller95,
       author = {{Keller}, C.~U. and {Johannesson}, A.},
        title = "{Speckle spectrography of extended objects.}",
      journal = {\aaps},
         year = 1995,
        month = may,
       volume = {110},
        pages = {565},
       adsurl = {https://ui.adsabs.harvard.edu/abs/1995A&AS..110..565K}
}

@ARTICLE{Noort22,
       author = {{van Noort}, M. and {Doerr}, H. -P.},
        title = "{Data reduction and restoration of spectropolarimetric microlensed hyperspectral imager data}",
      journal = {\aap},
         year = 2022,
        month = dec,
       volume = {668},
          eid = {A151},
        pages = {A151},
          doi = {10.1051/0004-6361/202243466},
       adsurl = {https://ui.adsabs.harvard.edu/abs/2022A&A...668A.151V}
}

@BOOK{IntroductiontoSpectropolarimetry,
       author = {{del Toro Iniesta}, Jose Carlos},
        title = "{Introduction to Spectropolarimetry}",
         year = 2007,
       adsurl = {https://ui.adsabs.harvard.edu/abs/2007insp.book.....D},
      publisher={Cambridge University Press}
}

@article{Goncalves02,  
    title={Fundamentals and applications of spectroscopic ellipsometry},  
    volume={25},  
    ISSN={0100-4042},  
    url={https://doi.org/10.1590/S0100-40422002000500015},  
    DOI={10.1590/S0100-40422002000500015}, 
    number={5},  
    journal={Química Nova},  
    publisher={Sociedade Brasileira de Química},    
    author={Gonçalves, Débora and Irene, Eugene A.},  
    year={2002},  
    month={Sep}, 
    pages={794–800} 
}

@ARTICLE{Rimmele20,
       author = {{Rimmele}, Thomas R. and {Warner}, Mark and {Keil}, Stephen L. and {Goode}, Philip R. and {Kn{\"o}lker}, Michael and {Kuhn}, Jeffrey R. and {Rosner}, Robert R. and {McMullin}, Joseph P. and {Casini}, Roberto and {Lin}, Haosheng and {W{\"o}ger}, Friedrich and {von der L{\"u}he}, Oskar and {Tritschler}, Alexandra and {Davey}, Alisdair and {de Wijn}, Alfred and {Elmore}, David F. and {Fehlmann}, Andr{\'e} and {Harrington}, David M. and {Jaeggli}, Sarah A. and {Rast}, Mark P. and {Schad}, Thomas A. and {Schmidt}, Wolfgang and {Mathioudakis}, Mihalis and {Mickey}, Donald L. and {Anan}, Tetsu and {Beck}, Christian and {Marshall}, Heather K. and {Jeffers}, Paul F. and {Oschmann}, Jacobus M. and {Beard}, Andrew and {Berst}, David C. and {Cowan}, Bruce A. and {Craig}, Simon C. and {Cross}, Eric and {Cummings}, Bryan K. and {Donnelly}, Colleen and {de Vanssay}, Jean-Benoit and {Eigenbrot}, Arthur D. and {Ferayorni}, Andrew and {Foster}, Christopher and {Galapon}, Chriselle Ann and {Gedrites}, Christopher and {Gonzales}, Kerry and {Goodrich}, Bret D. and {Gregory}, Brian S. and {Guzman}, Stephanie S. and {Guzzo}, Stephen and {Hegwer}, Steve and {Hubbard}, Robert P. and {Hubbard}, John R. and {Johansson}, Erik M. and {Johnson}, Luke C. and {Liang}, Chen and {Liang}, Mary and {McQuillen}, Isaac and {Mayer}, Christopher and {Newman}, Karl and {Onodera}, Brialyn and {Phelps}, LeEllen and {Puentes}, Myles M. and {Richards}, Christopher and {Rimmele}, Lukas M. and {Sekulic}, Predrag and {Shimko}, Stephan R. and {Simison}, Brett E. and {Smith}, Brett and {Starman}, Erik and {Sueoka}, Stacey R. and {Summers}, Richard T. and {Szabo}, Aimee and {Szabo}, Louis and {Wampler}, Stephen B. and {Williams}, Timothy R. and {White}, Charles},
        title = "{The Daniel K. Inouye Solar Telescope - Observatory Overview}",
      journal = {\solphys},
         year = 2020,
        month = dec,
       volume = {295},
       number = {12},
          eid = {172},
        pages = {172},
          doi = {10.1007/s11207-020-01736-7},
       adsurl = {https://ui.adsabs.harvard.edu/abs/2020SoPh..295..172R}
}

@INPROCEEDINGS{Rubio06,
       author = {{Bellot Rubio}, L.~R.},
        title = "{Stokes Inversion Techniques: Recent Advances and New Challenges}",
    booktitle = {Solar Polarization 4},
         year = 2006,
       editor = {{Casini}, R. and {Lites}, B.~W.},
       series = {Astronomical Society of the Pacific Conference Series},
       volume = {358},
        month = dec,
        pages = {107},
          doi = {10.48550/arXiv.astro-ph/0601483},
archivePrefix = {arXiv},
       eprint = {astro-ph/0601483},
 primaryClass = {astro-ph},
       adsurl = {https://ui.adsabs.harvard.edu/abs/2006ASPC..358..107B}
}

@ARTICLE{Suetterlin00,
       author = {{S{\"u}tterlin}, P. and {Wiehr}, E.},
        title = "{Applying speckle masking to spectra}",
      journal = {\solphys},
         year = 2000,
        month = may,
       volume = {194},
       number = {1},
        pages = {35-42},
          doi = {10.1023/A:1005273130450},
       adsurl = {https://ui.adsabs.harvard.edu/abs/2000SoPh..194...35S}
}

@article{Shenstone29,
author = {A.G. Shenstone and H.A. Blair},
title = {LXXXIII. A method for the magnetic analysis of a spectrum by means of unresolved Zeeman patterns, and its application to Ag II },
journal = {The London, Edinburgh, and Dublin Philosophical Magazine and Journal of Science},
volume = {8},
number = {52},
pages = {765--771},
year = {1929},
publisher = {Taylor \& Francis},
doi = {10.1080/14786441108564938},
URL = {https://doi.org/10.1080/14786441108564938}
}

@phdthesis{laverick2019phd,
  author    = {Mike Laverick},
  title     = {Fundamental atomic data deduced using stellar spectra},
  school    = {Faculty of Science, KU Leuven},
  year      = {2019},
  address   = {Leuven, Belgium},
  url       = {https://fys.kuleuven.be/ster/pub/phd-mike-laverick/mike-laverick-thesis-final.pdf}
}

@ARTICLE{Balthasar84,
       author = {{Balthasar}, H.},
        title = "{Asymmetries and Wavelengths of Solar Spectral Lines and the Solar Rotation Determined from Fourier-Transform Spectra}",
      journal = {\solphys},
         year = 1984,
        month = jul,
       volume = {93},
       number = {2},
        pages = {219-241},
          doi = {10.1007/BF02270836},
       adsurl = {https://ui.adsabs.harvard.edu/abs/1984SoPh...93..219B}
}

@ARTICLE{Solanki85,
       author = {{Solanki}, S.~K. and {Stenflo}, J.~O.},
        title = "{Models of solar magnetic fluxtubes - Constraints imposed by Fe I and II lines}",
      journal = {\aap},
         year = 1985,
        month = jul,
       volume = {148},
       number = {1},
        pages = {123-132},
       adsurl = {https://ui.adsabs.harvard.edu/abs/1985A&A...148..123S}
}

@ARTICLE{ToroIniesta16,
       author = {{del Toro Iniesta}, Jose Carlos and {Ruiz Cobo}, Basilio},
        title = "{Inversion of the radiative transfer equation for polarized light}",
      journal = {Living Reviews in Solar Physics},
         year = 2016,
        month = nov,
       volume = {13},
       number = {1},
          eid = {4},
        pages = {4},
          doi = {10.1007/s41116-016-0005-2},
archivePrefix = {arXiv},
       eprint = {1610.10039},
 primaryClass = {astro-ph.SR},
       adsurl = {https://ui.adsabs.harvard.edu/abs/2016LRSP...13....4D}
}

@ARTICLE{deWijn20,
       author = {{De Wijn}, Alfred and {K{\"o}nnen}, Gunther},
        title = "{Venus Transitis the Solar Corona}",
      journal = {\skytel},
         year = 2020,
        month = oct,
       volume = {140},
       number = {4},
        pages = {11},
       adsurl = {https://ui.adsabs.harvard.edu/abs/2020S&T...140d..11D}
}

@ARTICLE{2008A&A...489..429V,
       author = {{van Noort}, M.~J. and {Rouppe van der Voort}, L.~H.~M.},
        title = "{Stokes imaging polarimetry using image restoration at the Swedish 1-m solar telescope}",
      journal = {\aap},
         year = 2008,
        month = oct,
       volume = {489},
       number = {1},
        pages = {429-440},
          doi = {10.1051/0004-6361:200809959},
archivePrefix = {arXiv},
       eprint = {0805.4296},
 primaryClass = {astro-ph},
       adsurl = {https://ui.adsabs.harvard.edu/abs/2008A&A...489..429V}
}

@INPROCEEDINGS{Lites01,
       author = {{Lites}, B.~W. and {Elmore}, D.~F. and {Streander}, K.~V.},
        title = "{The Solar-B Spectro-Polarimeter}",
    booktitle = {Advanced Solar Polarimetry -- Theory, Observation, and Instrumentation},
         year = 2001,
       editor = {{Sigwarth}, Michael},
       series = {Astronomical Society of the Pacific Conference Series},
       volume = {236},
        month = jan,
        pages = {33},
       adsurl = {https://ui.adsabs.harvard.edu/abs/2001ASPC..236...33L}
}

@ARTICLE{Kosugi07,
       author = {{Kosugi}, T. and {Matsuzaki}, K. and {Sakao}, T. and {Shimizu}, T. and {Sone}, Y. and {Tachikawa}, S. and {Hashimoto}, T. and {Minesugi}, K. and {Ohnishi}, A. and {Yamada}, T. and {Tsuneta}, S. and {Hara}, H. and {Ichimoto}, K. and {Suematsu}, Y. and {Shimojo}, M. and {Watanabe}, T. and {Shimada}, S. and {Davis}, J.~M. and {Hill}, L.~D. and {Owens}, J.~K. and {Title}, A.~M. and {Culhane}, J.~L. and {Harra}, L.~K. and {Doschek}, G.~A. and {Golub}, L.},
        title = "{The Hinode (Solar-B) Mission: An Overview}",
      journal = {\solphys},
         year = 2007,
        month = jun,
       volume = {243},
       number = {1},
        pages = {3-17},
          doi = {10.1007/s11207-007-9014-6},
       adsurl = {https://ui.adsabs.harvard.edu/abs/2007SoPh..243....3K}
}

@ARTICLE{Tsuneta08,
       author = {{Tsuneta}, S. and {Ichimoto}, K. and {Katsukawa}, Y. and {Nagata}, S. and {Otsubo}, M. and {Shimizu}, T. and {Suematsu}, Y. and {Nakagiri}, M. and {Noguchi}, M. and {Tarbell}, T. and {Title}, A. and {Shine}, R. and {Rosenberg}, W. and {Hoffmann}, C. and {Jurcevich}, B. and {Kushner}, G. and {Levay}, M. and {Lites}, B. and {Elmore}, D. and {Matsushita}, T. and {Kawaguchi}, N. and {Saito}, H. and {Mikami}, I. and {Hill}, L.~D. and {Owens}, J.~K.},
        title = "{The Solar Optical Telescope for the Hinode Mission: An Overview}",
      journal = {\solphys},
         year = 2008,
        month = jun,
       volume = {249},
       number = {2},
        pages = {167-196},
          doi = {10.1007/s11207-008-9174-z},
archivePrefix = {arXiv},
       eprint = {0711.1715},
 primaryClass = {astro-ph},
       adsurl = {https://ui.adsabs.harvard.edu/abs/2008SoPh..249..167T}
}

@ARTICLE{Scharmer02b,
       author = {{Scharmer}, G{\"o}ran B. and {Gudiksen}, Boris V. and {Kiselman}, Dan and {L{\"o}fdahl}, Mats G. and {Rouppe van der Voort}, Luc H.~M.},
        title = "{Dark cores in sunspot penumbral filaments}",
      journal = {\nat},
         year = 2002,
        month = nov,
       volume = {420},
       number = {6912},
        pages = {151-153},
          doi = {10.1038/nature01173},
       adsurl = {https://ui.adsabs.harvard.edu/abs/2002Natur.420..151S}
}

@ARTICLE{Adjabshirzadeh83,
       author = {{Adjabshirzadeh}, A. and {Koutchmy}, S.},
        title = "{Photometric analysis of sunspot umbral dots. III Spectrophotometry and preliminary model of a 2-component umbra}",
      journal = {\aap},
         year = 1983,
        month = jun,
       volume = {122},
       number = {1-2},
        pages = {1-8},
       adsurl = {https://ui.adsabs.harvard.edu/abs/1983A&A...122....1A}
}

@ARTICLE{Asplund00,
       author = {{Asplund}, M.},
        title = "{Line formation in solar granulation. III. The photospheric Si and meteoritic Fe abundances}",
      journal = {\aap},
         year = 2000,
        month = jul,
       volume = {359},
        pages = {755-758},
          doi = {10.48550/arXiv.astro-ph/0005322},
archivePrefix = {arXiv},
       eprint = {astro-ph/0005322},
 primaryClass = {astro-ph},
       adsurl = {https://ui.adsabs.harvard.edu/abs/2000A&A...359..755A}
}

@ARTICLE{Asplund21,
       author = {{Asplund}, M. and {Amarsi}, A.~M. and {Grevesse}, N.},
        title = "{The chemical make-up of the Sun: A 2020 vision}",
      journal = {\aap},
         year = 2021,
        month = sep,
       volume = {653},
          eid = {A141},
        pages = {A141},
          doi = {10.1051/0004-6361/202140445},
archivePrefix = {arXiv},
       eprint = {2105.01661},
 primaryClass = {astro-ph.SR},
       adsurl = {https://ui.adsabs.harvard.edu/abs/2021A&A...653A.141A}
}

@ARTICLE{Borrero02,
       author = {{Borrero}, J.~M. and {Bellot Rubio}, L.~R.},
        title = "{A two-component model of the solar photosphere from the inversion of spectral lines}",
      journal = {\aap},
         year = 2002,
        month = apr,
       volume = {385},
        pages = {1056-1072},
          doi = {10.1051/0004-6361:20020176},
       adsurl = {https://ui.adsabs.harvard.edu/abs/2002A&A...385.1056B}
}

@ARTICLE{Borrero03,
       author = {{Borrero}, J.~M. and {Bellot Rubio}, L.~R. and {Barklem}, P.~S. and {del Toro Iniesta}, J.~C.},
        title = "{Accurate atomic parameters for near-infrared spectral lines}",
      journal = {\aap},
         year = 2003,
        month = jun,
       volume = {404},
        pages = {749-762},
          doi = {10.1051/0004-6361:20030548},
       adsurl = {https://ui.adsabs.harvard.edu/abs/2003A&A...404..749B}
}

@ARTICLE{Biermann41,
       author = {{Biermann}, L.},
        title = "{Der gegenw{\"a}rtige Stand der Theorie konvektiver Sonnenmodelle}",
      journal = {Vierteljahresschrift der Astronomischen Gesellschaft},
         year = 1941,
        month = jan,
       volume = {76},
        pages = {194-200},
       adsurl = {https://ui.adsabs.harvard.edu/abs/1941VAG....76..194B}
}

@ARTICLE{Bharti09,
       author = {{Bharti}, Lokesh and {Joshi}, Chandan and {Jaaffrey}, S.~N.~A. and {Jain}, Rajmal},
        title = "{Spectropolarimetery of umbral fine structures from Hinode: evidence for magnetoconvection}",
      journal = {\mnras},
         year = 2009,
        month = feb,
       volume = {393},
       number = {1},
        pages = {65-70},
          doi = {10.1111/j.1365-2966.2008.14203.x},
archivePrefix = {arXiv},
       eprint = {0811.1722},
 primaryClass = {astro-ph},
       adsurl = {https://ui.adsabs.harvard.edu/abs/2009MNRAS.393...65B}
}

@ARTICLE{Bharti10,
       author = {{Bharti}, L. and {Beeck}, B. and {Sch{\"u}ssler}, M.},
        title = "{Properties of simulated sunspot umbral dots}",
      journal = {\aap},
         year = 2010,
        month = jan,
       volume = {510},
          eid = {A12},
        pages = {A12},
          doi = {10.1051/0004-6361/200913328},
archivePrefix = {arXiv},
       eprint = {0911.5068},
 primaryClass = {astro-ph.SR},
       adsurl = {https://ui.adsabs.harvard.edu/abs/2010A&A...510A..12B}
}

@ARTICLE{Borrero11,
       author = {{Borrero}, Juan M. and {Ichimoto}, Kiyoshi},
        title = "{Magnetic Structure of Sunspots}",
      journal = {Living Reviews in Solar Physics},
         year = 2011,
        month = sep,
       volume = {8},
       number = {1},
          eid = {4},
        pages = {4},
          doi = {10.12942/lrsp-2011-4},
archivePrefix = {arXiv},
       eprint = {1109.4412},
 primaryClass = {astro-ph.SR},
       adsurl = {https://ui.adsabs.harvard.edu/abs/2011LRSP....8....4B}
}

@INPROCEEDINGS{Brault1978,
       author = {{Brault}, J.~W.},
        title = "{Solar Fourier Transform Spectroscopy}",
    booktitle = {Future solar optical observations needs and constraints},
         year = 1978,
       editor = {{Godoli}, G.},
       volume = {106},
        month = jan,
        pages = {33},
       adsurl = {https://ui.adsabs.harvard.edu/abs/1978fsoo.conf...33B}
}

@INPROCEEDINGS{Cao10,
       author = {{Cao}, Wenda and {Gorceix}, Nicolas and {Coulter}, Roy and {Coulter}, Aaron and {Goode}, Philip R.},
        title = "{First light of the 1.6 meter off-axis New Solar Telescope at Big Bear Solar Observatory}",
    booktitle = {Ground-based and Airborne Telescopes III},
         year = 2010,
       editor = {{Stepp}, Larry M. and {Gilmozzi}, Roberto and {Hall}, Helen J.},
       series = {Society of Photo-Optical Instrumentation Engineers (SPIE) Conference Series},
       volume = {7733},
        month = jul,
          eid = {773330},
        pages = {773330},
          doi = {10.1117/12.856611},
       adsurl = {https://ui.adsabs.harvard.edu/abs/2010SPIE.7733E..30C}
}

@ARTICLE{Chevalier16,
       author = {{Chevalier}, S.},
        title = "{Tache solaire No 568}",
      journal = {Annales de l'Observatoire astronomique de Z{\^o}-s{\`e} (Chine},
         year = 1916,
        month = jan,
       volume = {9},
        pages = {B29},
       adsurl = {https://ui.adsabs.harvard.edu/abs/1916AnOAZ...9B..29C}
}

@ARTICLE{Choudhuri86,
       author = {{Choudhuri}, A.~R.},
        title = "{The Dynamics of Magnetically Trapped Fluids. I. Implications for Umbral Dots and Penumbral Grains}",
      journal = {\apj},
         year = 1986,
        month = mar,
       volume = {302},
        pages = {809},
          doi = {10.1086/164042},
       adsurl = {https://ui.adsabs.harvard.edu/abs/1986ApJ...302..809C}
}

@BOOK{TheSun,
       author = {{Stix}, Michael},
        title = "{The Sun. an Introduction}",
         year = 1991,
    publisher = "Springer",
       adsurl = {https://ui.adsabs.harvard.edu/abs/1991sun..book.....S}
}

@book{AstronOptics,
    title = {Astronomical optics},
    author = {Schroeder, Daniel J.},
    address = {San Diego},
    publisher = {Academic Press},
    year = {2000},
    pages = {352},
    isbn = {9780080499512},
    url = {https://learning.oreilly.com/library/view/~/9780126298109/?ar}
}

@ARTICLE{Danielson64,
       author = {{Danielson}, Robert E.},
        title = "{The Structure of Sunspot Umbras. I. Observations.}",
      journal = {\apj},
         year = 1964,
        month = jan,
       volume = {139},
        pages = {45},
          doi = {10.1086/147738},
       adsurl = {https://ui.adsabs.harvard.edu/abs/1964ApJ...139...45D}
}

@ARTICLE{Frutiger00,
       author = {{Frutiger}, C. and {Solanki}, S.~K. and {Fligge}, M. and {Bruls}, J.~H.~M.~J.},
        title = "{Properties of the solar granulation obtained from the inversion of low spatial resolution spectra}",
      journal = {\aap},
         year = 2000,
        month = jun,
       volume = {358},
        pages = {1109-1121},
       adsurl = {https://ui.adsabs.harvard.edu/abs/2000A&A...358.1109F}
}

@phdthesis{Frutiger00phd,
  title        = {Inversions of Zeeman split Stokes profiles},
  author       = {{Frutiger}, C.},
  year         = 2000,
  month        = {March},
  note         = {Naturwissenschaften ETH Zürich, Nr. 13896},
  type         = {PhD thesis},
  doi          = {10.3929/ethz-a-004040479},
  url          = {http://hdl.handle.net/20.500.11850/144840},
  publisher    = {ETH Zürich},
  school       = {Institute of Astronomy, ETH Zürich}
}

@ARTICLE{Gingerich71,
       author = {{Gingerich}, O. and {Noyes}, R.~W. and {Kalkofen}, W. and {Cuny}, Y.},
        title = "{The Harvard-Smithsonian reference atmosphere}",
      journal = {\solphys},
         year = 1971,
        month = jul,
       volume = {18},
       number = {3},
        pages = {347-365},
          doi = {10.1007/BF00149057},
       adsurl = {https://ui.adsabs.harvard.edu/abs/1971SoPh...18..347G}
}

@ARTICLE{Holzreuter15,
       author = {{Holzreuter}, R. and {Solanki}, S.~K.},
        title = "{Three-dimensional non-LTE radiative transfer effects in Fe i lines. III. Line formation in magneto-hydrodynamic atmospheres}",
      journal = {\aap},
         year = 2015,
        month = oct,
       volume = {582},
          eid = {A101},
        pages = {A101},
          doi = {10.1051/0004-6361/201526373},
archivePrefix = {arXiv},
       eprint = {1507.04224},
 primaryClass = {astro-ph.SR},
       adsurl = {https://ui.adsabs.harvard.edu/abs/2015A&A...582A.101H}
}

@ARTICLE{Holzreuter13,
       author = {{Holzreuter}, R. and {Solanki}, S.~K.},
        title = "{Three-dimensional non-LTE radiative transfer effects in Fe i lines. II. Line formation in 3D radiation hydrodynamic simulations}",
      journal = {\aap},
         year = 2013,
        month = oct,
       volume = {558},
          eid = {A20},
        pages = {A20},
          doi = {10.1051/0004-6361/201322135},
archivePrefix = {arXiv},
       eprint = {1308.2490},
 primaryClass = {astro-ph.SR},
       adsurl = {https://ui.adsabs.harvard.edu/abs/2013A&A...558A..20H}
}

@ARTICLE{Holzreuter12,
       author = {{Holzreuter}, R. and {Solanki}, S.~K.},
        title = "{Three-dimensional non-LTE radiative transfer effects in Fe I lines. I. Flux sheet and flux tube geometries}",
      journal = {\aap},
         year = 2012,
        month = nov,
       volume = {547},
          eid = {A46},
        pages = {A46},
          doi = {10.1051/0004-6361/201219477},
archivePrefix = {arXiv},
       eprint = {1209.0559},
 primaryClass = {astro-ph.SR},
       adsurl = {https://ui.adsabs.harvard.edu/abs/2012A&A...547A..46H}
}

@inproceedings{Kurucz09,
   title={Including All the Lines},
   url={http://dx.doi.org/10.1063/1.3250087},
   DOI={10.1063/1.3250087},
   booktitle={AIP Conference Proceedings},
   publisher={AIP},
   author={Kurucz, Robert L. and Hubeny, Ivan and Stone, James M. and MacGregor, Keith and Werner, Klaus},
   year={2009},
   pages={43–51} }

@INPROCEEDINGS{Kurucz18,
       author = {{Kurucz}, R.~L.},
        title = "{Including All the Lines: Data Releases for Spectra and Opacities through 2017}",
    booktitle = {Workshop on Astrophysical Opacities},
         year = 2018,
       series = {Astronomical Society of the Pacific Conference Series},
       volume = {515},
        month = aug,
        pages = {47},
       adsurl = {https://ui.adsabs.harvard.edu/abs/2018ASPC..515...47K}
}

@ARTICLE{Leka97,
       author = {{Leka}, K.~D.},
        title = "{The Vector Magnetic Fields and Thermodynamics of Sunspot Light Bridges: The Case for Field-free Disruptions in Sunspots}",
      journal = {\apj},
         year = 1997,
        month = jul,
       volume = {484},
       number = {2},
        pages = {900-919},
          doi = {10.1086/304363},
       adsurl = {https://ui.adsabs.harvard.edu/abs/1997ApJ...484..900L}
}

@article{Hilland24,
  author    = {Rainer Hilland and Andreas Christen},
  title     = {A Systematic Investigation of the Applicability of Taylor’s Hypothesis in an Idealized Surface Layer},
  journal   = {Boundary-Layer Meteorology},
  volume    = {190},
  number    = {5},
  pages     = {22},
  year      = {2024},
  month     = {April},
  issn      = {1573-1472},
  doi       = {10.1007/s10546-024-00861-1},
  url       = {https://doi.org/10.1007/s10546-024-00861-1},
}

@inproceedings{SaintJacques2000,
author = {David Saint-Jacques and John E. Baldwin},
title = {{Taylor's hypothesis: good for nuts}},
volume = {4006},
booktitle = {Interferometry in Optical Astronomy},
editor = {Pierre J. Lena and Andreas Quirrenbach},
organization = {International Society for Optics and Photonics},
publisher = {SPIE},
pages = {951 -- 962},
year = {2000},
doi = {10.1117/12.390175},
URL = {https://doi.org/10.1117/12.390175}
}

@ARTICLE{Glindemann2000,
       author = {{Glindemann}, A. and {Hippler}, S. and {Berkefeld}, T. and {Hackenberg}, W.},
        title = "{Adaptive Optics on Large Telescopes}",
      journal = {Experimental Astronomy},
         year = 2000,
        month = apr,
       volume = {10},
       number = {1},
        pages = {5-47},
          doi = {10.1023/A:1008116831367},
       adsurl = {https://ui.adsabs.harvard.edu/abs/2000ExA....10....5G}
}

@ARTICLE{Yadav18,
       author = {{Yadav}, Rahul and {Mathew}, Shibu K.},
        title = "{Physical Properties of Umbral Dots Observed in Sunspots: A Hinode Observation}",
      journal = {\solphys},
         year = 2018,
        month = apr,
       volume = {293},
       number = {4},
          eid = {54},
        pages = {54},
          doi = {10.1007/s11207-018-1272-y},
       adsurl = {https://ui.adsabs.harvard.edu/abs/2018SoPh..293...54Y}
}

@ARTICLE{Kilcik20,
       author = {{Kilcik}, Ali and {Sarp}, Volkan and {Yurchyshyn}, Vasyl and {Rozelot}, Jean-Pierre and {Ozguc}, Atila},
        title = "{Physical Characteristics of Umbral Dots Derived from a High-Resolution Observations}",
      journal = {\solphys},
         year = 2020,
        month = apr,
       volume = {295},
       number = {4},
          eid = {58},
        pages = {58},
          doi = {10.1007/s11207-020-01618-y},
archivePrefix = {arXiv},
       eprint = {2004.07687},
 primaryClass = {astro-ph.SR},
       adsurl = {https://ui.adsabs.harvard.edu/abs/2020SoPh..295...58K}
}

@ARTICLE{Calisir23,
       author = {{Calisir}, M. Ali and {Yazici}, H. Tayfun and {Kilcik}, Ali and {Yurchyshyn}, Vasyl},
        title = "{Relationships Between Physical Parameters of Umbral Dots Measured for 12 Sunspot Umbras with the Goode Solar Telescope}",
      journal = {\solphys},
         year = 2023,
        month = sep,
       volume = {298},
       number = {9},
          eid = {103},
        pages = {103},
          doi = {10.1007/s11207-023-02198-3},
       adsurl = {https://ui.adsabs.harvard.edu/abs/2023SoPh..298..103C}
}

@ARTICLE{Ortiz10,
       author = {{Ortiz}, A. and {Bellot Rubio}, L.~R. and {Rouppe van der Voort}, L.},
        title = "{Downflows in Sunspot Umbral Dots}",
      journal = {\apj},
         year = 2010,
        month = apr,
       volume = {713},
       number = {2},
        pages = {1282-1291},
          doi = {10.1088/0004-637X/713/2/1282},
archivePrefix = {arXiv},
       eprint = {1003.1897},
 primaryClass = {astro-ph.SR},
       adsurl = {https://ui.adsabs.harvard.edu/abs/2010ApJ...713.1282O}
}

@ARTICLE{Taylor38,
       author = {{Taylor}, G.~I.},
        title = "{The Spectrum of Turbulence}",
      journal = {Proceedings of the Royal Society of London Series A},
         year = 1938,
        month = feb,
       volume = {164},
       number = {919},
        pages = {476-490},
          doi = {10.1098/rspa.1938.0032},
       adsurl = {https://ui.adsabs.harvard.edu/abs/1938RSPSA.164..476T}
}

@ARTICLE{Wilson96,
       author = {{Wilson}, R.~W. and {Jenkins}, C.~R.},
        title = "{Adaptive optics for astronomy: theoretical performance and limitations}",
      journal = {\mnras},
         year = 1996,
        month = jan,
       volume = {278},
       number = {1},
        pages = {39-61},
          doi = {10.1093/mnras/278.1.39},
       adsurl = {https://ui.adsabs.harvard.edu/abs/1996MNRAS.278...39W}
}

@ARTICLE{MiHI,
       author = {{van Noort}, M. and {Bischoff}, J. and {Kramer}, A. and {Solanki}, S.~K. and {Kiselman}, D.},
        title = "{A prototype of a microlensed hyperspectral imager for solar observations}",
      journal = {\aap},
         year = 2022,
        month = dec,
       volume = {668},
          eid = {A149},
        pages = {A149},
          doi = {10.1051/0004-6361/202243464},
       adsurl = {https://ui.adsabs.harvard.edu/abs/2022A&A...668A.149V}
}

@ARTICLE{Zhao24,
       author = {{Zhao}, Jie and {Yu}, Fu and {Zhu}, Xiaoshuai and {Yang}, Xu and {Su}, Jiangtao and {Schmieder}, Brigitte and {Li}, Hui and {Cao}, Wenda},
        title = "{Filamentary Structures Formation around a Magnetic Pore Using High-resolution Observations of the Goode Solar Telescope}",
      journal = {\apj},
         year = 2024,
        month = sep,
       volume = {973},
       number = {1},
          eid = {33},
        pages = {33},
          doi = {10.3847/1538-4357/ad5c3a},
       adsurl = {https://ui.adsabs.harvard.edu/abs/2024ApJ...973...33Z}
}

@article{kuridze25,
doi = {10.3847/2041-8213/add470},
url = {https://dx.doi.org/10.3847/2041-8213/add470},
year = {2025},
month = {may},
publisher = {The American Astronomical Society},
volume = {985},
number = {1},
pages = {L23},
author = {Kuridze, David and Wöger, Friedrich and Uitenbroek, Han and Rempel, Matthias and Tritschler, Alexandra and Rimmele, Thomas and Fischer, Catherine and Steiner, Oskar},
title = {The Striated Solar Photosphere Observed at 0.03" Resolution},
journal = {The Astrophysical Journal Letters}
}

@ARTICLE{Zhang13,
       author = {{Zhang}, Y. and {Ichimoto}, K.},
        title = "{Properties of sunspot penumbral grains observed with Hinode}",
      journal = {\aap},
         year = 2013,
        month = dec,
       volume = {560},
          eid = {A77},
        pages = {A77},
          doi = {10.1051/0004-6361/201118119},
       adsurl = {https://ui.adsabs.harvard.edu/abs/2013A&A...560A..77Z}
}

@ARTICLE{Harvey73,
       author = {{Harvey}, K. and {Harvey}, J.},
        title = "{Observations of Moving Magnetic Features near Sunspots}",
      journal = {\solphys},
         year = 1973,
        month = jan,
       volume = {28},
       number = {1},
        pages = {61-71},
          doi = {10.1007/BF00152912},
       adsurl = {https://ui.adsabs.harvard.edu/abs/1973SoPh...28...61H}
}

@INPROCEEDINGS{Munjiba24,
       author = {{Munjiba}, M.~M. and {Singh}, Prakhar and {Adithya}, H.~N. and {Padinhatteeri}, Sreejith and {Athiray}, P.~S.},
        title = "{Study of Solar Atmospheric Heating due to Non-Flaring Active Region}",
    booktitle = {42nd meeting of the Astronomical Society of India (ASI)},
         year = 2024,
       volume = {42},
        month = jan,
          eid = {P235},
        pages = {P235},
       adsurl = {https://ui.adsabs.harvard.edu/abs/2024asi..confP.235M}
}

@INPROCEEDINGS{Lites72,
       author = {{Lites}, B.~W. and {Athay}, R.~G.},
        title = "{FeI Ionization and Excitation Equilibrium in the Solar Atmosphere.}",
    booktitle = {Bulletin of the American Astronomical Society},
         year = 1972,
       volume = {4},
        month = mar,
        pages = {212},
       adsurl = {https://ui.adsabs.harvard.edu/abs/1972BAAS....4..212L}
}

@ARTICLE{Bharti16,
       author = {{Bharti}, L. and {Quintero Noda}, C. and {Joshi}, C. and {Rakesh}, S. and {Pandya}, A.},
        title = "{Fine structures at pore boundary}",
      journal = {\mnras},
         year = 2016,
        month = oct,
       volume = {462},
       number = {1},
        pages = {L93-L97},
          doi = {10.1093/mnrasl/slw130},
       adsurl = {https://ui.adsabs.harvard.edu/abs/2016MNRAS.462L..93B}
}

@book{boreman2001,
    author = {{Boreman}, G. D.},
    title = { Modulation Transfer Function in Optical and Electro-Optical Systems},
    publisher = {SPIE PRESS} ,
    year = 2001 
}

@ARTICLE{Jess23,
       author = {{Jess}, David B. and {Grant}, Samuel D.~T. and {Bate}, William and {Liu}, Jiajia and {Jafarzadeh}, Shahin and {Keys}, Peter H. and {Vieira}, Lu{\'\i}s E.~A. and {Dal Lago}, Alisson and {Guarnieri}, Fernando L. and {Christian}, Damian J. and {Gilliam}, Doug and {Banerjee}, Dipankar},
        title = "{The Fibre Resolved OpticAl and Near-Ultraviolet Czerny-Turner Imaging Spectropolarimeter (FRANCIS)}",
      journal = {\solphys},
         year = 2023,
        month = dec,
       volume = {298},
       number = {12},
          eid = {146},
        pages = {146},
          doi = {10.1007/s11207-023-02237-z},
archivePrefix = {arXiv},
       eprint = {2312.00122},
 primaryClass = {astro-ph.SR},
       adsurl = {https://ui.adsabs.harvard.edu/abs/2023SoPh..298..146J}
}

@ARTICLE{Noort12,
       author = {{van Noort}, M.},
        title = "{Spatially coupled inversion of spectro-polarimetric image data. I. Method and first results}",
      journal = {\aap},
         year = 2012,
        month = dec,
       volume = {548},
          eid = {A5},
        pages = {A5},
          doi = {10.1051/0004-6361/201220220},
archivePrefix = {arXiv},
       eprint = {1210.4636},
 primaryClass = {astro-ph.IM},
       adsurl = {https://ui.adsabs.harvard.edu/abs/2012A&A...548A...5V}
}

@ARTICLE{Noort13,
       author = {{van Noort}, M. and {Lagg}, A. and {Tiwari}, S.~K. and {Solanki}, S.~K.},
        title = "{Peripheral downflows in sunspot penumbrae}",
      journal = {\aap},
         year = 2013,
        month = sep,
       volume = {557},
          eid = {A24},
        pages = {A24},
          doi = {10.1051/0004-6361/201321073},
archivePrefix = {arXiv},
       eprint = {1308.0466},
 primaryClass = {astro-ph.SR},
       adsurl = {https://ui.adsabs.harvard.edu/abs/2013A&A...557A..24V}
}

@article{Noort24,
       author = {{van Noort}, M. and {H{\"o}lken}, J. and {Doerr}, H.-P. and {Chae}, J. and {Cao}, W. and {Gorceix}, N. and {Kang}, J. and {Ahn}, K. and {Solanki}, S.~K.},
        title = "{Large-aperture diffraction-limited spectro-polarimetry with FISS-SP}",
      journal = {\aap},
         year = 2025,
        month = dec,
       volume = {704},
          eid = {A215},
        pages = {A215},
          doi = {10.1051/0004-6361/202556041},
       adsurl = {https://ui.adsabs.harvard.edu/abs/2025A&A...704A.215V}
}

@ARTICLE{Milic18,
       author = {{Mili{\'c}}, I. and {van Noort}, M.},
        title = "{Spectropolarimetric NLTE inversion code SNAPI}",
      journal = {\aap},
         year = 2018,
        month = sep,
       volume = {617},
          eid = {A24},
        pages = {A24},
          doi = {10.1051/0004-6361/201833382},
archivePrefix = {arXiv},
       eprint = {1806.08134},
 primaryClass = {astro-ph.SR},
       adsurl = {https://ui.adsabs.harvard.edu/abs/2018A&A...617A..24M}
}

@ARTICLE{Parker79,
       author = {{Parker}, E.~N.},
        title = "{Sunspots and the physics of magnetic flux tubes. IX. Umbral dots and longitudinal overstability.}",
      journal = {\apj},
         year = 1979,
        month = nov,
       volume = {234},
        pages = {333-347},
          doi = {10.1086/157501},
       adsurl = {https://ui.adsabs.harvard.edu/abs/1979ApJ...234..333P}
}

@ARTICLE{Rempel11,
       author = {{Rempel}, Matthias and {Schlichenmaier}, Rolf},
        title = "{Sunspot Modeling: From Simplified Models to Radiative MHD Simulations}",
      journal = {Living Reviews in Solar Physics},
         year = 2011,
        month = dec,
       volume = {8},
       number = {1},
          eid = {3},
        pages = {3},
          doi = {10.12942/lrsp-2011-3},
       adsurl = {https://ui.adsabs.harvard.edu/abs/2011LRSP....8....3R}
}

@ARTICLE{Riethmueller13,
       author = {{Riethm{\"u}ller}, T.~L. and {Solanki}, S.~K. and {van Noort}, M. and {Tiwari}, S.~K.},
        title = "{Vertical flows and mass flux balance of sunspot umbral dots}",
      journal = {\aap},
         year = 2013,
        month = jun,
       volume = {554},
          eid = {A53},
        pages = {A53},
          doi = {10.1051/0004-6361/201321075},
archivePrefix = {arXiv},
       eprint = {1305.1164},
 primaryClass = {astro-ph.SR},
       adsurl = {https://ui.adsabs.harvard.edu/abs/2013A&A...554A..53R}
}

@ARTICLE{Riethmueller08,
       author = {{Riethm{\"u}ller}, T.~L. and {Solanki}, S.~K. and {Lagg}, A.},
        title = "{Stratification of Sunspot Umbral Dots from Inversion of Stokes Profiles Recorded by Hinode}",
      journal = {\apjl},
         year = 2008,
        month = may,
       volume = {678},
       number = {2},
        pages = {L157},
          doi = {10.1086/588636},
archivePrefix = {arXiv},
       eprint = {0805.4324},
 primaryClass = {astro-ph},
       adsurl = {https://ui.adsabs.harvard.edu/abs/2008ApJ...678L.157R}
}

@INPROCEEDINGS{Rutten88,
       author = {{Rutten}, Robert J.},
        title = "{The NLTE formation of iron lines in the solar photosphere}",
    booktitle = {IAU Colloq. 94: Physics of Formation of FE II Lines Outside LTE},
         year = 1988,
       editor = {{Viotti}, Roberto and {Vittone}, Alberto and {Friedjung}, Michael},
       series = {Astrophysics and Space Science Library},
       volume = {138},
        month = jan,
        pages = {185-210},
          doi = {10.1007/978-94-009-4023-9_23},
       adsurl = {https://ui.adsabs.harvard.edu/abs/1988ASSL..138..185R}
}

@ARTICLE{Shchukina01,
       author = {{Shchukina}, Nataliya and {Trujillo Bueno}, Javier},
        title = "{The Iron Line Formation Problem in Three-dimensional Hydrodynamic Models of Solar-like Photospheres}",
      journal = {\apj},
         year = 2001,
        month = apr,
       volume = {550},
       number = {2},
        pages = {970-990},
          doi = {10.1086/319789},
       adsurl = {https://ui.adsabs.harvard.edu/abs/2001ApJ...550..970S}
}

@ARTICLE{Sinjan24,
       author = {{Sinjan}, J. and {Solanki}, S.~K. and {Hirzberger}, J. and {Riethm{\"u}ller}, T.~L. and {Przybylski}, D.},
        title = "{Magnetograms underestimate even unipolar magnetic flux nearly everywhere on the solar disk}",
      journal = {\aap},
         year = 2024,
        month = oct,
       volume = {690},
          eid = {A341},
        pages = {A341},
          doi = {10.1051/0004-6361/202450267},
archivePrefix = {arXiv},
       eprint = {2407.08368},
 primaryClass = {astro-ph.SR},
       adsurl = {https://ui.adsabs.harvard.edu/abs/2024A&A...690A.341S}
}

@INPROCEEDINGS{Schlueter58,
       author = {{Schl{\"u}ter}, A. and {Temesv{\'a}ry}, S.},
        title = "{The Internal Constitution of Sunspots}",
    booktitle = {Electromagnetic Phenomena in Cosmical Physics},
         year = 1958,
       editor = {{Lehnert}, B.},
       volume = {6},
        month = jan,
        pages = {263},
       adsurl = {https://ui.adsabs.harvard.edu/abs/1958IAUS....6..263S}
}

@ARTICLE{Suetterlin96,
       author = {{S{\"u}tterlin}, P. and {Schr{\"o}ter}, E.~H. and {Muglach}, K.},
        title = "{Polarimetry of Solar Pores}",
      journal = {\solphys},
         year = 1996,
        month = mar,
       volume = {164},
       number = {1-2},
        pages = {311-320},
          doi = {10.1007/BF00146643},
       adsurl = {https://ui.adsabs.harvard.edu/abs/1996SoPh..164..311S}
}

@ARTICLE{Socas-Navarro04,
       author = {{Socas-Navarro}, H. and {Mart{\'\i}nez Pillet}, V. and {Sobotka}, M. and {V{\'a}zquez}, M.},
        title = "{The Thermal and Magnetic Structure of Umbral Dots from the Inversion of High-Resolution Full Stokes Observations}",
      journal = {\apj},
         year = 2004,
        month = oct,
       volume = {614},
       number = {1},
        pages = {448-456},
          doi = {10.1086/423487},
       adsurl = {https://ui.adsabs.harvard.edu/abs/2004ApJ...614..448S}
}

@ARTICLE{Socas-Navarro21,
       author = {{Socas-Navarro}, H. and {Asensio Ramos}, A.},
        title = "{Exploring the Sun's upper atmosphere with neural networks: Reversed patterns and the hot wall effect}",
      journal = {\aap},
         year = 2021,
        month = aug,
       volume = {652},
          eid = {A78},
        pages = {A78},
          doi = {10.1051/0004-6361/202140424},
       adsurl = {https://ui.adsabs.harvard.edu/abs/2021A&A...652A..78S}
}

@ARTICLE{Solanki06,
       author = {{Solanki}, Sami K. and {Inhester}, Bernd and {Sch{\"u}ssler}, Manfred},
        title = "{The solar magnetic field}",
      journal = {Reports on Progress in Physics},
         year = 2006,
        month = mar,
       volume = {69},
       number = {3},
        pages = {563-668},
          doi = {10.1088/0034-4885/69/3/R02},
archivePrefix = {arXiv},
       eprint = {1008.0771},
 primaryClass = {astro-ph.SR},
       adsurl = {https://ui.adsabs.harvard.edu/abs/2006RPPh...69..563S}
}

@ARTICLE{Schuessler06,
       author = {{Sch{\"u}ssler}, M. and {V{\"o}gler}, A.},
        title = "{Magnetoconvection in a Sunspot Umbra}",
      journal = {\apjl},
         year = 2006,
        month = apr,
       volume = {641},
       number = {1},
        pages = {L73-L76},
          doi = {10.1086/503772},
archivePrefix = {arXiv},
       eprint = {astro-ph/0603078},
 primaryClass = {astro-ph},
       adsurl = {https://ui.adsabs.harvard.edu/abs/2006ApJ...641L..73S}
}

@ARTICLE{Stein00,
       author = {{Stein}, Robert F. and {Nordlund}, {\r{A}}ke},
        title = "{Realistic Solar Convection Simulations}",
      journal = {\solphys},
         year = 2000,
        month = mar,
       volume = {192},
        pages = {91-108},
          doi = {10.1023/A:1005260918443},
       adsurl = {https://ui.adsabs.harvard.edu/abs/2000SoPh..192...91S}
}

@ARTICLE{Smitha20,
       author = {{Smitha}, H.~N. and {Holzreuter}, R. and {van Noort}, M. and {Solanki}, S.~K.},
        title = "{The influence of NLTE effects in Fe I lines on an inverted atmosphere. I. 6301 {\r{A}} and 6302 {\r{A}} lines formed in 1D NLTE}",
      journal = {\aap},
         year = 2020,
        month = jan,
       volume = {633},
          eid = {A157},
        pages = {A157},
          doi = {10.1051/0004-6361/201937041},
archivePrefix = {arXiv},
       eprint = {1912.07007},
 primaryClass = {astro-ph.SR},
       adsurl = {https://ui.adsabs.harvard.edu/abs/2020A&A...633A.157S}
}

@ARTICLE{Smitha21,
       author = {{Smitha}, H.~N. and {Holzreuter}, R. and {van Noort}, M. and {Solanki}, S.~K.},
        title = "{The influence of NLTE effects in Fe I lines on an inverted atmosphere. II. 6301 {\r{A}} and 6302 {\r{A}} lines formed in 3D NLTE}",
      journal = {\aap},
         year = 2021,
        month = mar,
       volume = {647},
          eid = {A46},
        pages = {A46},
          doi = {10.1051/0004-6361/202039107},
archivePrefix = {arXiv},
       eprint = {2101.00506},
 primaryClass = {astro-ph.SR},
       adsurl = {https://ui.adsabs.harvard.edu/abs/2021A&A...647A..46S}
}

@INPROCEEDINGS{Tritschler09,
       author = {{Tritschler}, A.},
        title = "{Structure and Dynamics of Sunspots (Keynote)}",
    booktitle = {The Second Hinode Science Meeting: Beyond Discovery-Toward Understanding},
         year = 2009,
       editor = {{Lites}, B. and {Cheung}, M. and {Magara}, T. and {Mariska}, J. and {Reeves}, K.},
       series = {Astronomical Society of the Pacific Conference Series},
       volume = {415},
        month = dec,
        pages = {339},
          doi = {10.48550/arXiv.0903.1300},
archivePrefix = {arXiv},
       eprint = {0903.1300},
 primaryClass = {astro-ph.SR},
       adsurl = {https://ui.adsabs.harvard.edu/abs/2009ASPC..415..339T}
}

@ARTICLE{Vukadinovic24,
       author = {{Vukadinovi{\'c}}, D. and {Smitha}, H.~N. and {Korpi-Lagg}, A. and {van Noort}, M. and {Castellanos Dur{\'a}n}, J.~S. and {Solanki}, S.~K.},
        title = "{globin: A spectropolarimetric inversion code for the coupled inference of atomic line parameters}",
      journal = {\aap},
         year = 2024,
        month = jun,
       volume = {686},
          eid = {A262},
        pages = {A262},
          doi = {10.1051/0004-6361/202347752},
archivePrefix = {arXiv},
       eprint = {2404.03291},
 primaryClass = {astro-ph.SR},
       adsurl = {https://ui.adsabs.harvard.edu/abs/2024A&A...686A.262V}
}

@ARTICLE{Watanabe12,
       author = {{Watanabe}, Hiroko and {Bellot Rubio}, Luis R. and {de la Cruz Rodr{\'\i}guez}, Jaime and {Rouppe van der Voort}, Luc},
        title = "{Temporal Evolution of Velocity and Magnetic Field in and around Umbral Dots}",
      journal = {\apj},
         year = 2012,
        month = sep,
       volume = {757},
       number = {1},
          eid = {49},
        pages = {49},
          doi = {10.1088/0004-637X/757/1/49},
archivePrefix = {arXiv},
       eprint = {1207.6006},
 primaryClass = {astro-ph.SR},
       adsurl = {https://ui.adsabs.harvard.edu/abs/2012ApJ...757...49W}
}

@ARTICLE{Wiegelmann14,
       author = {{Wiegelmann}, Thomas and {Thalmann}, Julia K. and {Solanki}, Sami K.},
        title = "{The magnetic field in the solar atmosphere}",
      journal = {\aapr},
         year = 2014,
        month = nov,
       volume = {22},
          eid = {78},
        pages = {78},
          doi = {10.1007/s00159-014-0078-7},
archivePrefix = {arXiv},
       eprint = {1410.4214},
 primaryClass = {astro-ph.SR},
       adsurl = {https://ui.adsabs.harvard.edu/abs/2014A&ARv..22...78W}
}

@ARTICLE{Ryabchikova15,
       author = {{Ryabchikova}, T. and {Piskunov}, N. and {Kurucz}, R.~L. and {Stempels}, H.~C. and {Heiter}, U. and {Pakhomov}, Yu and {Barklem}, P.~S.},
        title = "{A major upgrade of the VALD database}",
      journal = {\physscr},
         year = 2015,
        month = may,
       volume = {90},
       number = {5},
          eid = {054005},
        pages = {054005},
          doi = {10.1088/0031-8949/90/5/054005},
       adsurl = {https://ui.adsabs.harvard.edu/abs/2015PhyS...90e4005R}
}

@ARTICLE{Pakhomov19,
       author = {{Pakhomov}, Yu. V. and {Ryabchikova}, T.~A. and {Piskunov}, N.~E.},
        title = "{Hyperfine Splitting in the VALD Database of Spectral-line Parameters}",
      journal = {Astronomy Reports},
         year = 2019,
        month = dec,
       volume = {63},
       number = {12},
        pages = {1010-1021},
          doi = {10.1134/S1063772919120047},
archivePrefix = {arXiv},
       eprint = {1911.03189},
 primaryClass = {astro-ph.IM},
       adsurl = {https://ui.adsabs.harvard.edu/abs/2019ARep...63.1010P}
}

@ARTICLE{Bigot06,
       author = {{Bigot}, L. and {Th{\'e}venin}, F.},
        title = "{Atomic oscillator strengths in the spectral domain of Gaia}",
      journal = {\mnras},
         year = 2006,
        month = oct,
       volume = {372},
       number = {2},
        pages = {609-614},
          doi = {10.1111/j.1365-2966.2006.10701.x},
       adsurl = {https://ui.adsabs.harvard.edu/abs/2006MNRAS.372..609B}
}

@ARTICLE{Heiter21,
       author = {{Heiter}, U. and {Lind}, K. and {Bergemann}, M. and {Asplund}, M. and {Mikolaitis}, {\v{S}}. and {Barklem}, P.~S. and {Masseron}, T. and {de Laverny}, P. and {Magrini}, L. and {Edvardsson}, B. and {J{\"o}nsson}, H. and {Pickering}, J.~C. and {Ryde}, N. and {Bayo Ar{\'a}n}, A. and {Bensby}, T. and {Casey}, A.~R. and {Feltzing}, S. and {Jofr{\'e}}, P. and {Korn}, A.~J. and {Pancino}, E. and {Damiani}, F. and {Lanzafame}, A. and {Lardo}, C. and {Monaco}, L. and {Morbidelli}, L. and {Smiljanic}, R. and {Worley}, C. and {Zaggia}, S. and {Randich}, S. and {Gilmore}, G.~F.},
        title = "{Atomic data for the Gaia-ESO Survey}",
      journal = {\aap},
         year = 2021,
        month = jan,
       volume = {645},
          eid = {A106},
        pages = {A106},
          doi = {10.1051/0004-6361/201936291},
archivePrefix = {arXiv},
       eprint = {2011.02049},
 primaryClass = {astro-ph.IM},
       adsurl = {https://ui.adsabs.harvard.edu/abs/2021A&A...645A.106H}
}

@MISC{SocasNavarro15,
       author = {{Socas-Navarro}, H.},
        title = "{NICOLE: NLTE Stokes Synthesis/Inversion Code}",
 howpublished = {Astrophysics Source Code Library, record ascl:1508.002},
         year = 2015,
        month = aug,
          eid = {ascl:1508.002},
       adsurl = {https://ui.adsabs.harvard.edu/abs/2015ascl.soft08002S}
}

@ARTICLE{MODEST24,
       author = {{Castellanos Dur{\'a}n}, J.~S. and {Milanovic}, N. and {Korpi-Lagg}, A. and {L{\"o}ptien}, B. and {van Noort}, M. and {Solanki}, S.~K.},
        title = "{The MODEST catalog of depth-dependent spatially coupled inversions of sunspots observed by Hinode/SOT-SP}",
      journal = {\aap},
         year = 2024,
        month = jul,
       volume = {687},
          eid = {A218},
        pages = {A218},
          doi = {10.1051/0004-6361/202449269},
archivePrefix = {arXiv},
       eprint = {2403.06960},
 primaryClass = {astro-ph.SR},
       adsurl = {https://ui.adsabs.harvard.edu/abs/2024A&A...687A.218C}
}

@ARTICLE{Solanki88,
       author = {{Solanki}, S.~K. and {Steenbock}, W.},
        title = "{NLTE effects in solar magnetic fluxtubes}",
      journal = {\aap},
         year = 1988,
        month = jan,
       volume = {189},
       number = {1-2},
        pages = {243-253},
       adsurl = {https://ui.adsabs.harvard.edu/abs/1988A&A...189..243S}
}

@ARTICLE{Stenflo73,
       author = {{Stenflo}, J.~O.},
        title = "{Magnetic-Field Structure of the Photospheric Network}",
      journal = {\solphys},
         year = 1973,
        month = sep,
       volume = {32},
       number = {1},
        pages = {41-63},
          doi = {10.1007/BF00152728},
       adsurl = {https://ui.adsabs.harvard.edu/abs/1973SoPh...32...41S}
}

@ARTICLE{Danilovic16,
       author = {{Danilovic}, S. and {van Noort}, M. and {Rempel}, M.},
        title = "{Internetwork magnetic field as revealed by two-dimensional inversions}",
      journal = {\aap},
         year = 2016,
        month = sep,
       volume = {593},
          eid = {A93},
        pages = {A93},
          doi = {10.1051/0004-6361/201527842},
archivePrefix = {arXiv},
       eprint = {1607.00772},
 primaryClass = {astro-ph.SR},
       adsurl = {https://ui.adsabs.harvard.edu/abs/2016A&A...593A..93D}
}

@ARTICLE{Orozco07,
       author = {{Orozco Su{\'a}rez}, David and {Bellot Rubio}, Luis R. and {Del Toro Iniesta}, Jose Carlos and {Tsuneta}, Saku and {Lites}, Bruce and {Ichimoto}, Kiyoshi and {Katsukawa}, Yukio and {Nagata}, Shin'ichi and {Shimizu}, Toshifumi and {Shine}, Richard A. and {Suematsu}, Yoshinori and {Tarbell}, Theodore D. and {Title}, Alan M.},
        title = "{Strategy for the Inversion of Hinode Spectropolarimetric Measurements in the Quiet Sun}",
      journal = {\pasj},
         year = 2007,
        month = nov,
       volume = {59},
        pages = {S837},
          doi = {10.1093/pasj/59.sp3.S837},
archivePrefix = {arXiv},
       eprint = {0709.2033},
 primaryClass = {astro-ph},
       adsurl = {https://ui.adsabs.harvard.edu/abs/2007PASJ...59S.837O}
}

@article{Barnes04,
    author = {Barnes, J. R.},
    title = {Deconvolution of high-resolution spectra as a signal-to-noise ratio enhancement method},
    journal = {Monthly Notices of the Royal Astronomical Society},
    volume = {348},
    number = {4},
    pages = {1295-1300},
    year = {2004},
    month = {03},
    issn = {0035-8711},
    doi = {10.1111/j.1365-2966.2004.07451.x},
    url = {https://doi.org/10.1111/j.1365-2966.2004.07451.x},
    eprint = {https://academic.oup.com/mnras/article-pdf/348/4/1295/2960257/348-4-1295.pdf},
}

@INPROCEEDINGS{Allende98,
       author = {{Allende Prieto}, Carlos and {Ruiz Cobo}, Basilio and {Garcia Lopez}, Ramon J.},
        title = "{An Inversion Technique to Derive Model Photospheres in Late-Type Stars from High-Resolution Spectroscopy: The Sun}",
    booktitle = {Cool Stars, Stellar Systems, and the Sun},
         year = 1998,
       editor = {{Donahue}, Robert A. and {Bookbinder}, Jay A.},
       series = {Astronomical Society of the Pacific Conference Series},
       volume = {154},
        month = jan,
        pages = {813},
          doi = {10.48550/arXiv.astro-ph/9710067},
archivePrefix = {arXiv},
       eprint = {astro-ph/9710067},
 primaryClass = {astro-ph},
       adsurl = {https://ui.adsabs.harvard.edu/abs/1998ASPC..154..813A}
}

@ARTICLE{DiazBaso25,
       author = {{D{\'\i}az Baso}, C.~J. and {Mili{\'c}}, I. and {Rouppe van der Voort}, L. and {Schlichenmaier}, R.},
        title = "{Spectral resolution effects on the information content in solar spectra}",
      journal = {\aap},
         year = 2025,
        month = jan,
       volume = {693},
          eid = {A272},
        pages = {A272},
          doi = {10.1051/0004-6361/202452331},
archivePrefix = {arXiv},
       eprint = {2409.13677},
 primaryClass = {astro-ph.SR},
       adsurl = {https://ui.adsabs.harvard.edu/abs/2025A&A...693A.272D}
}

@article{ZERNIKE1934,
    title = {Beugungstheorie des schneidenver-fahrens und seiner verbesserten form, der phasenkontrastmethode},
    journal = {Physica},
    volume = {1},
    number = {7},
    pages = {689-704},
    year = {1934},
    issn = {0031-8914},
    doi = {https://doi.org/10.1016/S0031-8914(34)80259-5},
    url = {https://www.sciencedirect.com/science/article/pii/S0031891434802595},
    author = {Zernike, {Frits}}
}

@article{Arlt2020,
  author = {Arlt, Rainer and Vaquero, José M.},
  title = {Historical sunspot records},
  journal = {Living Reviews in Solar Physics},
  volume = {17},
  issue = {1},
  pages = {1},
  year = {2020},
  doi = {10.1007/s41116-020-0023-y},
  publisher = {Springer Nature},
  issn = {1614-4961},
  url = {https://doi.org/10.1007/s41116-020-0023-y},
  date = {2020-02-26}
}

@book{Frisch22,
title = {Radiative Transfer : An Introduction to Exact and Asymptotic Methods},
author = {Frisch, Hélène},
address = {Cham},
publisher = {Springer International Publishing},
year = {2022},
edition = {1st ed. 2022},
pages = {593},
isbn = {9783030952471},
url = {https://doi.org/10.1007/978-3-030-95247-1?nosfx=y}
}

@book{shurcliff62,
	address = {Cambridge, Massachusetts},
	title = {Polarized {Light}, production and use},
	language = {english},
	publisher = {Harvard University Press},
	author = {Shurcliff, William A.},
	year = {1962}
}

@ARTICLE{Gandorfer99,
       author = {{Gandorfer}, Achim M.},
        title = "{Ferroelectric retarders as an alternative to piezoelastic modulators for use in solar Stokes vector polarimetry}",
      journal = {Optical Engineering},
         year = 1999,
        month = aug,
       volume = {38},
        pages = {1402-1408},
          doi = {10.1117/1.602183},
       adsurl = {https://ui.adsabs.harvard.edu/abs/1999OptEn..38.1402G}
}

@INPROCEEDINGS{Pillet99,
       author = {{M{\'a}rtinez Pillet}, V. and {Collados}, M. and {S{\'a}nchez Almeida}, J. and {Gonz{\'a}lez}, V. and {Cruz-Lopez}, A. and {Manescau}, A. and {Joven}, E. and {Paez}, E. and {Diaz}, J. and {Feeney}, O. and {S{\'a}nchez}, V. and {Scharmer}, G. and {Soltau}, D.},
        title = "{LPSP \& TIP: Full Stokes Polarimeters for the Canary Islands Observatories}",
    booktitle = {High Resolution Solar Physics: Theory, Observations, and Techniques},
         year = 1999,
       editor = {{Rimmele}, T.~R. and {Balasubramaniam}, K.~S. and {Radick}, R.~R.},
       series = {Astronomical Society of the Pacific Conference Series},
       volume = {183},
        month = jan,
        pages = {264},
       adsurl = {https://ui.adsabs.harvard.edu/abs/1999ASPC..183..264M}
}

@article{Yang24,
author = {Zihao Yang  and Hui Tian  and Steven Tomczyk  and Xianyu Liu  and Sarah Gibson  and Richard J. Morton  and Cooper Downs },
title = {Observing the evolution of the Sun’s global coronal magnetic field over 8 months},
journal = {Science},
volume = {386},
number = {6717},
pages = {76-82},
year = {2024},
doi = {10.1126/science.ado2993},
URL = {https://www.science.org/doi/abs/10.1126/science.ado2993},
eprint = {https://www.science.org/doi/pdf/10.1126/science.ado2993}}

@ARTICLE{Gizon05,
       author = {{Gizon}, Laurent and {Birch}, Aaron C.},
        title = "{Local Helioseismology}",
      journal = {Living Reviews in Solar Physics},
         year = 2005,
        month = dec,
       volume = {2},
       number = {1},
          eid = {6},
        pages = {6},
          doi = {10.12942/lrsp-2005-6},
       adsurl = {https://ui.adsabs.harvard.edu/abs/2005LRSP....2....6G}
}

@ARTICLE{Tripathy23,
       author = {{Tripathy}, Sushanta},
        title = "{Seismology of active regions: Current status and perspectives}",
      journal = {Frontiers in Astronomy and Space Sciences},
         year = 2023,
        month = mar,
       volume = {10},
          eid = {1091777},
        pages = {1091777},
          doi = {10.3389/fspas.2023.1091777},
       adsurl = {https://ui.adsabs.harvard.edu/abs/2023FrASS..1091777T}
}

@book{collett05,
  author    = {Collett, Edward},
  title     = {Field Guide to Polarization},
  year      = {2005},
  publisher = {SPIE Press},
  address   = {Bellingham, Wash.},
  series    = {SPIE Field Guides},
  volume    = {5},
  isbn      = {0-8194-5868-6},
  doi       = {10.1117/3.626141},
  note      = {Includes bibliographical references and index. Spiral bound},
  language  = {English},
  pages     = {xiii+134}
}

@INPROCEEDINGS{Gisler03,
       author = {{Gisler}, Daniel and {Feller}, Alex and {Gandorfer}, Achim M.},
        title = "{Achromatic liquid crystal polarisation modulator}",
    booktitle = {Polarimetry in Astronomy},
         year = 2003,
       editor = {{Fineschi}, Silvano},
       series = {Society of Photo-Optical Instrumentation Engineers (SPIE) Conference Series},
       volume = {4843},
        month = feb,
        pages = {45-54},
          doi = {10.1117/12.458835},
       adsurl = {https://ui.adsabs.harvard.edu/abs/2003SPIE.4843...45G}
}

\chapter*{Acknowledgements}
\markboth{Acknowledgements}{Acknowledgements}
\addcontentsline{toc}{chapter}{Acknowledgements}

Looking back on my journey from training as a carpenter, through studying mathematics and philosophy, and pursuing a career in the space industry, it still feels surreal to now have this dissertation completed. And it would not have been written without the help of many people and institutions. \\

I was very lucky to have received funding and support within the framework of the International Max Planck Research School (IMPRS) for Solar System Science at the University of Göttingen and at TU Braunschweig hosted at the Max-Planck-Institut für Sonnensystemforschung (MPS). I would also like to thank the National Science Foundation (NSF) and New Jersey Institute of Technology (NJIT) for the granted Goode Solar Telescope (GST) telescope time and the Korea Astronomy and Space Science Institute and the Seoul National University for the generous access to the FISS instrument (This project was supported by US NSF-AGS grants 2309939 and 2408174). Further, I wish to express my thanks to the Sunrise-III project for additional conference and travel funding. It was a pleasure to work, learn and study at so many interesting and different places as the MPS clean rooms, the Big Bear Solar Observatory (BBSO), or the ESRANGE space center in Kiruna.

This dissertation project also benefited from the high performance computing (HPC) capabilities at MPS and the Gesellschaft f\"ur wissenschaftliche Datenverarbeitung mbH G\"ottingen (GWDG). 
I made a private donation to compensate for (some of) the CO\(_2\) emission created by the HPC usage for this thesis.
This work has made use of further infrastructure provided by other institutions, namly the VALD database, operated at Uppsala University, the Institute of Astronomy RAS in Moscow, and the University of Vienna.   
For the research I used NASA’s Astrophysics Data System (ADS) quite heavily.\\

As institutions usually don't just randomly spit out money I'd rather like to thank the people behind those decisions. So thank you Sami and Alex for having faith that a carpenter, with a degree in mathematics, starting a PhD in solar physics wasn’t a totally silly idea when giving me the opportunity to work on the Sunrise III project. But also to the other members of my thesis advisory committee (TAC), namely Stefan and Thorsten, for the good guidance and support. I owe great thanks to Michiel and HaPe for taking me aboard the FISS-SP project after the failed first launch of Sunrise III in Summer 2022. I also want to thank Wenda and Jongchul for their support, which enabled the FISS-SP project. Great thanks to the IMPRS coordinator, for all the help with the administrative hassles, formalities, and generally keeping the back of us PhD students free (by giving us extra work and hassles at the same time).  Finally, again to Stefan, for making it possible for me to give my own lectures at the University of G\"ottingen.

\newpage
The financial support might be necessary, but what really counts is the supervision, guidance, and moral support from supervisors, colleagues, and friends. I would therefore like to take this opportunity to conclude this project with a few personal words:

Sami, it has been a great pleasure to learn from your vast knowledge. I hope I have been able to learn from you to keep both details and context in view, and to better assess my own strengths and weaknesses. Alex, thank you for your ever-positive outlook and your ability to see and take the next step forward, no matter how challenging a situation may seem, and for meeting me as an equal from the very beginning. Michiel, thank you for your patience with my endless questions, your willingness to reconsider matters for the umpteenth time, and for the (sometimes wine-rich) philosophical conversations far beyond technical topics. HaPe, for all the vibration hunting fun and letting me fumble with the precious optics and the trust you had in me.

A big thanks to the whole Sunrise III team, especially Sami, Andi, Achim, Alex, and Tino for always pulling together as a team, even in the most devastating moments. Kiruna 2022 was an emotional roller-coaster, and each and everyone put the team first - or was it the olut? To Francisco for patient discussions and reviews. To Du\v{s}an for being such a pleasant room-mate for many weeks in Esrange. To Sebasti\'an for showing us this ``famous coffee place'' in \'Abeskovvu (Abisko). And finally, to Janek and Johannes, who became friends in Kiruna and stayed beyond while bringing my coffee addiction to the cinema screen.

To all the other nice people I met at MPS, rendering my time here as pleasant as it was. However, I don't just want to drop your names, the list would be simply too long and I hope you anyway know who you are. To the members of the AVG  for all the practical exercises when refining the Hainberg spectrograph or while maintaining the historic observatory.
To Salvatore for helpful comments to this thesis and lovely evenings spend cooking and dining together. 

Music is my way of keeping sanity and taking a break from work. I am therefore very grateful to have been so warmly welcomed by the Göttingen Lindy Hop and Ukulele communities. Special thanks to the band members of ``UkuLiesels'', ``MegaGauss'', and ``At some point (@\(\sum\cdot\))'' for so many shared songs, concerts, dances, and jam-sessions. 

Last, but as far from least as it can be, I owe a big thank you and an equally big apology to my dear friends. 
A thank you for putting up with my countless complaints and worries. And apologies for my rants and (extraordinarily good) puns. Andrea, Michael, and Cilia, I am grateful to have such great people in my life. 
But my warmest thanks belong to my family. To my mother for her constant support and for being an inspiring friend with shared values and actions. To my father, whom I miss for more than twenty years now and who wouldn't believe I am finishing a thesis in astrophysics (well, I don't). And, finally, to the wonderful Anna-Louise. I simply don’t have the words to express my feelings in a single sentence. Thank you for some of the happiest moments in my life in these past years living with you. We've been down, we've been up, but where we've not been is the future - let’s explore it together!\\[10pt]
\noindent
\textit{...ach ihr seid ja noch da, aber jetzt - abschalten!}
\begin{flushright}
Johannes H\"olken\\
\footnotesize{G\"ottingen, Oktober 2025}
\end{flushright}

\newpage

\end{document}